\pdfoutput=1
\documentclass[11pt,a4paper,twoside,openright]{report}
\usepackage[english]{ETHDAsfs}

\usepackage{pdfpages}
\usepackage{amsbsy}
\usepackage{amssymb}
\usepackage{graphicx}
\usepackage{natbib}
\usepackage{texab}
\usepackage{amsmath}
\usepackage{enumerate}
\usepackage{relsize}
\usepackage{color} 
\usepackage{listings}
\usepackage{xfrac} 
\usepackage[toc,page, titletoc]{appendix}
\usepackage{float} 
\usepackage{bbm} 
\usepackage{mathtools} 
\usepackage{multirow}
\usepackage{caption} 
\usepackage{subcaption} 
\usepackage{rotating} 
\usepackage{array} 

\definecolor{Mygrey}{gray}{0.75}
\definecolor{Cgrey}{gray}{0.4}
\numberwithin{equation}{subsection}

\begin{document}
\bibliographystyle{chicago}

\pagenumbering{roman}

\period{Spring 2017}
\dasatype{Master Thesis}
\students{Emiliano Díaz}
\mainreaderprefix{Adviser:}
\mainreader{Prof.\ Dr.\ Marloes Maathuis}
\submissiondate{March 19th 2017}
\title{Online deforestation detection}

\maketitle
\cleardoublepage

\markright{}
\vspace*{\stretch{1}}
\begin{center}
    To my family, my mother Alejandra, my father Manuel and my brother Manolo who supported me so much throughout the master program. To Dr. Markus Kalisch who always provided guidance and support during my studies at ETH. To my adviser Prof. Dr. Marloes Maathuis for all her patience and guidance in the whole master thesis process. 
\end{center}
\vspace*{\stretch{2}}


\newpage
\markboth{Abstract}{Abstract}
\chapter*{Abstract}

Deforestation detection using satellite images can make an important contribution to forest management. Current approaches can be broadly divided into those that compare two images taken at similar periods of the year and those that monitor changes by using multiple images taken during the growing season. The CMFDA algorithm described in \cite{CMFDA} is an algorithm that builds on the latter category by implementing a year-long, continuous, time-series based approach to monitoring images. This algorithm was developed for 30m resolution, 16-day frequency reflectance data from the Landsat satellite. In this work we adapt the algorithm to 1km, 16-day frequency reflectance data from the modis sensor aboard the Terra satellite. The CMFDA algorithm is composed of two submodels which are fitted on a pixel-by-pixel basis. The first estimates the amount of surface reflectance as a function of the day of the year. The second estimates the ocurrence of a deforestation event by comparing the last \emph{few} predicted and real reflectance values. For this comparison, the reflectance observations for six different bands are first combined into a \emph{forest} index. Real and predicted values of the \emph{forest} index are then compared and high absolute differences for consecutive observation dates are flagged as deforestation events. Our adapted algorithm also uses the two model framework. However, since the modis 13A2 dataset used, includes reflectance data for different spectral bands than those included in the Landsat dataset, we cannot construct the \emph{forest} index. Instead we propose two contrasting approaches: a multivariate and an index approach similar to that of CMFDA. In the first prediction errors (form first model) for selected bands are first compared against, band-specific, thresholds to produce one deforestation flag per band. The multiple deforestation flags are then combined using an \emph{or} rule to produce a general deforestation flag. In the second approach, as with the CMFDA algorithm, the reflectance observations for selected bands are combined into an index. We chose to use the local Mahalanobis distance of prediction errors for the selected bands as our index. This index will measure how atypical a given multivariate predicted error is therby helping us to detect when an intervention to the data generating mechanism has occurred, i.e. a deforestation event. We found that, in general, the multivariate approach obtained slightly better performance although the index approach, based on the Mahalanobis distance, was better at detecting deforestation early. Our training approach was different to that used  in \cite{CMFDA} in that the lower resolution of the reflectance data and the \emph{pseudo} ground-truth deforestation data used allowed us to select a much larger and diverse area including nine sites with different types of forest and deforestation, and training and prediction windows spanning 2003-2010. In \cite{CMFDA} reflectance and deforestation information from only one site and only the 2001-2003 period is used. This approach allowed us to make conclusions about how the methodology generalizes accross space (specifically pixels) and accross the day of the year. In the CMFDA and our adapted CMFDA methodology a single (possibly multivariate) threshold is applied to the prediction errors irrespective of the location or the time of the year. By comparing the results when thresholds were optimized on a site-by-site basis, to those when a single threshold was optimized for all nine sites we found that optimal thresholds do not translate accross sites, rather they display a local behavior. This is a direct consequence of the local behavior of the prediction error distibutions. This lead us to try to homogenize the error distributions accross space and time by applying transformations based on different observations and assumptions about the predicted error distributions and their dependence on time and space. However, our efforts in this sense did not improve performance leading us to recommend the implementation of the multivariate approach without transforming predicted errors.


\newpage
\tableofcontents
\newpage
\listoffigures
\newpage
\listoftables


\cleardoublepage
\pagenumbering{arabic}

\chapter{Introduction} \label{ch:intro}

\section{Deforestation problem} \label{s:defoProb}

The opening paragraph of \cite{FRA_FAO_2005} aptly and succintly describes the value of forests for our societies:
	
\begin{quotation}
	The contributions of forests to the well-being of humankind are extraordinarily vast and far-reaching. Forests play a fundamental role in combating rural poverty, ensuring food security and providing decent livelihoods; they offer promising mid-term green growth opportunities; and they deliver vital long-term ecosystem services, such as clean air and water, conservation of biodiversity and mitigation of climate change.	
\end{quotation}

According to \cite{FRA_FAO_2005} the total net forest on earth decreased at an annual net loss rate of 0.13 percent, from 4,128 million hectares in 1990, to 3,999 million hectares in 2015. However, the rate of annual net loss of forest has decreased from 0.18 percent in 1990-2000 period to 0.08 percent over the 2010-2015 period. 

According to \cite{InformeMexico} the total net forest in Mexico decreased at an annual net loss rate of 0.21 percent, well above the world average, from 69.8 million hectares in 1990, to 66.0 million hectares in 2015. As with the rest of the world, the rate of annual net loss of forest decreased markedly from 0.27 percent in the 1990-2000 period to 0.14 percent over the last the 2010-2015 period. 

It is probable that improvements in remote sensing technology and data processing techniques as well as in forest management policy have played a role in this improvement in the world and in Mexico. As our ability to acquire satellite images of the earth's surface with increasing resolution and frequency improves, the possibility of detecting deforestation events quickly and mitigating their scope and impact increases. The goal of this work is to adapt and implement to satellite images from Mexico, the method described in \cite{CMFDA}, called \emph{continuous monitoring of forest disturbance algorithm} (CMFDA) which seeks to detect deforestation events within a few weeks of their occurrance. 

Various deforestation detection algorithms have been developed. They can be split into those that are based on the comparison of images from two dates from similar periods of the year (see for example  \citealt{Healey} or \citealt{Masek}) and those that use images from multiple dates during the growing season (see for example  \citealt{Kennedy} or  \citealt{Vogelmann}). The CMFDA algorithm is more akin to the second type of algorithm with the difference that it monitors the images of forests throughout the year allowing it to detect deforestation at any time. 

The CMFDA algorithm consists of several steps geared toward estimating two models: a \emph{reflectance model} in which reflectance for each spectral band (the intensity of each \emph{colour}) is modeled as a function of the day of the year; and a \emph{deforestation model} which models the occurrence of a deforestation event as a function of the size of recent reflectance prediction errors. The idea is that if observed reflectance deviates a lot from model predicted reflectance it is likely due to a landcover change (in this case deforestation). In this work we adapt the CMFDA algorithm, which was developed for 30m resolution Landsat satellite reflectance data, to 1km resolution reflectance data from the modis sensor aboard the Terra satellite. The two main differences between the two sets of data is the resolution, the former has 30m resolution, the latter 1km resolution; and, in the case of the spectral bands available, the former has six optical bands, while the latter only four (plus two vegetation indices). These two differences determine the most important part of the adaptations developed in this work. 

To detect important differences between observed and model predicted reflectance values in a way that is most useful for deforestation detection, the CMFDA algorithm prescribes the construction of a \emph{forest} index using all six optical bands available for the Landsat data. Since the modis 13A2 dataset used in this work only includes four spectral bands we cannot build this \emph{forest} index. Instead we implement two different approaches, a \emph{multivariate} and an \emph{index} approach. In the multivariate approach a selection of the reflectance bands are first predicted and if the prediction error of each exceeds a certain threshold, particular to each band, then a band flag is triggered. The flags for the selection of bands are then combined using an \emph{or} rule to obtain the general deforestation flag. In the index approach, as is the case in CMFDA, the selection of reflectance bands are first combined to form a reflectance index, and the prediction error for the index then checked against a threshold. In contrast to CMFDA we use a data-driven approach based on the Mahalanobis distance to construct our index. 

The fact that we work with much lower resolution reflectance data allows us to use  a more extensive study period and study area than that used in \cite{CMFDA}. This, in turn, will allow us to make conclusions about how methodology generalizes to sites with different characteristics and accross time. 

\section{Structure} \label{s:structure}

The report is organized as follows: 

In chapter \ref{ch:data}  we explore the data used in this work. In section \ref{s:landuse} we explore the deforestation information used to build our binary deforestation response variable. This is the response variable of the deforestation model, mentioned above. Section \ref{s:refl} is an exploration of the satellite image time series dataset which we use to detect deforestation. The dataset includes the reflectance data which is the response variable for the reflectance model, mentioned above. This data set is used both in the development of the deforestation methododology that is the subject of this work and in its implementation. We try to understand the measurement and processing steps that are carried out to transform raw radiance data into estimated surface reflectance data. We also explore the sun-sensor geometry and quality conditions, such as the presence of cloud, shadow or snow, under which the surface reflectance data was recorded. Although many of the observations made will not be exploited in this work, they give good undestanding of the data and are necessary to aid future improvements. 

Chapter \ref{ch:meth} describes the CMFDA methodology, including adaptations made to it, and shows the results of training and testing. Section \ref{s:CMFDA} describes the CMFDA algorithm in detail. The idea is that by following the image of a given location, specifically a pixel, where we know there is forest through time, we can detect when the image changes and we can flag this as deforestation. More specifically CMFDA consists of two models. The first is a \emph{reflectance model}, in which the surface reflectance values (one for each band or \emph{colour}) that make up a forest pixel are explained as a function of the day of the year. There are six optical bands and each one is modeled in this way. The second is a \emph{deforestation model}, in which deforestation or non-deforestation is explained as a function of the difference between real observed reflectance values and the predicted reflectance values obtained using the first model. Predicted and real values for each of the six bands are first combined into a real and predicted \emph{forest} index, then the real and predicted values are compared. There are many other details such as what to do when clouds contaminate the measurement of reflectance, for a given pixel, with noise. These are fully described in this section.  The CMFDA algorithm was developed with and for 30m resolution Landsat satellite images. In this work we work with 1km resolution Terra satellite images. In section \ref{s:adapt} we describe adaptations made to the algorithm so that it may work with this data. Since the data set we use does not include reflectance information for the same bands we cannot construct the \emph{forest} index of the CMFDA algorithm. In section \ref{s:adapt} we leave open the issue of which of the available bands we should use and how we should combine them. As in CMFDA, the real and predicted values of the different bands are compared with a thresholding function that flags when the distance is too large, but the specific form will be left unspecified: do we combine bands into an index first and then threshold their prediction error as in the CMFDA algorithm, or  threshold the prediction errors of all bands first and then combine using an \emph{and}/\emph{or} rule? If we use an index which one do we use given we cannot construct the forest index? How do we train thresholding functions in each case? These issues are resolved in sections \ref{s:studyArea}-\ref{s:mahl}. In section \ref{s:studyArea} we select the study area such that sufficient deforestation occurs there during our prediction period (2005-2010) and it is diverse enough in terms of the types of deforestation that occur. Nine 25 by 25km sites are chosen. We choose the two bands that will be used in our adapted algorithm based on an exploratory analysis of the behavior of the different reflectance bands for deforested and non deforested pixels in these areas. We also identify which bands are useful for each type of deforestation based on this analysis. In section \ref{s:gridSearch} we implement the adapted CMFDA methodology on a site-by-site basis using for each site only the band appropriate for that site, based on the knowledge of the type of deforestation that occurred there. The performance results of this section will serve only as a benchmark as this version of the methodology is not implementable given the use of prior knowledge and the need to obtain local thresholds. In sections \ref{s:simAn} and \ref{s:mahl} we test two approaches that \emph{are} implementable and which specify the form of the thresholding function to use and how to train it, thereby completing the specification of the adapted CMFDA algorithm which we left open in section \ref{s:adapt}. In section \ref{s:simAn} we test an approach where the predicted errors for the bands selected in section \ref{s:studyArea} are thresholded separateley first and the resulting band-wise flags then combined into a general deforestation flag using an \emph{or} rule. The multivariate thresholding level is trained using simulated annealing. In section \ref{s:mahl} we first combine reflectance values for the different selected bands into an index and then threshold the predicted errors of that index. The index used is the local Mahalanobis distance of the prediction errors for the selected bands. This index takes into account the covariance of prediction errors for the different selected bands. The covariance is assumed to be a function of both location and the time of the year for estimation purposes.   

The approach described in chapter \ref{ch:meth} is a pixel-by-pixel modeling approach. In chapter \ref{ch:modhet} we analyze whether there is a pattern to how the different components of the model vary accross pixels and through time. In section \ref{s:coef} we explore whether the coefficients of the reflectance model for each pixel show a spatial pattern. In section \ref{s:var} we focus on how the prediction error distributions vary accross pixels and time. This is very important as the eventual goal of the methodology is to use a single thresholding function on the reflectance model prediction errors of any pixel at any time. In sections \ref{s:homStdr}-\ref{s:homECDF} we try to improve the adapted methodolgy described in chapter \ref{ch:meth} by finding a way to homogenize the distribution of prediction errors accross pixels and time of the year, so that we can apply a single thresholding function with improved performance. In section \ref{s:homStdr} we try to standardize the errors by dividing them by their standard deviation. We estimate the standard deviation under different assumptions about the space and time dependence of errors. In section \ref{s:homECDF} we try to homogenize the distribution of errors by transforming them using the empirical cumulative distribution function (ecdf). We, again, estimate the ecdf under different assumptions about the space and time dependence of errors.

In chapter \ref{ch:imp} we describe how to implement the methodology taking into account all the considerations and findings of chapters \ref{ch:meth} and \ref{ch:modhet}. 

In chapter \ref{ch:Summary} we summarize our findings and give a list of possible improvements to the algorithm and directions for future research.

\chapter{Data description and exploration} \label{ch:data}

In this chapter  we explore the data used to train and test the algorithm. As was mentioned in chapter \ref{ch:intro} the CMFDA consists of reflectance model and a deforestation model. 

In section \ref{s:landuse} the data used to construct the deforestation flag is explored. The deforestation flag is the response variable of the \emph{deforestation model}. The land-use classification for 2005 and 2010 from the North American Land-Change Monitoring System (NALCMS) described in \cite{landUse}, was used to build a response variable for deforestation. Firstly the classification model used to produce this data is described:  input data, statistical techniques and ground truth data that were used to develop it. The land-use classification of the model for 2005 and 2010 is then explored, including the amount of net forest loss that occurred in this period.

In section \ref{s:refl} we study the modis 13A2 dataset. This data set includes \emph{quality}, \emph{sun-sensor geometry}, \emph{surface reflectance} and \emph{vegetation index} variables. The \emph{surface relectance} variables are the response variables for the \emph{reflectance model}. We first give a detailed account of the measurement and data processing that goes into transforming the raw radiance data obtained by the modis sensor aboard the Terra into gridded, composited surface reflectance data. The particular circumstances in which a reflectance value for a given pixel was recorded  (angle between sun, surface and sensor, presence of clouds, shadows or snow, etc) and the processing involved (whether the reflectance value was atmosphere corrected or not) are described by the quality and sun-sensor geometry variables. We explore their distribution and behavior accross pixels and time. Although a lot of this analysis is not used further in this work we include it as we think it helps the understanding of the underlying physical causal processes involved. This in turn can inform future improvements to the methodology described in this work.

	\section{Land use classification} \label{s:landuse}
	
	The land-use classification model developed by NALCMS is now described. The input data used in \cite{landUse} to generate the land-use classification for Mexico consists of:
	
	\begin{itemize}
	 \item Modis/Terra top-of-atmosphere reflectance data (bands 1-7) at 250 metre spatial and 10-day temporal resolution over Mexico for 2005-2006 and 2010-2011,
	 \item a digital elevation model at 250 metre spatial resolution and derivatives (slope and aspect maps) and 
	 \item climate variables: monthly average minimum, mean, and maximum temperatures, total days of precipitation per year, total precipitation and total evaporation in the 1970-2000 period. 
	 \end{itemize}
	 
	 Reflectance variables were pre-processed using resampling, downscaling, compositing and denoising procedures described in \cite{landUse} to produce filtered 10-day composites. They were then averaged on a pixel-by-pixel basis to aggregate to 12 monthly composites at 250 metre resolution. Decision-trees together with boosting were used to perform classification. Two levels of classification were used, Level I, consisting of 12 classes, and Level II, which is more detailed, consisting of 19 classes. Table \ref{tabClasses}, based on table 20.1 of \cite{landUse} lists the different land-use categories. 
	 
	 \begin{table}[H] 
	 \begin{tabular}{|l|l|l|}
		 \hline
	 \textbf{Level I} & \textbf{Level II} & \textbf{Forest/Other} \\ 
	 \hline 
	 \multirow{2}{*}{1. Needleleaf forest} & 1.Temperate or subpolar forest                        & \multirow{2}{*}{Forest} \\ 
	                                       & 2. Subpolar taiga needleleaf forest                   &						  \\
	 \hline 
	 \multirow{3}{*}{2. Broadleaf forest}  & 3. Tropical or subtropical broadleaf evergreen forest & \multirow{3}{*}{Forest}  \\
	                                       & 4. Tropical or subtropical broadleaf deciduous forest &                          \\
										   & 5. Temperate or subpolar broadleaf deciduous forest   &                         \\
     \hline										   
	 3. Mixed forest                       & 6. Mixed forest   									   & Forest                  \\		
	 \hline
     \multirow{2}{*}{4. Shrubland}         & 7. Tropical or subtropical shrubland 				   & \multirow{2}{*}{Other}  \\	                   										   & 8. Temperate or subpolar shrubland                    &                         \\				   	
     \hline
     \multirow{2}{*}{5. Grassland}         & 9. Tropical or subtropical grassland 				   & \multirow{2}{*}{Other}  \\	                   										   & 10. Temperate or subpolar grassland                   &                          \\				   	
     \hline
     \multirow{4}{*}{6. Linchen/Moss}      & 11. Subpolar or polar shrubland-lichen-moss 		   & \multirow{3}{*}{Other}  \\	                   	  								    & 12. Subpolar or polar grassland-lichen-moss                  &                           \\				   	
	                                       & 13. Subpolar or polar barren-lichen-moss              &                              \\				   	
     \hline
	 7. Wetland                            & 14. Wetland   									       & Other                  \\		
	 \hline
	 8. Cropland                           & 15. Cropland   									   & Other                  \\		
	 \hline
	 9. Barren land                        & 16. Barren land   									   & Other                  \\		
	 \hline
	 10. Urban and built-up                & 17. Urban and built-up   							   & Other                  \\		
	 \hline
	 11. Water                             & 18. Water   									       & Other                  \\		
	 \hline
	 12. Snow and ice                      & 19. Snow and ice   								   & Other                  \\		
	 \hline									   	
	 \end{tabular}\\
	 \caption{NALCMS land-use classes}
	 \label{tabClasses}
	 \end{table}
	 
	 To train the decision trees ground-truth data from previous studies carried out by Mexican government agencies and academic institutions was used (see table 20.4 of \citealt{landUse}). The output of the model is a land-use classification, according to the categories described in table \ref{tabClasses}, for 2005 and 2010 at 250m resolution.
	 
\begin{figure}[H]  
  
	   \begin{subfigure}{.5\textwidth}
	     \centering
	 	\includegraphics[width=1\textwidth]{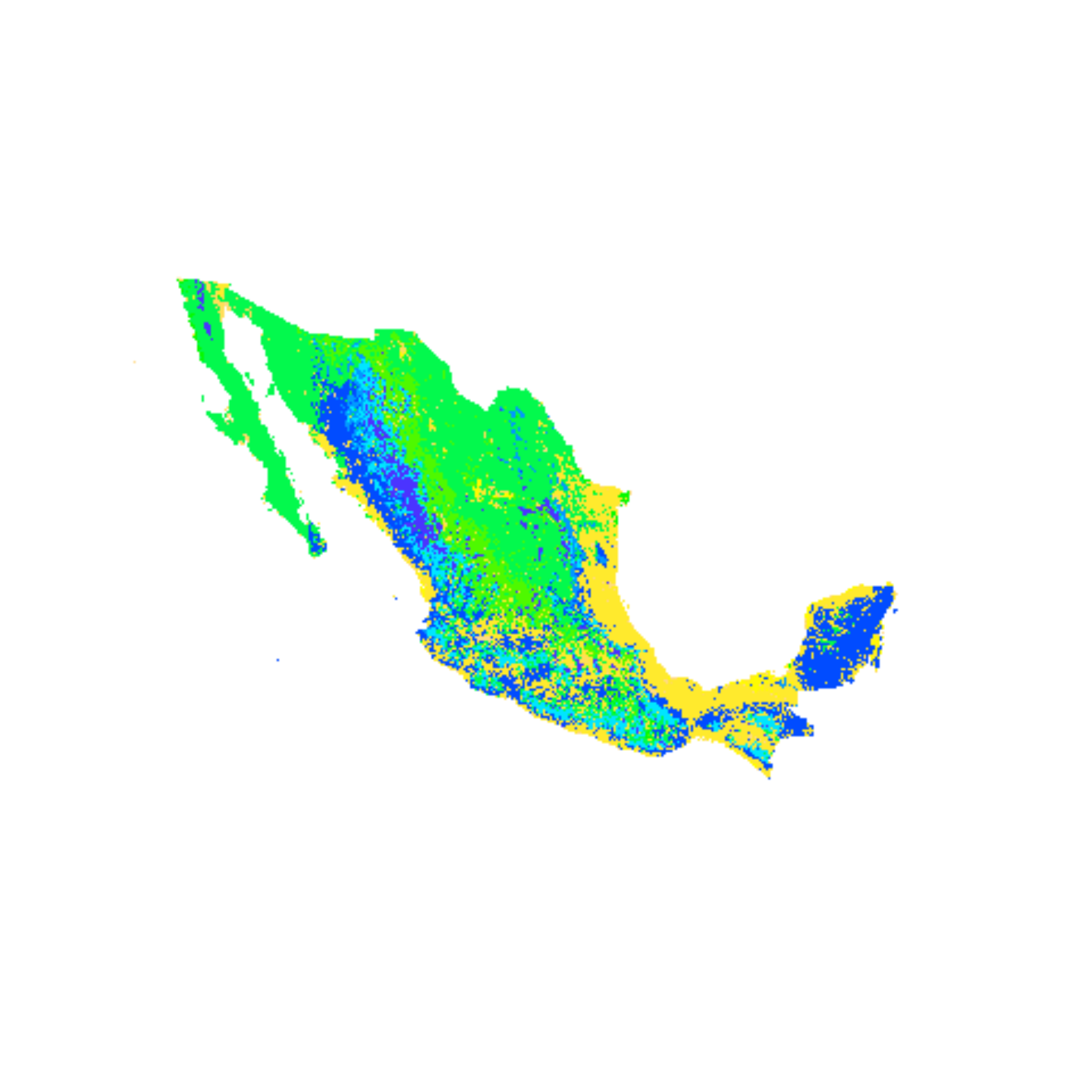} 
	     \caption{Mexico, 2005}
	     \label{fig:sfig1a}
	   \end{subfigure}%
	   \begin{subfigure}{.5\textwidth}
	     \centering
	     \includegraphics[width=1\textwidth]{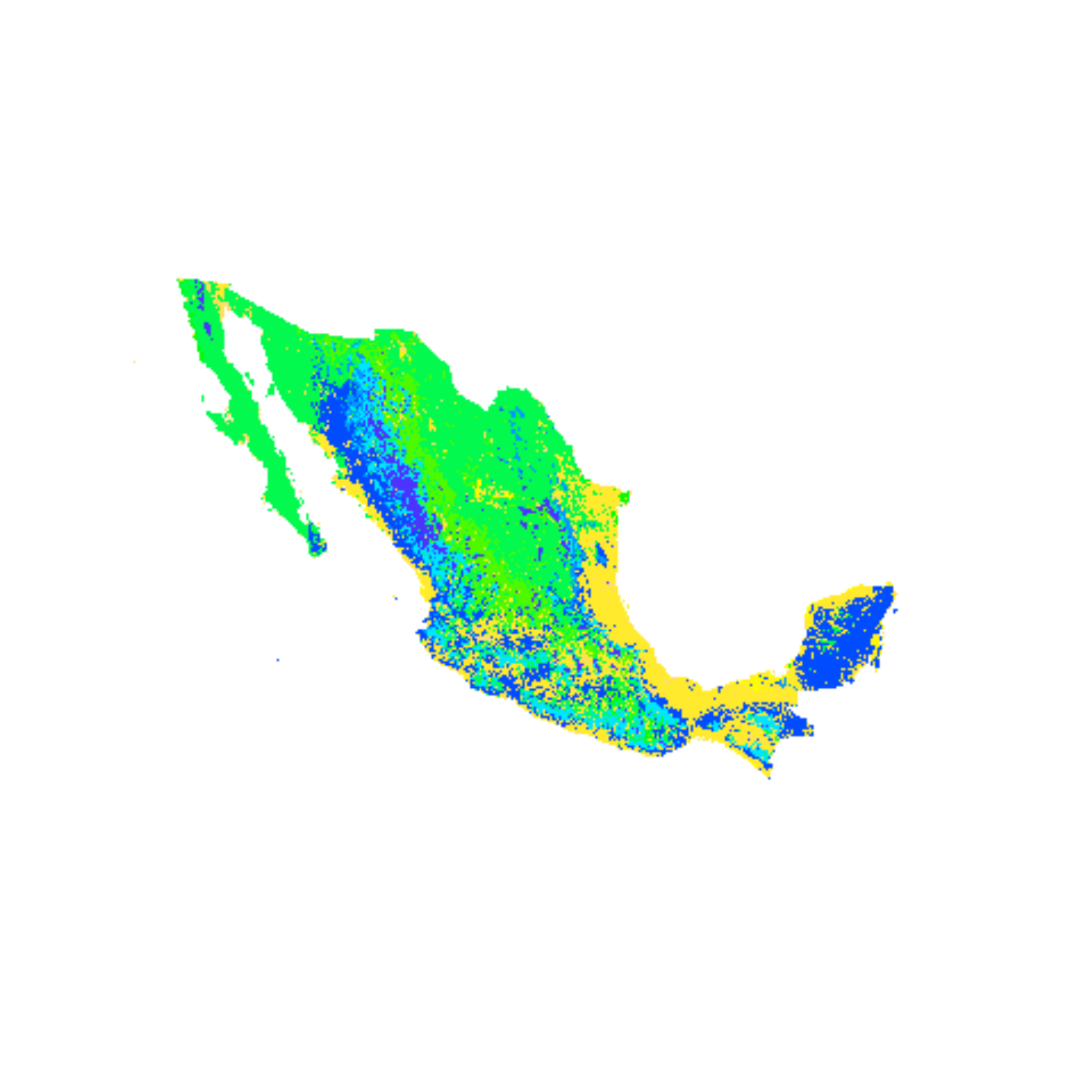}
	     \caption{Mexico, 2010}
	     \label{fig:sfig1b}
	   \end{subfigure}
	   \begin{subfigure}{.4\textwidth}
	     \centering
	     \includegraphics[width=1\textwidth]{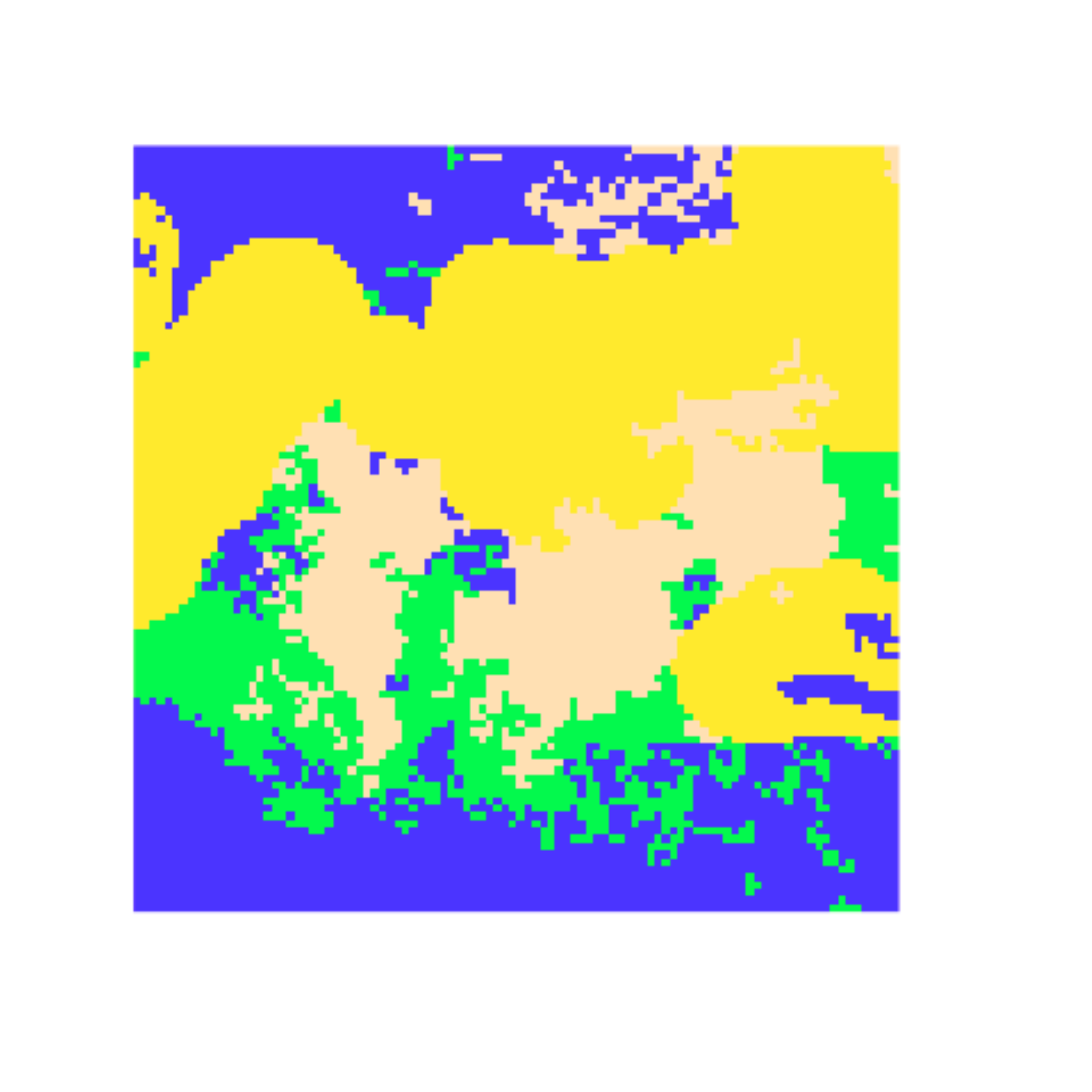}
	     \caption{25 km$^2$ close-up of location in south-west Coahuila, 2005}
	     \label{fig:sfig1c}
	   \end{subfigure}
	   \begin{subfigure}{.4\textwidth}
	     \centering
	     \includegraphics[width=1\textwidth]{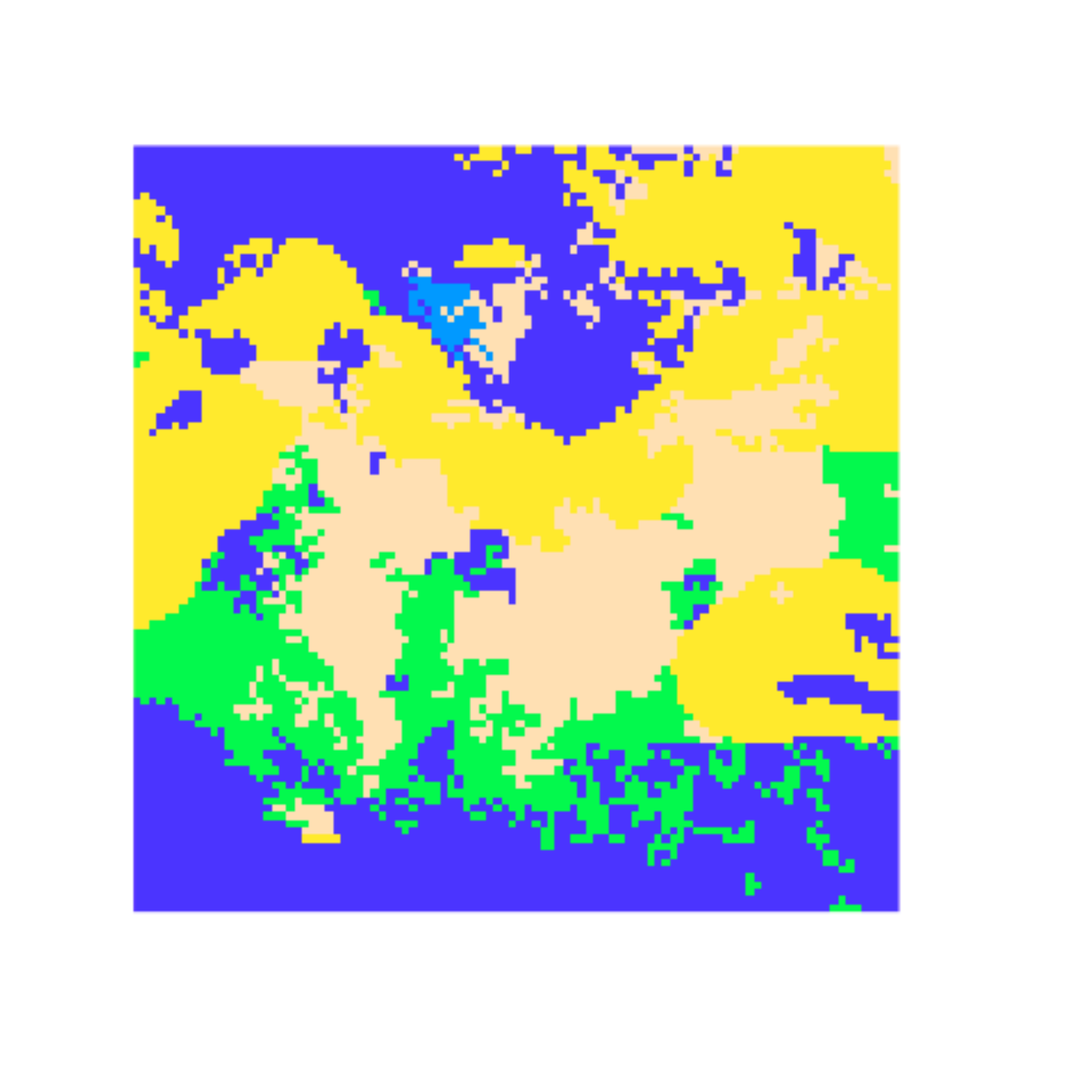}
	     \caption{25 km$^2$  close-up of location in south-west Coahuila, 2010}
	     \label{fig:sfig1d}
	   \end{subfigure}
	   \begin{subfigure}{.4\textwidth}
	     \centering
	     \includegraphics[width=1\textwidth]{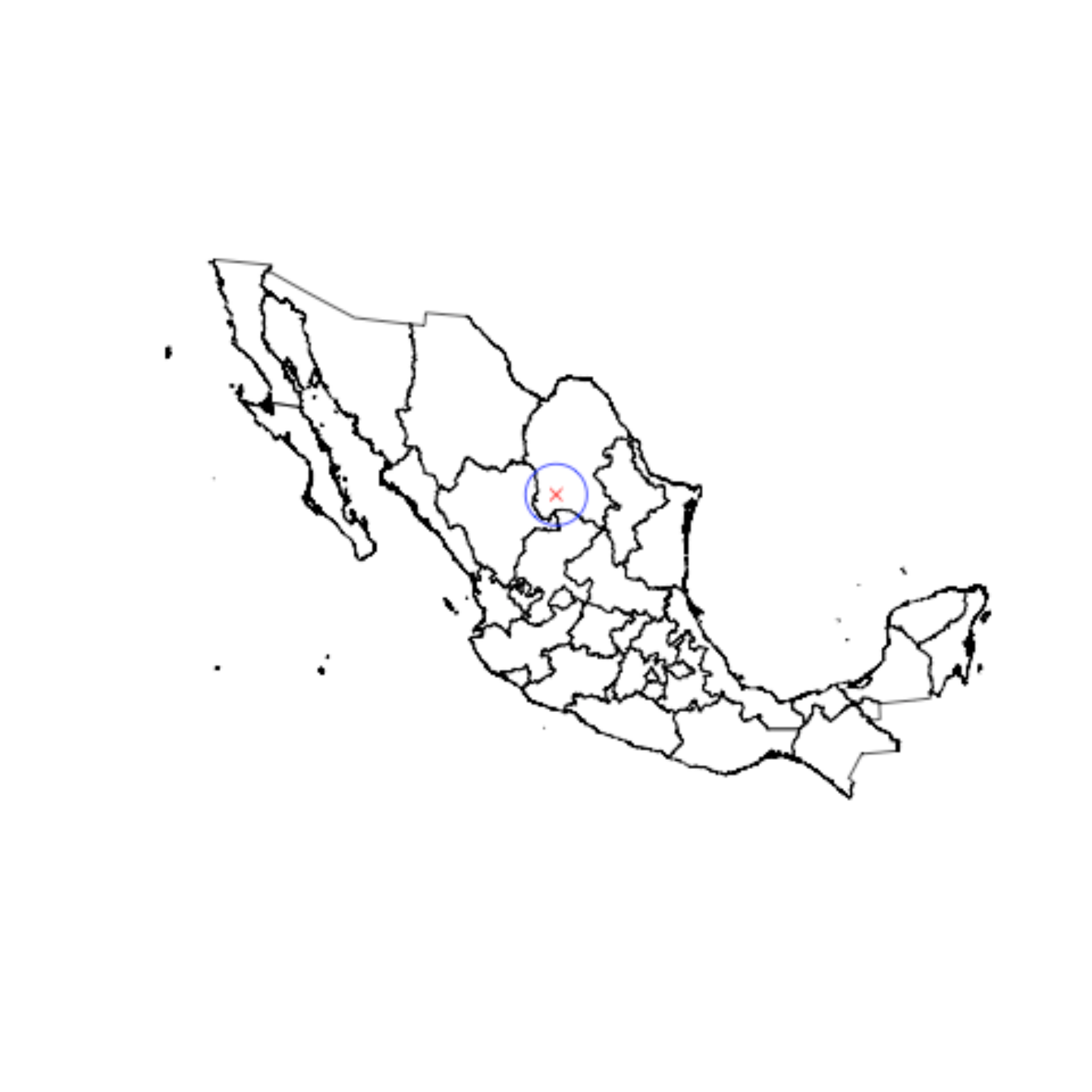}
	     \caption{Close-up location in south-west Coahuila}
	     \label{fig:sfig1e}
	   \end{subfigure}
	   \begin{subfigure}{.4\textwidth}
	     \centering
	     \includegraphics[width=1\textwidth]{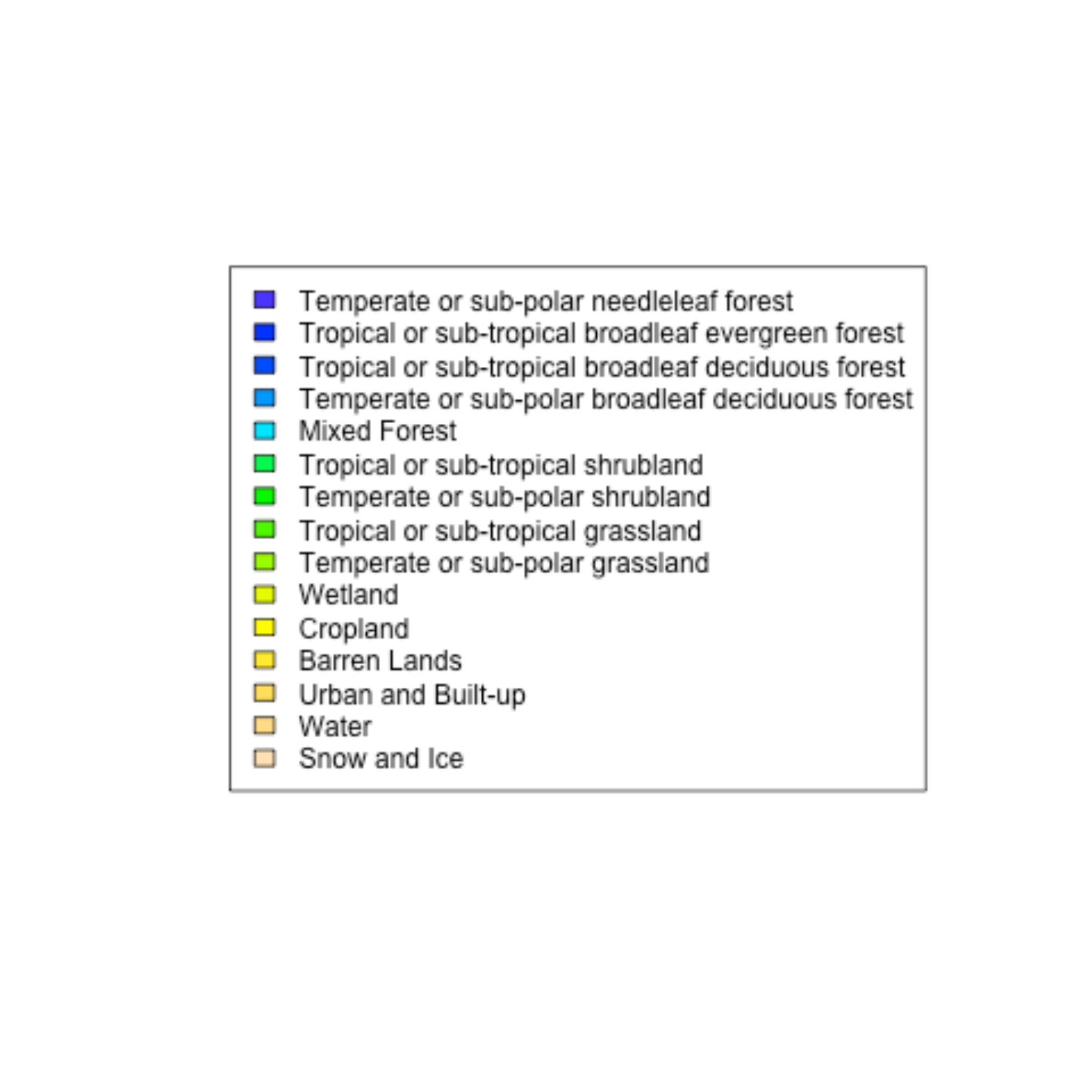}
	     \caption{Land use legend}
	     \label{fig:sfig1f}
	   \end{subfigure}	   
	   \caption[NALCMS land-use classification of Mexico, 2005 and 2010 ]
	   {NALCMS land-use classification of Mexico, 2005 and 2010}
	   \label{fig:landUseMex}
	 \end{figure}

	 Figure \ref{fig:landUseMex} shows the land-use in Mexico and in a location in south-west Coahuila, in 2005 and 2010, according to the land-use classification model. 
	 
\begin{figure}[H]
	  \centering
	  \includegraphics[width=.5\textwidth]{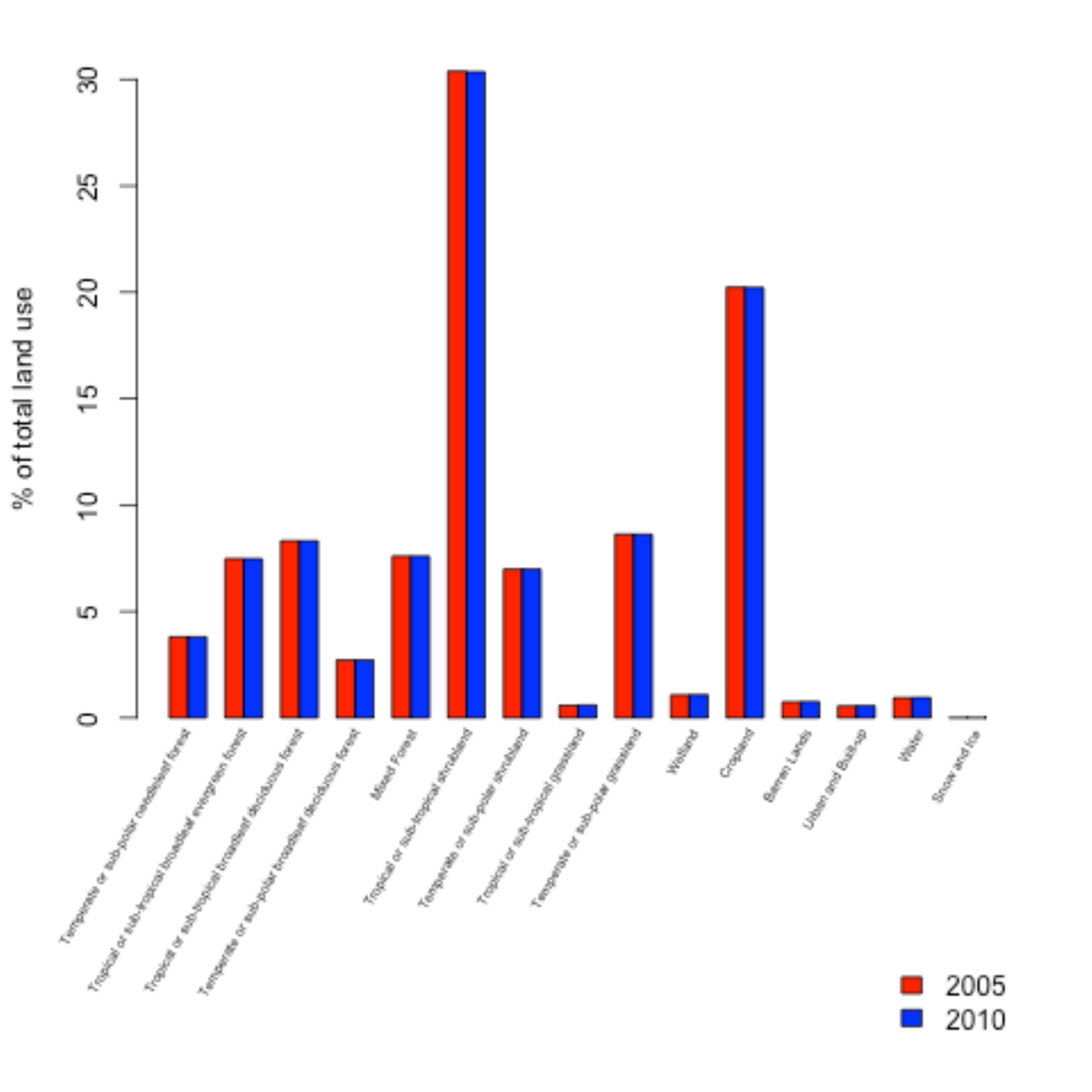} 
	  \caption[NALCMS land-use distribution of Mexico, 2005 and 2010]
	  {NALCMS land-use distribution of Mexico, 2005 and 2010}
	  \label{fig:landUseDist}
	\end{figure}

	 From figure \ref{fig:landUseDist} and table \ref{tab:landUseDist} we can see that the change in land use between 2005 and 2010 was marginal according to the NALCMS classification. Table \ref{tab:deforest} shows only 0.032\% of land classified as forest in 2005 is classified as \emph{non-forest} in 2010. This contrasts with the figure from \cite{InformeMexico} which estimatates it at 0.87\%. By choosing study areas with high deforestation, such that the new land-use type is widespread, we hope the classification model has reasonable accuracy: the idea is that this new type of land-use will be easily detected by the classification model since it is not spatially isolated.
	 
	\begin{center}
		\footnotesize\addtolength{\tabcolsep}{-2pt}
\begin{table}[ht]
\centering
\begingroup\tiny
\begin{tabular}{rrrrrrrrrrrrrrrr}
  & \begin{sideways} Temperate or sub-polar needleleaf forest \end{sideways} & \begin{sideways} Tropical or sub-tropical broadleaf evergreen forest \end{sideways} & \begin{sideways} Tropical or sub-tropical broadleaf deciduous forest \end{sideways} & \begin{sideways} Temperate or sub-polar broadleaf deciduous forest \end{sideways} & \begin{sideways} Mixed Forest \end{sideways} & \begin{sideways} Tropical or sub-tropical shrubland \end{sideways} & \begin{sideways} Temperate or sub-polar shrubland \end{sideways} & \begin{sideways} Tropical or sub-tropical grassland \end{sideways} & \begin{sideways} Temperate or sub-polar grassland \end{sideways} & \begin{sideways} Wetland \end{sideways} & \begin{sideways} Cropland \end{sideways} & \begin{sideways} Barren Lands \end{sideways} & \begin{sideways} Urban and Built-up \end{sideways} & \begin{sideways} Water \end{sideways} & \begin{sideways} Snow and Ice \end{sideways} \\ 
  \hline
2005 & 3.797 & 7.477 & 8.316 & 2.709 & 7.595 & 30.384 & 6.980 & 0.579 & 8.623 & 1.071 & 20.232 & 0.743 & 0.552 & 0.941 & 0.001 \\ 
   \hline
2010 & 3.797 & 7.474 & 8.313 & 2.708 & 7.594 & 30.354 & 6.981 & 0.593 & 8.617 & 1.082 & 20.217 & 0.756 & 0.561 & 0.952 & 0.001 \\ 
   \hline
\end{tabular}
\endgroup
\caption[NALCMS land-use distribution of Mexico, 2005 and 2010.]{NALCMS land-use distribution of Mexico, 2005 and 2010.} 
\label{tab:landUseDist}
\end{table}

	\end{center}

	\begin{center}
		\footnotesize\addtolength{\tabcolsep}{-2pt}
\begin{table}[ht]
\centering
\begin{tabular}{rr}
 Non-forest loss & forest loss \\ 
  \hline
99.968 & 0.032 \\ 
   \hline
\end{tabular}
\caption[\% net forest loss from 2005 to 2010 in Mexico according to NALCMS model.]{\% net forest loss from 2005 to 2010 in Mexico according to NALCMS model.} 
\label{tab:deforest}
\end{table}

	\end{center}
	
	As is mentioned above, the output of the NALCMS land-use distribution model is at 250m resolution.  A deforestation flag at 1km resolution is needed for the adapted CMFDA algorithm since the reflectance information that is used is at this coarser resolution. To construct a deforestation flag with 1km resolution  a deforestation flag with 250m resolution was first constructed. Each 1 by 1km pixel consists of sixteen 250 by 250m resolution pixels. To construct a deforestation flag at 1km the number of 250m resolution pixels within each 1km resolution pixel, that changed from forest to non-forest in the 2005-2010 period, were counted. The 1km resolution deforestation flag consists of assigning the value 1 to any 1km resolution where the number of 250m resolution deforested pixels was at least one. 
	

	\section{Reflectance data} \label{s:refl}
	
	The modis 13A2 dataset is now described. This data includes the reflectance variables that will be used as response variables in the \emph{reflectance} model of the CMFDA algorithm. The measurement and data-processing steps necessary to transform raw radiance data into surface reflectance data is first described. 

\begin{figure}[H]
	  \centering
	  \includegraphics[width=0.8\textwidth]{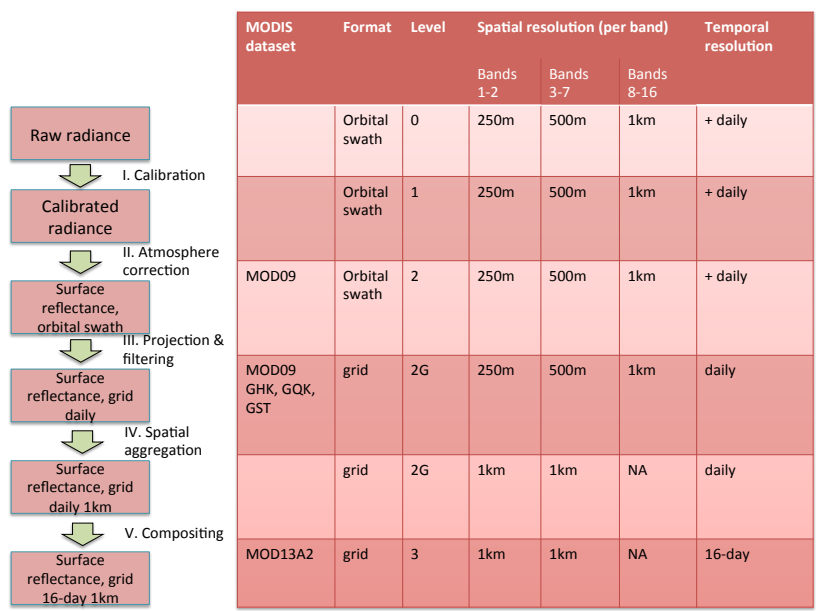} 
	  \caption[MOD13A2 data set production]
	  {MOD13A2 data set production}
	  \label{fig:mod13a2}
	\end{figure}

	We used the 13A2 Vegetation Index dataset \citep{modis} from the modis sensor aboard the Terra satellite. Figure \ref{fig:mod13a2}, based on information from the \citet{userGuide_mod09} and \citet{userGuide_VI}, shows how the 13A2 reflectance data is constructed. In fact, this section is based on \citet{userGuide_mod09}, \citet{userGuide_VI} and on an exploration of the sample from the \cite{modis} dataset.
	
	The level of a given satellite dataset refers to the amount of processing:
	
	\begin{enumerate}
		\item \textbf{Level 0:} Raw satellite feeds (radiance data). Data in orbital swath format.
		\item \textbf{Level 1:} Radiometrically calibrated satellite radiance data. Data in orbital swath format.
		\item \textbf{Level 2:} Atmospheric correction to yield estimation of surface reflectance. Data in orbital swath format.
		\item \textbf{Level 3:} Geolocation and temporal aggregation (compositing/averaging). Data in grid format. 
		\item \textbf{Level 4:} Additional processing. 
	\end{enumerate}	
	
	An orbital swath is the area that a satellite \emph{sees} as it travels along its orbit. The orbital swath format consists of radiance or reflectance values  associated to \emph{granules} which are five minute sections of the \emph{swath}, or seen path, as opposed to the grid format, where the radiance or reflectance data is associated to a grid cell of a two-dimensional coordinate system.  Level 2G data consists of level 2 data that has been geolocated but no temporal aggregation has yet been applied to it.  We now give a brief description of the processes carried out to transform the level 0 radiance data obtained by the modis aboard the Terra satellite into the MOD13A2 Vegetation Index data set. 
	
	\begin{enumerate}[I.]
		
		\item \textbf{Calibration:} Level 0 data collected by modis aboard Terra satellite, is geolocated to 1A data and then calibrated to Level 1B radiance data. Bands 1-2 are measured at 250 metre resolution, bands 1-7 at 500m resolution and bands 1-16 at 1km resolution. There is possibly more than one observation per day, especially for pixels near the poles where the near-pole orbits of the Terra come closer together.

		\item \textbf{Atmosphere correction:} Level 1B calibrated radiance data is transformed into level 2 surface reflectance and corrected for the effects of atmospheric gases and aerosols. First radiance data is scaled and divided by the cosine of the solar zenith angle to obtain a top-of-atomosphere reflectance value. Then ground level surface reflectance is estimated by correcting for atmospheric scattering and absorption.  Correction involves many different variables including atmospheric intrinsic reflectance, gaseous transmission, atmospheric transmission, spherical albedo, surface pressure, ozone, water vapor and aerosol optical thickness.

		\item \textbf{Projection \& filtering} Data is projection onto a grid and temporally aggregated into one daily observation per pixel to obtain daily level 2G data.

		\item  \textbf{Spatial aggregation}: Bands 1-2 from 250m to 1km and bands 3-7 from 500m to 1km are spatially aggregated. This is done by the modis aggregation algorithm which takes into account the fact that the original granule observations may intersect more than one grid cell at any given resolution.

		\item \textbf{Compositing}: This involves aggregating the daily reflectance data to produce a time series with a 16-day frequency. This is done in two steps. First the data is pre-composited, or aggregated, into 8-day frequency using a set of filters based on quality, cloud presence, and viewing angle. The more cloud-contamination and  residual atmospheric contamination the poorer the quality of the observation. Also, the further away from a nadir sensor view (target directly beneath sensor) the poorer the quality of the pixel. The second step, called compositing, involves computing the NDVI for both 8-day periods and picking the best observation based on highest NDVI (Maximum Value Composite) and other quality assurance conditions. Note that the compositing procedure means that  there can be spatial discontinuities in the reflectance values because adjacent pixels may be measurements from different days (within the same 16-day window) and different sun-sensor-pixel geometries.
				
	\end{enumerate}	
	
	The contents of modis 13A2 dataset is now described. It is composed of four types of variables: quality, sun-sensor geometry, surface reflectance and vegetation index variables. Specifically, the 13A2 data set includes the variables of table \ref{tab:MODISvars}:
	
 \begin{table}[H] 
	\tiny 
 \begin{tabular}{|p{4cm}|l|l|l|l|l|l|l|}
	 \hline
 \textbf{Variable}                              & \textbf{Type}                        & \textbf{band} & \textbf{wavelength (nm)} & \textbf{units} & \textbf{missing flag} & \textbf{min} & \textbf{max} \\ 
 \hline                                                                                                                                              
 pixel reliability                              & \multirow{2}{*}{quality}             &    &           & rank            & 255   &   0    &  3      \\
 VI Quality                                     &                                      &    &           & flag            & 65535 &   0    &  65534  \\
 \hline                                                                                                                                              
 view zenith angle                              & \multirow{4}{*}{sun-sensor geometry} &    &           & degree          & -100  &   -90  & 90      \\
 sun zenith angle                               &                                      &    &           & degree          & -100  &   -90  & 90      \\
 relative azimuth angle                         &                                      &    &           & degree          & -40   &   -360 & 360     \\
 composite day                                  &                                      &    &           & day of the year & -1    &    1   & 366     \\
 \hline                                                                                                                                              
 red reflectance                                & \multirow{4}{*}{reflectance}         & 1  & 620-670   & \%              & -0.1  &    0   & 1       \\
 near infra-red (NIR) reflectance               &                                      & 2  & 841-876   & \%              & -0.1  &    0   & 1       \\
 blue reflectance                               &                                      & 3  & 459-479   & \%              & -0.1  &    0   & 1       \\
 middle infra-red (MIR) reflectance             &                                      & 7  & 2105-2155 & \%              & -0.1  &    0   & 1       \\
 \hline                                                                                                                                              
 normalized difference vegetation index (NDVI)  & \multirow{2}{*}{vegetation indices}  &    &           & index           & -0.3  &  -0.2  & 1       \\
 enhanced vegetation index (EVI)                &                                      &    &           & index           & -0.3  &  -0.2  & 1       \\
 \hline
 \end{tabular}\\
 \caption{MODIS 13A2 variables}
 \label{tab:MODISvars}
 \end{table}
 
All 12 variables are available for:
 
 \begin{itemize}
  \item 343 time observations at 16 day intervals from 2000-02-18 to 2015-01-01, and
  \item data for the whole of Mexico at 1km resolution which corresponds to 7,128,000  pixels per date-variable. 
 \end{itemize}
 
 The behavior of these variables accross time and different types of land-cover will be explored in order to get a better understanding of the different circumstances under which surface reflectance is measured and to get a first sense of the behavior of the surface reflectance and vegetation index variables. Before providing a more detailed description and exploration of each variable we describe the sample used to explore the data. The sample consists of:

\begin{itemize}
 \item Only pixels within Mexico which showed no change in landcover from 2005 to 2010,
 \item Only pixels found at least 1km away from pixels of other classes to mitigate misclassification at borders,
 \item Samples of 100 pixels for each of the 14 significant classes present in Mexico (snow and ice pixels not sampled), and, 
 \item 1,400 pixels sampled in total.
\end{itemize}

\begin{figure}[H]
	  \centering
	  \includegraphics[width=1\textwidth]{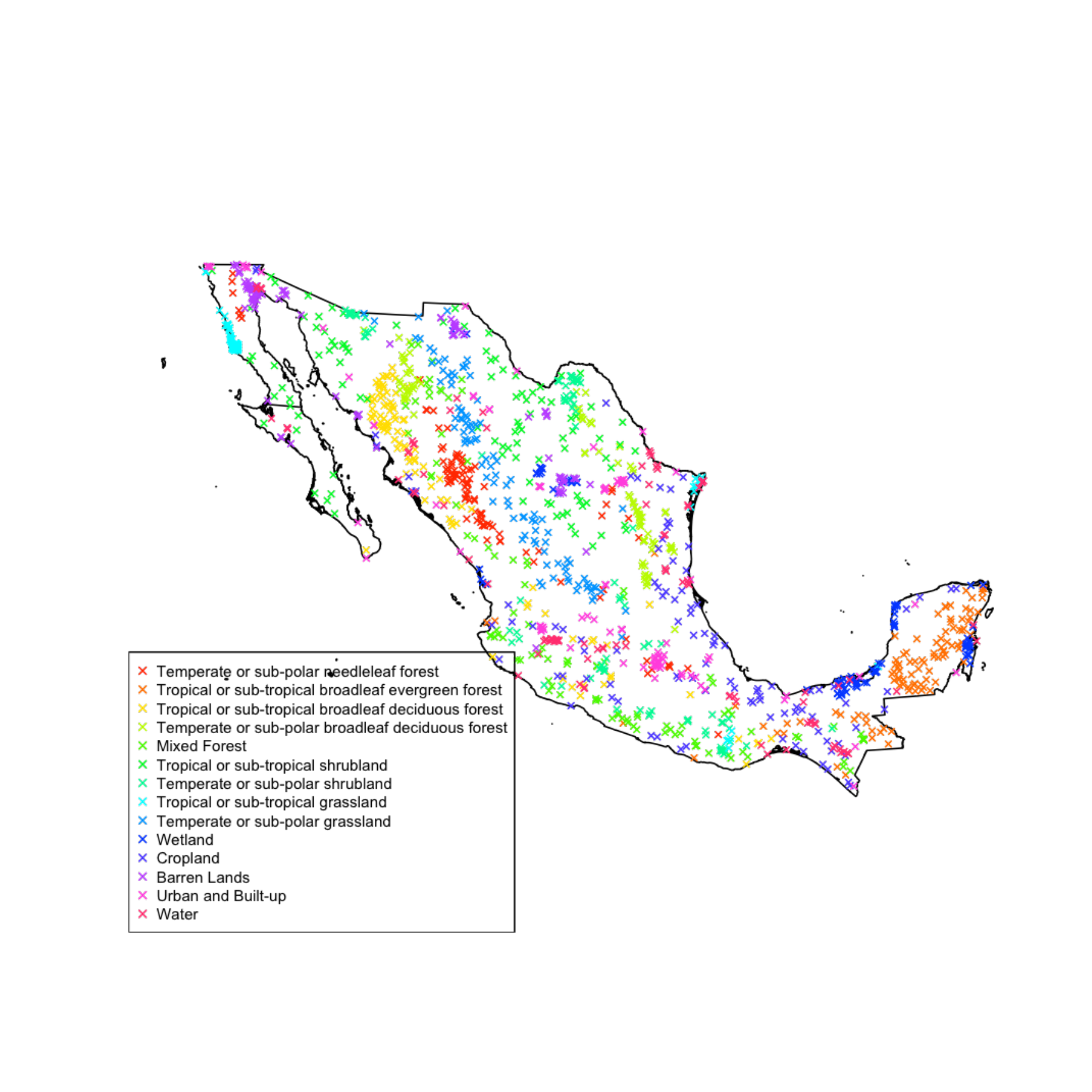} 
	  \caption[Sample used to explore data]
	  {Sample used to explore data}
	  \label{fig:sample}
	\end{figure}

Figure \ref{fig:sample} shows the spatial distribution of the sample. 

The behavior of the different variables is now explored. Although some of the analysis in the rest of the chapter will not be explicitly used in the adapted CMFDA algorithm, it provides a better understanding of the remote-sensing process and may aid future improvements to the algorithm.

\subsection{Quality}

Two variables describe the quality of the data, pixel reliability and VI Quality. In the adapted CMFDA algorithm only the pixel reliability variable is used to filter observations with bad quality. However, VI Quality gives a more detailed description of the different sources of noise in the data. Pixel reliability gives an overall assessment of the quality of the reflectance observations. The quality of each pixel-date observation is assessed and takes the values shown in table \ref{tab:pixRel}:

 \begin{center} 

 \begin{tabular}{|l|l|l|}
	 \hline
 \textbf{Value} & \textbf{Summary} & \textbf{Description}  \\ 
 \hline                                                                                                                                              
 -1 & Fill/No data  &  Not processed                            \\
\hline                                                                                       
 0  & Good data     &  Use with confidence                      \\
 \hline                                                                                       
 1  & Marginal data &  Useful, but look at other QA information \\
\hline                                                                                       
 2  & Snow/Ice      &  Target covered with snow/ice             \\
\hline                                                                                       
 3  & Cloudy        &  Target not visible, covered with cloud   \\
\hline                                                                                       
 \end{tabular}\\
 \captionof{table}{Pixel reliability values}
  \label{tab:pixRel}
 \end{center}

	\begin{center}
		\footnotesize\addtolength{\tabcolsep}{-2pt}
\begin{table}[ht]
\centering
\begin{tabular}{rrrrr}
 Good data & Marginal data & Snow/ice & Cloudy & Not processed \\ 
  \hline
65.519 & 30.677 & 0.027 & 3.203 & 0.574 \\ 
   \hline
\end{tabular}
\caption[\% pixel reliability in exploration sample.]{\% pixel reliability in exploration sample.} 
\label{tab:pixRelDist}
\end{table}

	\end{center}
	
	As table \ref{tab:pixRelDist} shows over 95\% of the data is classified as either \emph{good data} or \emph{marginal data}. By exploring the VI Quality variable the types of noise affecting data flagged as \emph{marginal data} will be shown.

	\begin{figure}[H]
	  \centering
	  \includegraphics[width=.8\textwidth]{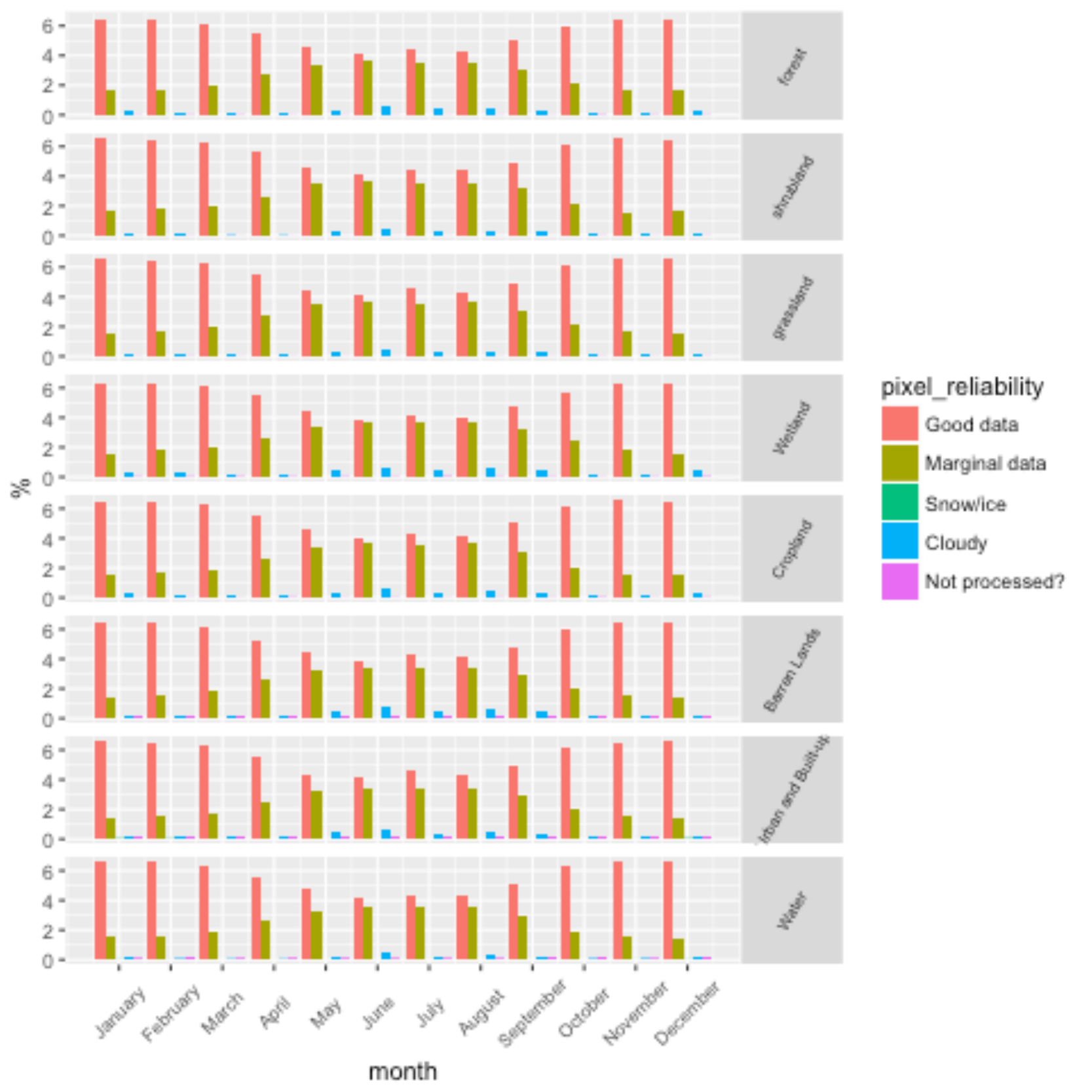} 
	  \caption[Pixel reliability by month and land use type]
	  {Pixel reliability by month and land use type}
	  \label{fig:pixRel_month}
	\end{figure}
	
	Figure \ref{fig:pixRel_month} shows that the quality of the data drops during the months from April to September coinciding with the rainy season in many parts of Mexico. 
	
	Although the pixel realiability variable will be used to filter noisy data before fitting the \emph{reflectance} model of the CMFDA algorithm,  the VI Quality variable is explored in order to understand what kinds of noise may be affecting observations flagged with different pixel realiability values.

The variable VI Quality encodes several flags (in a 16-bit  integer) detailing various aspects of the quality of the reflectance observations. The quality of each pixel-date observation is assessed in detail. Table \ref{MODISvars}, based on table 5 from \cite{userGuide_VI}, shows the flags encoded.

 \begin{center} 
	\tiny 
 \begin{tabular}{|l|l|l|l|}
	 \hline
 \textbf{Inverse position in 16-bit} & \textbf{variable encoded} & \textbf{values} & \textbf{description} \\ 
 \hline                                                                                                                                              
 \multirow{4}{*}{0-1}  & \multirow{4}{*}{MODLAND QA} & 00 & VI produced, good quality       \\
                       &                             & 01 & VI produced, but check other QA \\
                       &                             & 10 & pixel produced, but most probably cloudy \\
                       &                             & 11 & pixel not produced due to other reasons other than clouds \\					   
 \hline                                                                                                 
 \multirow{11}{*}{2-5} & \multirow{11}{*}{VI usefulness} & 0000 & highest quality       \\
                       &                                 & 0001 & lower quality         \\
                       &                                 & 0010 & decreasing quality    \\
                       &                                 & 0100 & decreasing quality    \\					   
                       &                                 & 1000 & decreasing quality    \\
					   &                                 & 1001 & decreasing quality    \\
					   &                                 & 0101 & decreasing quality    \\
					   &                                 & 1100 & lowest quality        \\					   					
					   &                                 & 1101 & quality so low that it is not useful \\					  
					   &                                 & 1110 & L1B data faulty \\
					   &                                 & 1111 & Not useful for any other reason/ not processed \\					 										    					   
 \hline
 \multirow{4}{*}{6-7}  & \multirow{4}{*}{Aerosol quantity} & 00 & Climatology       \\
                       &                                   & 01 & Low \\
                       &                                   & 10 & Average \\
                       &                                   & 11 & High  \\					   
 \hline
 \multirow{2}{*}{8}    & \multirow{2}{*}{Adjacent cloud detected} & 1 & Yes       \\
                       &                                          & 0 & No        \\
 \hline
 \multirow{2}{*}{9}    & \multirow{2}{*}{Atmosphere BRDF correction performed} & 1 & Yes       \\
                       &                                                       & 0 & No        \\
 \hline
 \multirow{2}{*}{10}    & \multirow{2}{*}{Mixed clouds} & 1 & Yes       \\
                       &                                & 0 & No        \\
 \hline
 \multirow{4}{*}{11-13}  & \multirow{4}{*}{Land/water flag}  & 000 & Shallow ocean                \\
                         &                                   & 001 & Land (nothing else but land) \\
                         &                                   & 010 & Ocean coastlines and lake shorelines \\
                         &                                   & 011 & Shallow inland water          \\					   
                         &                                   & 100 & Ephemeral water          \\
                         &                                   & 101 & Deep inland water          \\
                         &                                   & 110 & Moderate or continental ocean          \\					   
                         &                                   & 111 & Deep ocean         \\					   						 
 \hline
 \multirow{2}{*}{14}    & \multirow{2}{*}{Possible snow/ice} & 1 & Yes       \\
                       &                                     & 0 & No        \\
 \hline
 \multirow{2}{*}{15}    & \multirow{2}{*}{Possible shadow} & 1 & Yes       \\
                       &                                   & 0 & No        \\
 \hline
 \end{tabular}\\
 \captionof{table}{VI Quality variables}
  \label{MODISvars}
 \end{center}
 
 Tables \ref{tab:modland}-\ref{tab:possShadow} show the distribution of the VI Quality variables for the sample described previously. 

	\begin{center}
		\footnotesize\addtolength{\tabcolsep}{-2pt}
\begin{table}[H]
	\tiny
\centering
\begin{tabular}{p{3cm}|p{3cm}|p{3cm}|p{3cm}}
 VI produced, good quality & VI produced, but check other QA & Pixel produced, but most probably cloudy & Pixel not produced due to other reasons than clouds \\ 
  \hline
92.215 & 0.024 & 7.184 & 0.577 \\ 
   \hline
\end{tabular}
\caption[MODLAND distribution]{MODLAND distribution} 
\label{tab:modland}
\end{table}

	\end{center}
	
	Table \ref{tab:modland} specifies whether the quality of the data was good enough to estimate surface reflectances and vegetation indices (VI produced), only surface reflectances (pixel produced) or neither. For over 92\% of the data both types of variables are estimated and for over 97\% the surface reflectances are estimated. 

	\begin{center}
		\footnotesize\addtolength{\tabcolsep}{-2pt}
\begin{table}[H]
	\tiny
\centering
\begin{tabular}{r|r|r|r|r|r|r|r}
 2 & 3 & 4 & 5 & 6 & 7 & 8 & 12 \\ 
  \hline
0.375 & 93.426 & 0.641 & 3.051 & 0.029 & 2.274 & 0.065 & 0.139 \\ 
   \hline
\end{tabular}
\caption[Usefulness distribution]{Usefulness distribution} 
\label{tab:usefulness}
\end{table}

	\end{center}
	
	The \emph{usefuleness} variable is supposed to take 11 values however table \ref{tab:usefulness} suggests 12 possible values. We simply took lower values to mean higher quality, although, we did not use this variable for any part of the methodology. 
	
	\begin{center}
		\footnotesize\addtolength{\tabcolsep}{-2pt}
\begin{table}[H]
	\tiny
\centering
\begin{tabular}{r|r|r|r}
 Climatology & Low & Average & High \\ 
  \hline
94.565 & 4.861 & 0.000 & 0.574 \\ 
   \hline
\end{tabular}
\caption[Aerosol quantity]{Aerosol quantity}
\label{tab:aerosol} 
\end{table}

	\end{center}
	
	Table \ref{tab:aerosol} shows that over 94\% of the reflectance data is corrected for the presence of aerosols in the atmosphere (\emph{Climatology} value). 
	
	\begin{center}
		\footnotesize\addtolength{\tabcolsep}{-2pt}
\begin{table}[H]
	\tiny
\centering
\begin{tabular}{r|r}
 0 & 1 \\ 
  \hline
77.911 & 22.089 \\ 
   \hline
\end{tabular}
\caption[Adjacent cloud flag]{Adjacent cloud flag} 
\label{tab:adjacentCloud}
\end{table}

	\end{center}
	
	Table \ref{tab:adjacentCloud} shows that around 22\% of the reflectance observations were taken when a cloud was detected for an adjacent pixel. While a cloud was not detected for the pixel associated to the observation, the fact that cloud was detected in an adjacent pixel increases the possibility that the reflectance observation is affected by cloudy conditions.
	
	\begin{center}
		\footnotesize\addtolength{\tabcolsep}{-2pt}
\begin{table}[H]
	\tiny
\centering
\begin{tabular}{r|r}
 0 & 1 \\ 
  \hline
23.605 & 76.395 \\ 
   \hline
\end{tabular}
\caption[Atmosphere BRDF correction flag]{Atmosphere BRDF correction flag} 
\label{tab:AtmBRDF}
\end{table}

	\end{center}
	
	Table \ref{tab:AtmBRDF} shows that over 76\% of the reflectance data was adjusted for atmospheric effects. 
	
	\begin{center}
		\footnotesize\addtolength{\tabcolsep}{-2pt}
\begin{table}[H]
	\tiny
\centering
\begin{tabular}{r|r}
 0 & 1 \\ 
  \hline
99.125 & 0.875 \\ 
   \hline
\end{tabular}
\caption[Mixed cloud flag]{Mixed cloud flag} 
\label{tab:mixedCloud}
\end{table}

	\end{center}
	
	Table \ref{tab:mixedCloud} shows that in less than 1\% of the observations were taken when a cloud with \emph{mixed} phase was detected.
	
	\begin{center}
		\footnotesize\addtolength{\tabcolsep}{-2pt}
\begin{table}[H]
	\tiny
\centering
\begin{tabular}{p{1.8cm}|p{1.8cm}|p{1.8cm}|p{1.8cm}|p{1.8cm}|p{1.8cm}|p{1.8cm}|p{1.8cm}}
 Shallow ocean & Land (Nothing else but land) & Ocean coastlines and lake shorelines & Shallow inland water & Ephemeral water & Deep inland water & Moderate or continental ocean & Deep ocean \\ 
  \hline
0.092 & 62.387 & 16.664 & 6.336 & 6.114 & 2.614 & 2.459 & 3.333 \\ 
   \hline
\end{tabular}
\caption[Land/water categories]{Land/water categories} 
\label{tab:landWater}
\end{table}

	\end{center}
	
	Table \ref{tab:landWater} shows that the Land/water flag allows us to determine whether the surface where the reflectance observation was estimated is land or water. For over 62\% of the observations the type of surface was land. 
	
	\begin{center}
		\footnotesize\addtolength{\tabcolsep}{-2pt}
\begin{table}[H]
	\tiny
\centering
\begin{tabular}{r|r}
 0 & 1 \\ 
  \hline
96.223 & 3.777 \\ 
   \hline
\end{tabular}
\caption[Ice/snow flag]{Ice/snow flag} 
\label{tab:iceSnow}
\end{table}

	\end{center}
	
	Table \ref{tab:iceSnow} shows that ice or snow was detected on the surface for less than 4\% of the observations. 
	
	\begin{center}
		\footnotesize\addtolength{\tabcolsep}{-2pt}
\begin{table}[H]
	\tiny
\centering
\begin{tabular}{r|r}
 0 & 1 \\ 
  \hline
71.708 & 28.292 \\ 
   \hline
\end{tabular}
\caption[Possible shadow flag]{Possible shadow flag} 
\label{tab:possShadow}
\end{table}

	\end{center}
	
	Finally, table \ref{tab:possShadow} shows that over 28\% of observations were taken at times and locations where nearby clouds could have cast a shadow over the surface. The relationship between the VI Quality flags and the pixel reliability variable is now explored. 
	
	Data flagged as \emph{good data} by the pixel reliability variable all have desirable flags in the VI Quality variables (\emph{VI produced, good quality} or \emph{VI produced but check other QA} for MODLAND, \emph{Climatology} for aerosol quantity, no adjacent clouds, atmosphere BRDF correction, no mixed cloud, no ice or snow and \emph{land} or \emph{ocean coastlines and lake shorelines} as land/water category), so this data has no discernible flaws based on the quality flags. Table \ref{tab:pixRel_VI} shows the distribution of data flagged as \emph{marginal data} by \emph{pixel reliability}  with respect to VI Quality variables:

	 \begin{center} 
		 \tiny
	 \begin{tabular}{|p{12cm}|l|}
		 \hline
	 \textbf{undesirable flags} & \textbf{\%}  \\ 
	 \hline                                                                                                      
	 none &  9.672  \\
	\hline                                               
	 land flag $\neq \{$ land, ocean coastlines and lake shorelines $\}$ &  0.0122\\
	 \hline                                               
	 land flag $\neq \{$ land, ocean coastlines and lake shorelines $\}$, not atmosphere BRDF corrected and possible shadow &  4.598 \\
	\hline                                               
	 land flag $\neq \{$ land, ocean coastlines and lake shorelines $\}$, adjacent cloud and possible shadow  &  17.595 \\
	\hline                                               
	 not atmosphere BRDF corrected, possible shadow and adjacent cloud &  33.459 \\
	\hline
	 land flag $\neq \{$ land, ocean coastlines and lake shorelines $\}$ not atmosphere BRDF corrected, adjacent cloud and possible shadow &  11.684 \\
	\hline
	 land flag $\neq \{$ land, ocean coastlines and lake shorelines $\}$ not atmosphere BRDF corrected, aerosol $\neq$ \emph{climatology} and possible shadow & 1.376 \\
	\hline
	  MODLAND $=$ \emph{pixel produced but probably cloudy} , land flag $\neq \{$ land, ocean coastlines and lake shorelines $\}$  and possible shadow & 2.897   \\
	\hline
	  MODLAND $=$ \emph{pixel produced but probably cloudy} , land flag $\neq \{$ land, ocean coastlines and lake shorelines $\}$, not atmosphere BRDF corrected and possible shadow &  0.057 \\
	\hline
	  MODLAND $=$ \emph{pixel produced but probably cloudy} , land flag $\neq \{$ land, ocean coastlines and lake shorelines $\}$, aerosol $\neq$ \emph{climatology} and possible shadow &  1.898 \\
	\hline
	  MODLAND $=$ \emph{pixel produced but probably cloudy} , land flag $\neq \{$ land, ocean coastlines and lake shorelines $\}$, aerosol $\neq$ \emph{climatology}, possible shadow and mixed cloud & 0.001  \\
	\hline
	  MODLAND $=$ \emph{pixel produced but probably cloudy} , land flag $\neq \{$ land, ocean coastlines and lake shorelines $\}$, aerosol $\neq$ \emph{climatology}, possible shadow and not atmosphere BRDF corrected & 3.645   \\
	\hline
	  MODLAND $=$ \emph{pixel produced but probably cloudy}, land flag $\neq \{$ land, ocean coastlines and lake shorelines $\}$, aerosol $\neq$ \emph{climatology}, not atmosphere BRDF corrected and possible shadow & 13.103  \\
	\hline
	  MODLAND $=$ \emph{pixel produced but probably cloudy}, land flag $\neq \{$ land, ocean coastlines and lake shorelines $\}$, aerosol $\neq$ \emph{climatology}, not atmosphere BRDF corrected, possible shadow and adjacent cloud & 0.002  \\
	\hline
	 \end{tabular}\\
	 \captionof{table}{Distribution of \emph{Marginal data} according to VI Quality flags}
	 \label{tab:pixRel_VI}
	 \end{center}
	
	As we can see from table \ref{tab:pixRel_VI} the data flagged as \emph{marginal data} does have several problems. Most significantly 21.604 \% of \emph{marginal data} is flagged as \emph{probably cloudy} and 90.316\% is flagged as having a \emph{possible shadow}. 

\subsection{Sun-sensor geometry}

The Terra satellite has a sun-synchronous near-polar orbit which means it orbits the earth from north to south and then south to north in such a way that it always passes over a given point of the planet's surface at the same local solar time. Although, to be clear, it does not pass over all points at the same local solar time. This is useful since it allows for relatively constant illumination of a given point accross different time observations. 

Sun-sensor geometry variables describe the geometric circumstances in which radiance data was recorded for any given pixel at any given date. This is important because radiance data is converted into reflectance data using the sun-zenith angle. Additionally, the view zenith angle is used in the compositing process to obtain the highest quality measurement. This means these variables could potentially be useful in modeling reflectance. For this reason, even though these variables are not used in the adapted CMFDA algorithm, an exporation of their behavior is included here. Figure \ref{fig:sunSensorGeom} describes the three sun-sensor geometry variables:

	\begin{figure}[H]
	  \centering
	  \includegraphics[width=.7\textwidth]{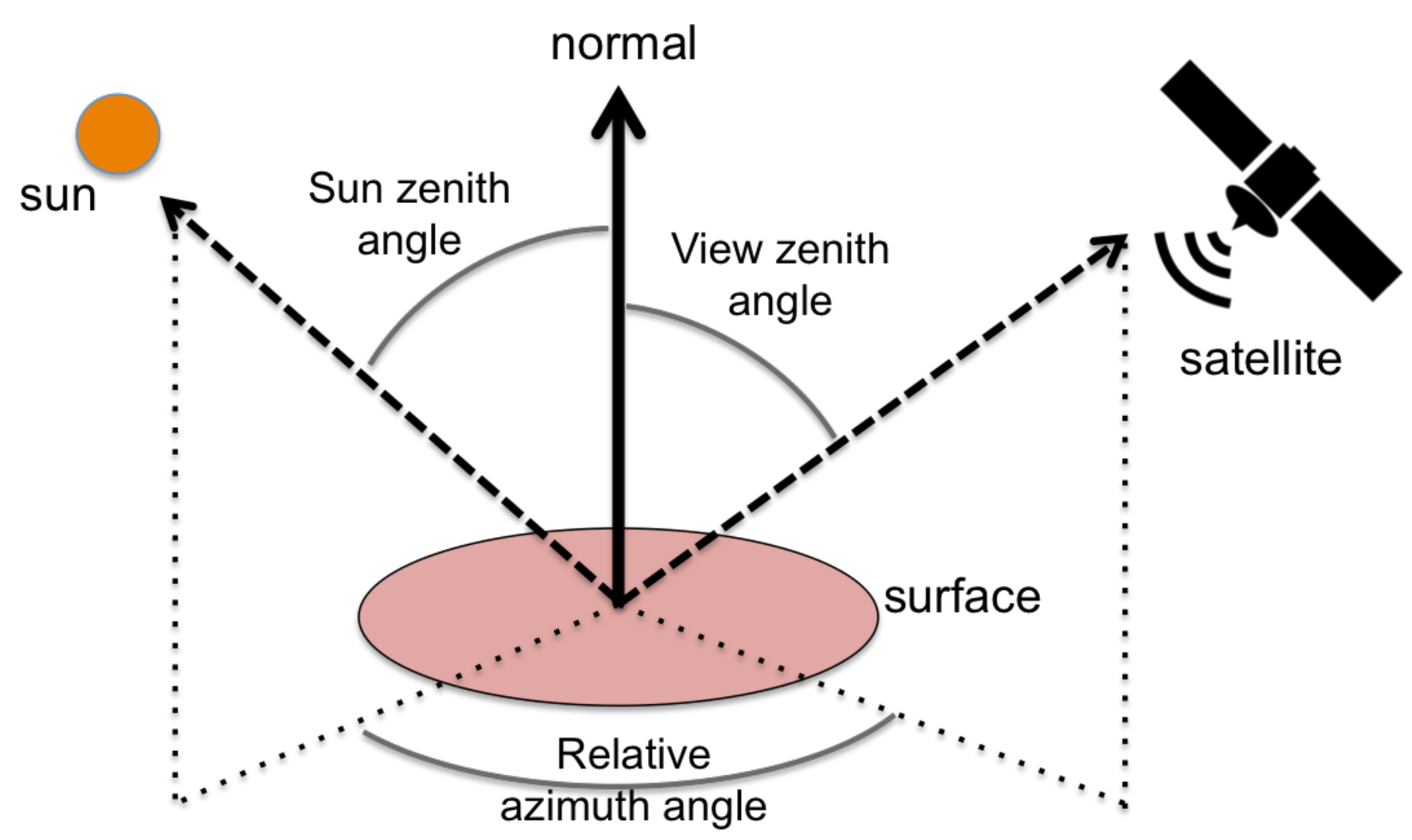} 
	  \caption[Sun-sensor geometry]
	  {Sun-sensor geometry}
	  \label{fig:sunSensorGeom}
	\end{figure}

	\begin{figure}[H]
	  \centering
	  \includegraphics[width=0.4\textwidth]{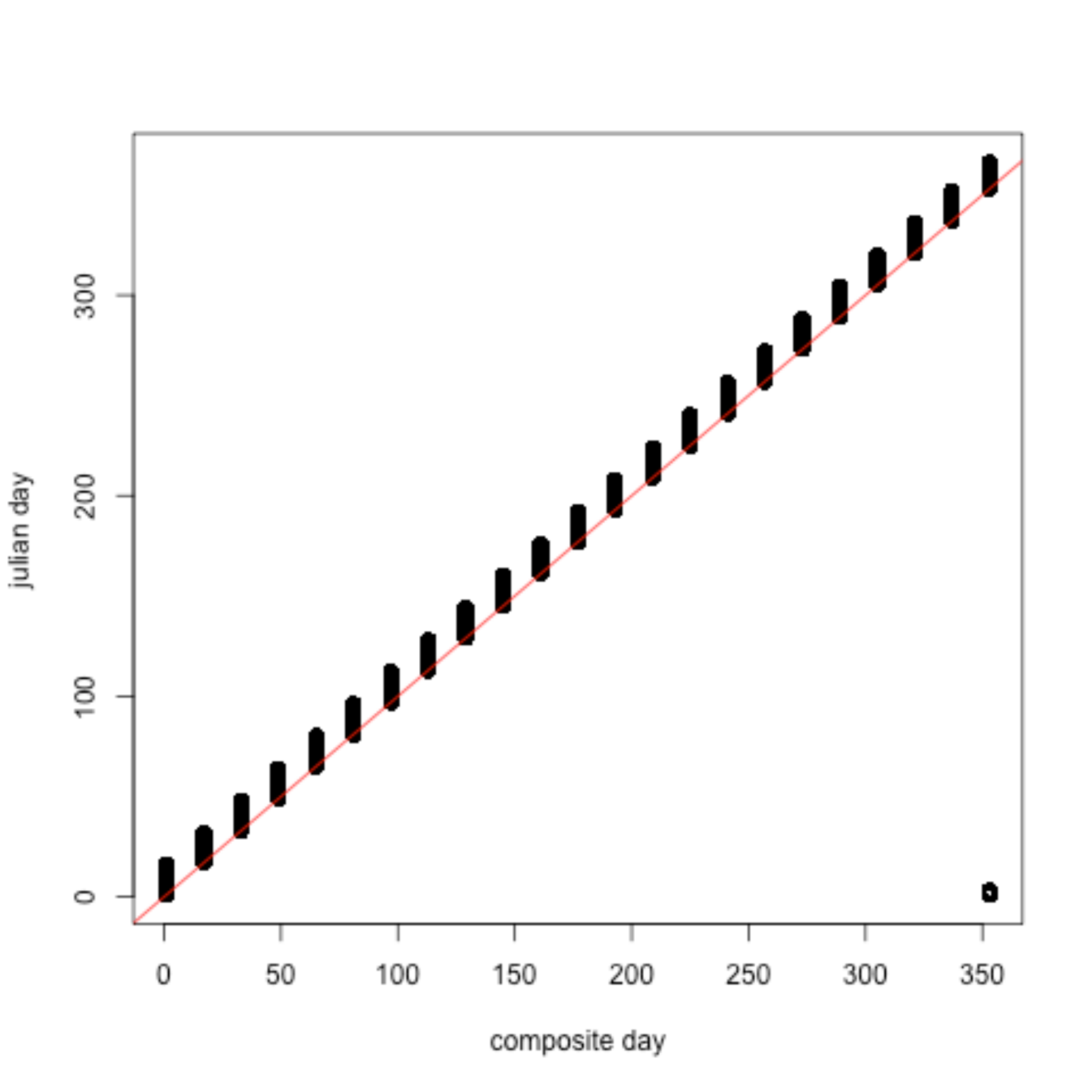} 
	  \caption[Composite day of the year vs. julian day]
	  {Composite day of the year vs. julian day}
	  \label{fig:compJulianDay}
	\end{figure}

	As figure \ref{fig:compJulianDay} shows, the composite day of the year, the day to which a pixel observation actually corresponds, can be 0-15 days after the julian-day, the date of the image file. 

	
 \begin{figure}[H]  
 
   \begin{subfigure}{.5\textwidth}
     \centering
 	\includegraphics[width=1\textwidth]{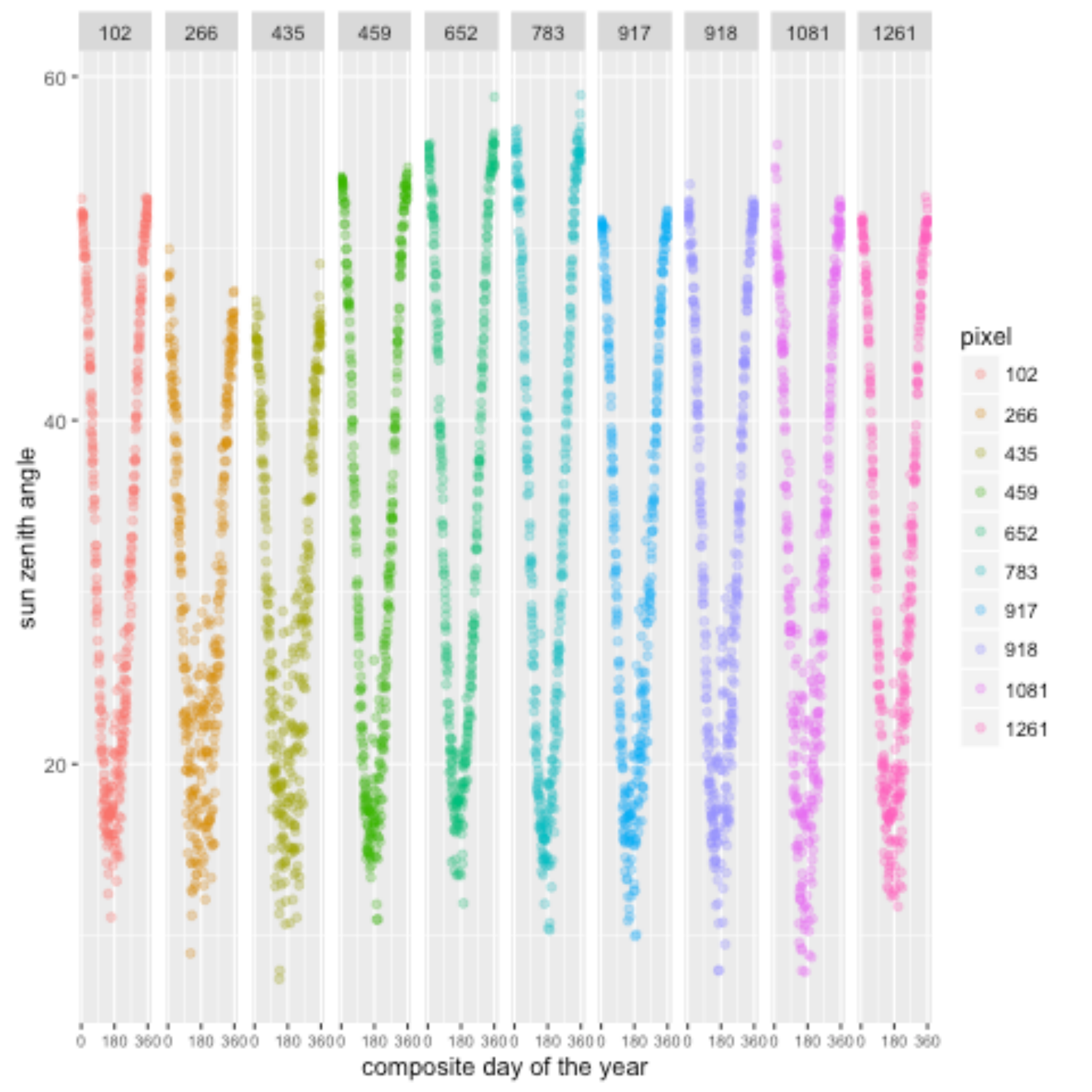} 
     \caption{by composite day and pixel}
     \label{fig:sunzenith_pix}
   \end{subfigure}%
   \begin{subfigure}{.5\textwidth}
     \centering
     \includegraphics[width=1\textwidth]{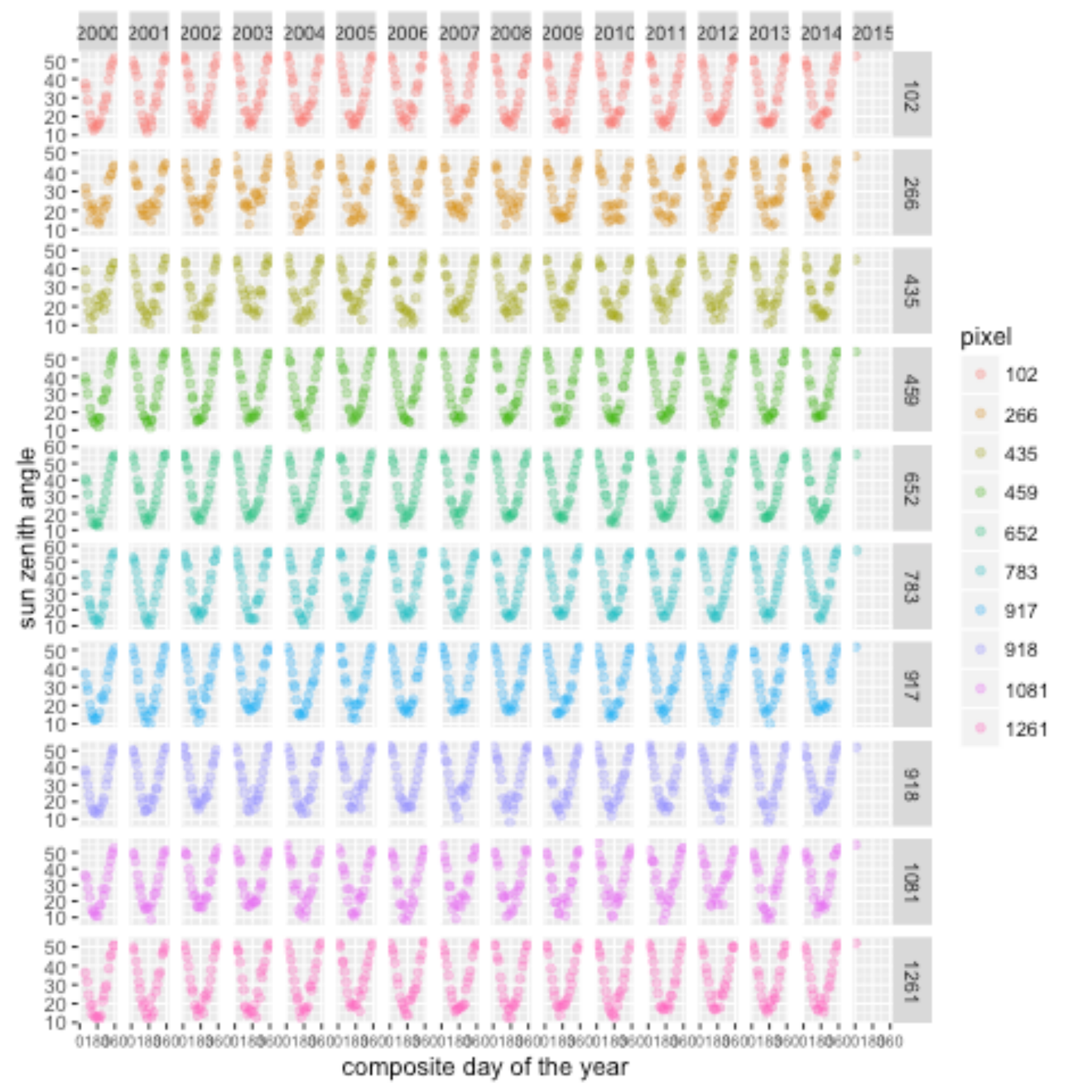}
     \caption{by composite day, pixel and year}
     \label{fig:sunzenith_year}
   \end{subfigure}
   \caption[Sun zenith angle for 10 sampled pixels for 2000-2015 period]
   {Sun zenith angle for 10 sampled pixels for 2000-2015 period}
   \label{fig:sunzenith}
 \end{figure}

	Figure \ref{fig:sunzenith} shows there is a clear yearly seasonal pattern to the sun-zenith angle associated to the reflectance readings of each pixel. In the summer months, when the sun is highest in the sky, the angle is lowest.  There is not much variation accross years or within months due to the sun-synchronous nature of Terra's orbit. There is some variance in the summer months, possibly because the presence of more clouds means that the time of day when a quality pixel measurement is available varies more.  
	
	\begin{figure}[H]
      \begin{subfigure}{.5\textwidth}
        \centering
    	\includegraphics[width=1\textwidth]{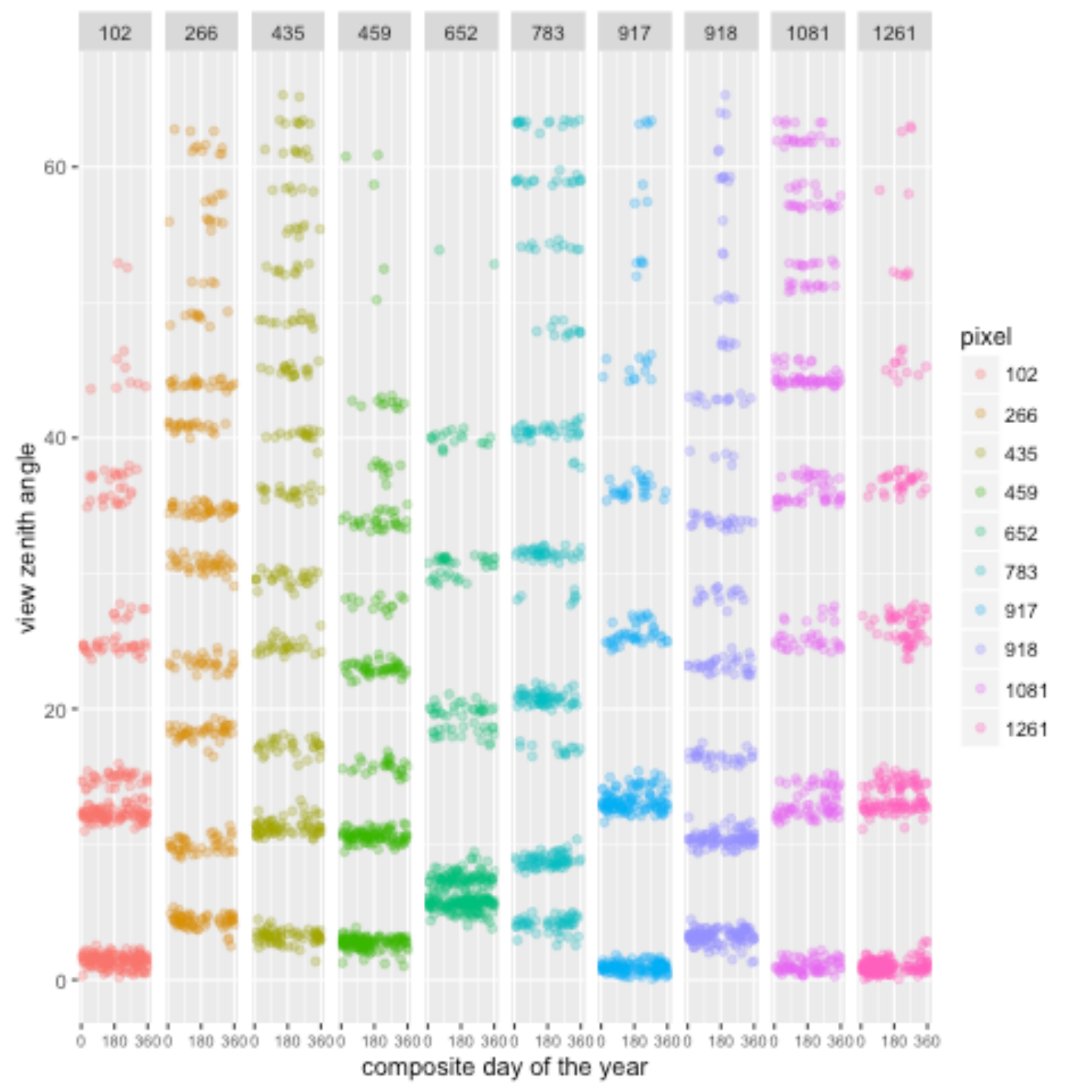} 
        \caption{by composite day and pixel}
        \label{fig:viewzenith_pix}
      \end{subfigure}%
      \begin{subfigure}{.5\textwidth}
        \centering
        \includegraphics[width=1\textwidth]{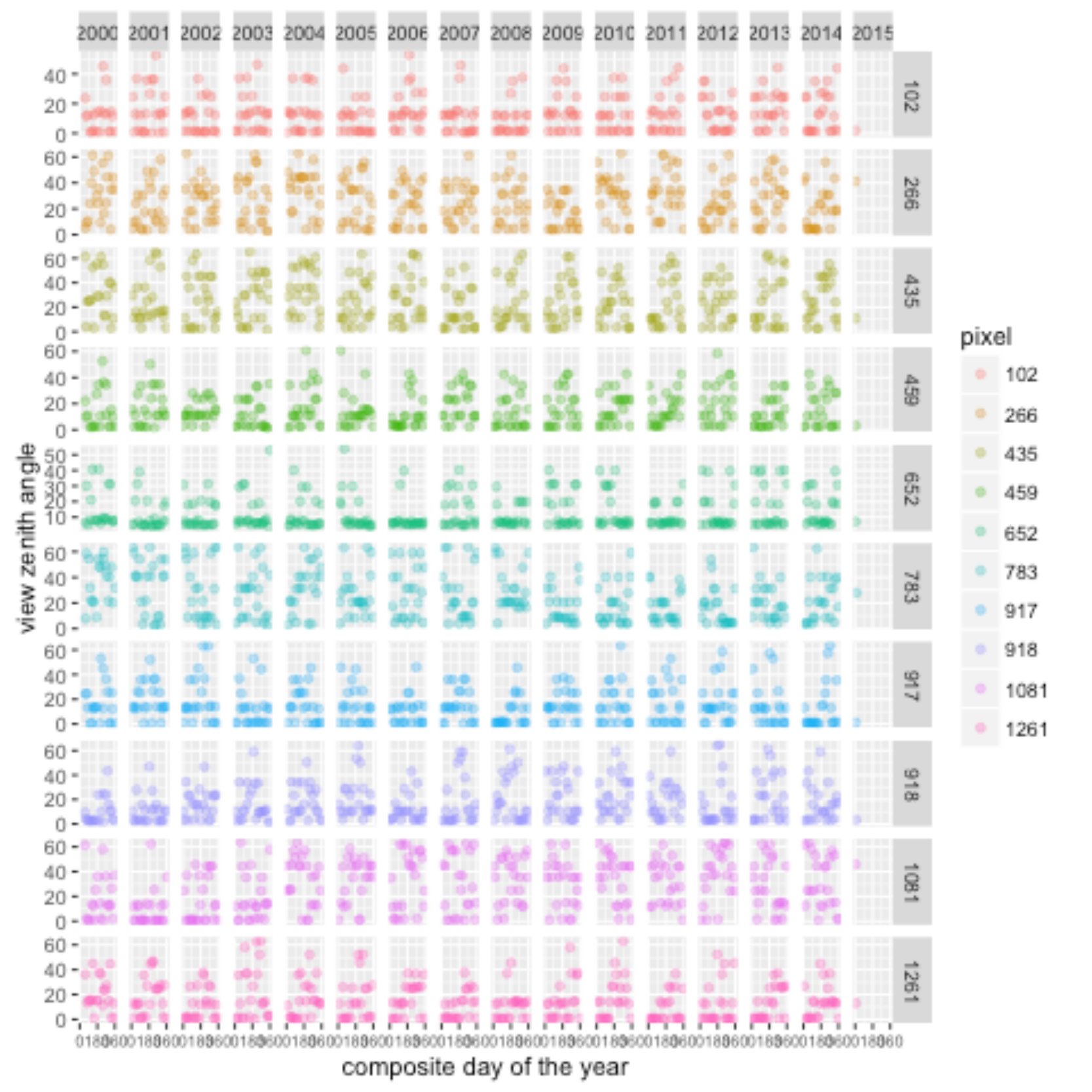}
        \caption{by composite day, pixel and year}
        \label{fig:viewzenith_year}
      \end{subfigure}
      \caption[View zenith angle for 10 sampled pixels for 2000-2015 period]
      {View zenith angle for 10 sampled pixels for 2000-2015 period}
      \label{fig:viewzenith}
    \end{figure}
	
	Figure \ref{fig:viewzenith} shows there is no obvious seasonality to the view-zenith angle associated to the reflectance readings of each pixel, rather they seem to occur at discrete angle increments. Most of the observations are at lower angles. The few observation at higher view-zenith angles are concentrated in the summer months, probably due to the fact that increased cloud cover in these months forces the compositing algorithm to use more extreme off-nadir observations. 
	
	\begin{figure}[H]
      \begin{subfigure}{.5\textwidth}
        \centering
    	\includegraphics[width=1\textwidth]{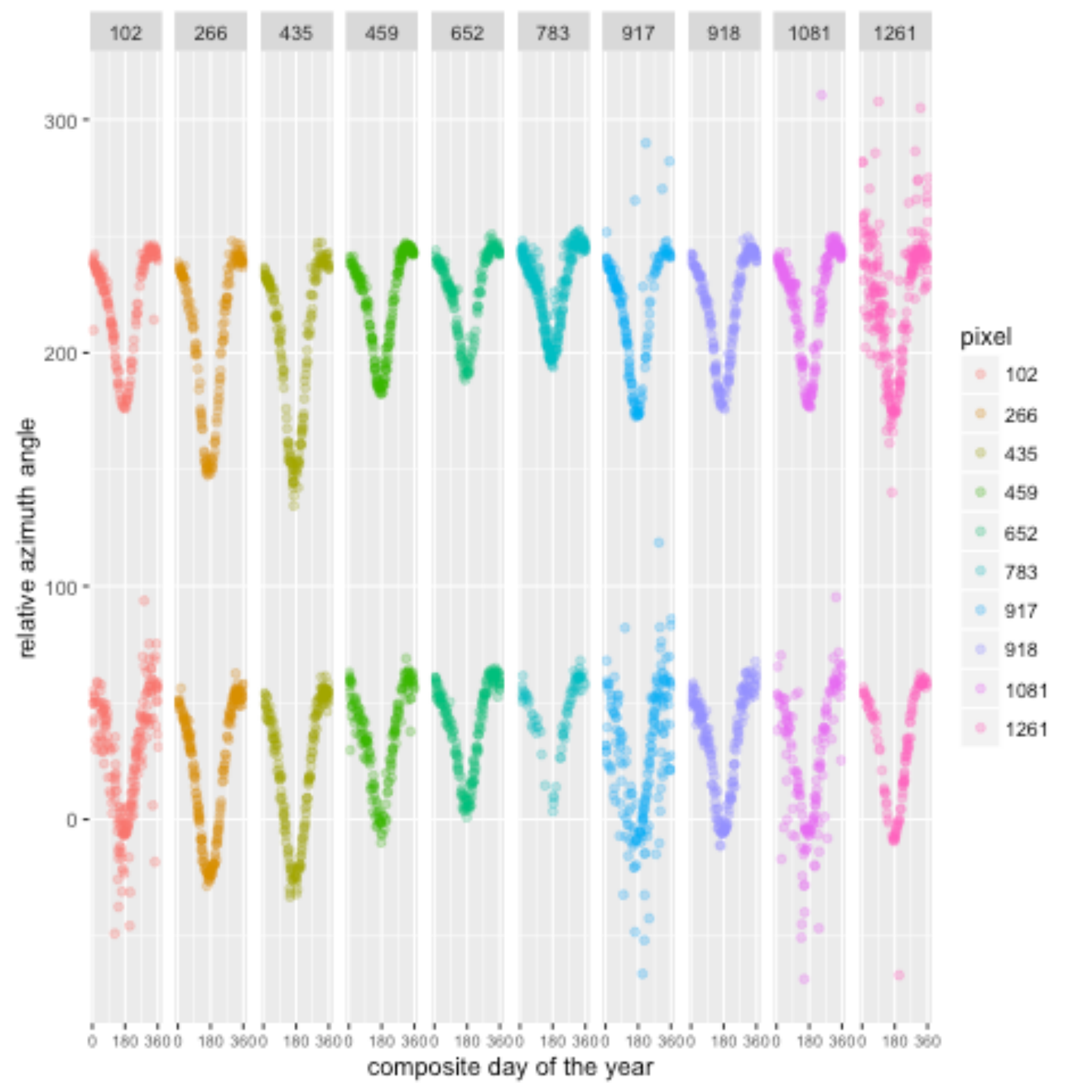} 
        \caption{by composite day and pixel}
        \label{fig:relAzi_pix}
      \end{subfigure}%
      \begin{subfigure}{.5\textwidth}
        \centering
        \includegraphics[width=1\textwidth]{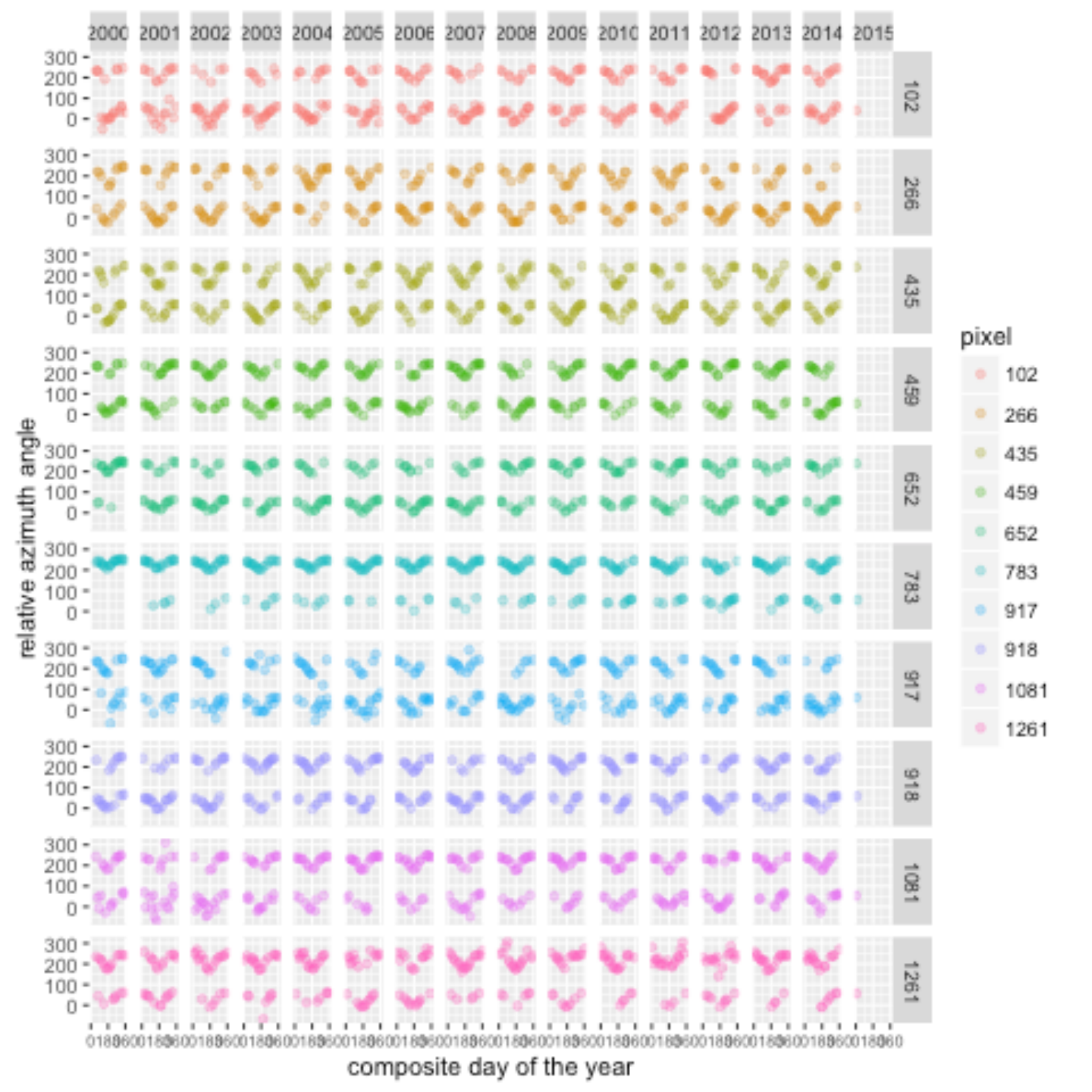}
        \caption{by composite day, pixel and year}
        \label{fig:relAzi_year}
      \end{subfigure}
      \caption[Relative azimuth angle for 10 sampled pixels for 2000-2015 period]
      {Relative azimuth angle for 10 sampled pixels for 2000-2015 period}
      \label{fig:relAzi}
    \end{figure}
	
	Figure \ref{fig:relAzi} shwos that the relative azimuth angle of observations associated to the reflectance readings of each pixel occur in two similar \emph{paths} approximately 180 degrees apart. The part of the path on which an observation lies appears to be related to the sun-zenith angle. However, which one of the two paths an observation belongs to is probably related to the location of the satellite with respect to the path formed by the projection of the target-sun segment, on earth's surface. We apply the following transformation to the relative-azimuth angle in order to \emph{join} the two paths:
	
	\begin{align}
		g(\theta_{ra}) =      
		\begin{cases}
      	  	\theta_{ra}, &  \theta_{ra} < 125 \\
			\theta_{ra} - 180, & \theta_{ra}  \geq 125 
		\end{cases}
	\end{align}

Where $\theta_{ra}$ is the relative-azimuth angle. Now observe the relationship between $g(\theta_{ra})$ and, on the one hand, the composite day of the year, and on the other, the sun-zenith angle. 

	\begin{figure}[H]
      \begin{subfigure}{.5\textwidth}
        \centering
    	\includegraphics[width=1\textwidth]{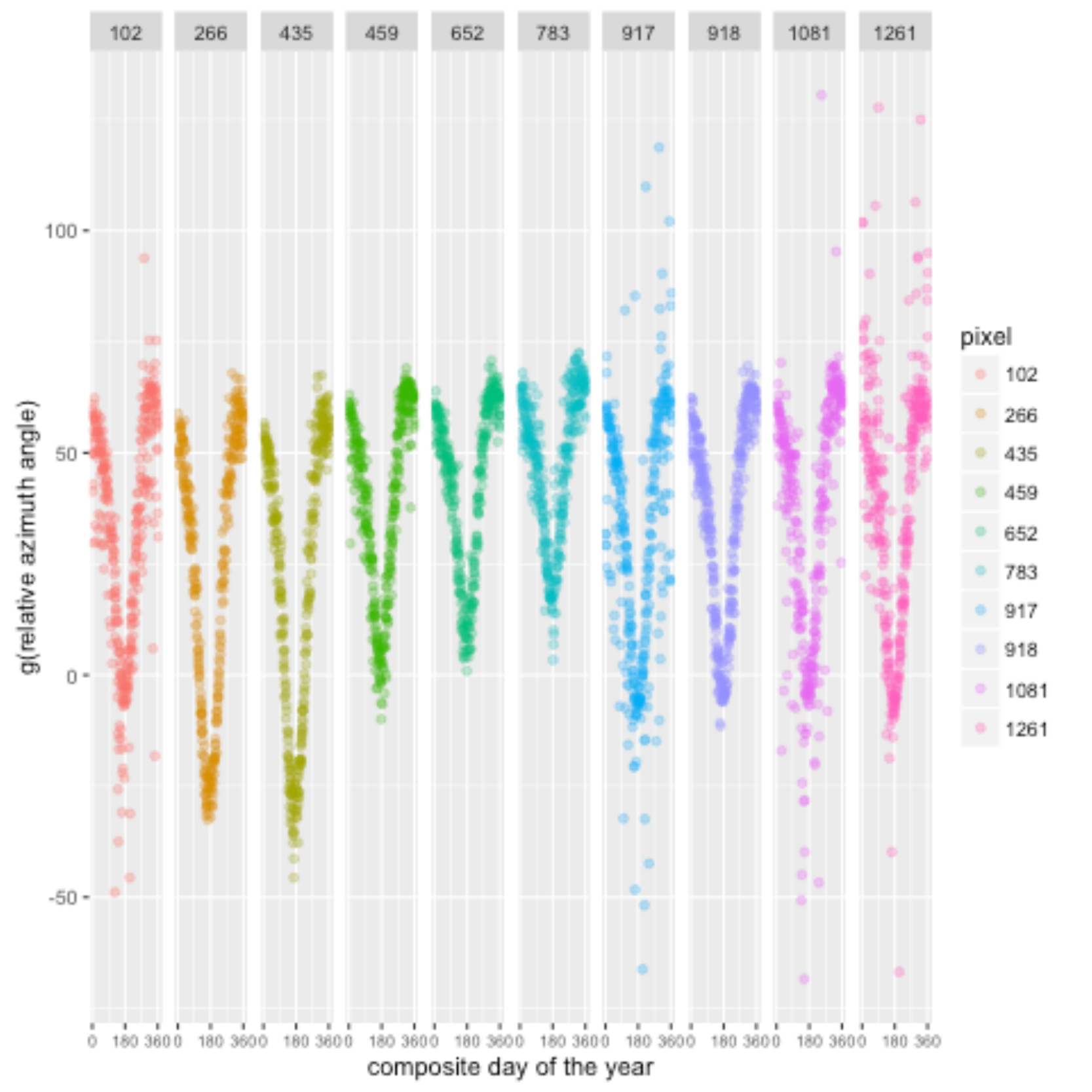} 
        \caption{by composite day and pixel}
        \label{fig:relAziTr}
      \end{subfigure}%
      \begin{subfigure}{.5\textwidth}
        \centering
        \includegraphics[width=1\textwidth]{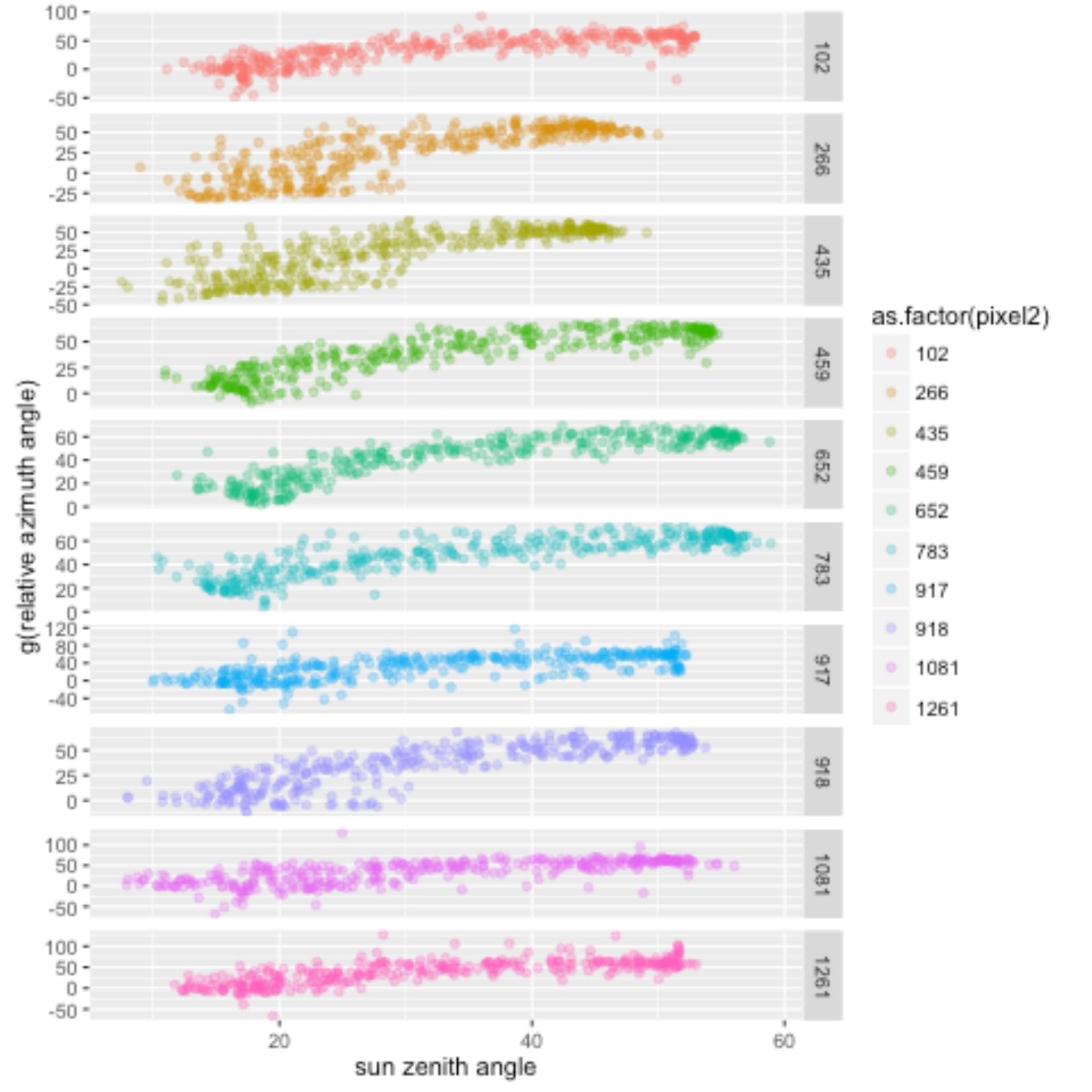}
        \caption{by sun-zenith angle and pixel}
        \label{fig:relAziTr_sunZe}
      \end{subfigure}
      \caption[Transformed relative azimuth angle for 10 sampled pixels for 2000-2015 period]
      {Transformed relative azimuth angle for 10 sampled pixels for 2000-2015 period}
      \label{fig:relAziTr}
    \end{figure}
	
	As figure \ref{fig:relAziTr} shows, the transformed relative azimuth angle seems to depend quadratically on the sun-zenith angle. 

\subsection{Reflectance} \label{ss:reflectance}

Surface reflectance is the amount of light reflected by a surface. It is a ratio of surface radiance to surface irradiance, so it is unitless, and typically has values between 0.0 and 1.0. The atmospheric correction algorithm described in section \ref{s:refl} results in values typically between -0.1 and 16. 

Since part of the CMFDA algorithm consists of modeling surface reflectance the distributions of the four reflectance variables is plotted. It is also of interest to distinguish the difference in the reflectance patterns related to the type of landcover so the median reflectance is plotted against the month of the year for the different types of landcover.

	
	\begin{figure}[H]
      \begin{subfigure}{.5\textwidth}
        \centering
    	\includegraphics[width=0.8\textwidth]{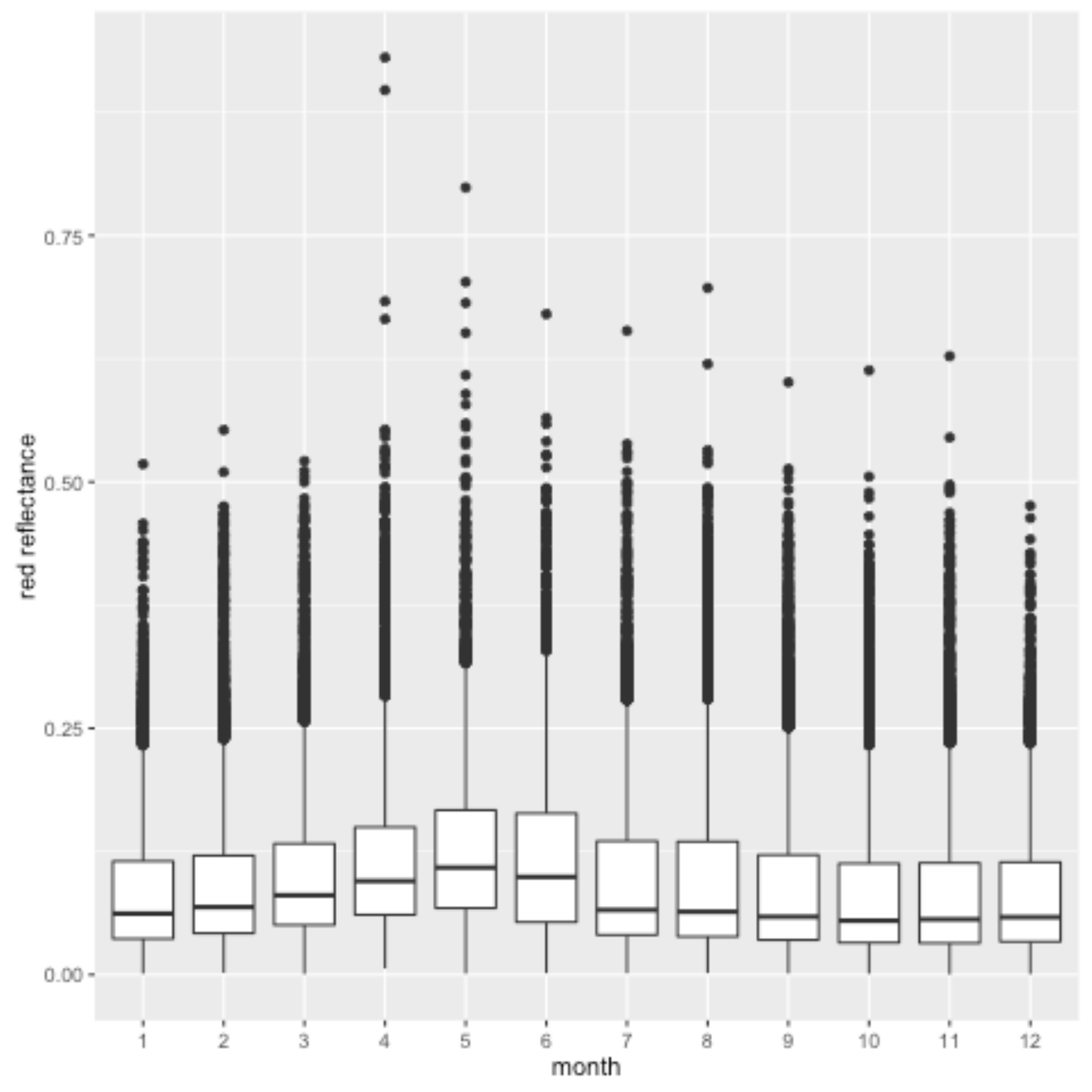} 
        \caption{box plot by month}
        \label{fig:redRef_box}
      \end{subfigure}%
      \begin{subfigure}{.5\textwidth}
        \centering
        \includegraphics[width=0.8\textwidth]{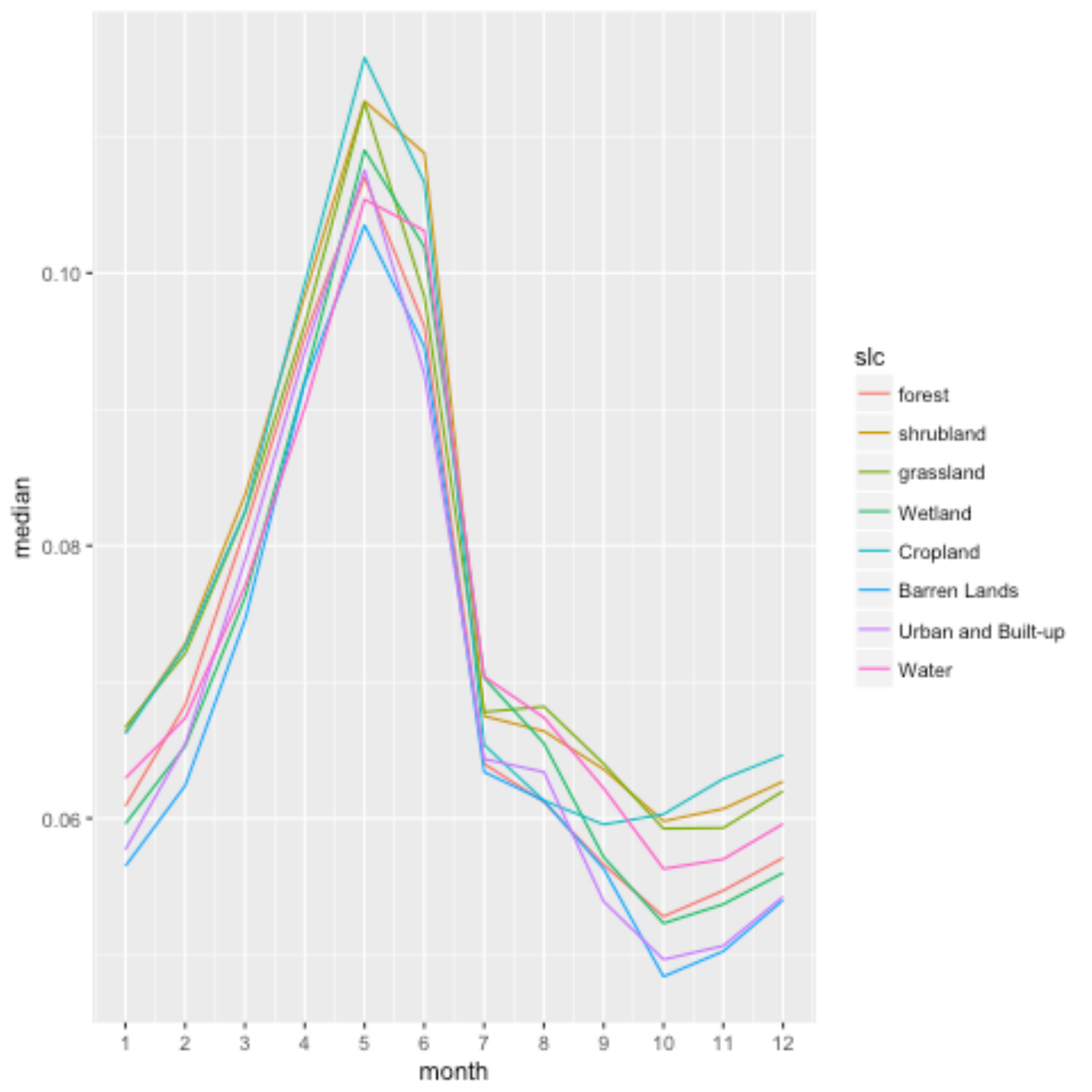}
        \caption{median by month and land cover type}
        \label{fig:redRef_med}
      \end{subfigure}
      \caption[Red reflectance]
      {Red reflectance}
      \label{fig:redRef}
    \end{figure}
	
	Figure \ref{fig:redRef} shows that red reflectance has a seasonal pattern which peaks in May and June. It has markedly a positively-skewed distribution. Green photosynthetically active pigments absorb most of the red and blue visible spectrum, however it seems only possible to distinguish grassland from the rest of the landcover types by its yearly red-reflectance median values.  Forest landcover can only be distinguished from some of the other landcovers, such as \emph{urban and built up}. 
	
	
	\begin{figure}[H]
      \begin{subfigure}{.5\textwidth}
        \centering
    	\includegraphics[width=0.8\textwidth]{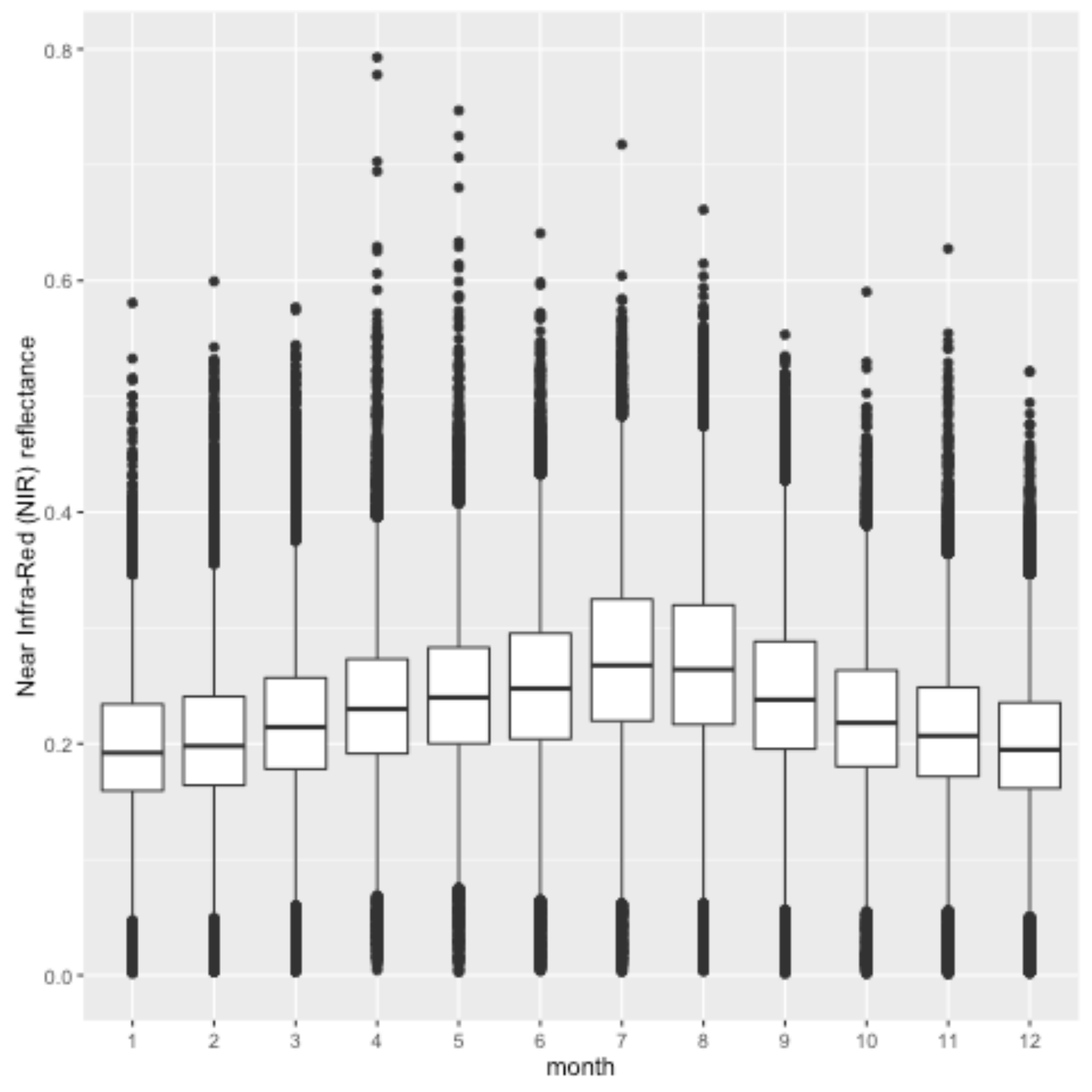} 
        \caption{box plot by month}
        \label{fig:NIRRef_box}
      \end{subfigure}%
      \begin{subfigure}{.5\textwidth}
        \centering
        \includegraphics[width=0.8\textwidth]{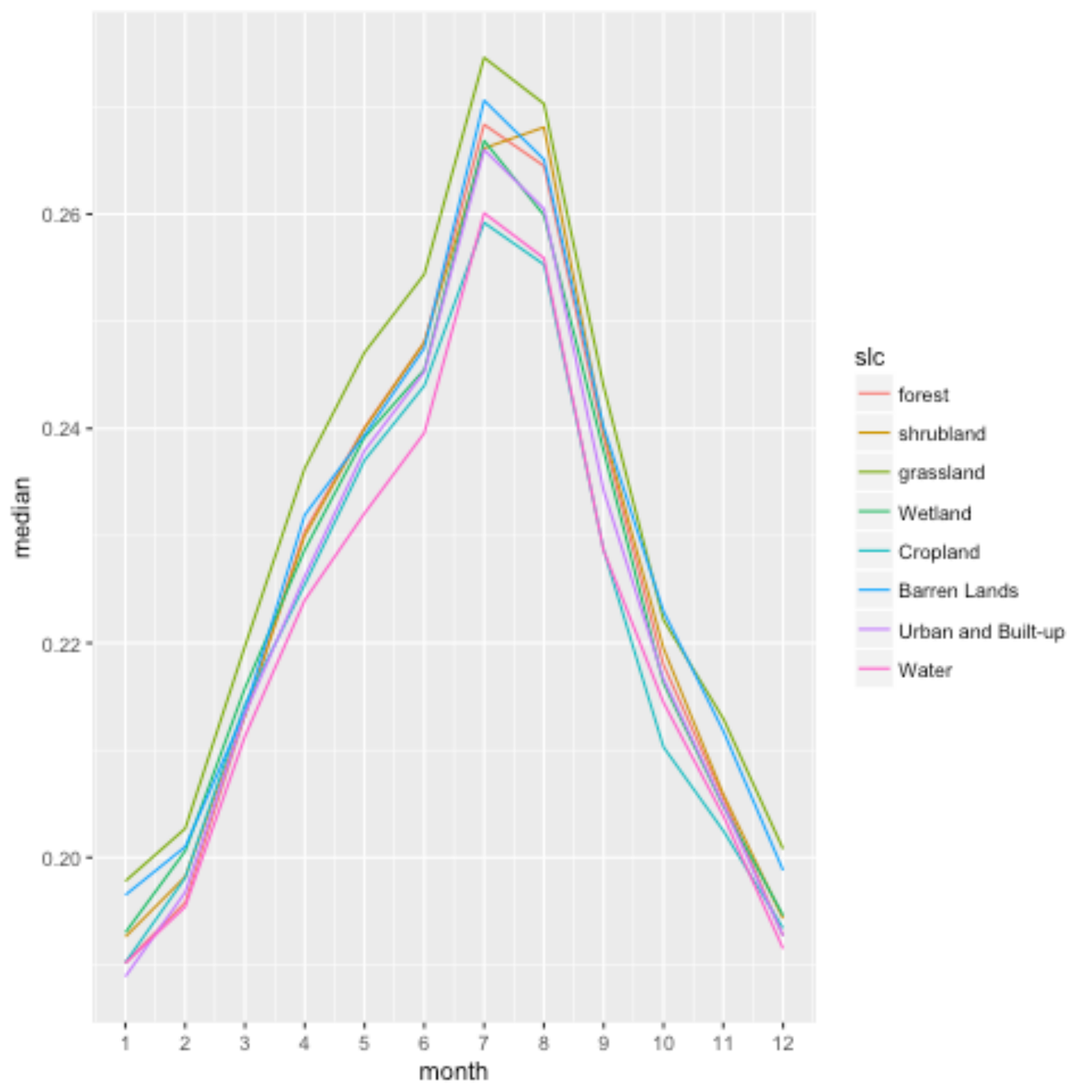}
        \caption{median by month and land cover type}
        \label{fig:NIRRef_med}
      \end{subfigure}
      \caption[Near Infra-Red (NIR) reflectance]
      {Near Infra-Red (NIR) reflectance}
      \label{fig:NIRRef}
    \end{figure}
	
	Figure \ref{fig:NIRRef} shows that NIR reflectance has a seasonal pattern which peaks a little later than red reflectance in July and August. It has a slightly positively-skewed distribution. Green photosynthetically active pigments reflect or transmit most of the NIR spectrum. The strong increase in NIR refelectance of cropland landcovers from August to December distinguishes it from other landcover NIR reflectance patterns. This is possibly due to spring/summer crops such as corn which reach its critical growing stages in the months of August and September.
	
	
	\begin{figure}[H]
      \begin{subfigure}{.5\textwidth}
        \centering
    	\includegraphics[width=0.8\textwidth]{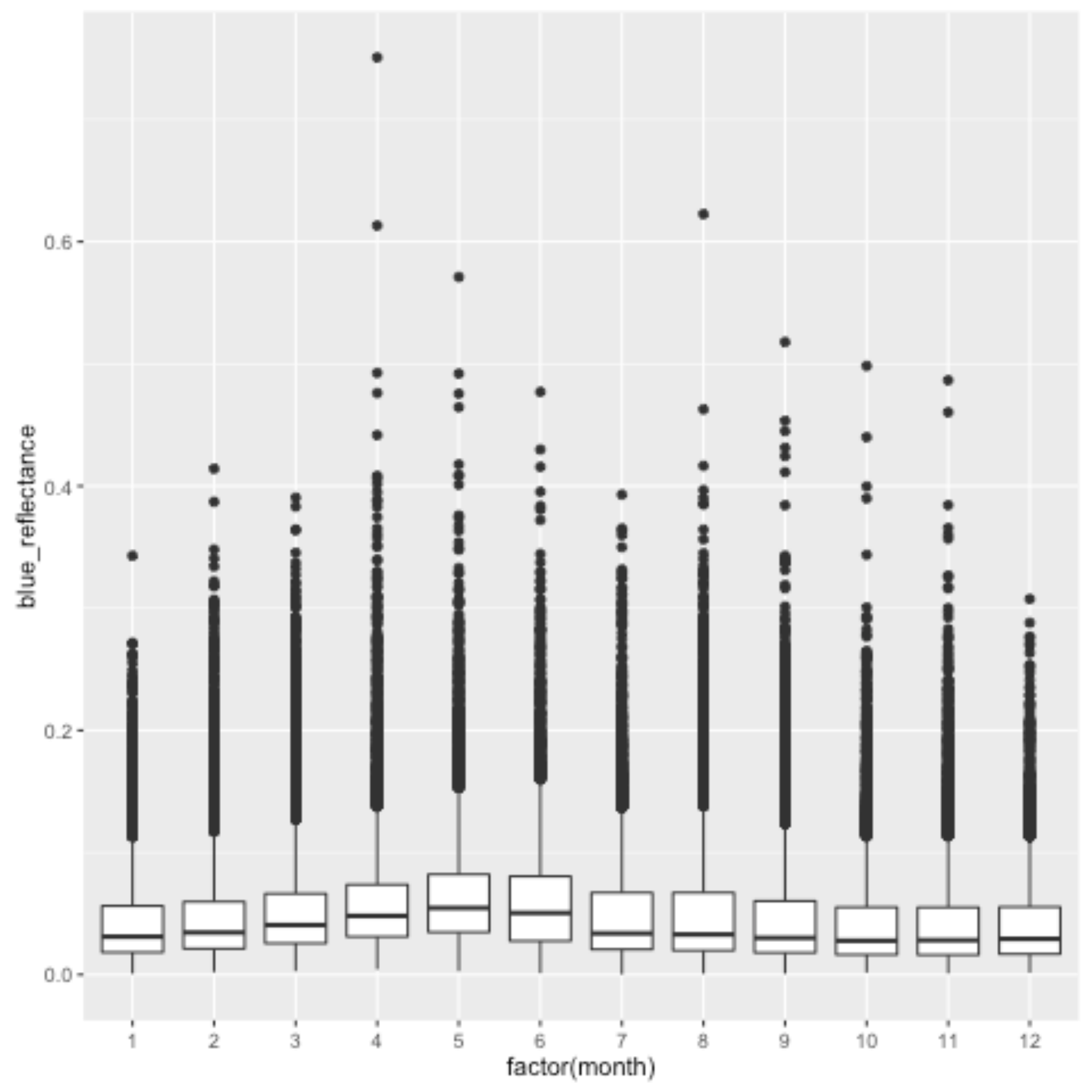} 
        \caption{box plot by month}
        \label{fig:blueRef_box}
      \end{subfigure}%
      \begin{subfigure}{.5\textwidth}
        \centering
        \includegraphics[width=0.8\textwidth]{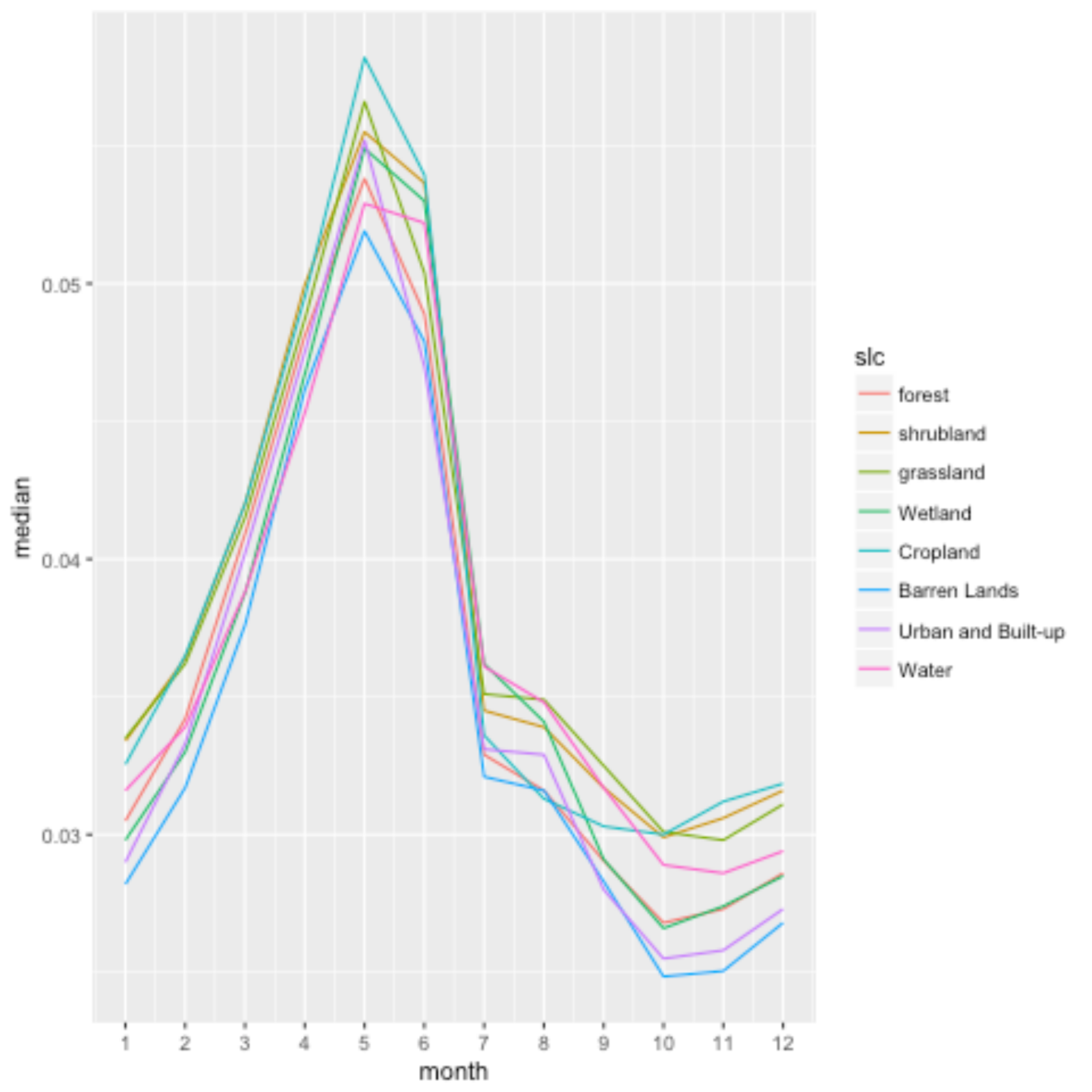}
        \caption{median by month and land cover type}
        \label{fig:blueRef_med}
      \end{subfigure}
      \caption[Blue reflectance]
      {Blue reflectance}
      \label{fig:blueRef}
    \end{figure}
	
	Figure \ref{fig:blueRef} shows that blue reflectance has a seasonal pattern similar to that of red reflectance peaking in May and June. Its distribution also has a markedly positive skew. 
	
	
	\begin{figure}[H]
      \begin{subfigure}{.5\textwidth}
        \centering
    	\includegraphics[width=0.8\textwidth]{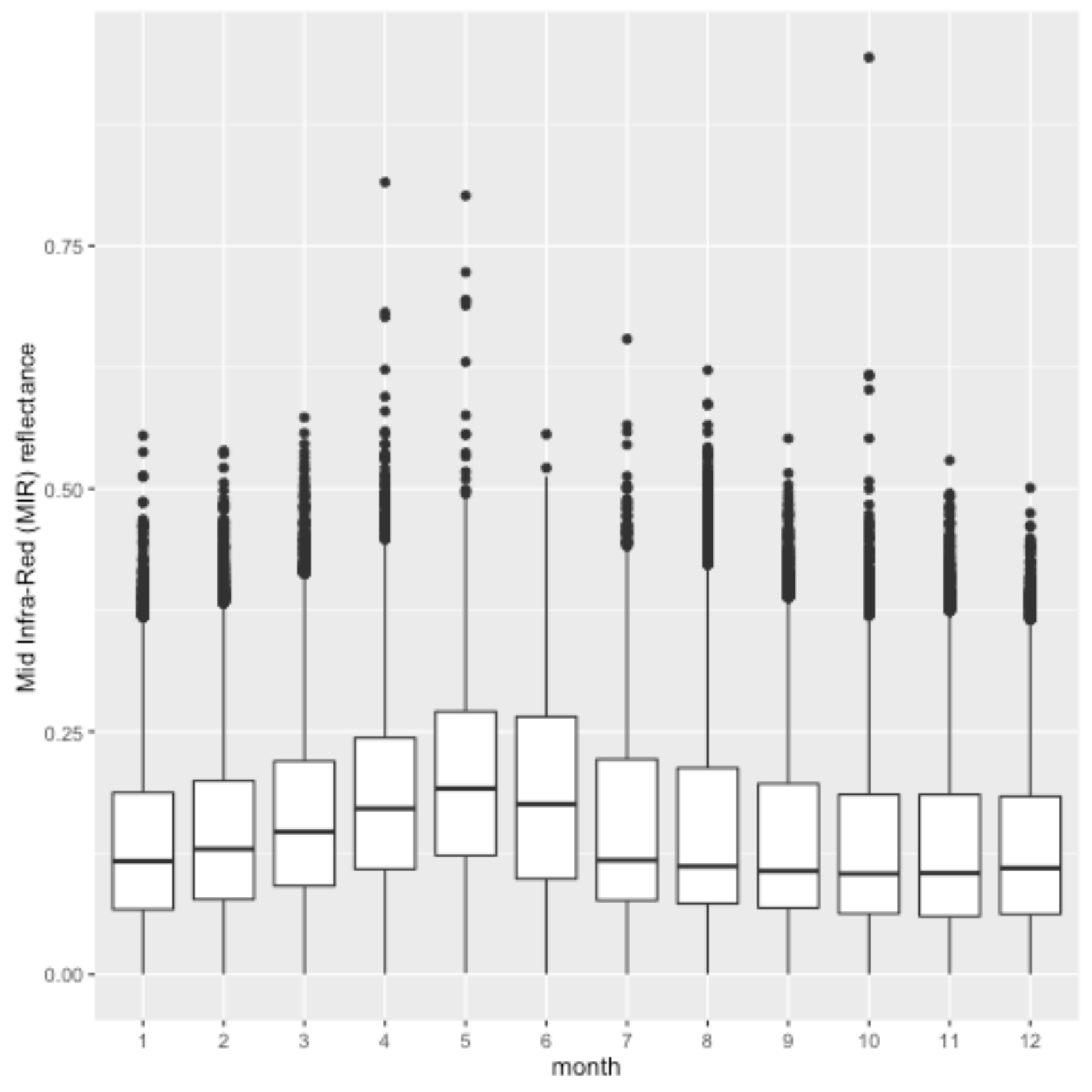} 
        \caption{box plot by month}
        \label{fig:MIRRef_box}
      \end{subfigure}%
      \begin{subfigure}{.5\textwidth}
        \centering
        \includegraphics[width=0.8\textwidth]{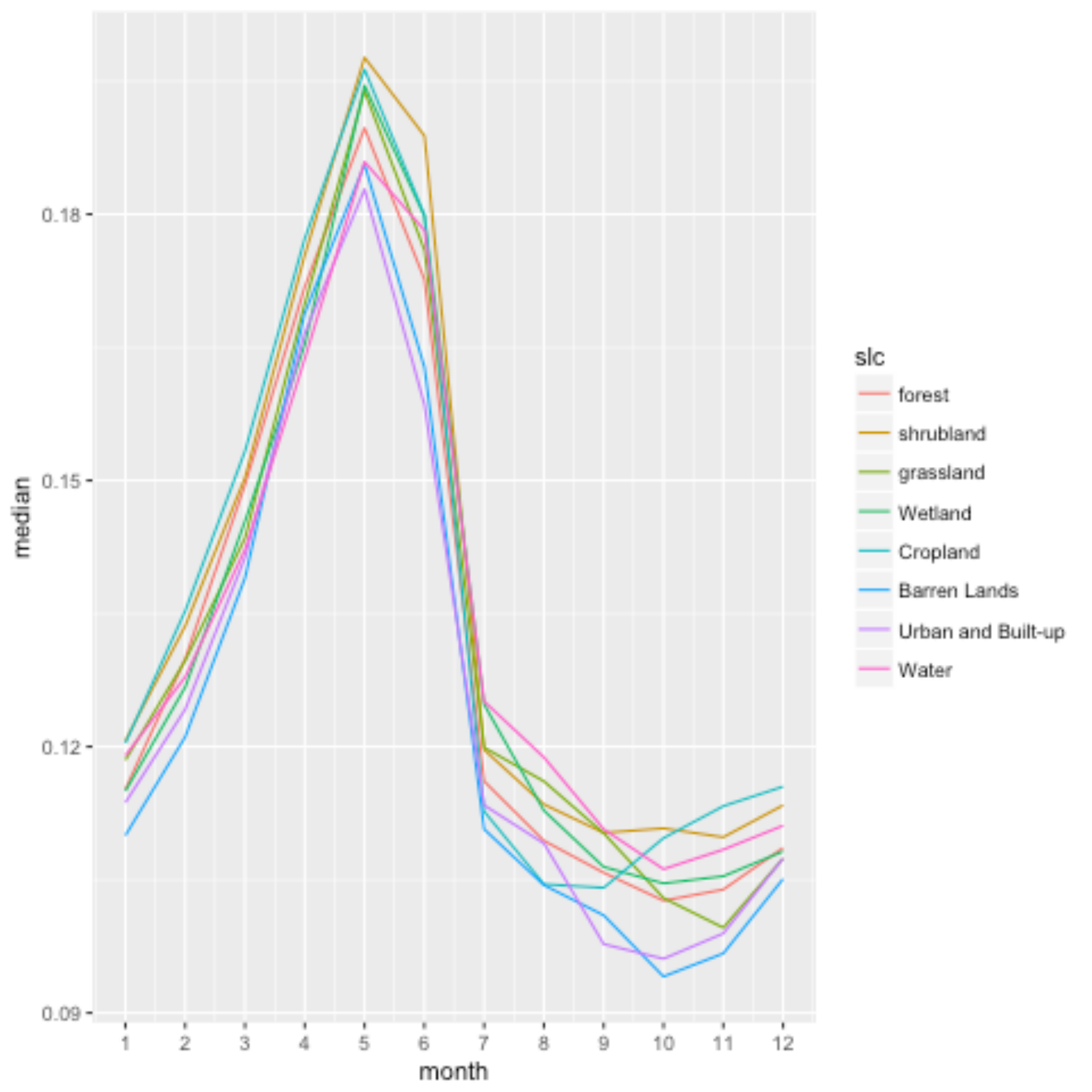}
        \caption{median by month and land cover type}
        \label{fig:MIRRef_med}
      \end{subfigure}
      \caption[Mid Infra-Red (MIR) reflectance]
      {Mid Infra-Red (MIR) reflectance}
      \label{fig:MIRRef}
    \end{figure}

	Figure \ref{fig:MIRRef} shows that MIR reflectance has a seasonal pattern that also peaks in May and June. It is slightly positively skewed similarly to NIR reflectance.
	
	In general, all the reflectance variables are positively skewed, meaning it might be advisable to try transformations of the reflectance variables when constructing the \emph{reflectance} model of the CMFDA algorithm.  By exploring the global (for all sampled pixels accross Mexico) median reflectance pattern as function of time it is not clear that the different landcovers can be distinguished. This is probably due to the fact that local factors, such as the local solar time at which observations were taken, are confounded with the landcover signal. This shows the importance of incorporating the sun-sensor geometry variables, which include information about the local solar time of measurements, into the \emph{reflectance} model. We did not explore these possibilities further but include the analysis to aid possible future improvements.

\subsection{Vegetation Indices}

	Vegetation indices measure the amount of vegetation activity at land surface. They are constructed so that reflectance signal from vegetation can be best distinguished from reflectance signal from other types of landcover. This is achieved by combining the reflectance of two or more wave bands, often red reflectance (0.6 - 0.7 micrometres) and NIR reflectance (0.7-1.1 micrometres).

	The theoretical basis for vegetation indices is derived from the spectral reflectance signatures of leaves. Reflectance in the visible spectrum is low since photosynthetically active pigments absorb most blue and red wavelengths. In contrast most of the near infra-red radiation (NIR) is reflected or transmitted. Accordingly, the contrast between red an NIR reflectance is a good proxy for vegetation amount with more contrast over targets with a higher amount of vegetation. 
	
	The NDVI is a function of the NIR and red reflectance designed to standardize values between -1 and 1. It is expressed as:
	
	\begin{align}
		I_{ND} = \frac{\rho_{2}-\rho_{1}}{\rho_{2}+\rho_{1}} 
	\end{align}	
	
	where,
	\begin{itemize}
		\item $I_{ND}$ is the NDVI,
		\item $\rho_{1}$ is red reflectance and 
		\item $\rho_{2}$ is NIR reflectance.
	\end{itemize}	
	
			Being a ratio, $I_{ND}$ can reduce several types of band-correlated noise from differnt sources: variations in direct/diffuse irradiance, presence of clouds and shadows, different sun and view angles, different topography and atmospheric conditions and instrument error. An important disadvantage of ratio-based indices is that their non-linearity leads to insensitivities to vegetation variation in certain cases. Additionally, it does not account for variation unrelated to vegetation amount such as additive atmospheric (path radiance) effects, canopy-background interactions and canopy bidirectional reflectance anisotropies. 
		
			The Enhanced Vegetation Index (EVI) has improved sensitivity in areas with high amounts of vegetation and better vegetation monitoring capacity in general. This is due to the decoupling of soil and atmospheric influences from the vegetation signal.  Atmospheric effects are reduced  by estimating the atmosphere influence level with a weighted difference in blue and red reflectances.  The effect of soil signal is accounted for through the estimation of a canopy background adjustment parameter. The EVI is expressed as: 

	 \begin{align}
	  I_{E}  =
	    \begin{cases}
	      G\frac{\rho_2-\rho_1}{\rho_2 + C_1 \rho_1 - C_2 \rho_3 + L}, &  \rho_3 < 0.2 \\
	      G\frac{\rho_2-\rho_1}{\rho_2 + 2.4 \rho_1 + L}, & \rho_3 \geq 0.2 
	    \end{cases}
	\end{align}	
		where NIR, 
		\begin{itemize}
			\item $I_{E}$ is the EVI,
			\item $\rho_{3}$ is blue reflectance,
			\item  $L$ is the canopy background adjustment parameter,
			\item $C_1$ and $C_2$ are the weights to estimate the atmospheric influence, 
			\item $G$ is a gain or scaling factor, and,  
			\item the coefficients adopted for the MODIS EVI algorithm are, $L=1$, $C_1=6$, $C_2=7.5$, and $G=2.5$.
		\end{itemize}
	
		The formula for $I_{E}$ is different where the blue refelectance, $\rho_3$, is greater than $0.2$ because these rare $\rho_3$ values, caused by bright targets such as heavy clouds, snow and ice, lead to extremely high $I_E$ values.

	In this work the NDVI and EVI variables are treated as additional spectral bands with which to distinguish between landcover types. This means they may also be used as response variables in the \emph{reflectance} model of the CMFDA algorithm. For this reason the distribution of the two vegetation indices are plotted. As with the reflectance variables, it is of interest to distinguish the difference in the vegetation index patterns related to the type of landcover. For this reason the median NDVI and EVI are plotted against the month of the year for the different types of landcover.

	
	\begin{figure}[H]
      \begin{subfigure}{.5\textwidth}
        \centering
    	\includegraphics[width=0.8\textwidth]{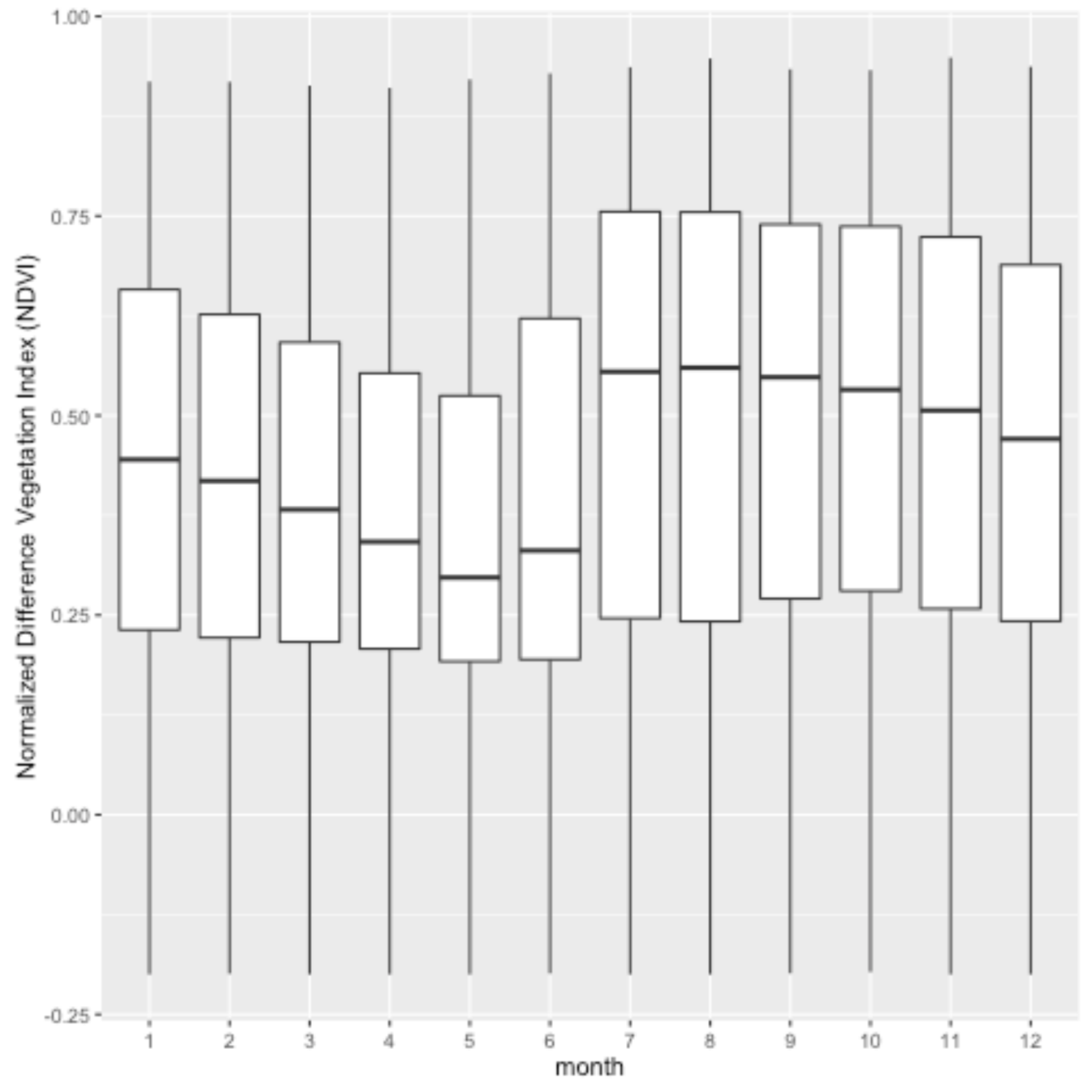} 
        \caption{box plot by month}
        \label{fig:NDVI_box}
      \end{subfigure}%
      \begin{subfigure}{.5\textwidth}
        \centering
        \includegraphics[width=0.8\textwidth]{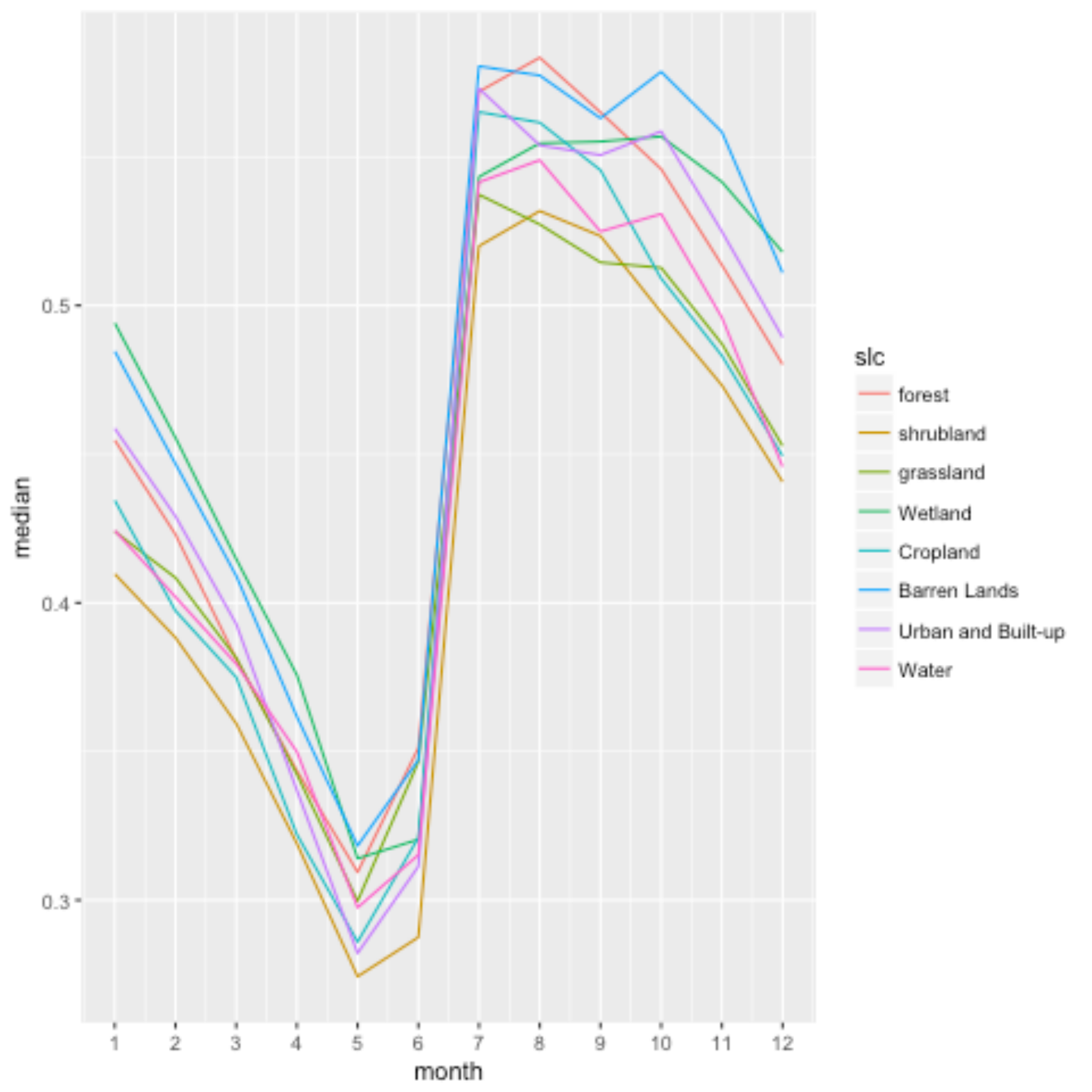}
        \caption{median by month and land cover type}
        \label{fig:NDVI_med}
      \end{subfigure}
      \caption[Normalized Difference Vegetation Index (NDVI)]
      {Normalized Difference Vegetation Index (NDVI)}
      \label{fig:NDVI}
    \end{figure}
	
		The shift in peak between red and NIR reflectance means that NDVI bottoms around May and June and jumps drastically to its peak in July and August, as is shown in \ref{fig:NDVI}.


	\begin{figure}[H]
      \begin{subfigure}{.5\textwidth}
        \centering
    	\includegraphics[width=0.8\textwidth]{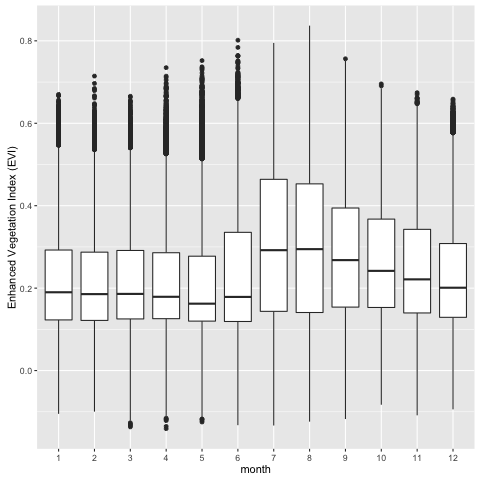} 
        \caption{box plot by month}
        \label{fig:EVI_box}
      \end{subfigure}%
      \begin{subfigure}{.5\textwidth}
        \centering
        \includegraphics[width=0.8\textwidth]{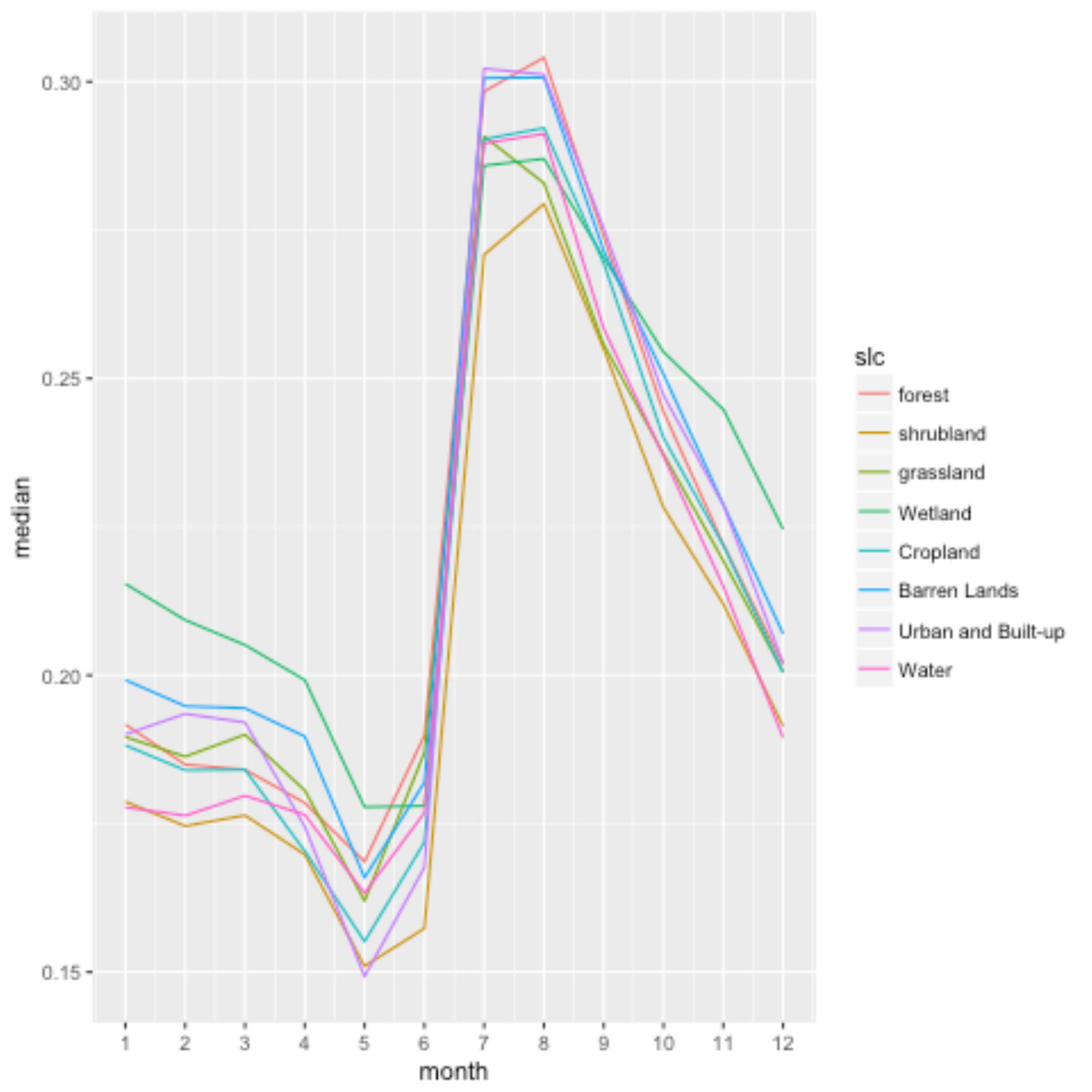}
        \caption{median by month and land cover type}
        \label{fig:EVI_med}
      \end{subfigure}
      \caption[Enhanced Vegetation Index (EVI)]
      {Enhanced Vegetation Index (EVI)}
      \label{fig:EVI}
    \end{figure}
	
	Figure \ref{fig:EVI} shows that just as in the case of NDVI, the EVI bottoms around May and June and jumps drastically to its peak in July and August.

\chapter{Methodology} \label{ch:meth} 

		To detect deforestation we follow the continuous monitoring of forest disturbance algorithm  (CMFDA) described in \cite{CMFDA} which consists of two basic steps involving the estimation of a \emph{reflectance model} and a \emph{deforestation model}:
		
		\begin{enumerate}[1.]
			\item \textbf{Reflectance model}: Model the surface reflectance $y$ as a function of the day of the year:
			
				\begin{align} \label{modelRef}
					y_{b,p,t} = g_{b,p}(d) + \epsilon_{b,p,t},
				\end{align}	
				
				where $b \in \{1,2,..,B\}$, $p \in \mathcal{S}$,$t \in \mathcal{T}$ and $d \in \{1,2,...,366\}$ refer to the band, pixel, date and day of the year respectively.

			\item \textbf{Deforestation model}: Then model the deforestation event $z_{p,t} \in \{0,1\}$ at pixel $p$ and date $t$ as a function of the prediction errors $\hat{\epsilon}_{b,p,t}=y_{b,p,t} - \hat{g}_{b,p}(d)$ :
			
				\begin{align} \label{modelDefo}
					z_{p,t} = h_L( \{\hat{\epsilon}_{b,p,s}\}_{b \in \{1,...,B\}, s \in \{t, t-1,...,t-C+1\} } ) + e 
				\end{align}	
				
				where $h_L(\cdot) \in \{0,1\}$ is a thresholding function which basically checks if the prediction errors  $\hat{\epsilon}_{b,p,t}$, for the last $C$ consecutive observations is larger in magnitude than a (possibly multivariate) thershold $L$. The prediction errors for either, a certain band $b \in \{1,...,B\}$ or a certain subset of bands $b \in \mathcal{B} \subset \{1,...,B\}$ are checked against a threshold $L$. Alternatively, a function of the prediction errors for different bands $I(\{\hat{\epsilon}_{b,p,t}\}_{b \in \mathcal{B} \subset \{1,...,B\}})$ is checked against a threshold $L$.
				
		\end{enumerate}	 
		
		Recalling section \ref{ss:reflectance}, surface reflectance is the amount of light reflected by a surface. It is a ratio of surface radiance to surface irradiance. Top-of-atmosphere reflectance is the amount of light that reaches the top of earth's atmosphere. The CMFDA algorithm was designed for 30m resolution \textbf{top-of-atmosphere} reflectance data from the Landsat satellite. We adapt the method to work with the 1km resolution \textbf{surface} reflectance data from the modis sensor aboard the Terra satellite. 
		
		In section \ref{s:CMFDA} we give a detailed description of the CMFDA algorithm. In section \ref{s:adapt} we describe the changes made to the algorithm to adapt it to the modis reflectance data, leaving the specific form of the thresholding function $h_L$ that should be used to be determined in sections \ref{s:studyArea}-\ref{s:mahl}. In section \ref{s:studyArea} we firstly choose the nine 25 by 25km sites that will be used to train and test our adapted CMFDA algorithm. We then identify two important types of deforestation, \emph{forest to water} and \emph{forest to urban or cropland}, and determine which of the bands in the modis data set can be used to identify each type i.e. we identify the subset of bands $\mathcal{B} \subset \{1,...,B\}$ on which our function $h_L$ should depend. 
		
		In section \ref{s:gridSearch} we train sites separately according to the type of deforestation that occurred at each. This means that we train functions $h_{L_i}(\{\hat{\epsilon}_{b,p,t}\}_{b=i, s \in \{t, t-1,...,t-C+1\} })$ for $i \in \mathcal{B}=\{2,8\}$. Bands two and eight correspond to NIR reflectance and the NDVI. We also obtain our first performance results for the algorithm. These results will serve mostly as a benchmark since the algorithm is still not implementable given that we use prior knowledge of the type of deforestation that occurred to choose $i \in \mathcal{B}=\{2,8\}$.  
		
		In section \ref{s:simAn} we train a thresholding function $h_L(\cdot)$ which depends on the prediction errors of two surface reflectance bands, NIR reflectance and NDVI, and where $L$ is a multivariate threshold, i.e. $h_L := h_L( \{\hat{\epsilon}_{b,p,t}\}_{b \in \mathcal{B} \subset \{1,...,B\}, s \in \{t, t-1,...,t-C+1\} } )$. In section \ref{s:mahl} we train a thresholding function $h_L(\cdot)$ which depends on an index calculated with the prediction errors of the NIR reflectance and NDVI bands, i.e. $h_L := h_L(I(\{\hat{\epsilon}_{b,p,t}\}_{b \in \mathcal{B} \subset \{1,...,B\}}))$. 

		The approach followed in \cite{CMFDA} is a pixel-by-pixel modeling approach where the time dependence of the expected value of reflectance, $\mathbb{E}[y]$ ,  is built in to the model through the function $g_{b,p}(d)$, which as we will see in section \ref{s:CMFDA} is a linear combination of harmonic functions of the day of the year $d$. The possible spatial dependence between the functions for different pixels, $g_{b,p}(d)$ and $g_{b,s}(d)$ with $p \neq s$, is not explored in that work. Additionally, no specific assumptions about the distribution of the errors $\epsilon$ are made. In this work we assume, very generally, that $\epsilon$ has a distribution $D$ such that $\mathbb{E}[D]=0$ and $\mathbb{V}[D]=\sigma^2$. In chapter \ref{ch:modhet} we will explore the spatial dependence of $g_{b,p}(d)$ accross pixels and the spatial and time dependence of the variance of errors $\sigma^2$.

	\section{CMFDA algorithm} \label{s:CMFDA}

		The CMFDA described in \cite{CMFDA}  uses level 1T (T stands for terrain corrected) satellite images at 30 metres resolution from the Landsat Enhanced Thematic Mapper Plus (ETM+) and Landsat Thematic Mapper (TM) sensors onboard the Landsat 7 and Landsat 5, respectively. The dataset contains 6 optical bands of top-of-atmosphere (TOA) reflectance data (bands 1,2,3,4,5 and 7) and band 6 which is Brightness Temperature (BT). This data is available at a maximum temporal frequency of 8 days (when both the Landsat 5 and 7 are used). As table \ref{tab:sensorWaves} shows, the wavelength corresponding to each band is slightly different for each sensor (ETM+ and TM).

	 \begin{table}[H] 
	\begin{center}
	 \begin{tabular}{|l|l|l|}
		 \hline
	 \textbf{Band} & \textbf{TM wavelength ($\mu$ m)}  & \textbf{ETM+ wavelength ($\mu$ m)} \\ 
	 \hline                                                                                                                                    
	 1 & 0.45-0.52    &  0.45-0.515   \\
	\hline
	 2 & 0.52-0.60    &  0.525-0.605  \\
	\hline
	 3 & 0.63-0.69    &  0.63-0.69    \\
	\hline
	 4 & 0.76-0.90    &  0.75-0.90    \\
	\hline
	 5 & 1.55-1.75    &  1.55-1.75    \\
	\hline
	 6 & 10.40-12.50  & 10.40-12.50   \\
	\hline
	 7 & 2.08-2.35    &  2.09-2.35    \\
	\hline                                                                                       
	 \end{tabular}\\
	 \caption{Wavelengths for bands 1-7 on TM and ETM+ sensors}
	 \label{tab:sensorWaves}
	\end{center}
	 \end{table}

	\subsection{Development}
	
	The CMFDA methodology was developed using information from a 60 by 60 km site located in the Savannah River basin in Georgia and South Carolina. It consists of 7 steps. Steps i-v are steps necessary to estimate the \emph{reflection model} from \ref{modelRef} while steps vi-vii are geared toward estimating the \emph{deforestation model} from \ref{modelDefo}:

	\begin{enumerate}

		\item \textbf{Cloud detection A.} This includes single date cloud, cloud shadow and snow masking using the Fmask algorithm described in \cite{FMask} on top-of-atmosphere reflectance data. This is done for each available date in the 2001-2003 period. All pixels in an image are processed concurrently as the algorithm works on a \emph{scene} by \emph{scene} basis: whether a pixel is flagged as cloud, cloud shadow or snow depends on operations carried out over the whole image or scene. 

		\item \textbf{Cloud detection B.} This entails multi date cloud, cloud shadow and snow masking: the Fourier model below is fitted on \textbf{top-of-atmosphere} reflectance data using robust iteratively reweighted least squares (RIRLS) to reduce the influence of outliers caused by cloud, snow or shadow presence.   The model is fitted for all 6 optical bands: $b \in \{1,2,3,4,5,7\}$ as in \ref{tab:sensorWaves}. The model is fitted on a pixel-by-pixel basis using all available observations from three year period (2001-2003). Outliers are then identified by comparing observed with model predicted values and flagged as \emph{cloud, snow or shadow} or \emph{clear sky} pixels. The thresholds for outlier identification are not specified in \cite{CMFDA} nor is it specified how many or which time-series, corresponding to the various bands,  must include outliers for a given pixel-date for that pixel-date to be classified as \emph{cloud, snow or shadow}.

			\begin{align} \label{fourierModel}
						g_{b,p}(d) = \alpha_{0,b,p} &+ \sum_{i=1}^N\Bigg(\alpha_{i,b,p} \cos\Big(\frac{2\pi x}{iT}\Big) + \beta_{i,b,p} \sin\Big(\frac{2\pi x}{iT}\Big) \Bigg)\\ &+ \alpha_{N+1,b,p} \cos\Big(\frac{2\pi x}{0.5T}\Big) + \beta_{N+1,b,p} \sin\Big(\frac{2\pi x}{0.5T}\Big)
			\end{align}

			Where:
			\begin{itemize}
				\item $d \in \{1,2,...,366\}$ is the day of the year,
				\item $N$ is the largest inter-annual change that is taken into account, $N$ was was chosen to be 2, 
				\item $T$ is the number of days in the year, taken to be 366, 
				\item Coefficients with $i=2,3,...,N$ represent variation that occurs on cycles lasting $i$-years and mostly result from land cover change,
				\item $\alpha_{i,b}$ and $\beta_{i,b}$ for $i \in \{0,...,N\}$ are coefficients corresponding to the surface reflectance model for band $b$,
				\item Coefficients $i=N+1$ capture bimodal variations which mostly occur in agricultural areas due to double-cropping,
				\item Coefficients $i = 1$ capture BRDF and phenology effects which are what we are ultimately interested in, and,
				\item Coefficients $\alpha_{0,b}$ represents mean overall surface reflectance of band $b$. 
			\end{itemize}
			
	At the end of the two filtering steps, i and ii, it is possible to identify the set $\mathcal{C}_t$ of pixels wich have clear observations at date $t$, for all dates $t$ in training window $\mathcal{T}_j$, and prediction period $\mathcal{P}_j$. In this case $j \in \{1\}$ since there is only one training window, 2001-2002, and one prediction window, 2003. It is also possible to identify the set of dates $\mathcal{T}_{j,p} \subset \mathcal{T}_j$ and $\mathcal{P}_{j,p} \subset \mathcal{P}_j$ for which there are clear reflectance observations for pixel $p$ and training or prediction window $j$.

	\item \textbf{Atmosphere correction.} Apply atmosphere correction to top-of-atmosphere reflectance data to get surface reflectance data. The set of clear surface reflectance observations can now be constructed: for training windows $Y_{j,p}^\mathcal{T}=\{y_{b,p,t}\}_{b \in \{1,2,3,4,5,7\}, t \in \mathcal{T}_{j,p}}$; and for prediction windows $Y_{j,p}^\mathcal{P}=\{y_{b,p,t}\}_{b \in \{1,2,3,4,5,7\}, t \in \mathcal{P}_{j,p}}$; for every period $j$ and every pixel $p$ in Savannah river basin site $\mathcal{S}$.

	\item \textbf{Fourier model.} For every pixel $p \in \mathcal{S}$ use observations $Y_{j,p}^\mathcal{T}$ to fit same Fourier model as above using ordinary least squares (OLS). The model is fitted for all 6 bands, this time on \textbf{surface} reflectance data. Only data from a two-year 2001-2002 period is used so that 2003 can be used for prediction and training of deforestation thresholds. 

	\item \textbf{Forest mask.} Define stable forest mask $\mathcal{F}$ of pixels using model coefficients and domain knowledge:
		\begin{enumerate}[a.]
			\item Forests are observed to have high NDVI values 
			\item and low SWIR reflectance (restrictions on $\alpha_{0,b}$ values)
			\item No landcover change in training period (low values for  $\alpha_{i,b}$ and $\beta_{i,b}$ for $i=2,...,N$)
		\end{enumerate}

	\item \textbf{Prediction.} Predict values for 2003 for days when clear observations are available for forest pixels $p \in \mathcal{F}$. Use only annual and bi-annual  seasonality coefficients: i.e. force $\alpha_{i,b}=0$ and $\beta_{b,i}=0$ for $i =2,...,N$  (they should already be low due to forest mask requirement c). The linear estimate function with these forced coefficients is $g_{b,p}^0$.  Figure \ref{fig:trainScheme_CMFDA} illustrates the training and prediction windows. 
	
	\begin{figure}[H]
	  \centering
	  \includegraphics[width=.8\textwidth]{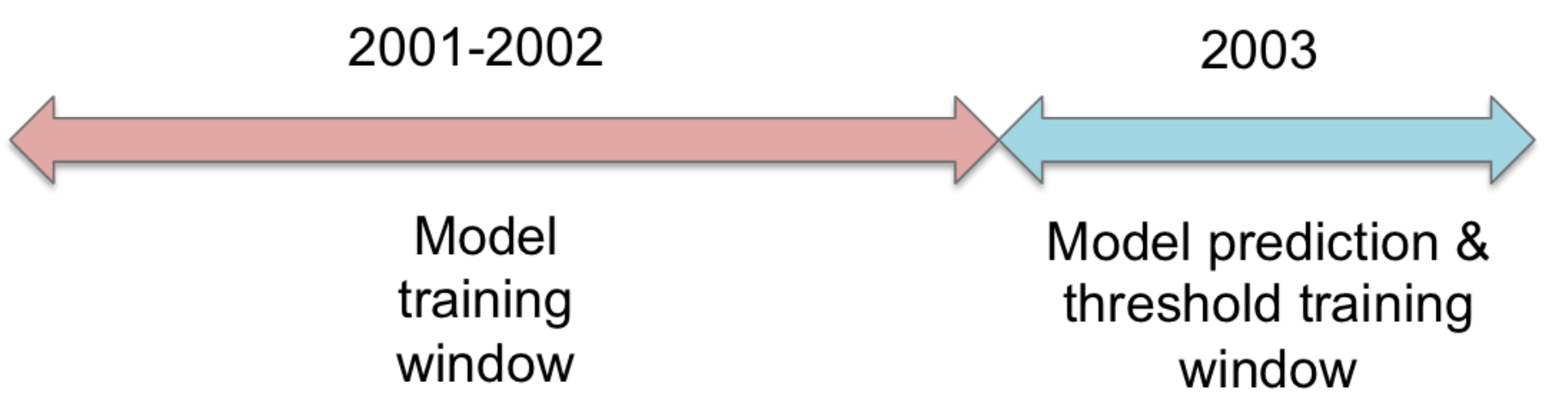} 
	  \caption[Model training, model prediction and threshold training scheme for CMFDA]
	  {Model training, model prediction and threshold training scheme for CMFDA}
	  \label{fig:trainScheme_CMFDA}
	\end{figure}
	
	The set of predicted surface reflectance observations can now be constructed: for prediction windows  $\hat{Y}_{j,p}^\mathcal{P}=\{\hat{y}_{b,p,t}=\hat{g}_{b,p}^0(d(t))\}_{b \in \{1,2,3,4,5,7\}, t \in \mathcal{P}_{j,p}}$; for every period $j$ and every pixel $p$ in forest mask $\mathcal{F} \subset \mathcal{S}$.

	\item \textbf{Change detection.}  The basic idea is to compare predicted and real values and when the difference is larger than a certain threshold $L$ flag the pixel as \emph{deforested}. However questions remain about:
	
	\begin{enumerate}
		\item Which band to use? Should all 6 optical bands be used concurrently in multivariate approach? If so, do all bands used need to show a large difference between the predicted and real values or just one? Should a \emph{summary} index be constructed to measure this difference?
		\item Should a single date (single-date, $C=1$) or several dates (multiple-date, $C>1$) be used when comparing real and predicted values?
	\end{enumerate}	
	
	In \cite{CMFDA} the index approach is used. After considering several indices they find the one with the best performance is the following \emph{forest} index:
	
	\begin{align}
		I_f = B - (G+W)
	\end{align}	
	
	Where, $B$, $G$, and $W$ are brightness, greenness, and wetness indices constructed from the six Landsat optical bands using the \emph{Tasseled Cap Transformation} (\citealt{Crist1} and \citealt{Crist2}). Both a single-date and multiple-date approach were tried. After training the single-date approach a threshold $L$ of 0.18 was found to optimize the performance measure. Training of the multiple-date approach, with threshold and number of consecutive violations of threshold ($C \in \{2,3,4,5\}$ were tried) as training parameters, yielded an optimal threshold of $L=0.12$ for $C=3$ consecutive violations. 
	
	Training of thresholds was done using deforestation data obtained by \emph{visual} inspection of Savannah river basin site during 2003. This was carried out by visually analysing landsat images at 30m resolution and with the aid of higher resolution images from Google Earth. Optimal thresholds were determined using a training-data hold-out sample scheme where the optimal threshold and consecutive number of violations were estimated using training data and the final performance measure evaluated on the hold-out sample. The performance measure used is the so-called \emph{producer's accuracy} ($a_p$) and \emph{user's accuracy} ($a_u$). The optimal threshold was defined as the threshold where producer's and user's accuracy are equal. Producer's and user's accuracy are defined with respect to the confusion matrix of table \ref{tab_confMat}. 
	
	\begin{table}[H]
		\begin{center}
	\begin{tabular}{cccc}
	predicted & Observed & & Total \\
	\hline
	  & 0 & 1 & \\
	\hline
	\hline
	0 & U & T & $W$ \\
	\hline
	1 & V & S & $R$ \\
	\hline
	Total & $N_0$ & $N_1$ & N \\
	\end{tabular}
	\caption{Confusion matrix showing observed and predicted deforestation (1 = deforestation).} 
	\label{tab_confMat}
	\end{center}
	\end{table}
	
	\begin{align}
		a_p &= \frac{S}{N_1} \\
		a_u &= \frac{S}{R}
	\end{align}	
	
	$N_1$ is fixed but as threshold goes up, $S$ and $R$, the number of deforested pixels correctly detected and the number of pixels flagged as deforested, respectively, both go down. This means that $a_p$, the producer's accuracy is a strictly decreasing function of the threshold. We can also argue that, in general, $a_u$, the user's accuracy will be an increasing function. First of all when the threshold is zero, $R=N$ and $R \geq S=T$, and when the threshold is really large $R=S=0$. This means that on average, per unit increase in threshold, $S$ decreases at a slower rate than $R$ so, on average $a_u$ increases. Additionally, if the index $I$ is any good at detecting deforestation, then as the difference between predicted and real $I$ increases the more likely a deforestation event occurred: setting increasing thresholds means we are more sure of our deforestation predictions and so $a_u$ goes up. \emph{Temporal accuracy}, the degree to which a deforestation event is detected at the time it actually occurred, was also evaluated in \cite{CMFDA}, although not optimized.

	\end{enumerate}
	
	\subsection{Implementation}
	
	Although it is not clear in \cite{CMFDA} exactly how CMFDA should be implemented going forward, it seems there are two possiblities:
	
	\begin{enumerate}
		\item \textbf{Sporadic retraining of Fourier model.} The Fourier model is only sporadically (every few years) retrained. New observations are only filtered for cloud, shadow or snow presence using the Fmask algorithm. Every time a new \emph{clear} observation becomes available, the predicted and real \emph{forest} index $I_f$ is computed, and the difference is checked against the threshold to see if a deforestation flag is triggered (either with single-date or multiple-date versions). The forest mask is constructed using the previous forest mask, which was constructed in last model training period, with pixels detected as having been deforested since omitted from mask. 
		\item \textbf{Online retraining of Fourier model}. Every time a new clear reflectance observation becomes available the model is re-trained using the last 2 years of clear observations (excluding new observation). All observations can be filtered for cloud, shadow or snow presence using, firstly, the Fmask algorithm and, secondly, outlier analysis based on robustley fitted Fourier model on top-of-atmosphere reflectance data. Filtered surface reflectance data can then be used to retrain model using OLS. A new forest mask can be constructed using retrained coefficients and checked for consistency against previous forest mask and pixels flagged as deforested since. Predictions are then made for date of new observation using retrained model, the real and predicted index $I$ computed, and the difference checked against the threshold. 
	\end{enumerate}	
		
	\section{Adaptation of algorithm} \label{s:adapt}
	
	The CMFDA algorithm was developed with and for 30m resolution Landsat satellite images. In this work we work with 1km resolution Terra satellite images. There are three main differences between the Landsat images used in \cite{CMFDA} and the Terra 13A2 data: the resolution is 30m for the former and 1km for the latter; the Landsat dataset used in \cite{CMFDA} consists of top-of-atmosphere reflectance whereas the 13A2 Terra dataset consists of surface reflectance; and, the spectral bands available are different in each dataset. In this section we describe the changes we made to the algorithm to adapt it to this new dataset. In essence these changes consist of a simplification of the algorithm.
	
	\subsection{Development}
	
	We developed a simplified version of CMFDA. It consists of 5 steps. Steps i-iv are steps necessary to estimate the \emph{reflection model} from \ref{modelRef} while steps iv-v are geared toward estimating the \emph{deforestation model} from \ref{modelDefo} (step iv involves both models):
	
		\begin{enumerate}

			\item \textbf{Cloud detection.} Instead of having two cloud, shadow and snow filtering steps we simply used the modis flag \emph{pixel reliability} to exclude outliers (we kept \emph{good} and \emph{marginal} data an excluded \emph{snow/icy} and \emph{cloudy} data). It is now possible to identify the set $\mathcal{C}_t$ of pixels wich have clear observations at date $t$, for all dates $t$ in training window $\mathcal{T}_j$, and prediction period $\mathcal{P}_j$. In this case $j \in \{1,2,...,5\}$ since there are five training windows (2003-2004, 2004-2005, 2005-2006, 2006-2007 and 2007-2008), and five prediction windows (2005, 2006, 2007, 2008 and 2009). It is also possible to identify the set of dates $\mathcal{T}_{j,p} \subset \mathcal{T}_j$ and $\mathcal{P}_{j,p} \subset \mathcal{P}_j$ for which there are clear reflectance observations for pixel $p$ and training or prediction window $j$.
			
			\item \textbf{Atmosphere correction.} Modis data is already surface reflectance as atmosphere correction has already been applied. The set of clear surface reflectance observations can now be constructed: for training windows $Y_{j,p}^\mathcal{T}=\{y_{b,p,t}\}_{b \in \{1,2,3,7,8,9\}, t \in \mathcal{T}_{j,p}}$; and for prediction windows $Y_{j,p}^\mathcal{P}=\{y_{b,p,t}\}_{b \in \{1,2,3,7,8,9\}, t \in \mathcal{P}_{j,p}}$; for every period $j$ and every pixel $p$.

			\item \textbf{Forest mask.} Instead of using the model itself to identify a forest mask $\mathcal{F}$, we simply used the landcover classification for  2005 and 2010 and assumed pixels classified as forest for both years were forest throughout. 

			\item \textbf{Fourier model and prediction.} For every pixel $p \in \mathcal{F}$ the model is fitted for all six available bands using observations $Y_{j,p}^\mathcal{T}$. The bands $\{1,2,3,7,8,9\}$ correspond to red reflectance, NIR reflectance, blue reflectance, MIR reflectance, NDVI and EVI. Since 2005-2010 is the period for which we have deforestation information for Mexico this is will be our prediction window. However, we retrained the model every year using a two year training window so that the prediction window for every model is one year. We added a 120 day period to the prediction window to make sure that any deforestation that occurs during the year can be detected using the multiple-date algorithm where the maximum number of consecutive violations $C$ of threshold, which we tried was six ($6*16=96<120$). Figure \ref{fig:trainScheme_adapt} shows the overlapping training and prediction window scheme which was used. 
			
			\begin{figure}[H]
			  \centering
			  \includegraphics[width=.8\textwidth]{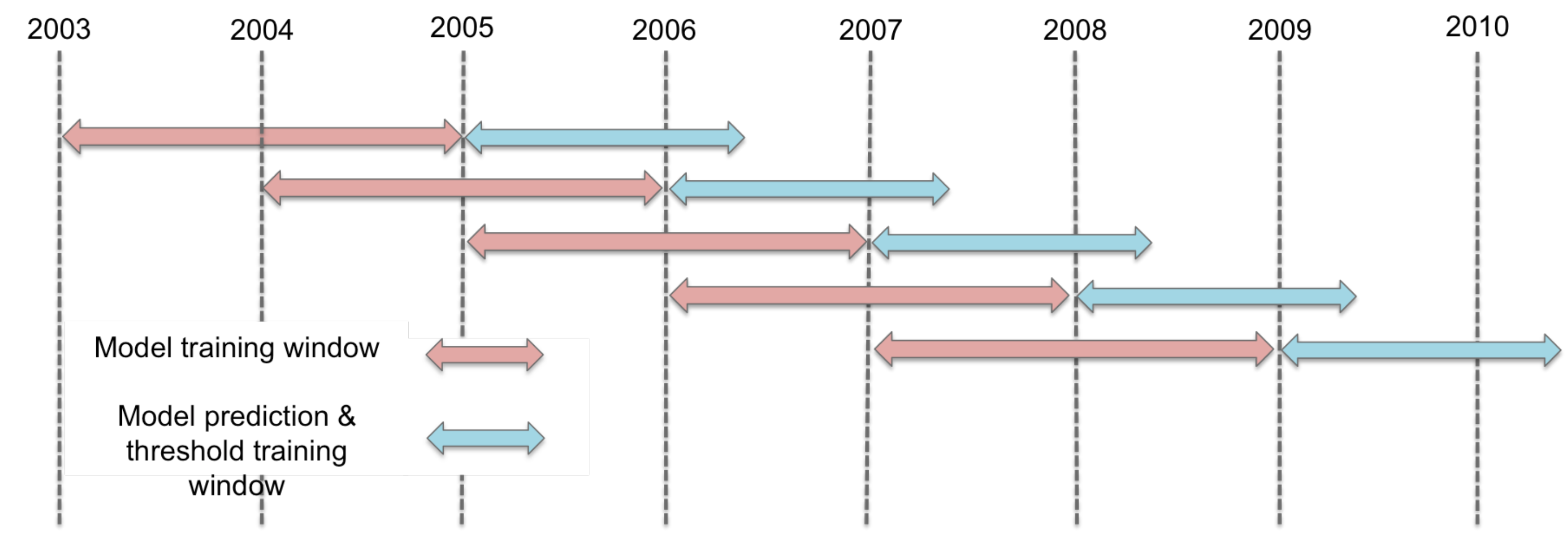} 
			  \caption[Model training, model prediction and threshold training scheme for adapted algorithm]
			  {Model training, model prediction and threshold training scheme for adapted algorithm}
			  \label{fig:trainScheme_adapt}
			\end{figure}
			
			We want to detect deforestation within one prediction window and not combine different prediction windows. If a deforestation event occurs toward the end of a given year and we used a one year prediction window we would need to combine two prediction windows to detect the deforestation, especially for multiple-date algorithms with a high number of consecutive violations of threshold. However this would mean that the second prediction window used a training window in which the deforestation event occurred which could affect model estimation adversely.

			\begin{figure}[H]
			  \centering
			  \includegraphics[width=.8\textwidth]{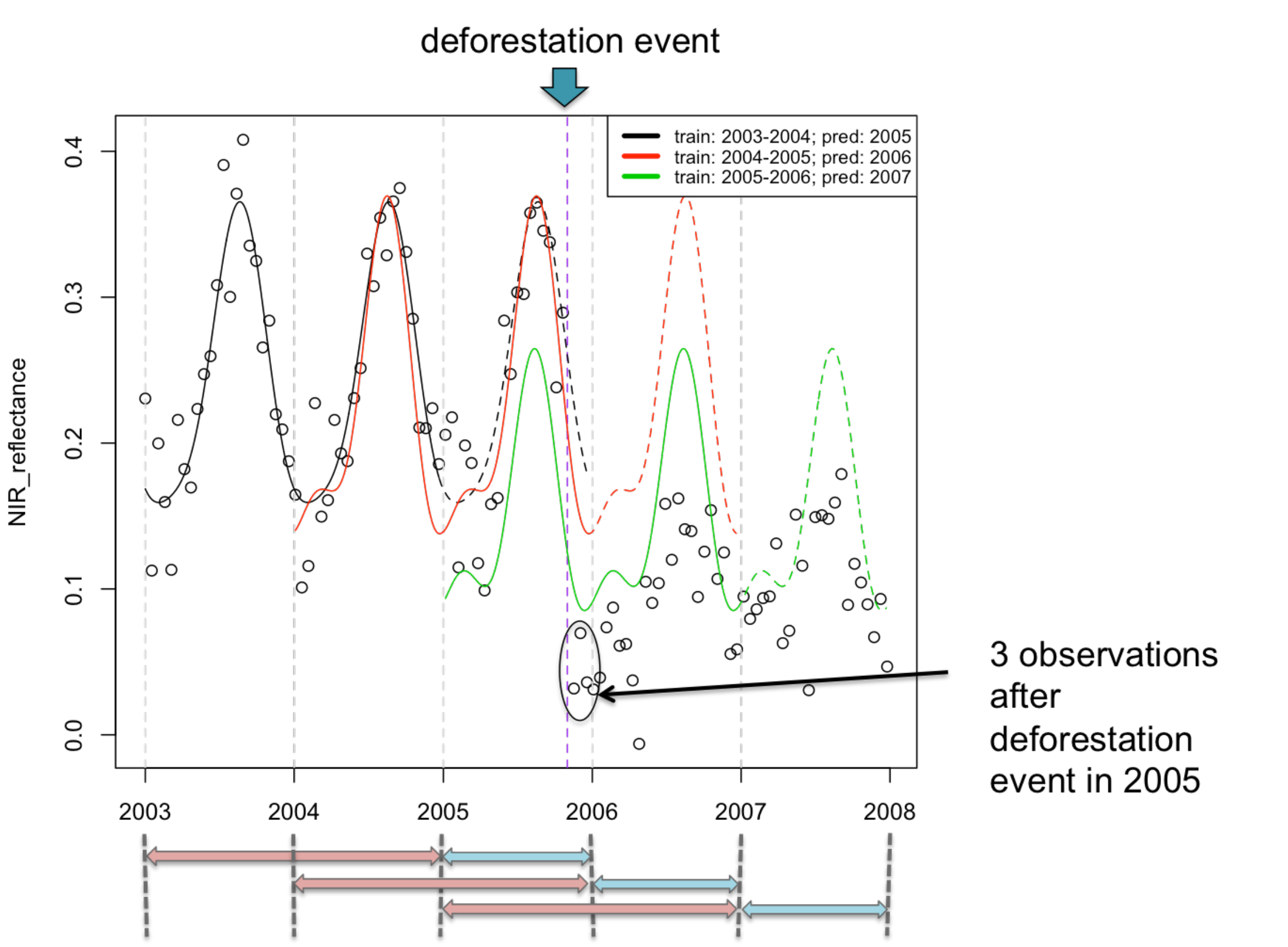} 
			  \caption[Simulated example explaining need for extended prediction window]
			  {Simulated example explaining need for extended prediction window}
			  \label{fig:overlap}
			\end{figure}
			
			In figure \ref{fig:overlap} we see that after the deforestation event which occurs in november 2005, there are only 3 observations left in the year which may be insufficient to detect the deforestation. Since the prediction windows are not extended in this example, we would need to use predictions using the the 2004-2005 training window. However, this means the model fit will be affected by the three \emph{outlying} observations circled in figure \ref{fig:overlap}. 
			
			When fitting models with inter-annual change coefficients ($\alpha_{i,b}$, $\beta_{i,b}$ for $i=2,...,N$) these turned out to have large values for most pixels. If we exclude these coefficents at prediction time, as is done in the CMFDA algorithm, the predictions are way off even when there is no landuse change. If we used $N=2$ and kept $i=2$ coefficients at prediction time, predictions are more reasonable however we chose to use $N=1$ since this model is more simple and performed similarly. Figure \ref{fig:fourier_day} day illustrates the similarity between the $N=1$ and $N=2$ models for a pixel in Nayarit.

			\begin{figure}[H]
			  \centering
			  \includegraphics[width=.5\textwidth]{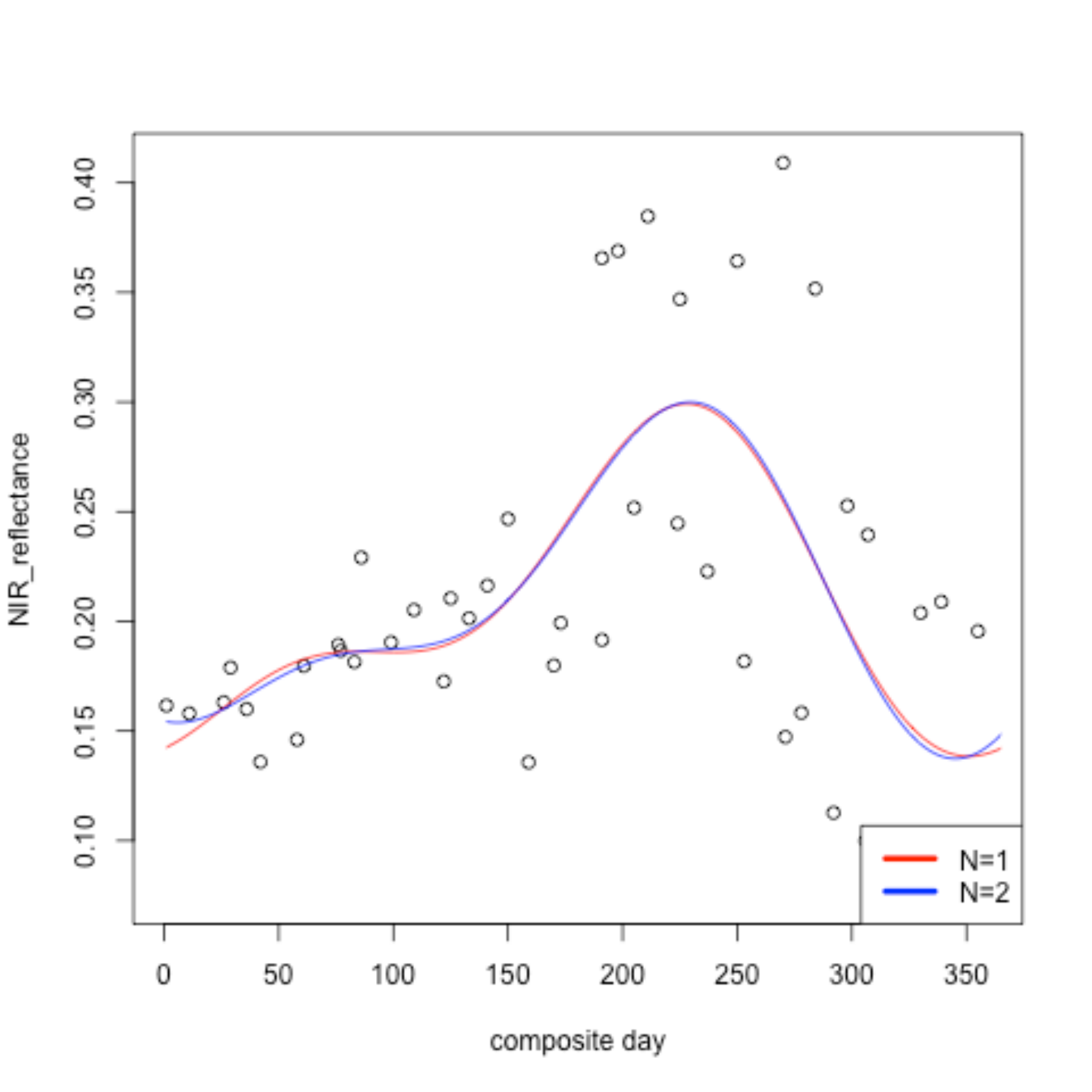} 
			  \caption[Example of fitted Fourier model for a pixel in Nayarit for $N=1$ and $N=2$]
			  {Example of fitted Fourier model for a pixel in Nayarit for $N=1$ and $N=2$}
			  \label{fig:fourier_day}
			\end{figure}
			
			Figure \ref{fig:fourier_date} exemplifies, training and prediction (figures \ref{fig:fourier_date_a},\ref{fig:fourier_date_c} and \ref{fig:fourier_date_e}) and monitoring of prediction errors (figures \ref{fig:fourier_date_b},\ref{fig:fourier_date_d} and \ref{fig:fourier_date_f}) with the different versions of the model discussed above:
			\begin{enumerate}[a.]
				\item $N=1$, i.e. no coefficients corresponding inter-annual variation: figures \ref{fig:fourier_date_a} and \ref{fig:fourier_date_b}; 
				\item $N=2$, i.e. coefficients corresponding to annual variation fitted during training, but forced to be zero during prediction: figures \ref{fig:fourier_date_c} and \ref{fig:fourier_date_d}; 
				\item $N=2$, i.e. coefficients corresponding to annual variation fitted during training, and included during prediction: figures \ref{fig:fourier_date_e} and \ref{fig:fourier_date_f}.
			\end{enumerate}	
			
			The figure illustrates that for the example shown, versions a. and c. work similarly. In version b. however prediction error is much higher because without the omitted coefficients, the model no longer fits the training data. In \cite{CMFDA} this was not the case because with the reflectance data used in that case, the fitted coefficients corresponding to annual variation were always very small.

	   	 \begin{figure}[H]  
  
	   	   \begin{subfigure}{.5\textwidth}
	   	     \centering
	   	 	\includegraphics[width=1\textwidth]{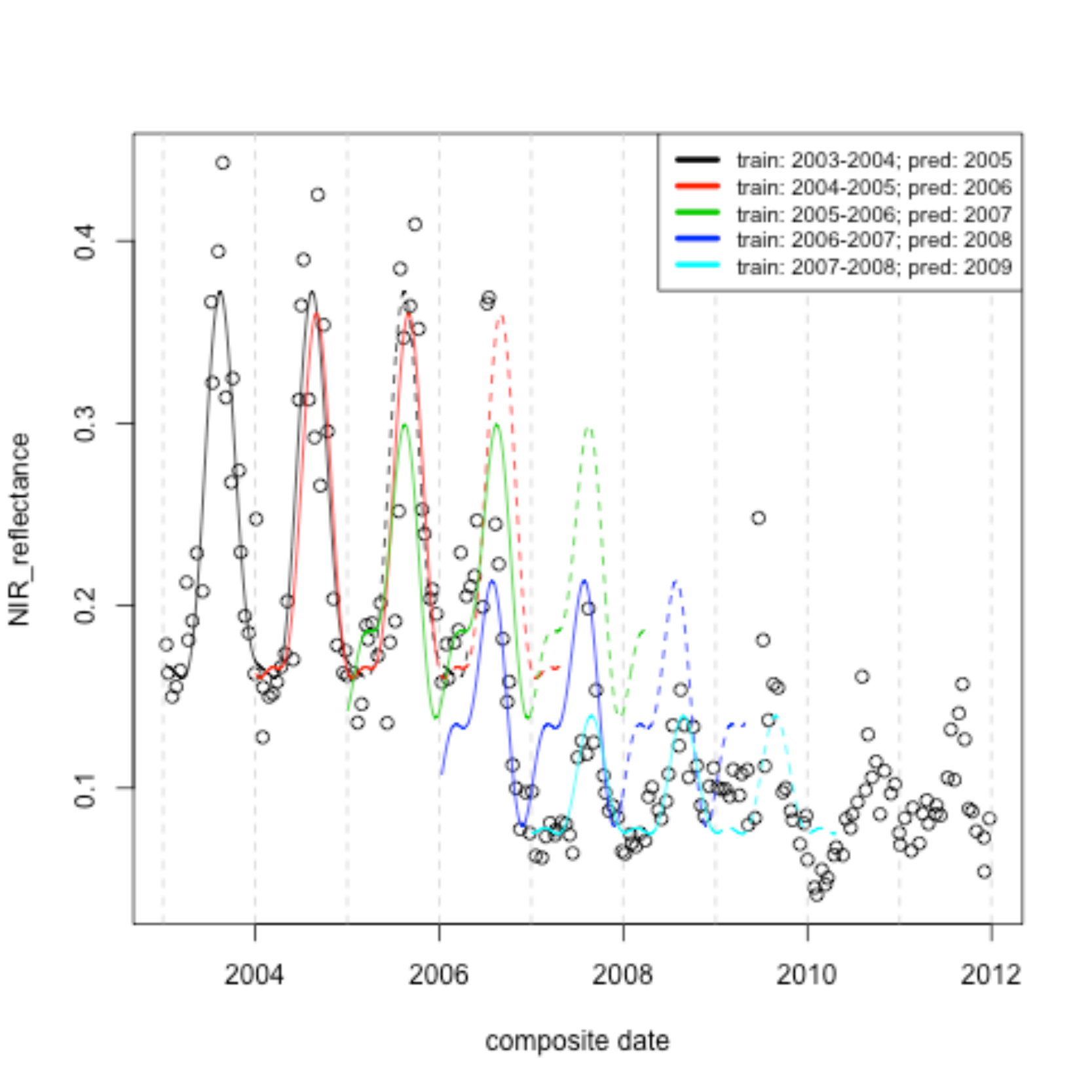} 
	   	     \caption{$N=1$, training and prediction}
	   	     \label{fig:fourier_date_a}
	   	   \end{subfigure}%
	   	   \begin{subfigure}{.5\textwidth}
	   	     \centering
	   	     \includegraphics[width=1\textwidth]{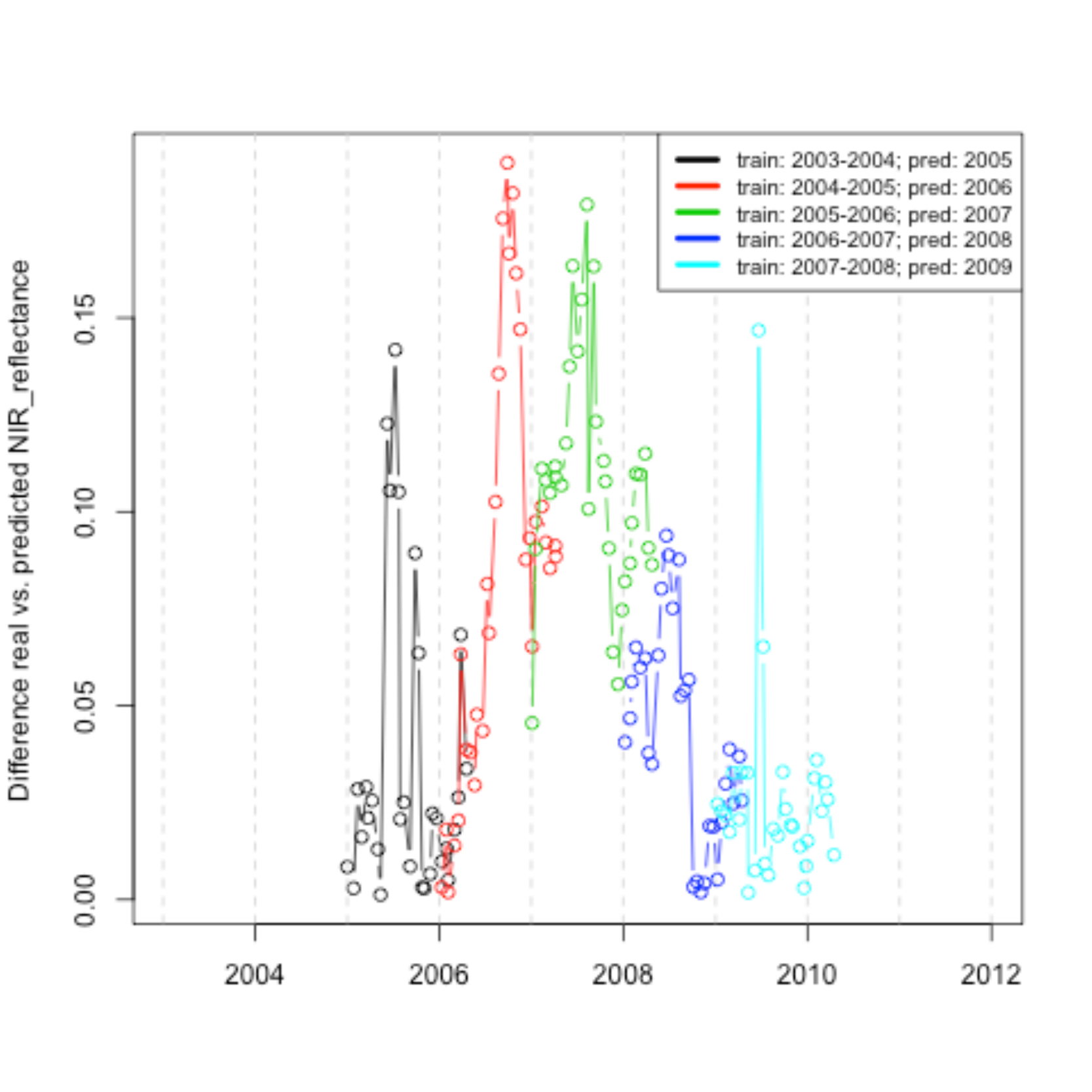}
	   	     \caption{$N=1$, difference between prediction and real values}
	   	     \label{fig:fourier_date_b}
	   	   \end{subfigure}
	   	   \begin{subfigure}{.4\textwidth}
	   	     \centering
	   	     \includegraphics[width=1\textwidth]{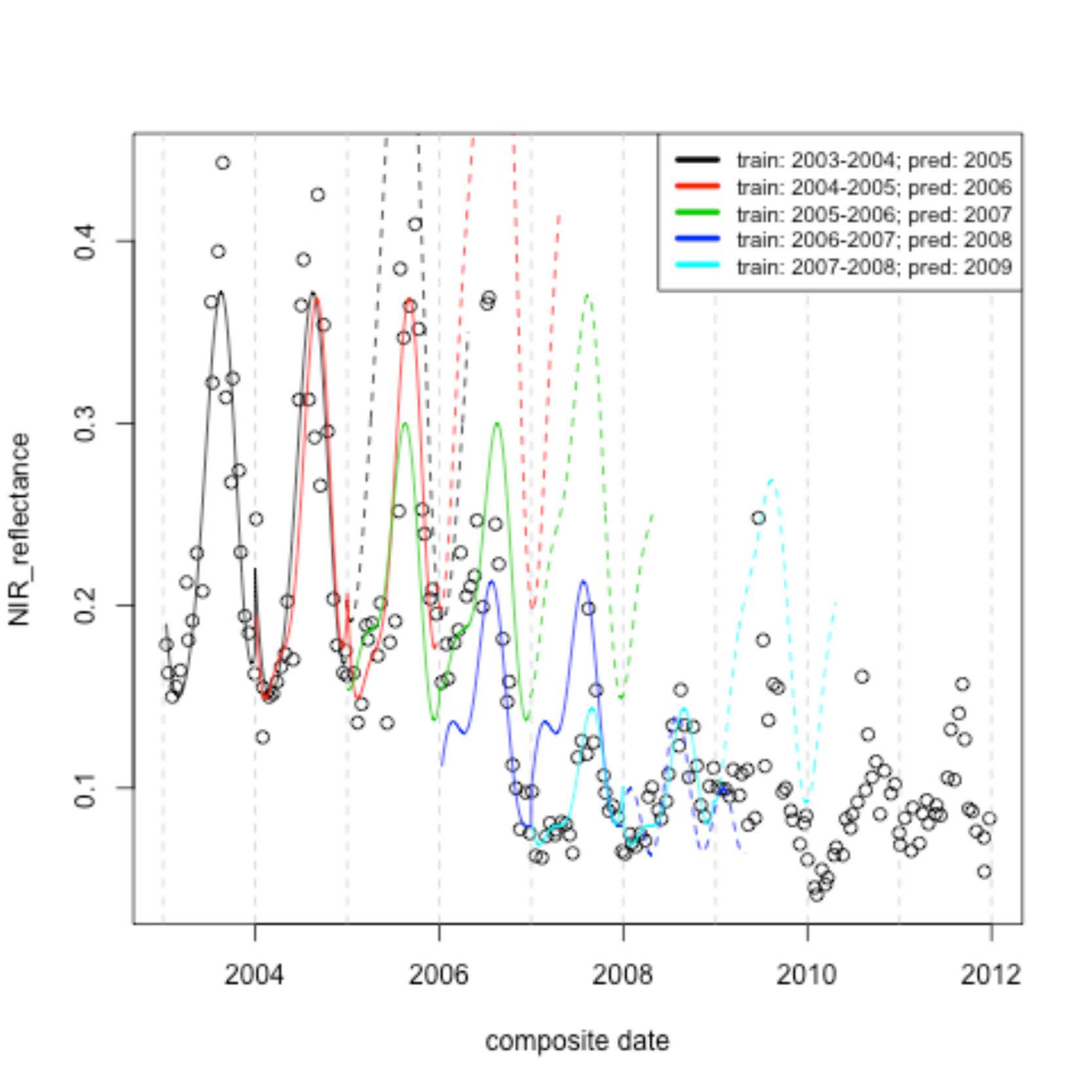}
	   	     \caption{$N=2$, training and prediction, $\alpha_2$ and $\beta_2$ fixed to zero for prediction}
	   	     \label{fig:fourier_date_c}
	   	   \end{subfigure}
	   	   \begin{subfigure}{.4\textwidth}
	   	     \centering
	   	     \includegraphics[width=1\textwidth]{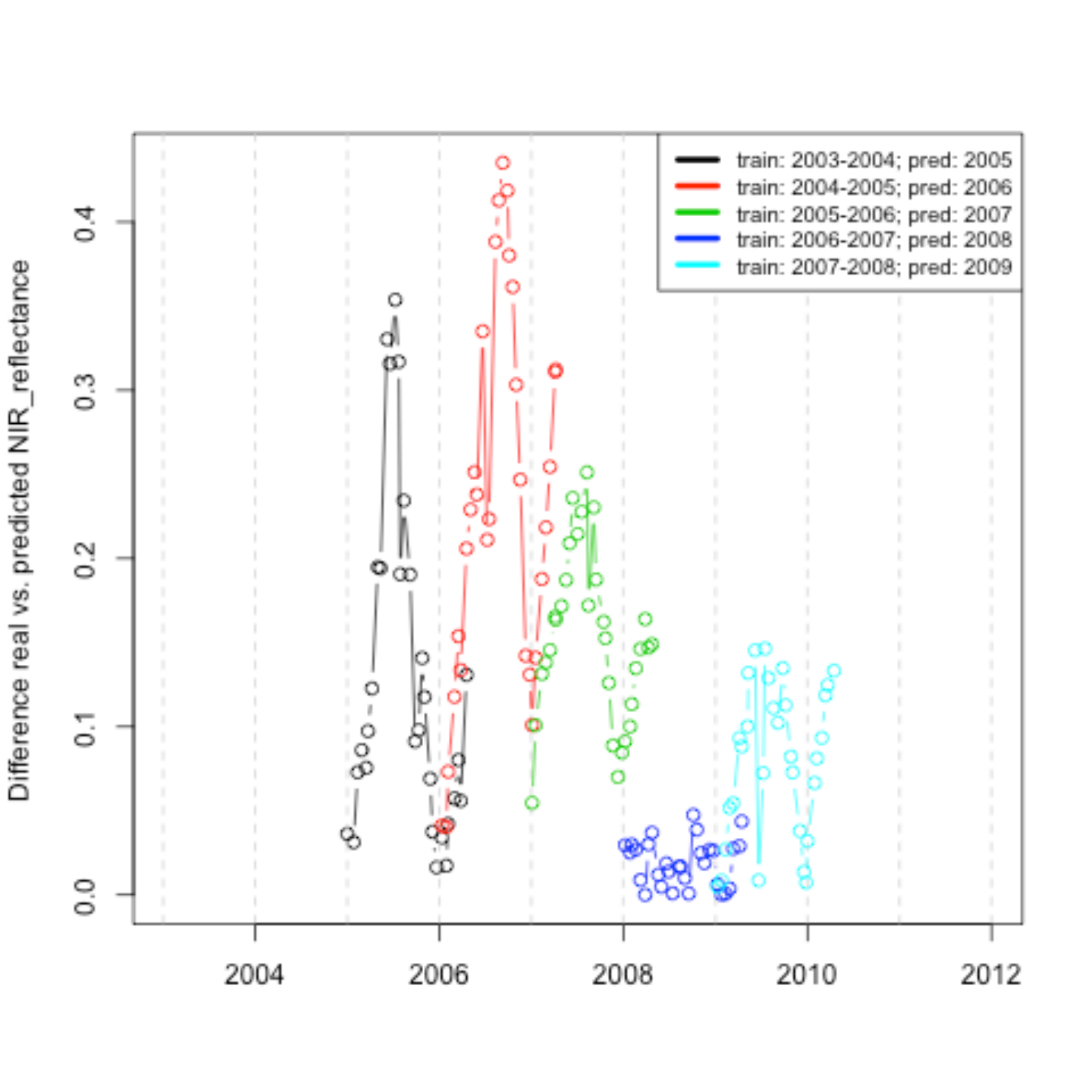}
	   	     \caption{$N=2$, difference between prediction and real values, $\alpha_2$ and $\beta_2$ fixed to zero for prediction}
	   	     \label{fig:fourier_date_d}
	   	   \end{subfigure}
	   	   \begin{subfigure}{.4\textwidth}
	   	     \centering
	   	     \includegraphics[width=1\textwidth]{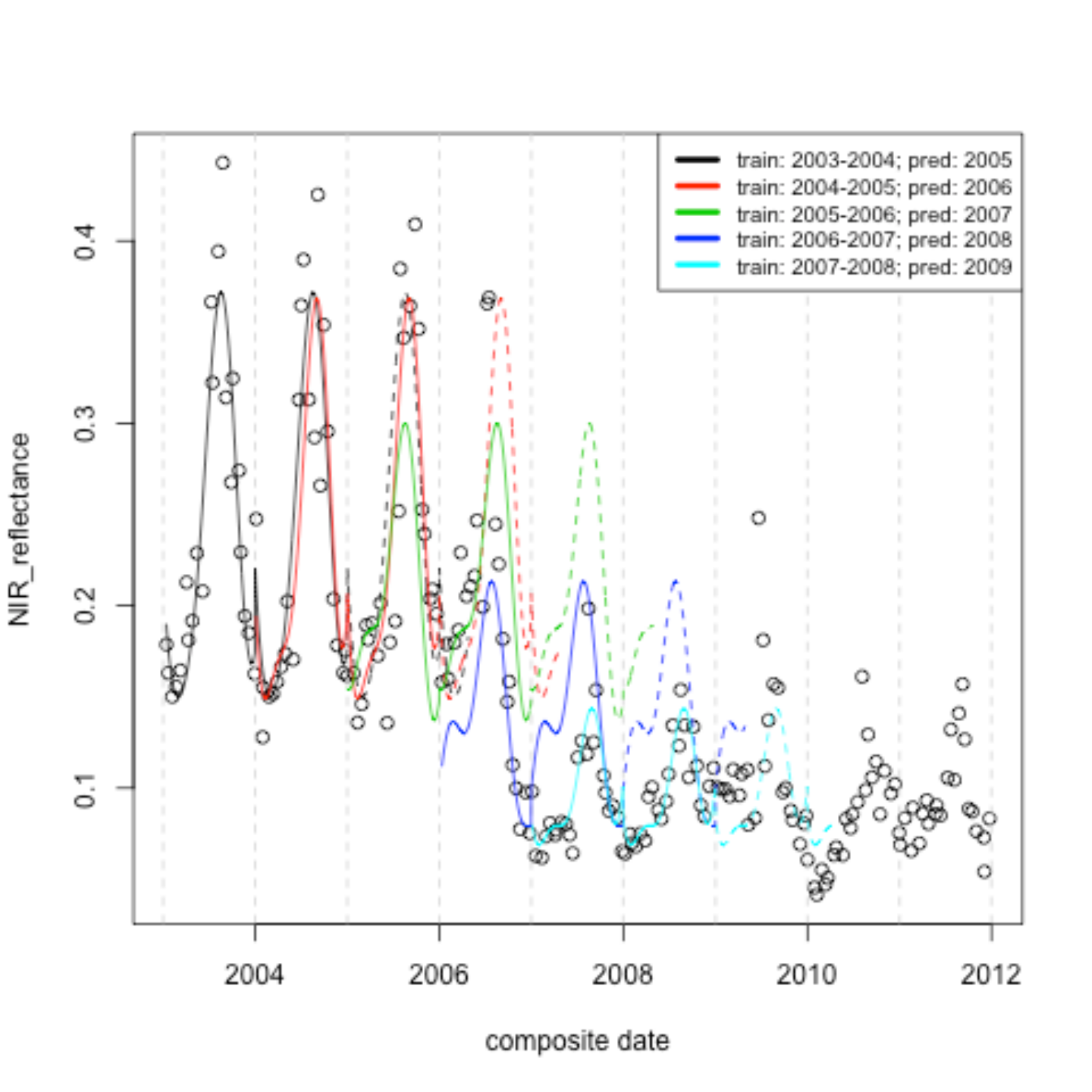}
	   	     \caption{$N=2$, training and prediction}
	   	     \label{fig:fourier_date_e}
	   	   \end{subfigure}
	   	   \begin{subfigure}{.4\textwidth}
	   	     \centering
	   	     \includegraphics[width=1\textwidth]{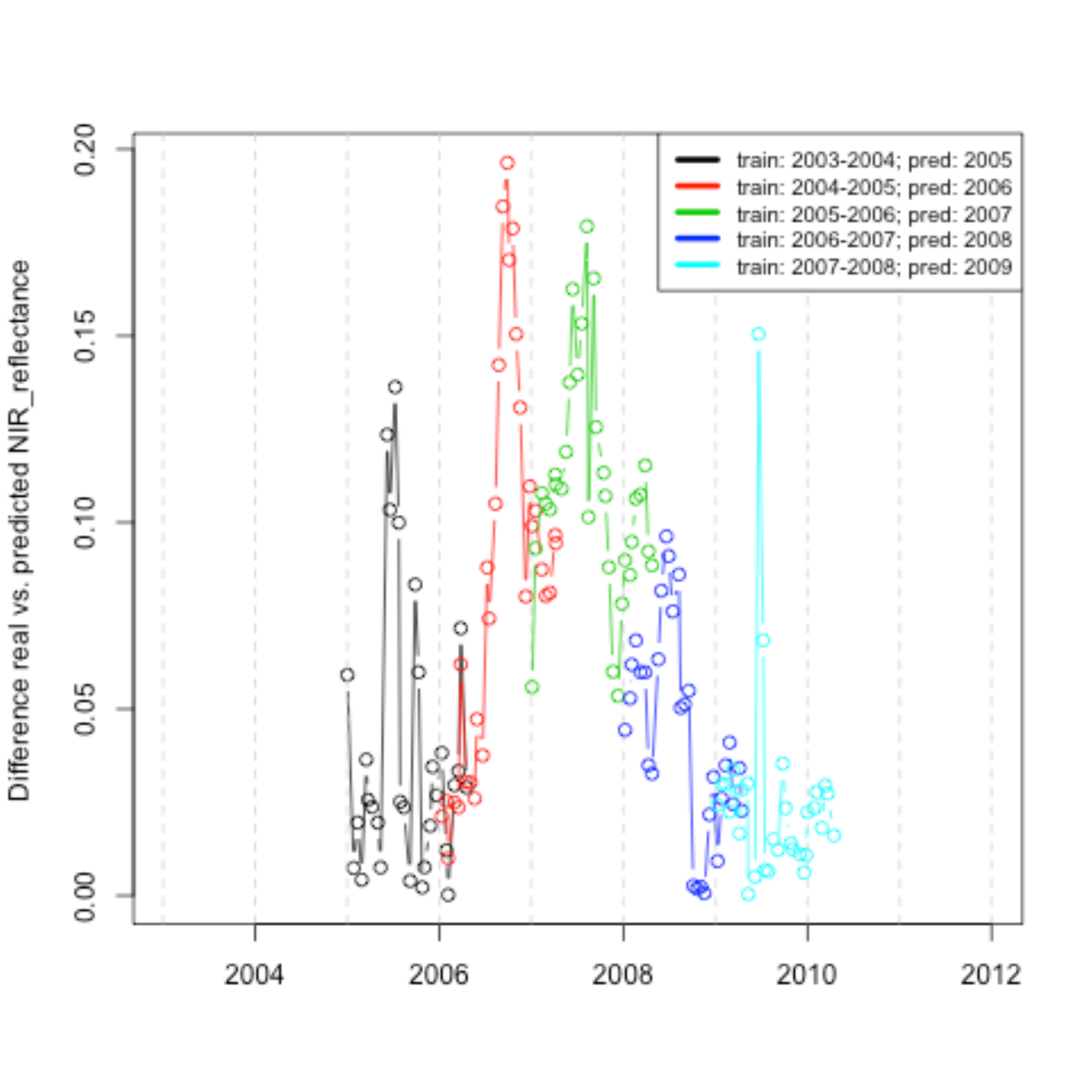}
	   	     \caption{$N=2$, difference between prediction and real values}
	   	     \label{fig:fourier_date_f}
	   	   \end{subfigure}	   
	   	   \caption[Training of, and prediction with, Fourier model]
	   	   {Training of, and prediction with, Fourier model}
	   	   \label{fig:fourier_date}
	   	 \end{figure}
		 
		 The set of predicted surface reflectance observations can now be constructed: for prediction windows  $\hat{Y}_{j,p}^\mathcal{P}=\{\hat{y}_{b,p,t}=\hat{g}_{b,p}(d(t))\}_{b \in \{1,2,3,7,8,9\}, t \in \mathcal{P}_{j,p}}$; for every period $j$ and every pixel $p$ in forest mask $\mathcal{F}$.
			
			\item \textbf{Change detection.} The modis 13A2 data does not include the same bands as the landsat ETM+ and TM datasets so we cannot use the same index as was used in \cite{CMFDA}. We chose to try two approaches in terms of the way we use the prediction errors of the available bands to detect change:
			
			\begin{enumerate}
				\item \textbf{Multivariate approach}. Use a subset $\mathcal{B} \subset \{1,2,3,7,8,9\}$ of the six available bands found to be helpful in detecting deforestation to detect change by training a multivariate threshold $L$. This requires the use of an optimization technique to train the different thresholds simultaneously. The thresholding rule can be \emph{strict} in that the prediction errors of all selected bands must exceed their threshold or \emph{lax} if flag is raised whenever any of the predicted errors exceeds its threshold.
				\item \textbf{Index approach}. Use a subset $\mathcal{B} \subset \{1,2,3,7,8,9\}$ of the six available bands found to be helpful in detecting deforestation to detect change by calculating the \emph{local} Mahalanobis distance of their predicted errors. In section \ref{s:mahl} we explain what we mean by \emph{local} Mahalanobis distance. This represents and index approach similar to the one used in \cite{CMFDA}.
			\end{enumerate}

			We applied a multiple-date approach and tried $C \in \{2,3,4,5,6\}$ consecutive violations of threshold using the \emph{lax} implementation. In other words we trained thresholds for different number of consecutive violations of that threshold. This means that the only training parameter was the threshold $L$ itself. The number of consecutive violations $C$ just specifies a variant of the algorithm. Based on some preliminary results we found that we obtained better performance if we applied the threshold to consecutive differences of the same sign. In the CMFDA algorithm differences between predicted and real values may alternate between positive and negative values and still trigger the deforestation flag. In this case only consecutive differences of the same sign may do so. 
			
			To train the thresholds we use the land-use classification for 2005 and 2010 from the NALCMS. As a performance measure we used the true skill statistic ($tss$). This is also defined with respect to the confusion matrix of table \ref{tab_confMat}.
			
			 \begin{align}
				 tss = \frac{S}{N_1} + \frac{U}{N_0}-1
			 \end{align} 
			
			The $tss$ takes into account the skill with which the algorithm detects correctly both \emph{deforestation} and \emph{non-deforestation} events. We do not have information on the timing of the deforestation events so cannot evaluate the temporal accuracy of the algorithm. In chapter \ref{ch:imp}, after determining whether to use multivariate or index appoach, the subject of implementation going forward will be broached. 
	
		\end{enumerate}

	\section{Study area selection} \label{s:studyArea}
	
	To train the model we chose nine 25 by 25km sites (625 1km pixels or 10,000 250m pixels). The criteria for choosing this sites was two-fold:
	
	\begin{enumerate}
		\item Choose sites with a high amount of deforestation in order to have data adequate for training thresholds, and,
		\item Choose sites with different types of land-use change. 
	\end{enumerate}	
	
	Table \ref{tab:sites} shows the sites chosen, the amount of deforestation in each and the predominant type of land-use change. 
	
	 \begin{table}[H] 
		 \tiny
	 \begin{tabular}{|p{2.2cm}|p{1.2cm}|p{4cm}|p{2cm}|p{1.5cm}|p{1.5cm}|p{1.5cm}|}
		 \hline
	 \textbf{Site name} & \textbf{State} & \textbf{Type of forest 2005 (top 90\%)} & \textbf{Land use deforested pixels 2010 (top 90\%)} & \textbf{Best time series} & \textbf{\# of forest pixels (at 250m resolution, out of 10,000) in 2005}   & \textbf{\# of deforested pixels (at 250m resolution, out of 10,000) 2005-2010}  \\ 
	 \hline                                                                                                                                      
	 Sonora232 &  Sonora & tropical or sub-tropical broadleaf deciduous forest & water, barren lands & NIR & 7,772 & 232  \\
	\hline 
	 Jalisco164 &  Jalisco & tropical or sub-tropical broadleaf deciduous forest, temperate or sub-polar broadleaf deciduous forest,  & water, wetland & NIR & 2,898 & 164  \\
	\hline
	 Nayarit151 &  Nayarit & tropical or sub-tropical broadleaf deciduous forest & water & NIR & 6,590 & 151  \\
	\hline
	 Nayarit109 &  Nayarit & tropical or sub-tropical broadleaf deciduous forest, mixed forest & water & NIR & 5,376 & 109  \\
	\hline
	\hline
	 Yucatan180 &  Yucatan & tropical or sub-tropical broadleaf evergreen forest & urban, cropland & NDVI & 6,862 & 180  \\
	\hline
	 QuintanaRoo100 & Quintana Roo & tropical or sub-tropical broadleaf evergreen forest & cropland, urban & NDVI & 4,386 & 100  \\
	\hline
	 Michoacan98 &  Michoacan & tropical or sub-tropical broadleaf deciduous forest, mixed forest, temperate or sub-polar broadleaf deciduous forest & urban, cropland, temperate or sub-polar shrubland, tropical or sub-tropical grassland & NDVI & 5,639 & 98  \\
	\hline
	 QuintanaRoo77 & Quintana Roo & tropical or sub-tropical broadleaf evergreen forest & cropland, urban & NDVI & 3,052 & 77  \\
	\hline
	 Sonora74 &  Sonora & tropical or sub-tropical broadleaf deciduous forest & barren lands, tropical or sub-tropical shrubland & NDVI & 8,878 & 74  \\
	\hline                                                                                      
	 \end{tabular}\\
	 \caption{Nine chosen sites for threshold training}
	 \label{tab:sites}
	 \end{table}
	
	As we will see in the analysis of each site, it appears land-use change from forest to water can be detected effectively using the NIR refelectance time series while land-use change from forest to urban or cropland can be better detected with the NDVI time series. It's important to notice that we don't know what type of land-use change will happen before hand. Although in the section \ref{s:gridSearch} we explore using the appropriate time series for each site, given the type of land-use change that happened, to have an algorithm we can use going forward, we will need to use both time series. In section \ref{s:simAn} we explore using multivariate time series and thresholds and in section \ref{s:mahl} we explore using an index approach similar to that used in \cite{CMFDA} based on the \emph{Mahalanobis distance}.  
		
In sections \ref{ss:sitesWater} and \ref{ss:sitesUrban} we illustrate the type, level and pattern of deforestation for one of the \emph{forest to water} sites, Sonora232, and for one of the \emph{forest to cropland or urban} sites, Yucatan180. 

		
		\subsection{Forest to water sites} \label{ss:sitesWater}
		
		\subsubsection{Sonora232}
		
\begin{figure}[H]  
  
   	   \begin{subfigure}{.5\textwidth}
   	     \centering
   	 	\includegraphics[width=0.7\textwidth]{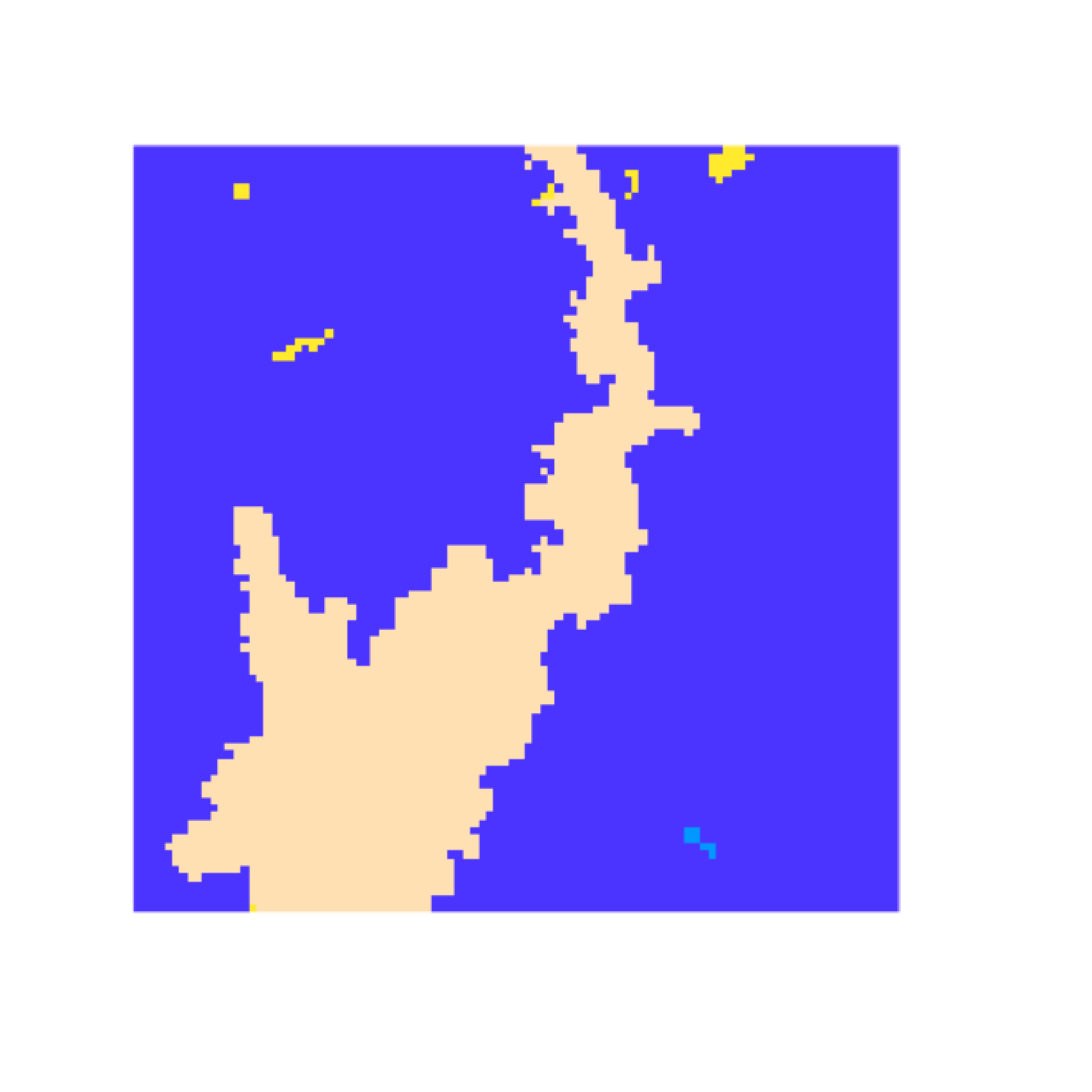} 
   	     \caption{Land-use 2005}
   	     \label{fig:sfig1a}
   	   \end{subfigure}%
   	   \begin{subfigure}{.5\textwidth}
   	     \centering
   	     \includegraphics[width=0.7\textwidth]{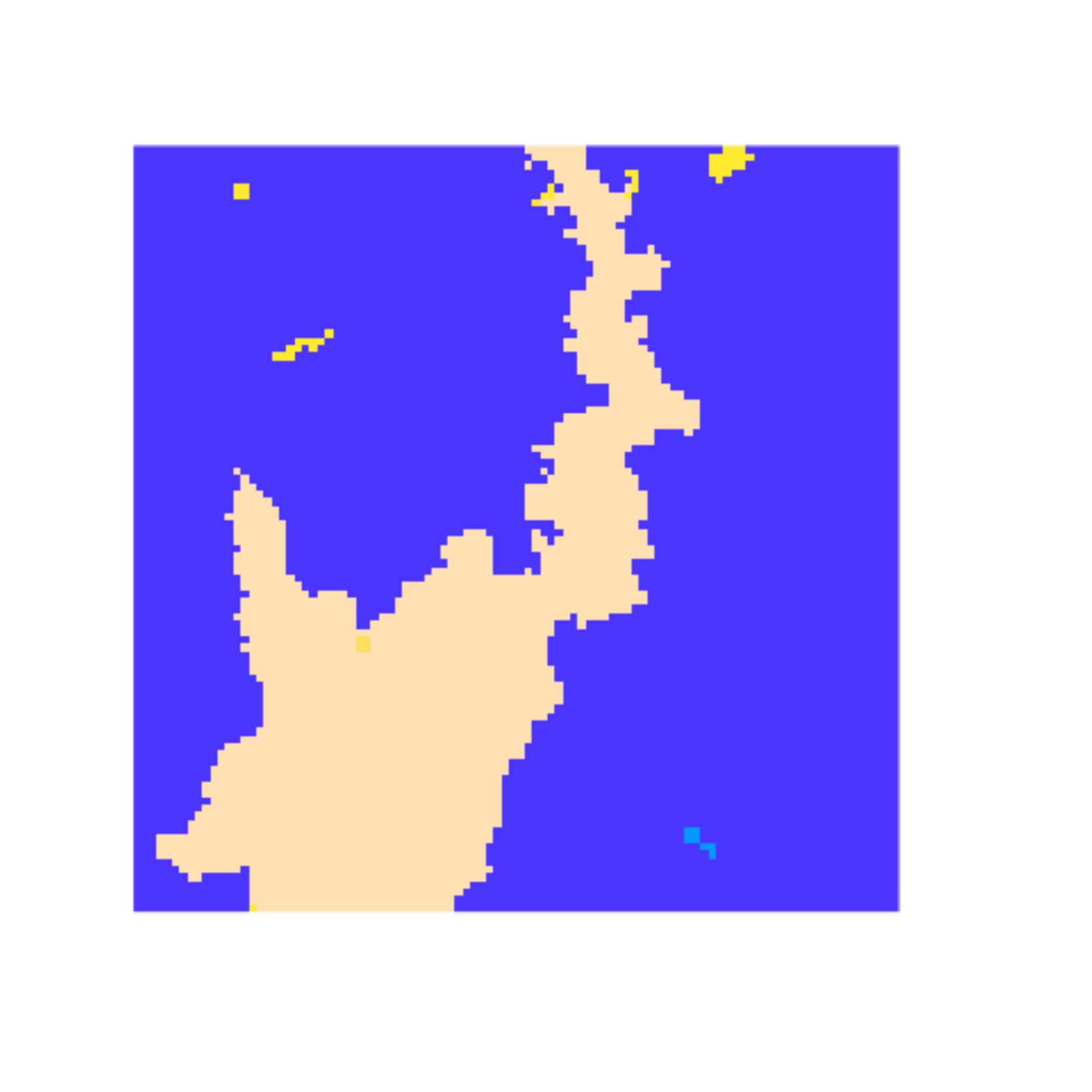}
   	     \caption{Land-use 2010}
   	     \label{fig:sfig1b}
   	   \end{subfigure}
   	   \begin{subfigure}{.4\textwidth}
   	     \centering
   	     \includegraphics[width=0.7\textwidth]{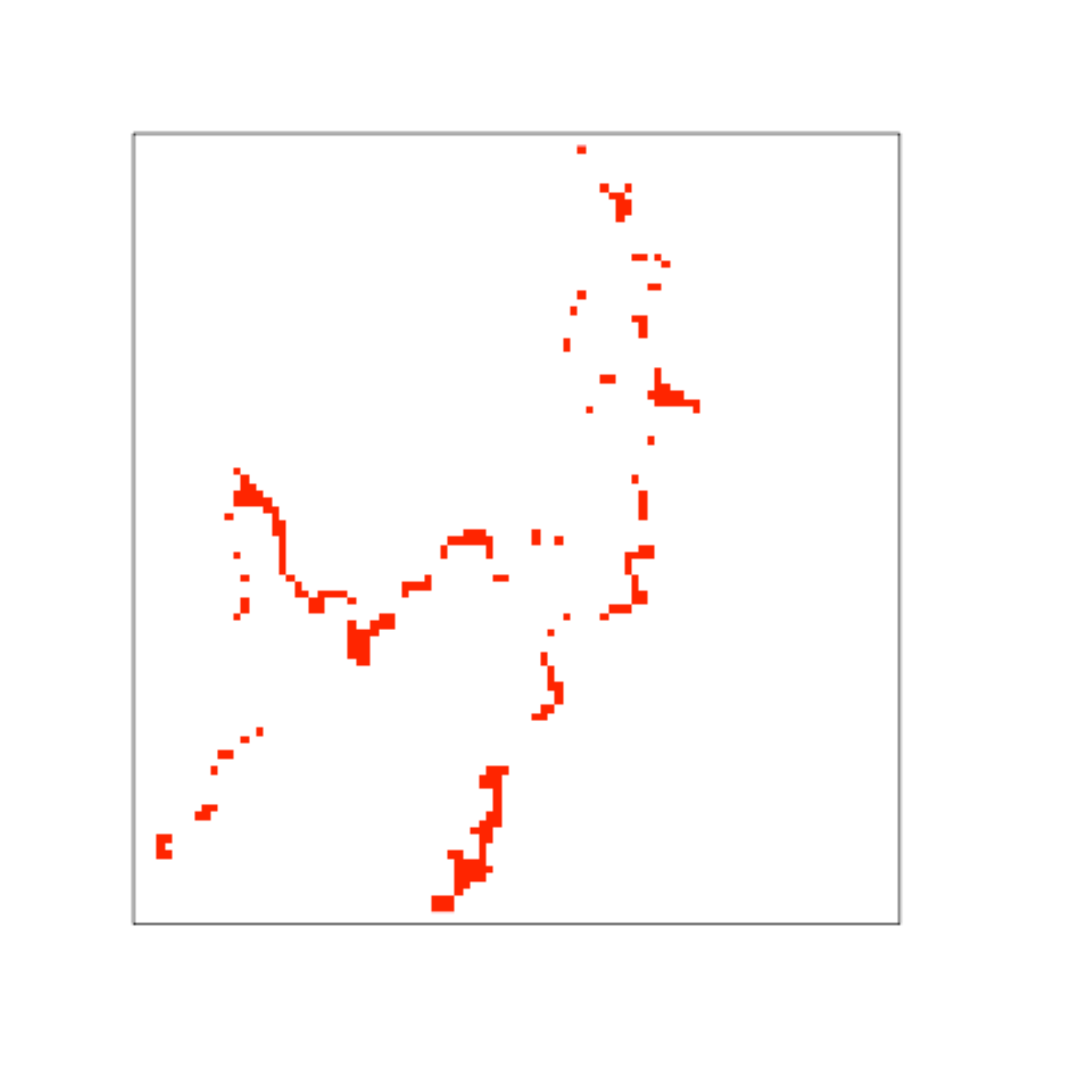}
   	     \caption{Deforestation 2005-2010}
   	     \label{fig:sfig1c}
   	   \end{subfigure}
   	   \begin{subfigure}{.4\textwidth}
   	     \centering
   	     \includegraphics[width=0.7\textwidth]{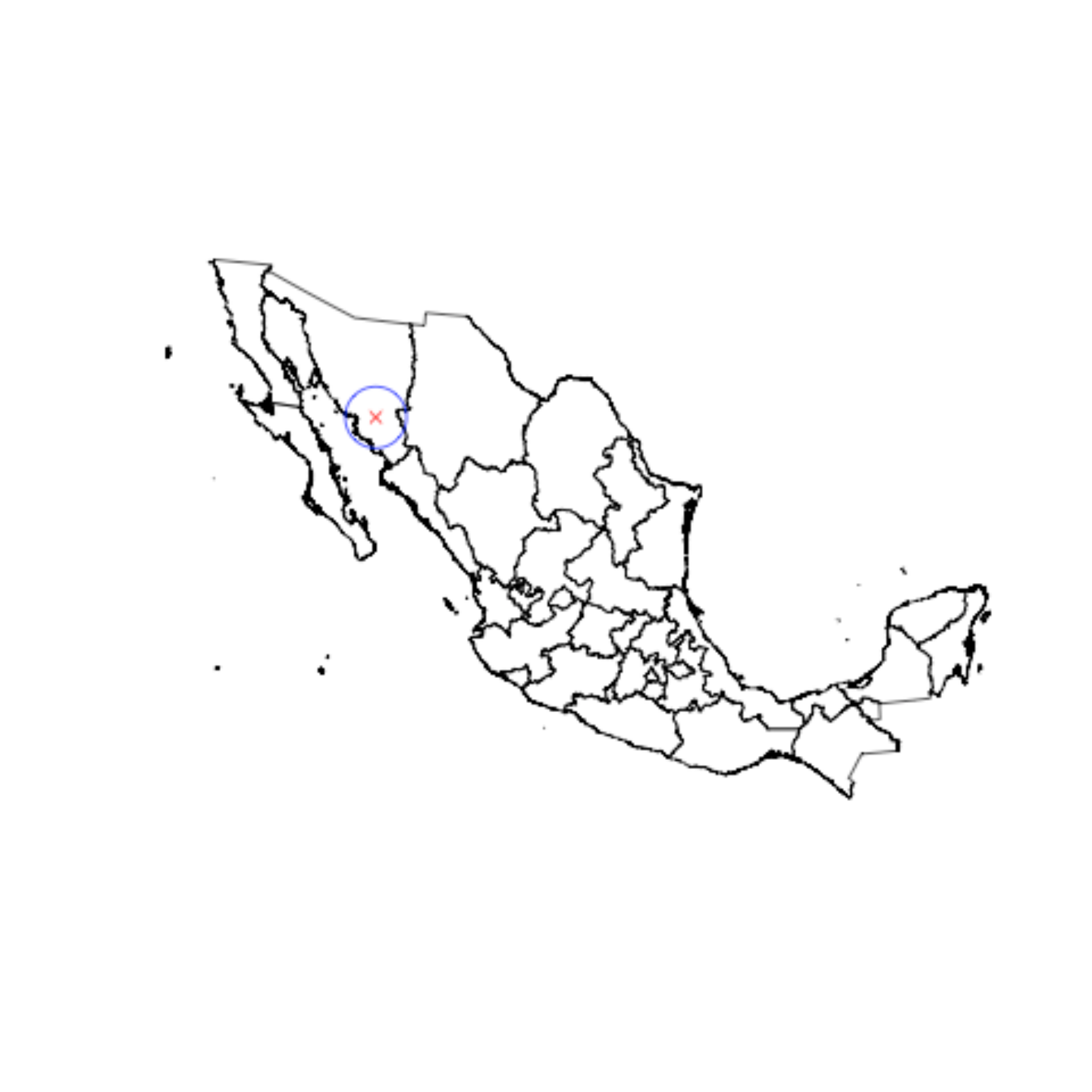}
   	     \caption{Location}
   	     \label{fig:sfig1d}
   	   \end{subfigure}   
      \begin{subfigure}{.5\textwidth}
        \centering
        \includegraphics[width=0.8\textwidth]{./2_Data/legend_landUse}
        \caption{Land use legend}
        \label{fig:legend}
      \end{subfigure}
   	   \caption[NALCMS land-use classification of Sonora232 site, 2005 and 2010 at 250m resolution.]
   	   {NALCMS land-use classification of Sonora232 site, 2005 and 2010 at 250m resolution.}
   	   \label{fig:landUseSonora232}
   	 \end{figure}

		Figure \ref{fig:landUseSonora232} illustrates the deforestation that occurred in the Sonora232 site located in the north-west of Mexico. It appears that a body of water expanded in the 2005-2010 period pushing back the forest edge. 
		
	 Table \ref{tab:landUseSonora232} confirms that the only type of deforestation that occurred at this site was from forest to water.

		\begin{center}
			\footnotesize\addtolength{\tabcolsep}{-2pt}
\begin{table}[ht]
\centering
\begingroup\tiny
\begin{tabular}{rrrrrrrrrrrrrrrrrrrrr}
  & \begin{sideways} Temperate or sub-polar needleleaf forest \end{sideways} & \begin{sideways} Sub-polar taiga needleleaf forest \end{sideways} & \begin{sideways} Tropical or sub-tropical broadleaf evergreen forest \end{sideways} & \begin{sideways} Tropical or sub-tropical broadleaf deciduous forest \end{sideways} & \begin{sideways} Temperate or sub-polar broadleaf deciduous forest \end{sideways} & \begin{sideways} Mixed Forest \end{sideways} & \begin{sideways} Tropical or sub-tropical shrubland \end{sideways} & \begin{sideways} Temperate or sub-polar shrubland \end{sideways} & \begin{sideways} Tropical or sub-tropical grassland \end{sideways} & \begin{sideways} Temperate or sub-polar grassland \end{sideways} & \begin{sideways} Sub-polar or polar shrubland-lichen-moss \end{sideways} & \begin{sideways} Sub-polar or polar grassland-lichen-moss \end{sideways} & \begin{sideways} Sub-polar or polar barren-lichen-moss \end{sideways} & \begin{sideways} Wetland \end{sideways} & \begin{sideways} Cropland \end{sideways} & \begin{sideways} Barren Lands \end{sideways} & \begin{sideways} Urban and Built-up \end{sideways} & \begin{sideways} Water \end{sideways} & \begin{sideways} Snow and Ice \end{sideways} & \begin{sideways} Total \end{sideways} \\ 
  \hline
Temperate or sub-polar needleleaf forest &    0 &    0 &    0 &    0 &    0 &    0 &    0 &    0 &    0 &    0 &    0 &    0 &    0 &    0 &    0 &    0 &    0 &    0 &    0 &    0 \\ 
  Sub-polar taiga needleleaf forest &    0 &    0 &    0 &    0 &    0 &    0 &    0 &    0 &    0 &    0 &    0 &    0 &    0 &    0 &    0 &    0 &    0 &    0 &    0 &    0 \\ 
  Tropical or sub-tropical broadleaf evergreen forest &    0 &    0 &    0 &    0 &    0 &    0 &    0 &    0 &    0 &    0 &    0 &    0 &    0 &    0 &    0 &    0 &    0 &    0 &    0 &    0 \\ 
  Tropical or sub-tropical broadleaf deciduous forest &    0 &    0 &    0 & 7540 &    0 &    0 &    0 &    0 &    0 &    0 &    0 &    0 &    0 &    0 &    0 &    4 &    0 &  228 &    0 & 7772 \\ 
  Temperate or sub-polar broadleaf deciduous forest &    0 &    0 &    0 &    0 &    0 &    0 &    0 &    0 &    0 &    0 &    0 &    0 &    0 &    0 &    0 &    0 &    0 &    0 &    0 &    0 \\ 
  Mixed Forest &    0 &    0 &    0 &    0 &    0 &    0 &    0 &    0 &    0 &    0 &    0 &    0 &    0 &    0 &    0 &    0 &    0 &    0 &    0 &    0 \\ 
   \hline
Tropical or sub-tropical shrubland &    0 &    0 &    0 &    0 &    0 &    0 &    7 &    0 &    0 &    0 &    0 &    0 &    0 &    0 &    0 &    0 &    0 &    0 &    0 &    7 \\ 
  Temperate or sub-polar shrubland &    0 &    0 &    0 &    0 &    0 &    0 &    0 &    0 &    0 &    0 &    0 &    0 &    0 &    0 &    0 &    0 &    0 &    0 &    0 &    0 \\ 
  Tropical or sub-tropical grassland &    0 &    0 &    0 &    0 &    0 &    0 &    0 &    0 &    0 &    0 &    0 &    0 &    0 &    0 &    0 &    0 &    0 &    0 &    0 &    0 \\ 
  Temperate or sub-polar grassland &    0 &    0 &    0 &    0 &    0 &    0 &    0 &    0 &    0 &    0 &    0 &    0 &    0 &    0 &    0 &    0 &    0 &    0 &    0 &    0 \\ 
  Sub-polar or polar shrubland-lichen-moss &    0 &    0 &    0 &    0 &    0 &    0 &    0 &    0 &    0 &    0 &    0 &    0 &    0 &    0 &    0 &    0 &    0 &    0 &    0 &    0 \\ 
  Sub-polar or polar grassland-lichen-moss &    0 &    0 &    0 &    0 &    0 &    0 &    0 &    0 &    0 &    0 &    0 &    0 &    0 &    0 &    0 &    0 &    0 &    0 &    0 &    0 \\ 
  Sub-polar or polar barren-lichen-moss &    0 &    0 &    0 &    0 &    0 &    0 &    0 &    0 &    0 &    0 &    0 &    0 &    0 &    0 &    0 &    0 &    0 &    0 &    0 &    0 \\ 
  Wetland &    0 &    0 &    0 &    0 &    0 &    0 &    0 &    0 &    0 &    0 &    0 &    0 &    0 &    0 &    0 &    0 &    0 &    0 &    0 &    0 \\ 
  Cropland &    0 &    0 &    0 &    0 &    0 &    0 &    0 &    0 &    0 &    0 &    0 &    0 &    0 &    0 &   43 &    0 &    0 &    0 &    0 &   43 \\ 
  Barren Lands &    0 &    0 &    0 &    0 &    0 &    0 &    0 &    0 &    0 &    0 &    0 &    0 &    0 &    0 &    0 &    0 &    0 &    0 &    0 &    0 \\ 
  Urban and Built-up &    0 &    0 &    0 &    0 &    0 &    0 &    0 &    0 &    0 &    0 &    0 &    0 &    0 &    0 &    0 &    0 &    0 &    0 &    0 &    0 \\ 
  Water &    0 &    0 &    0 &    0 &    0 &    0 &    0 &    0 &    0 &    0 &    0 &    0 &    0 &    0 &    0 &    0 &    0 & 2178 &    0 & 2178 \\ 
  Snow and Ice &    0 &    0 &    0 &    0 &    0 &    0 &    0 &    0 &    0 &    0 &    0 &    0 &    0 &    0 &    0 &    0 &    0 &    0 &    0 &    0 \\ 
   \hline
Total &    0 &    0 &    0 & 7540 &    0 &    0 &    7 &    0 &    0 &    0 &    0 &    0 &    0 &    0 &   43 &    4 &    0 & 2406 &    0 & 10000 \\ 
   \hline
\end{tabular}
\endgroup
\caption[NALCMS land-use distribution for site Sonora232, 2005 and 2010.]{NALCMS land-use distribution for site Sonora232, 2005 and 2010.} 
\label{tab:landUseSonora232}
\end{table}

		\end{center}
		
\begin{figure}[H]
		  \centering
		  \includegraphics[width=.5\textwidth]{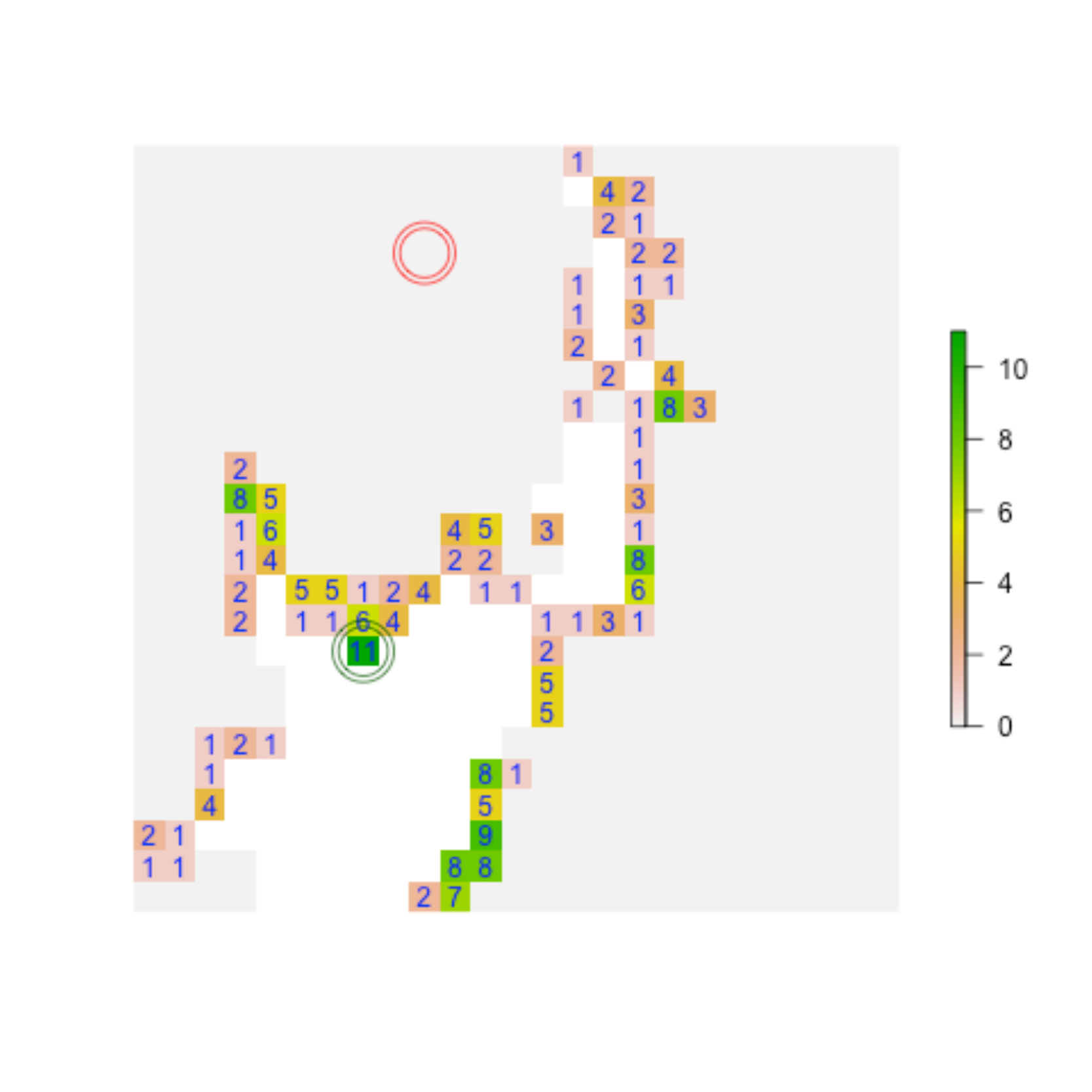} 
		  \caption[Deforestation in Sonora232 site at 1km resolution]
		  {Deforestation in Sonora232 site at 1km resolution: number of 250m pixels deforested out of 16}
		  \label{fig:defoSonora232}
		\end{figure}

\begin{figure}[H]
	      \begin{subfigure}{.5\textwidth}
	        \centering
	    	\includegraphics[width=1\textwidth]{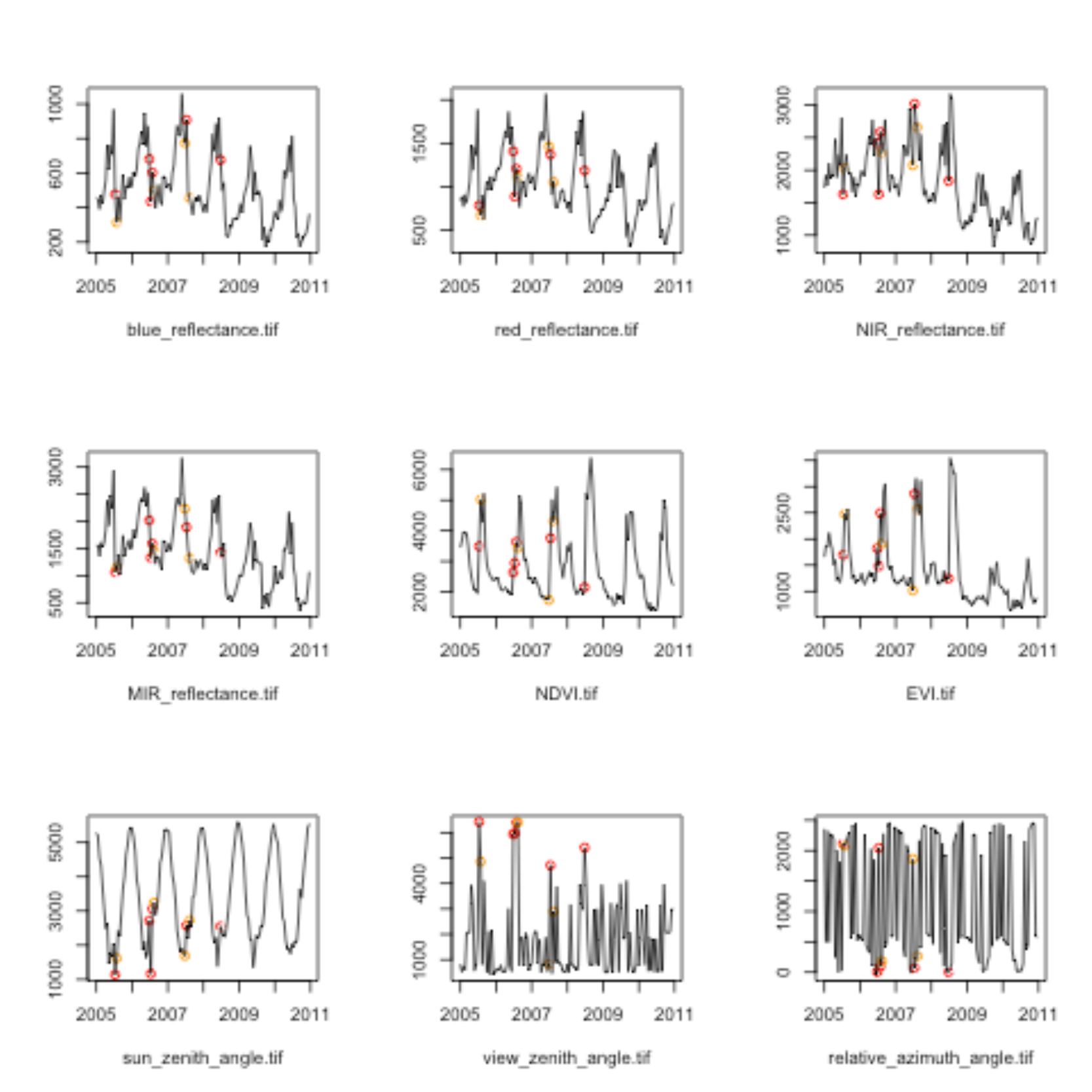} 
	        \caption{deforested pixel}
	        \label{fig:defo_pix}
	      \end{subfigure}%
	      \begin{subfigure}{.5\textwidth}
	        \centering
	        \includegraphics[width=1\textwidth]{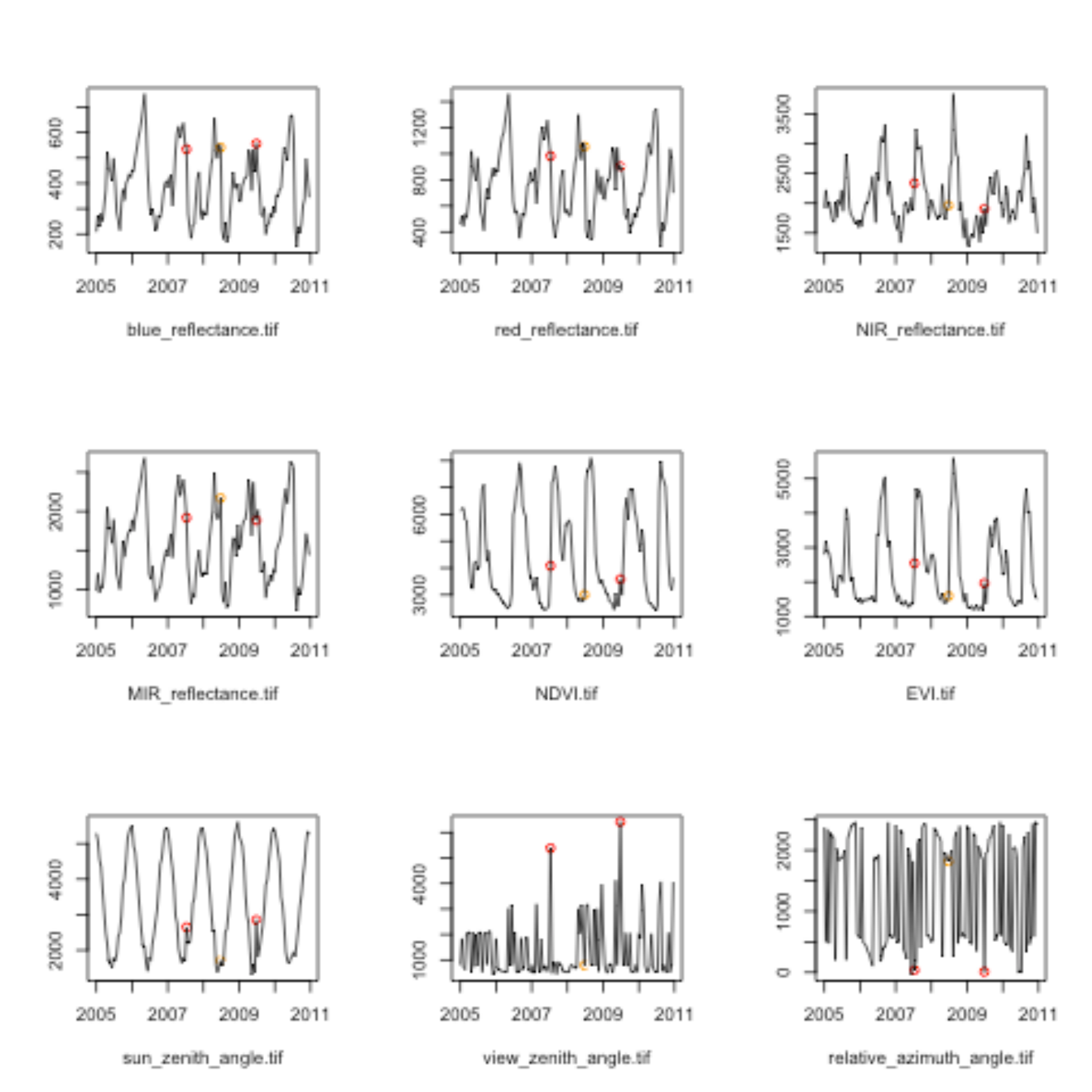}
	        \caption{non-deforested pixel}
	        \label{fig:non_defo_pix}
	      \end{subfigure}
	      \caption[Reflectance, vegetation indices and sun-sensor geometry time series for deforested and non-deforested pixel in Sonora232 site]
	      {Reflectance, vegetation indices and sun-sensor geometry time series for deforested and non-deforested pixel in Sonora 232 site, identified in figure \ref{fig:defoSonora232} with green and red circles respectively. Orange marked points are \emph{marginal data} and red marked points are \emph{cloudy} or \emph{not processed}}
	      \label{fig:timeSeriesSonora232}
	    \end{figure}

		Recalling section \ref{s:landuse}, the deforestation flag is constructed from a landuse classification model with a 250m resolution. Figure \ref{fig:defoSonora232} shows which pixels in the Sonora232 site were deforested and also the degree of deforestation: the number of 250m resolution deforested pixels, out of a possible 16, is labeled for deforested pixels. Additionally, two pixels, one deforested and one non-deforested, are identified with circles so that we may observe the behavior of their reflectance in the 2005-2010 period in figure \ref{fig:timeSeriesSonora232}. 
		
		In figure \ref{fig:timeSeriesSonora232} the sun-sensor geometry, surface reflectance and vegetation index time series are plotted for the 2005-2010 period for the two pixes identified in figure \ref{fig:defoSonora232}. We can see, by looking at the NIR reflectance time series, that the deforestation event for the deforested pixel occurred towards the end of 2008 or the beginning of 2009. 
		
		\subsection{Forest to urban or cropland sites} \label{ss:sitesUrban}
		
		\subsubsection{Yucatan180}
		
\begin{figure}[H]  
  
      	   \begin{subfigure}{.5\textwidth}
      	     \centering
      	 	\includegraphics[width=0.7\textwidth]{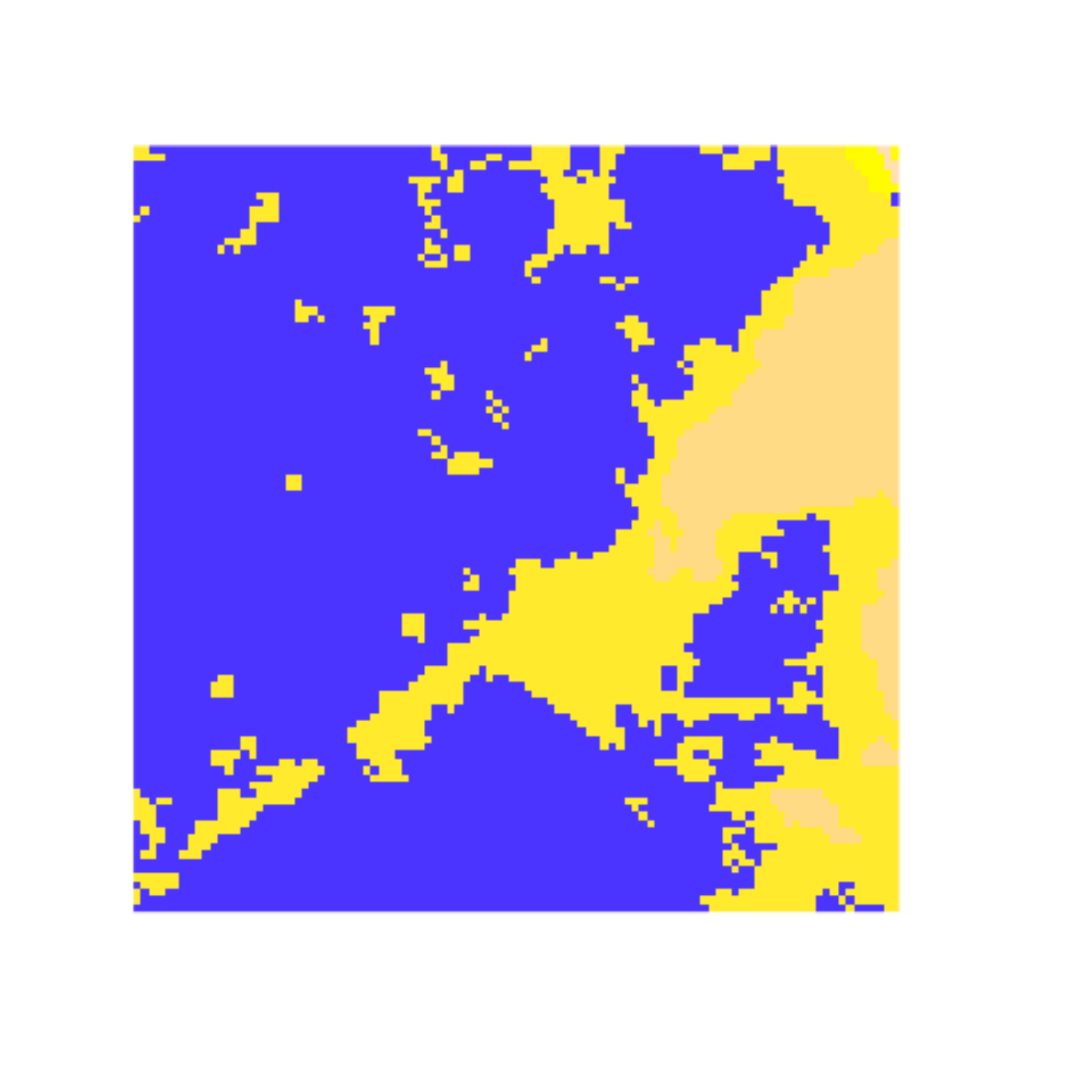} 
      	     \caption{Land-use 2005}
      	     \label{fig:sfig1a}
      	   \end{subfigure}%
      	   \begin{subfigure}{.5\textwidth}
      	     \centering
      	     \includegraphics[width=0.7\textwidth]{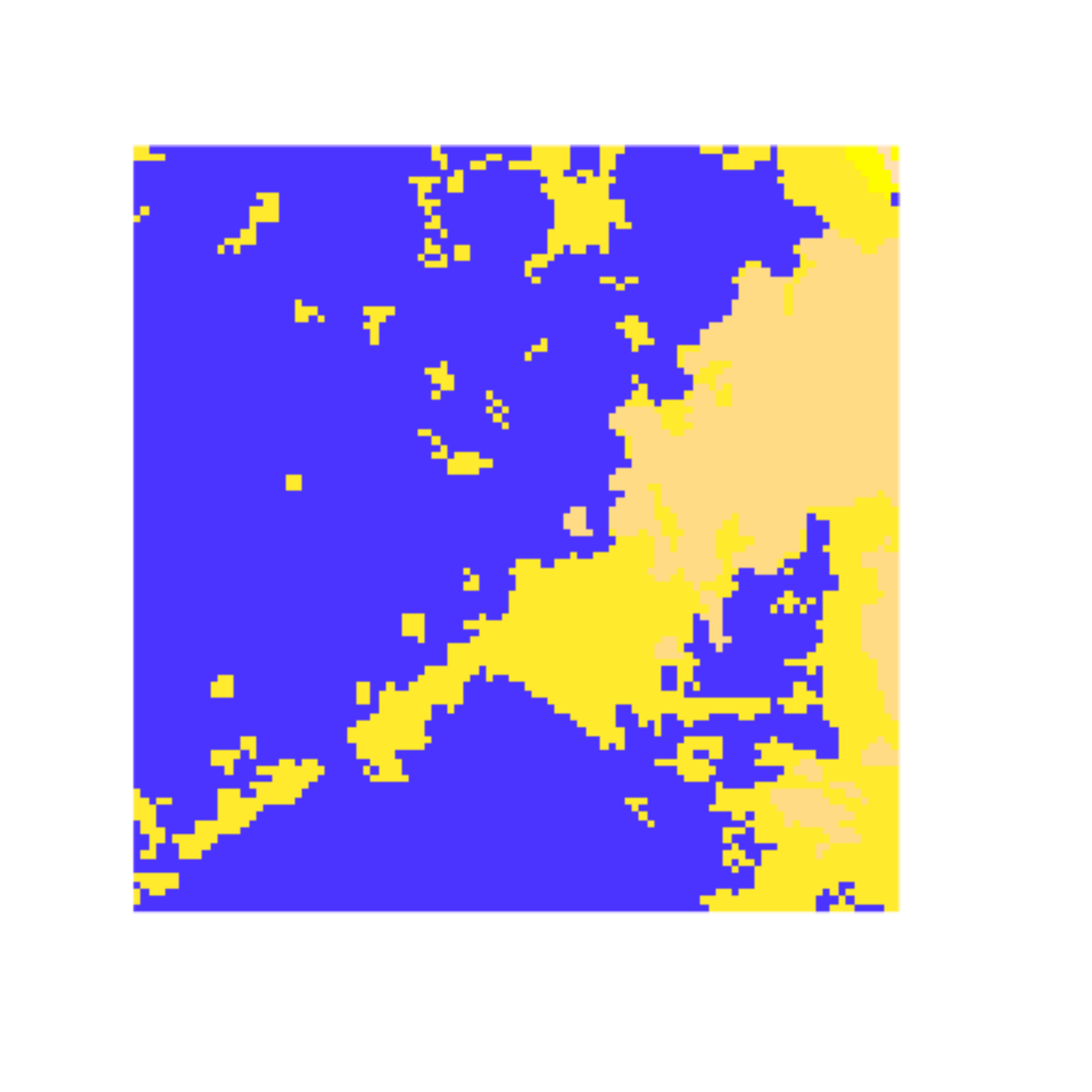}
      	     \caption{Land-use 2010}
      	     \label{fig:sfig1b}
      	   \end{subfigure}
      	   \begin{subfigure}{.4\textwidth}
      	     \centering
      	     \includegraphics[width=0.7\textwidth]{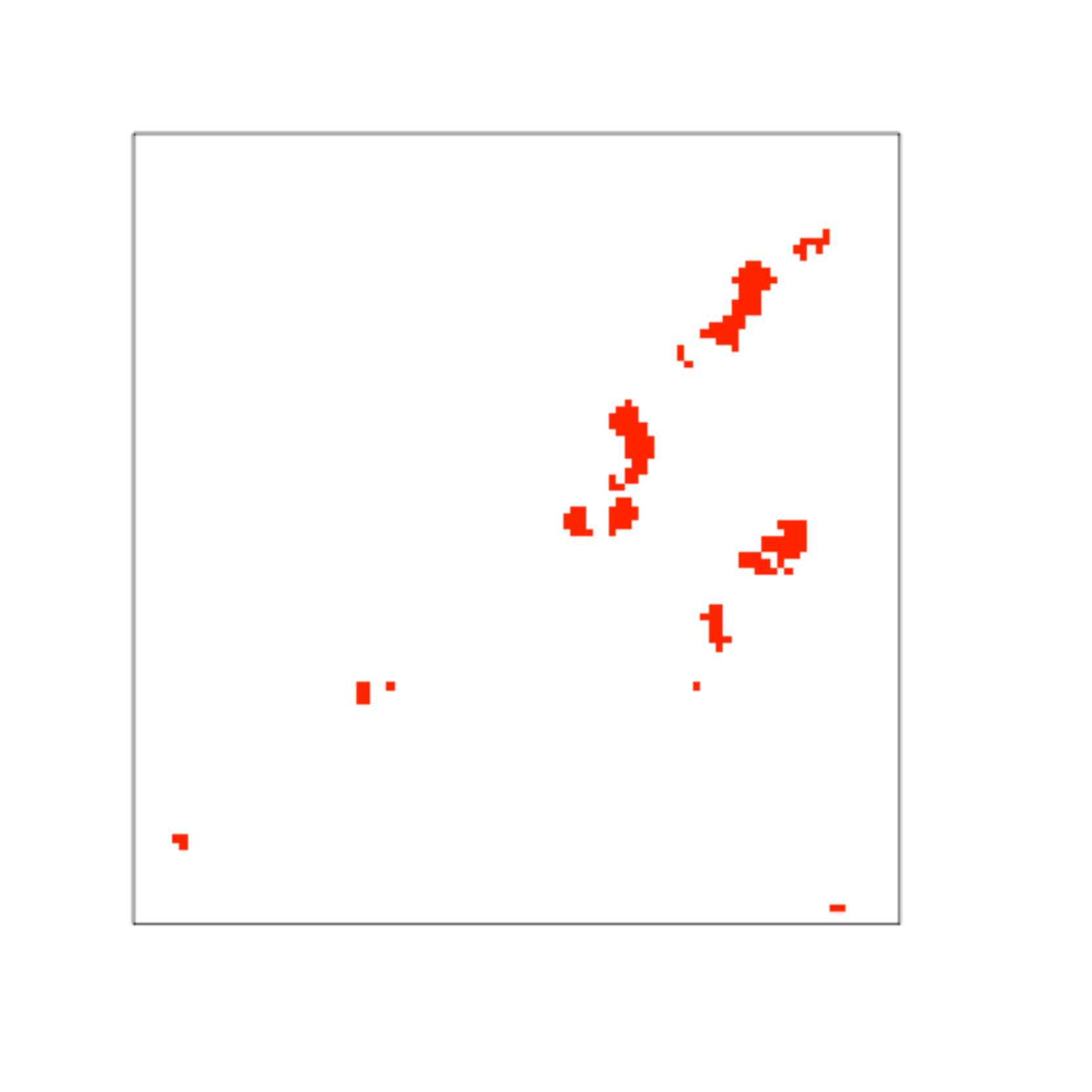}
      	     \caption{Deforestation 2005-2010}
      	     \label{fig:sfig1c}
      	   \end{subfigure}
      	   \begin{subfigure}{.4\textwidth}
      	     \centering
      	     \includegraphics[width=0.7\textwidth]{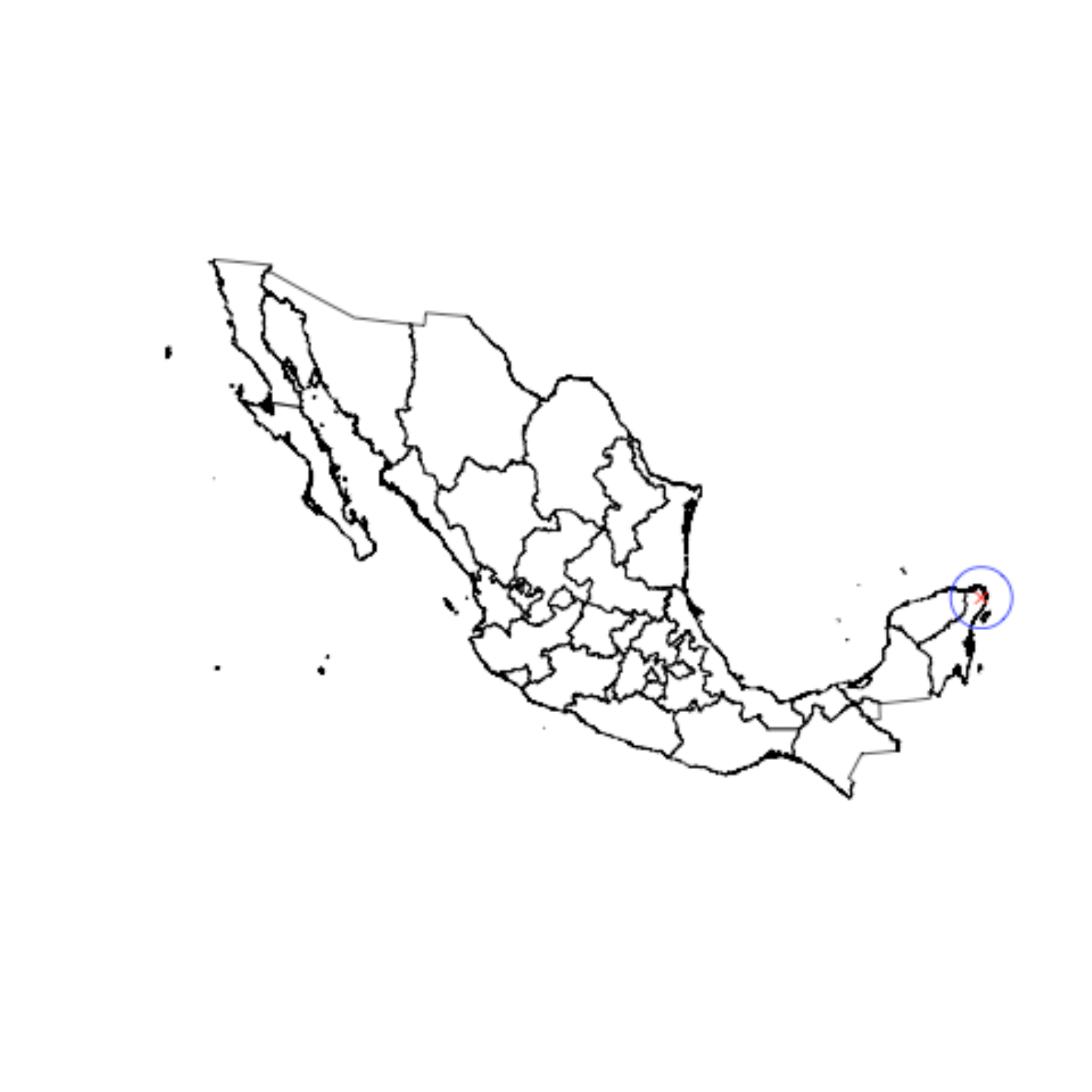}
      	     \caption{Location}
      	     \label{fig:sfig1d}
      	   \end{subfigure}   
         \begin{subfigure}{.5\textwidth}
           \centering
           \includegraphics[width=0.8\textwidth]{./2_Data/legend_landUse}
           \caption{Land use legend}
           \label{fig:legend}
         \end{subfigure}
      	   \caption[NALCMS land-use classification of Yucatan180 site, 2005 and 2010 at 250m resolution.]
      	   {NALCMS land-use classification of Yucatan180 site, 2005 and 2010 at 250m resolution.}
      	   \label{fig:landUseYucatan180}
      	 \end{figure}

		Figure \ref{fig:landUseYucatan180} illustrates the deforestation that occurred in the Yucatan180 site located in the south-east of Mexico, in the Yucatan penninsula. We can see that forest was cut down as the city expanded to the west and to the south. 
		
		Table \ref{tab:landUseYucatan180} shows that the forest that was lost in the 2005-2010 period was replaced by urban landuse and croplands. 
		
   		\begin{center}
   			\footnotesize\addtolength{\tabcolsep}{-2pt}
\begin{table}[ht]
\centering
\begingroup\tiny
\begin{tabular}{rrrrrrrrrrrrrrrrrrrrr}
  & \begin{sideways} Temperate or sub-polar needleleaf forest \end{sideways} & \begin{sideways} Sub-polar taiga needleleaf forest \end{sideways} & \begin{sideways} Tropical or sub-tropical broadleaf evergreen forest \end{sideways} & \begin{sideways} Tropical or sub-tropical broadleaf deciduous forest \end{sideways} & \begin{sideways} Temperate or sub-polar broadleaf deciduous forest \end{sideways} & \begin{sideways} Mixed Forest \end{sideways} & \begin{sideways} Tropical or sub-tropical shrubland \end{sideways} & \begin{sideways} Temperate or sub-polar shrubland \end{sideways} & \begin{sideways} Tropical or sub-tropical grassland \end{sideways} & \begin{sideways} Temperate or sub-polar grassland \end{sideways} & \begin{sideways} Sub-polar or polar shrubland-lichen-moss \end{sideways} & \begin{sideways} Sub-polar or polar grassland-lichen-moss \end{sideways} & \begin{sideways} Sub-polar or polar barren-lichen-moss \end{sideways} & \begin{sideways} Wetland \end{sideways} & \begin{sideways} Cropland \end{sideways} & \begin{sideways} Barren Lands \end{sideways} & \begin{sideways} Urban and Built-up \end{sideways} & \begin{sideways} Water \end{sideways} & \begin{sideways} Snow and Ice \end{sideways} & \begin{sideways} Total \end{sideways} \\ 
  \hline
Temperate or sub-polar needleleaf forest &    0 &    0 &    0 &    0 &    0 &    0 &    0 &    0 &    0 &    0 &    0 &    0 &    0 &    0 &    0 &    0 &    0 &    0 &    0 &    0 \\ 
  Sub-polar taiga needleleaf forest &    0 &    0 &    0 &    0 &    0 &    0 &    0 &    0 &    0 &    0 &    0 &    0 &    0 &    0 &    0 &    0 &    0 &    0 &    0 &    0 \\ 
  Tropical or sub-tropical broadleaf evergreen forest &    0 &    0 & 6682 &    0 &    0 &    0 &    0 &    0 &    0 &    0 &    0 &    0 &    0 &    0 &   34 &    0 &  146 &    0 &    0 & 6862 \\ 
  Tropical or sub-tropical broadleaf deciduous forest &    0 &    0 &    0 &    0 &    0 &    0 &    0 &    0 &    0 &    0 &    0 &    0 &    0 &    0 &    0 &    0 &    0 &    0 &    0 &    0 \\ 
  Temperate or sub-polar broadleaf deciduous forest &    0 &    0 &    0 &    0 &    0 &    0 &    0 &    0 &    0 &    0 &    0 &    0 &    0 &    0 &    0 &    0 &    0 &    0 &    0 &    0 \\ 
  Mixed Forest &    0 &    0 &    0 &    0 &    0 &    0 &    0 &    0 &    0 &    0 &    0 &    0 &    0 &    0 &    0 &    0 &    0 &    0 &    0 &    0 \\ 
   \hline
Tropical or sub-tropical shrubland &    0 &    0 &    0 &    0 &    0 &    0 &    0 &    0 &    0 &    0 &    0 &    0 &    0 &    0 &    0 &    0 &    0 &    0 &    0 &    0 \\ 
  Temperate or sub-polar shrubland &    0 &    0 &    0 &    0 &    0 &    0 &    0 &    0 &    0 &    0 &    0 &    0 &    0 &    0 &    0 &    0 &    0 &    0 &    0 &    0 \\ 
  Tropical or sub-tropical grassland &    0 &    0 &    0 &    0 &    0 &    0 &    0 &    0 &    0 &    0 &    0 &    0 &    0 &    0 &    0 &    0 &    0 &    0 &    0 &    0 \\ 
  Temperate or sub-polar grassland &    0 &    0 &    0 &    0 &    0 &    0 &    0 &    0 &    0 &    0 &    0 &    0 &    0 &    0 &    0 &    0 &    0 &    0 &    0 &    0 \\ 
  Sub-polar or polar shrubland-lichen-moss &    0 &    0 &    0 &    0 &    0 &    0 &    0 &    0 &    0 &    0 &    0 &    0 &    0 &    0 &    0 &    0 &    0 &    0 &    0 &    0 \\ 
  Sub-polar or polar grassland-lichen-moss &    0 &    0 &    0 &    0 &    0 &    0 &    0 &    0 &    0 &    0 &    0 &    0 &    0 &    0 &    0 &    0 &    0 &    0 &    0 &    0 \\ 
  Sub-polar or polar barren-lichen-moss &    0 &    0 &    0 &    0 &    0 &    0 &    0 &    0 &    0 &    0 &    0 &    0 &    0 &    0 &    0 &    0 &    0 &    0 &    0 &    0 \\ 
  Wetland &    0 &    0 &    0 &    0 &    0 &    0 &    0 &    0 &    0 &    0 &    0 &    0 &    0 &   27 &    0 &    0 &    0 &    0 &    0 &   27 \\ 
  Cropland &    0 &    0 &    0 &    0 &    0 &    0 &    0 &    0 &    0 &    0 &    0 &    0 &    0 &    0 & 1977 &    0 &  230 &    0 &    0 & 2207 \\ 
  Barren Lands &    0 &    0 &    0 &    0 &    0 &    0 &    0 &    0 &    0 &    0 &    0 &    0 &    0 &    0 &    0 &    0 &    0 &    0 &    0 &    0 \\ 
  Urban and Built-up &    0 &    0 &    0 &    0 &    0 &    0 &    0 &    0 &    0 &    0 &    0 &    0 &    0 &    0 &    0 &    0 &  897 &    0 &    0 &  897 \\ 
  Water &    0 &    0 &    0 &    0 &    0 &    0 &    0 &    0 &    0 &    0 &    0 &    0 &    0 &    0 &    0 &    0 &    0 &    7 &    0 &    7 \\ 
  Snow and Ice &    0 &    0 &    0 &    0 &    0 &    0 &    0 &    0 &    0 &    0 &    0 &    0 &    0 &    0 &    0 &    0 &    0 &    0 &    0 &    0 \\ 
   \hline
Total &    0 &    0 & 6682 &    0 &    0 &    0 &    0 &    0 &    0 &    0 &    0 &    0 &    0 &   27 & 2011 &    0 & 1273 &    7 &    0 & 10000 \\ 
   \hline
\end{tabular}
\endgroup
\caption[NALCMS land-use distribution for site Yucatan180, 2005 and 2010.]{NALCMS land-use distribution for site Yucatan180, 2005 and 2010.} 
\label{tab:landUseYucatan180}
\end{table}

   		\end{center}
		
\begin{figure}[H]
   		  \centering
   		  \includegraphics[width=.5\textwidth]{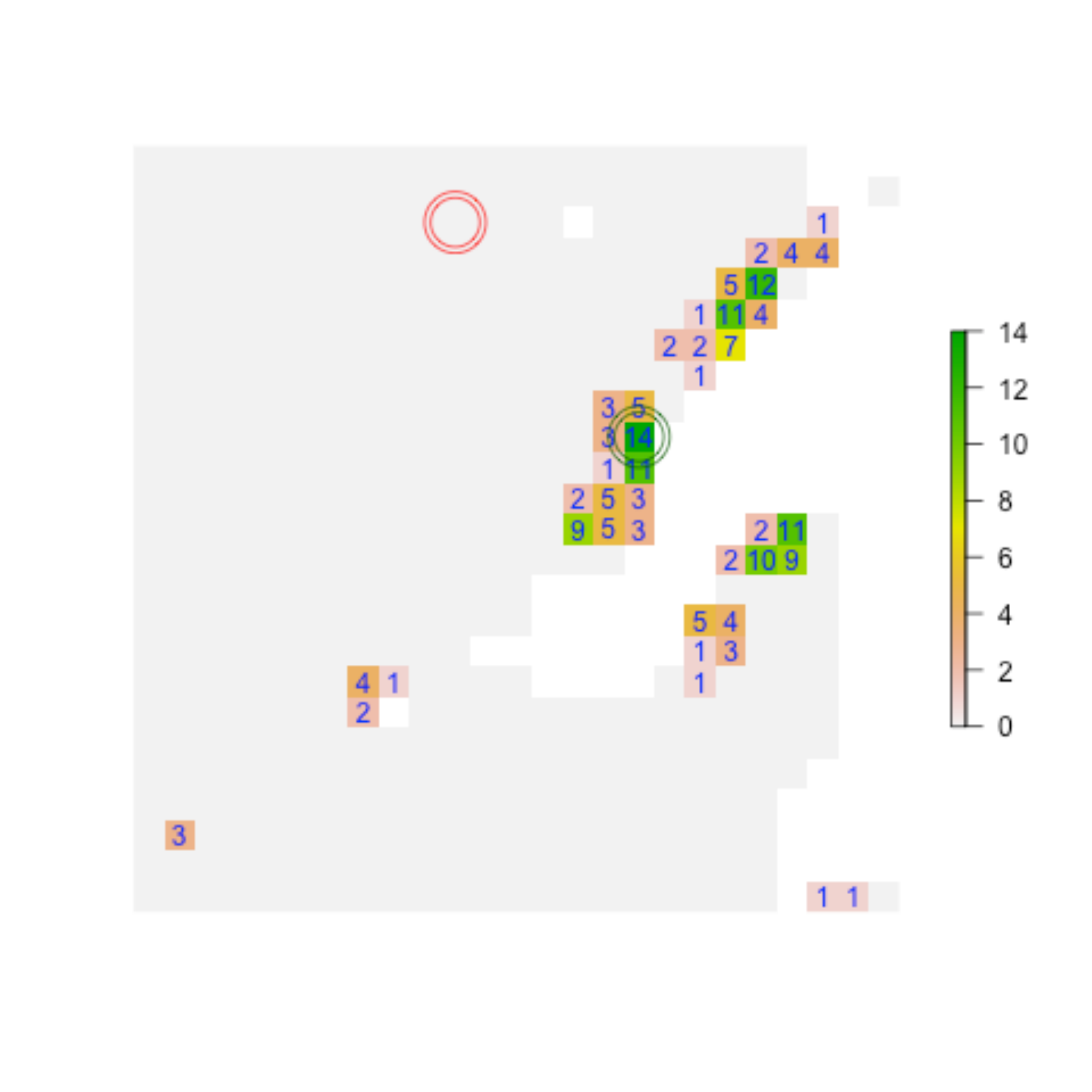} 
   		  \caption[Deforestation in Yucatan180 site at 1km resolution]
   		  {Deforestation in Yucatan180 site at 1km resolution: number of 250m pixels deforested out of 16}
   		  \label{fig:defoYucatan180}
   		\end{figure}

\begin{figure}[H]
   	      \begin{subfigure}{.5\textwidth}
   	        \centering
   	    	\includegraphics[width=1\textwidth]{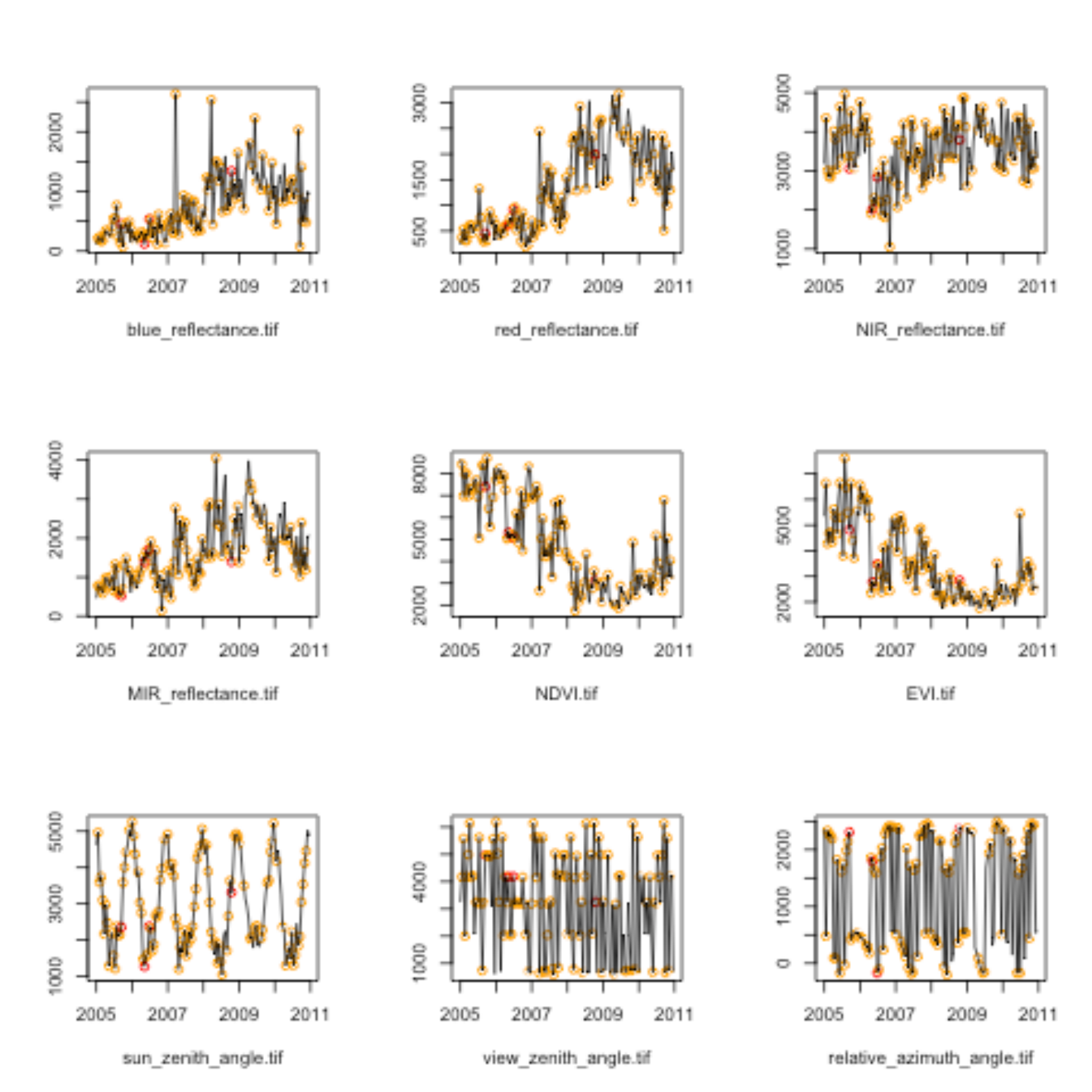} 
   	        \caption{deforested pixel}
   	        \label{fig:defo_pix}
   	      \end{subfigure}%
   	      \begin{subfigure}{.5\textwidth}
   	        \centering
   	        \includegraphics[width=1\textwidth]{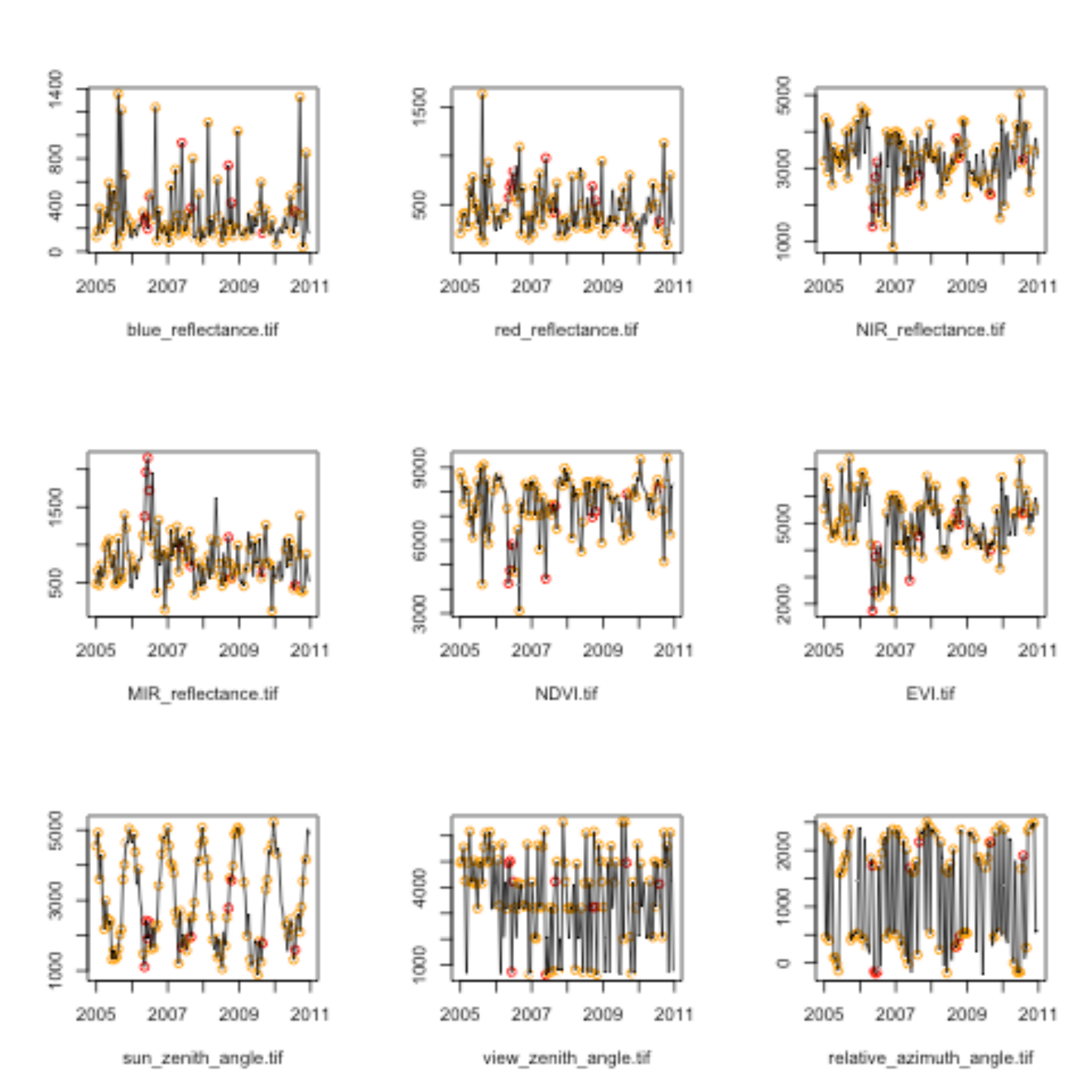}
   	        \caption{non-deforested pixel}
   	        \label{fig:non_defo_pix}
   	      \end{subfigure}
   	      \caption[Reflectance, vegetation indices and sun-sensor geometry time series for deforested and non-deforested pixel in Yucatan 180 site.]
   	      {Reflectance, vegetation indices and sun-sensor geometry time series for deforested and non-deforested pixel in Yucatan180 site, identified in figure \ref{fig:defoYucatan180} with green and red circles respectively. Orange marked points are \emph{marginal data} and red marked points are \emph{cloudy} or \emph{not processed}}
   	      \label{fig:timeSeriesYucatan180}
   	    \end{figure}

		 Figure \ref{fig:defoYucatan180} shows which pixels in the Yucatan180 site were deforested and also the degree of deforestation: the number of 250m resolution deforested pixels, out of a possible 16, is labeled for deforested pixels. Additionally, two pixels, one deforested and one non-deforested, are identified with circles so that we may observe the behavior of their reflectance in the 2005-2010 period in figure \ref{fig:timeSeriesYucatan180}. 
		
		In figure \ref{fig:timeSeriesYucatan180} the sun-sensor geometry, surface reflectance and vegetation index time series are plotted for the 2005-2010 period for the two pixes identified in figure \ref{fig:defoYucatan180}. We can see, by looking at the NDVI time series, that the deforestation event for the deforested pixel occurred during 2007.

	\section{Grid search results by site} \label{s:gridSearch} 
	
	In this section we will optimize the thresholds used on a site-by-site basis using information from table \ref{tab:sites} regarding what time series $b$ (i.e. which band) is appropriate. We want to maximize our utility function $tss$, which depends on the threshold $L$, the relevant time series $\hat{\epsilon}_{b,p,t}$, and on $z_p$, whether there was deforestation or not from 2005-2010, for pixel $p$:
	
	\begin{align} \label{utility_univariate}
		& tss(L; C, \{z_p\}_{p \in \mathcal{S}(p)},\{\hat{\epsilon}_{b(p),p,t}^j\}_{p \in \mathcal{S}(p),j \in \{1,...,5\},t \in \mathcal{P}(j,p)} ) =  \nonumber \\ 
		& \frac{S(L; C, \{\hat{\epsilon}_{b(p),p,t}^j\}_{p \in \mathcal{S}(p),j \in \{1,...,5\},t \in \mathcal{P}(j,p)} )}{N_1(\{z_p\}_{p \in \mathcal{S}(p)})} + \frac{U(L; C, \{\hat{\epsilon}_{b(p),p,t}^j\}_{p \in \mathcal{S}(p),j \in \{1,...,5\},t \in \mathcal{P}(j,p)} )}{N_0(\{z_p\}_{p \in \mathcal{S}(p)})}-1 = \nonumber \\
		& \frac{\sum_{p \in \mathcal{S}} z_p \bigg(\prod_{j=1}^5 \prod_{t \in \mathcal{P}_1(j,p)} h_L(\{\hat{\epsilon}_{b(p),p,s}^j\}_{s \in \{t, t+1, t+C-1\}}) \bigg)}{\sum_{p \in \mathcal{S}} z_p} + \nonumber \\
		& \frac{\sum_{p \in \mathcal{S}} (1-z_p) \bigg(1-\prod_{j=1}^5 \prod_{t \in \mathcal{P}_1(j,p)} h_L(\{\hat{\epsilon}_{b(p),p,s}^j\}_{s \in \{t, t+1, t+C-1\}}) \bigg)}{|\mathcal{S}|-\sum_{p \in \mathcal{S}} z_p} - 1
	\end{align}	
	
	where
	\begin{itemize}
		\item $b(p) \in \{2,8\}$ is the appropriate band (NIR or NDVI) for pixel $p$, according to table \ref{tab:sites} depending on which site pixel $p$ belongs to,
		\item $\mathcal{S}(p)$ is site to which pixel $p$ belongs,
		\item $\mathcal{P}(j,p)$ is the set of dates with clear reflectance observations for pixel $p$ and prediction \textbf{window}  $j$ i.e. as described in section \ref{s:adapt} and in figure \ref{fig:trainScheme_adapt}.
		\item $\mathcal{P}_1(j,p)$ is the set of dates with clear reflectance observations for pixel $p$ and  prediction \textbf{year} $j$,
		\item the super index $j$ indicates the training window $\mathcal{T}_{j,p}$ used to train the reflectance model parameters and obtain the errors $\hat{\epsilon}_{b,p,t}^j$, and
		\item the thresholding function $h_L(\cdot)$ with \textbf{univariate} threshold $L$ is defined as follows:
		
		\begin{align} \label{thrs_univariate}
			h_L(\{\hat{\epsilon}_{b(p),p,s}^j\}_{s \in \{t, t+1, t+C-1\}}) := \prod_{r=0}^{C-1} \mathbbm{1}_{\{\hat{\epsilon}_{b(p),p,t+r}^j > L\}} + \prod_{r=0}^{C-1} \mathbbm{1}_{\{\hat{\epsilon}_{b(p),p,t+r}^j < -L\}} 
		\end{align}	
	\end{itemize}	
	
	Tables \ref{tab:grdSrch_consecVaries} and \ref{tab:grdSrch_consecVaries_CV} show the optimized thresholds, optimal $tss$ values and user/producer accuracy obtained for each site and for different values of the parameter $C$ (2-6). Performance measures were estimated using the whole training sample and using cross validation respectively. 
	
	\begin{center}
		\footnotesize\addtolength{\tabcolsep}{-2pt}
\begin{table}[ht]
\centering
\begingroup\tiny
\begin{tabular}{ll|ccccc|ccccc|ccccc}
     &           &  \multicolumn{5}{|c|}{optimal threshold} &  \multicolumn{5}{|c|}{1-tss} &  \multicolumn{5}{|c|}{user/producer accuracy}\\	
	\hline
 site & variable & 2 & 3 & 4 & 5 & 6 & 2 & 3 & 4 & 5 & 6 & 2 & 3 & 4 & 5 & 6 \\ 
  \hline
Sonora232 & NIR\_reflectance & 0.09 & 0.08 & 0.08 & 0.06 & 0.05 & 83.42 & 70.45 & 58.79 & 52.87 & 59.23 & 25.61 & 31.51 & 46.59 & 54.21 & 48.75 \\ 
  Jalisco164 & NIR\_reflectance & 0.13 & 0.11 & 0.09 & 0.06 & 0.05 & 54.41 & 49.37 & 41.96 & 36.69 & 33.73 & 46.20 & 54.10 & 63.98 & 63.81 & 64.81 \\ 
  Nayarit151 & NIR\_reflectance & 0.13 & 0.10 & 0.08 & 0.06 & 0.04 & 35.74 & 25.97 & 23.92 & 43.78 & 39.37 & 50.07 & 63.77 & 50.00 & 60.15 & 60.15 \\ 
  Nayarit109 & NIR\_reflectance & 0.13 & 0.10 & 0.10 & 0.06 & 0.05 & 37.50 & 40.87 & 49.60 & 51.19 & 47.82 & 59.62 & 38.89 & 41.27 & 39.61 & 37.04 \\ 
   \hline
Yucatan180 & NDVI & 0.28 & 0.22 & 0.20 & 0.12 & 0.12 & 82.65 & 78.84 & 74.38 & 69.10 & 79.34 & 32.58 & 23.69 & 22.42 & 11.86 & 11.91 \\ 
  QuintanaRoo100 & NDVI & 0.24 & 0.18 & 0.17 & 0.09 & 0.09 & 76.47 & 83.21 & 81.80 & 77.42 & 81.41 & 9.13 & 5.72 & 21.64 & 14.58 & 15.15 \\ 
  Michoacan98 & NDVI & 0.21 & 0.17 & 0.13 & 0.11 & 0.09 & 72.71 & 59.11 & 58.78 & 57.99 & 53.77 & 13.56 & 20.02 & 23.39 & 24.05 & 32.39 \\ 
  QuintanaRoo77 & NDVI & 0.24 & 0.19 & 0.18 & 0.13 & 0.11 & 81.54 & 68.25 & 65.80 & 53.24 & 54.62 & 9.43 & 16.72 & 15.39 & 27.05 & 25.66 \\ 
  Sonora74 & NDVI & 0.24 & 0.22 & 0.25 & 0.17 & 0.22 & 49.83 & 58.98 & 60.68 & 46.61 & 60.51 & 28.64 & 36.67 & 38.18 & 45.00 & 31.76 \\ 
   \hline
\end{tabular}
\endgroup
\caption{optimal threshold, 1-tss and user/producer accuracy for each site (training data)} 
\label{tab:grdSrch_consecVaries}
\end{table}

	\end{center}

	\begin{center}
		\footnotesize\addtolength{\tabcolsep}{-2pt}
\begin{table}[ht]
\centering
\begingroup\tiny
\begin{tabular}{ll|ccccc|ccccc|ccccc}
&           &  \multicolumn{5}{|c|}{optimal threshold} &  \multicolumn{5}{|c|}{1-tss} &  \multicolumn{5}{|c|}{user/producer accuracy}\\		
\hline
 site & variable & 2 & 3 & 4 & 5 & 6 & 2 & 3 & 4 & 5 & 6 & 2 & 3 & 4 & 5 & 6 \\ 
  \hline
Sonora232 & NIR\_reflectance & 0.09 & 0.08 & 0.08 & 0.06 & 0.05 & 88.81 & 69.41 & 61.04 & 54.38 & 59.57 & 24.78 & 30.29 & 47.93 & 54.54 & 50.22 \\ 
  Jalisco164 & NIR\_reflectance & 0.13 & 0.11 & 0.09 & 0.06 & 0.05 & 58.36 & 52.56 & 43.31 & 46.41 & 34.00 & 49.97 & 55.12 & 63.88 & 66.64 & 62.69 \\ 
  Nayarit151 & NIR\_reflectance & 0.13 & 0.10 & 0.08 & 0.06 & 0.04 & 37.89 & 33.61 & 26.35 & 44.90 & 45.48 & 57.04 & 59.72 & 57.10 & 63.80 & 61.30 \\ 
  Nayarit109 & NIR\_reflectance & 0.13 & 0.10 & 0.10 & 0.06 & 0.05 & 47.82 & 49.21 & 53.57 & 64.29 & 52.58 & 60.00 & 41.81 & 42.64 & 40.05 & 37.18 \\ 
   \hline
Yucatan180 & NDVI & 0.28 & 0.22 & 0.20 & 0.12 & 0.12 & 85.98 & 87.23 & 80.86 & 72.64 & 83.79 & 32.84 & 24.55 & 23.54 & 12.16 & 10.81 \\ 
  QuintanaRoo100 & NDVI & 0.24 & 0.18 & 0.17 & 0.09 & 0.09 & 80.77 & 93.13 & 93.42 & 99.93 & 95.60 & 9.77 & 4.29 & 16.90 & 14.31 & 15.18 \\ 
  Michoacan98 & NDVI & 0.21 & 0.17 & 0.13 & 0.11 & 0.09 & 77.64 & 66.40 & 59.85 & 63.14 & 60.49 & 11.95 & 27.74 & 24.52 & 25.70 & 30.94 \\ 
  QuintanaRoo77 & NDVI & 0.24 & 0.19 & 0.18 & 0.13 & 0.11 & 100.78 & 82.64 & 65.71 & 57.76 & 60.18 & 16.24 & 26.52 & 12.41 & 40.72 & 26.29 \\ 
  Sonora74 & NDVI & 0.24 & 0.22 & 0.25 & 0.17 & 0.22 & 71.07 & 82.07 & 83.15 & 71.41 & 84.77 & 29.44 & 35.83 & 36.67 & 36.39 & 32.22 \\ 
   \hline
\end{tabular}
\endgroup
\caption[optimal threshold, 1-tss and user/producer accuracy for each site (cross validation)]{optimal threshold, 1-tss and user/producer accuracy for each site (cross validation)} 
\label{tab:grdSrch_consecVaries_CV}
\end{table}

	\end{center}
	
	Tables \ref{tab:grdSrch_consecVaries} and \ref{tab:grdSrch_consecVaries_CV} show that the performance of the algorithm, optimized on a site-by-site basis,  varies a lot between the different sites. More importantly, the optimal thresholds vary signifcantly for the \emph{forest to urban or cropland} sites. This will represent a complication since we eventually want to apply a chosen threshold uniformly to all of Mexico.

\begin{figure}[H]
      \begin{subfigure}{.5\textwidth}
        \centering
    	\includegraphics[width=1\textwidth]{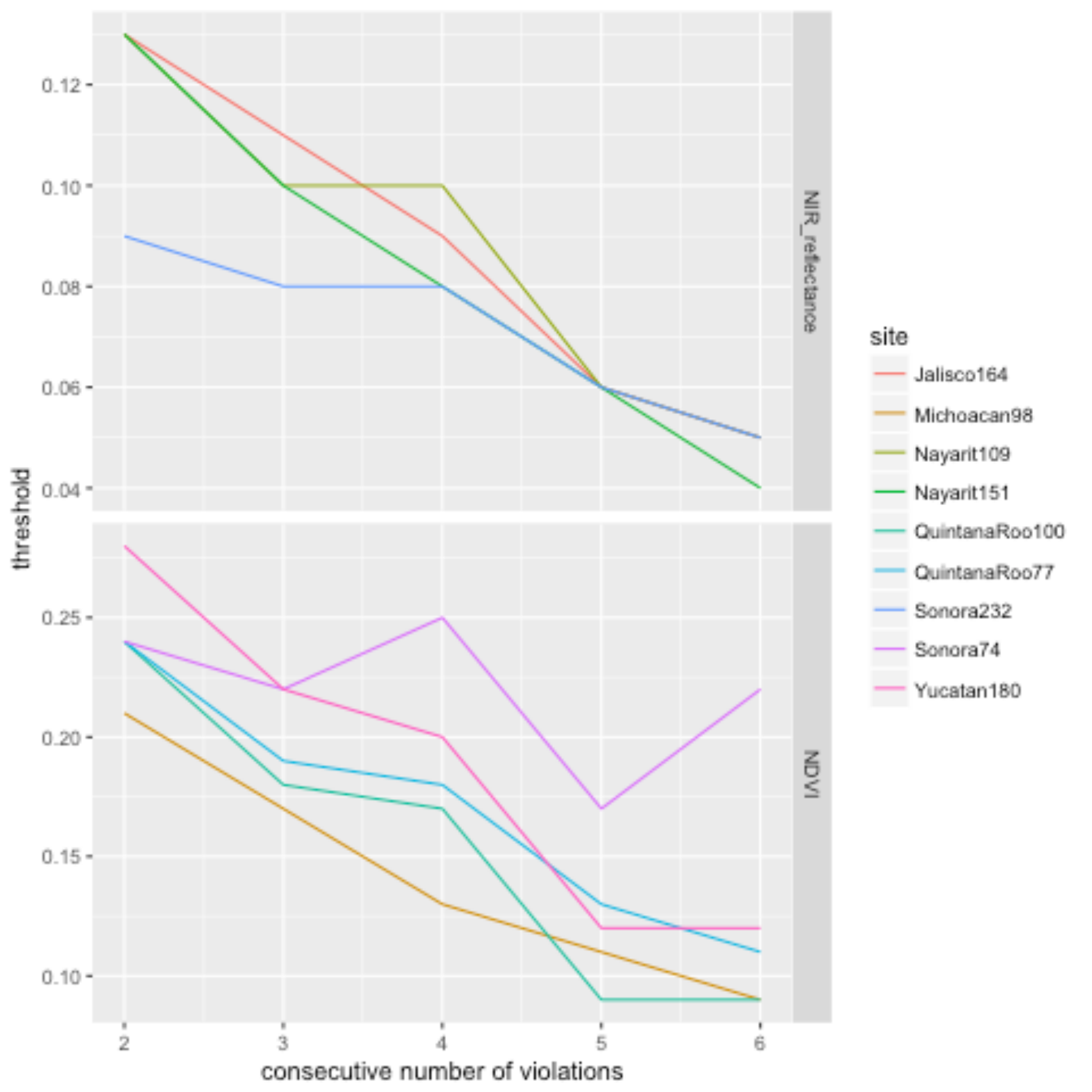} 
        \caption{Optimal thresholds by site and consecutive number of violations}
        \label{fig:opt_thrs}
      \end{subfigure}%
      \begin{subfigure}{.5\textwidth}
        \centering
        \includegraphics[width=1\textwidth]{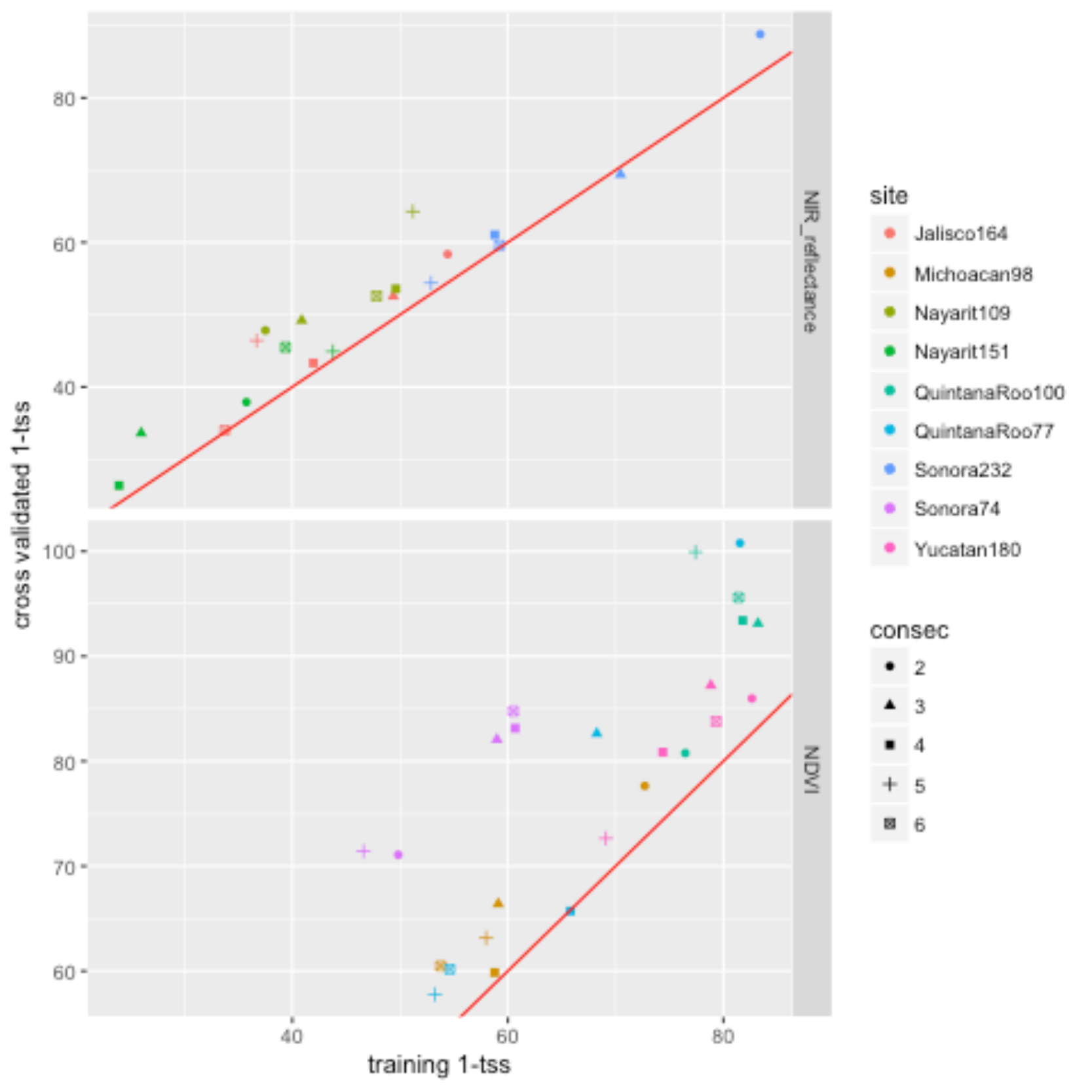}
        \caption{Training vs. cross-validated $1-tss$}
        \label{fig:train_cv}
      \end{subfigure}
      \caption[Comparison of optimal thresholds by site and comparison between training and cross validated results for grid-search]
      {Comparison of optimal thresholds by site and comparison between training and cross validated results for grid-search}
      \label{fig:grid_srch_train_cv}
    \end{figure}

	Figure \ref{fig:opt_thrs} confirms graphically that the optimal thresholds vary a lot between the different site especially for \emph{forest to urban or cropland sites}. Figure \ref{fig:train_cv} shows that the training algorithm overfits the threshold more for NDVI prediction errors since the cross-validated performance results are significantly higher than those when the thresholds were trained on the entire site sample. 
	
\begin{figure}[H]
	  \centering
	  \includegraphics[width=.5\textwidth]{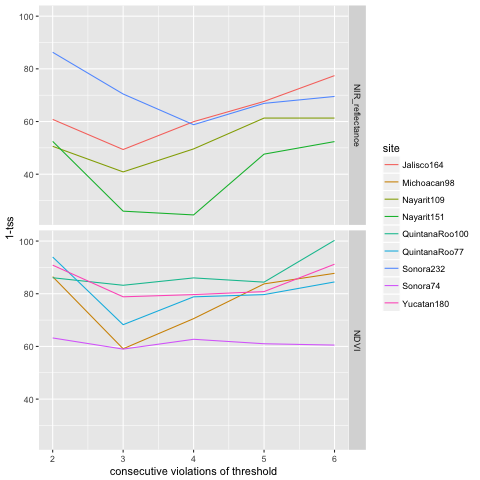} 
	  \caption[$1-tss$ results for each site trained separately. Results for optimal threshold for 3 consecutive violations.]
	  {$1-tss$ results for each site trained separately. Results for optimal threshold for 3 consecutive violations.}
	  \label{fig:grdSrch_consecFix}
	\end{figure}

	In order to visualize how the performance changes as we vary the parameter $C$, but while keeping the threshold fixed, we applied the optimal threshold found for $C=3$ accross the board, i.e. to values of $C$ in the 2-6 range. Table \ref{tab:grdSrch_consecFix} and figure \ref{fig:grdSrch_consecFix} show the results. 
	
	\begin{center}
		\footnotesize\addtolength{\tabcolsep}{-2pt}
\begin{table}[ht]
\centering
\begingroup\tiny
\begin{tabular}{lcl|ccccc}
&           &          & \multicolumn{5}{|c|}{\# of consecutive violations}\\
\hline	  
 site & threshold & variable & 2 & 3 & 4 & 5 & 6 \\ 
  \hline
Sonora232 & 0.080 & NIR\_reflectance & 86.293 & 70.448 & 58.785 & 66.891 & 69.523 \\ 
  Jalisco164 & 0.110 & NIR\_reflectance & 60.816 & 49.371 & 59.947 & 67.624 & 77.429 \\ 
  Nayarit151 & 0.100 & NIR\_reflectance & 52.462 & 25.974 & 24.540 & 47.619 & 52.381 \\ 
  Nayarit109 & 0.100 & NIR\_reflectance & 50.595 & 40.873 & 49.603 & 61.310 & 61.310 \\ 
   \hline
Yucatan180 & 0.220 & NDVI & 90.859 & 78.837 & 79.678 & 80.814 & 91.206 \\ 
  QuintanaRoo100 & 0.180 & NDVI & 86.074 & 83.214 & 86.014 & 84.406 & 100.268 \\ 
  Michoacan98 & 0.170 & NDVI & 86.486 & 59.106 & 70.547 & 83.742 & 87.748 \\ 
  QuintanaRoo77 & 0.190 & NDVI & 93.936 & 68.251 & 78.826 & 79.675 & 84.477 \\ 
  Sonora74 & 0.220 & NDVI & 63.220 & 58.983 & 62.712 & 61.017 & 60.508 \\ 
   \hline
\end{tabular}
\endgroup
\caption[1-tss for each site trained separately with time series based on type of land-use change ]{1-tss for each site trained separately with time series based on type of land-use change } 
\label{tab:grdSrch_consecFix}
\end{table}

	\end{center}
	
	It seems from the results shown in table \ref{tab:grdSrch_consecFix} and figure \ref{fig:grdSrch_consecFix} that the algorithm performs best for 3-4 consecutive violations of the threshold. 
	
	In sections \ref{ss:grdSrchWater} and \ref{ss:grdSrchUrban} we illustrate the grid-search optimization procedure used to determine the optimal thresholds and visualize the performance of the algorithm spatially  for one of the \emph{forest to water} sites, Sonora232, and for one of the \emph{forest to cropland or urban} sites, Yucatan180. 
	

		\subsection{Forest to water sites} \label{ss:grdSrchWater}

		\subsubsection{Sonora232}
		
\begin{figure}[H]
		  \centering
		  \includegraphics[width=.5\textwidth]{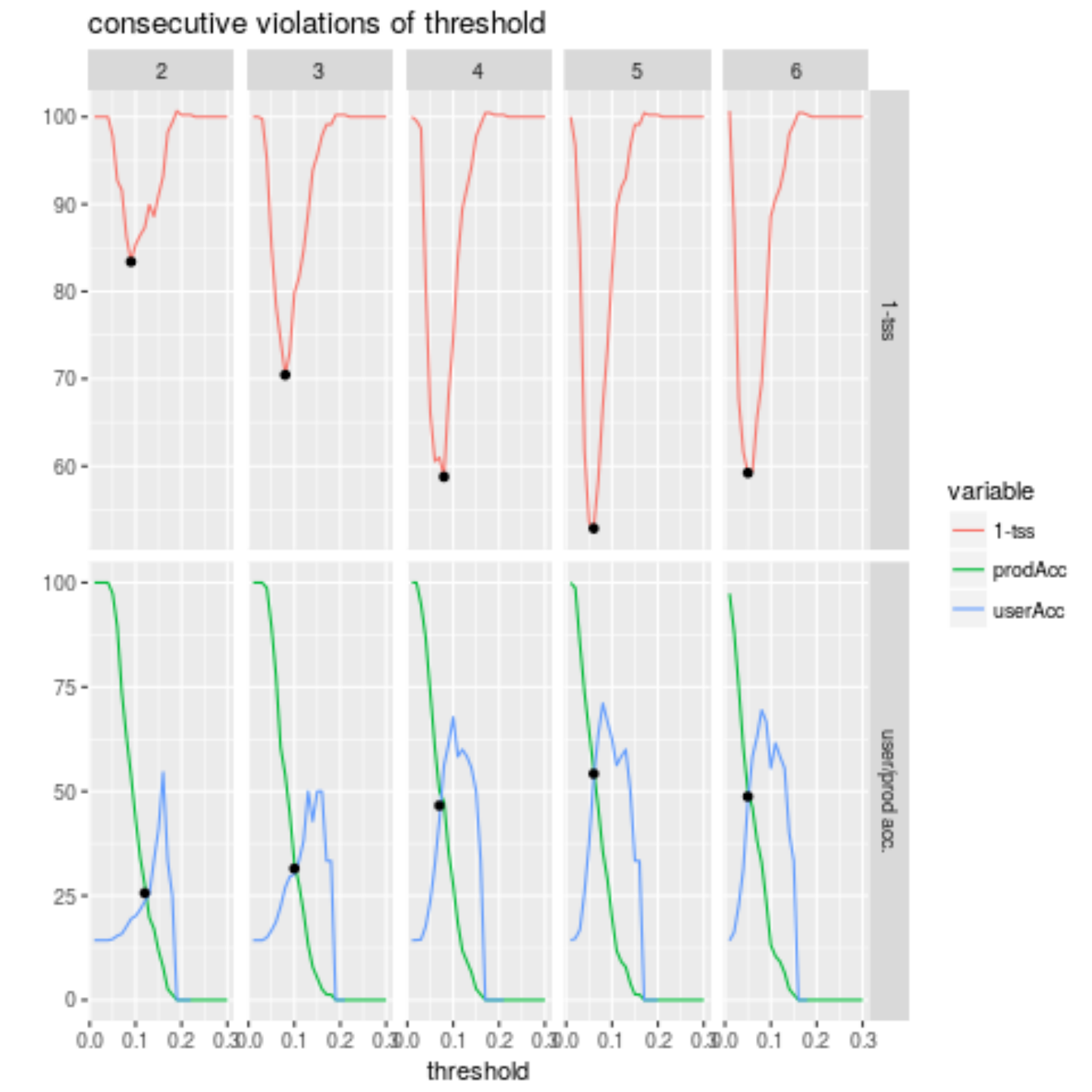} 
		  \caption[Grid search for optimal NIR reflectance threshold for Sonora232 site (training $1-tss$).]
		  {Grid search for optimal NIR reflectance threshold for Sonora232 site (training $1-tss$).}
		  \label{fig:grdSrch_Sonora232}
		\end{figure}

\begin{figure}[H]
	      \begin{subfigure}{.5\textwidth}
	        \centering
	    	\includegraphics[width=1\textwidth]{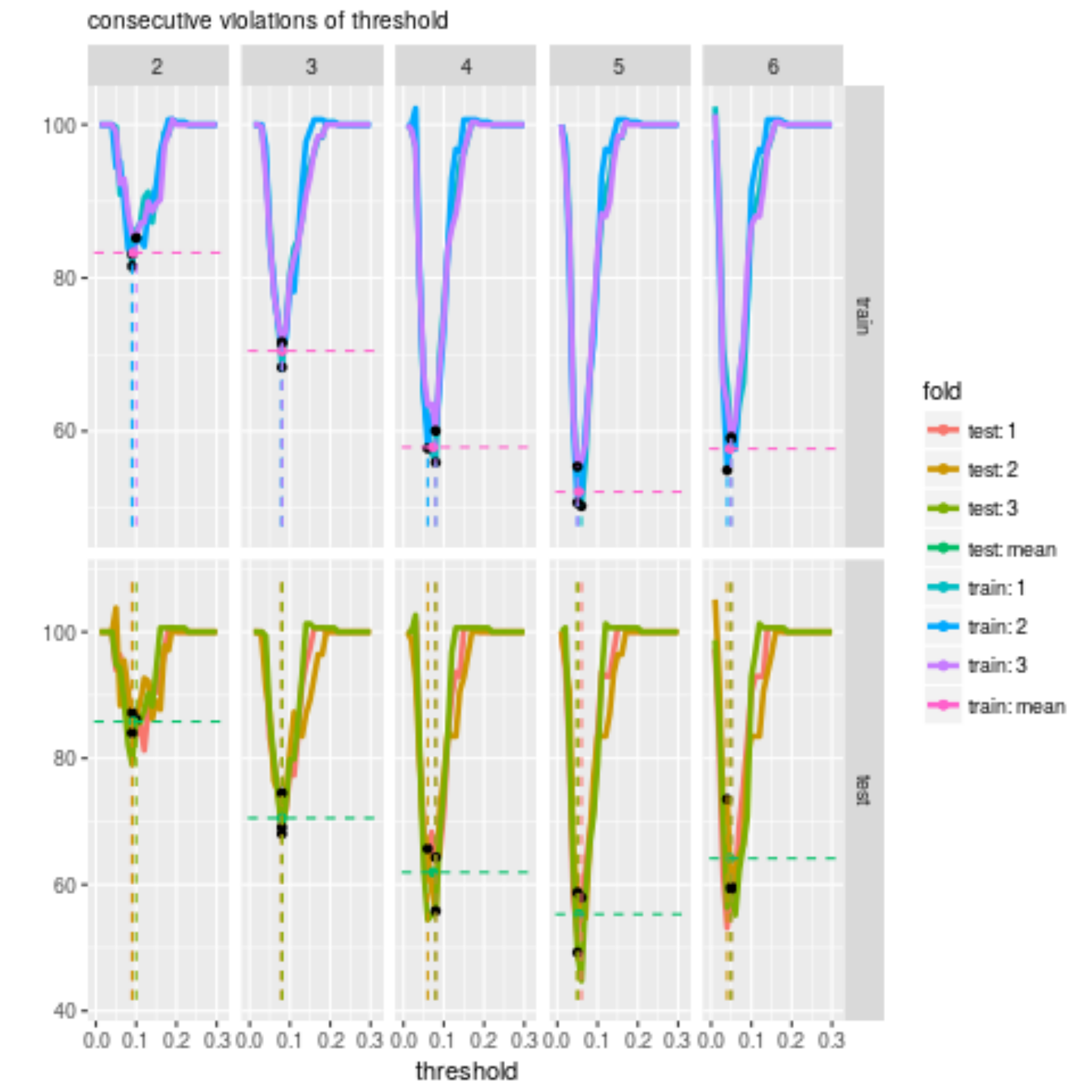} 
	        \caption{$1-tss$}
	        \label{fig:defo_pix}
	      \end{subfigure}%
	      \begin{subfigure}{.5\textwidth}
	        \centering
	        \includegraphics[width=1\textwidth]{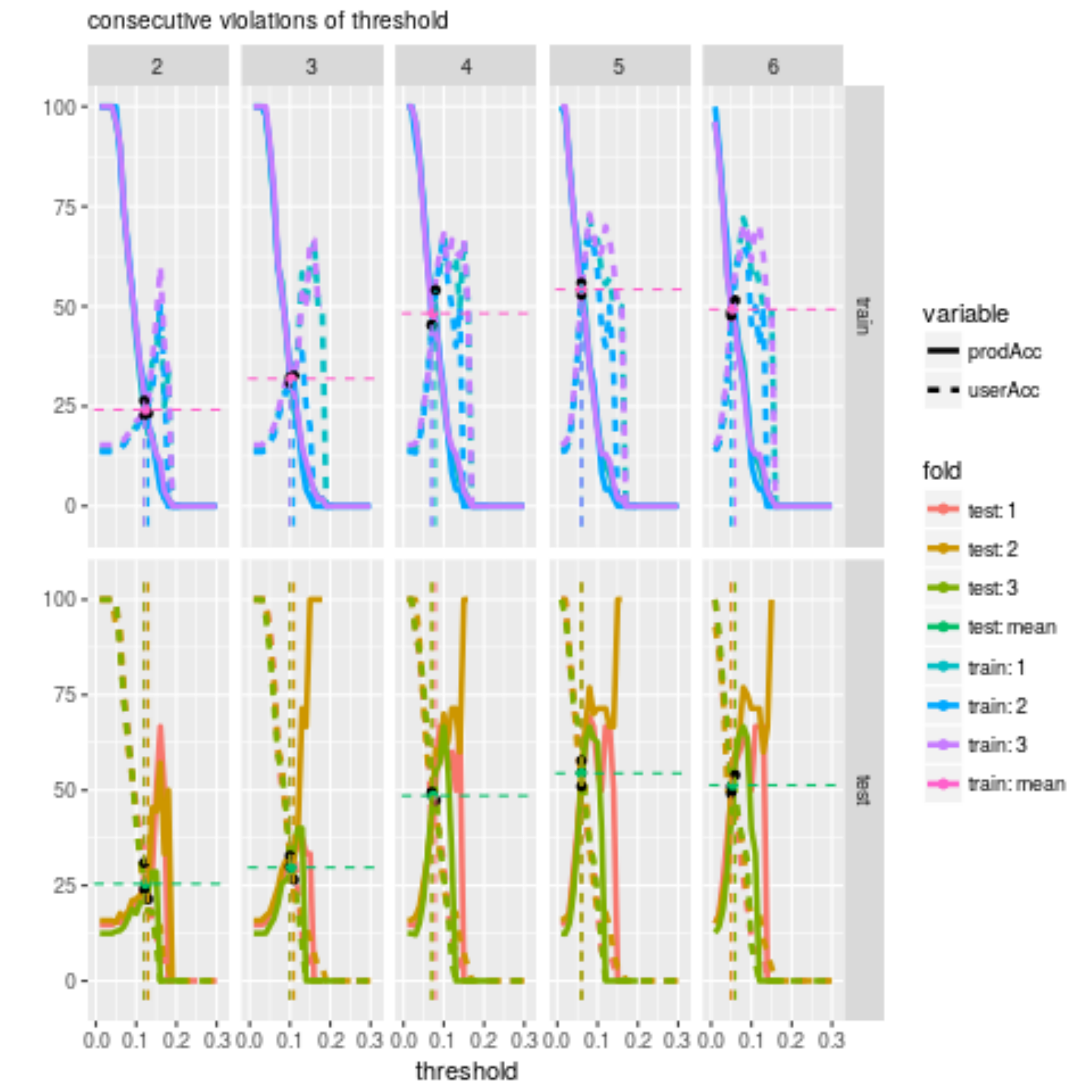}
	        \caption{$a_p$ and $a_u$}
	        \label{fig:non_defo_pix}
	      \end{subfigure}
	      \caption[Grid search for optimal NIR reflectance threshold for Sonora232 site (cross-validated $1-tss$, $a_p$ and $a_u$).]
	      {Grid search for optimal NIR reflectance threshold for Sonora232 site (cross-validated $1-tss$, $a_p$ and $a_u$).}
	      \label{fig:grid_srch_cv_Sonora232}
	    \end{figure}

		Figures \ref{fig:grdSrch_Sonora232} and \ref{fig:grid_srch_cv_Sonora232} illustrate the grid-search procedure used to find the optimal NIR reflectance prediction error threshold, for different values of $C$, in the Sonora232 site. 

		Figures \ref{fig:grdSrch_Sonora232} and \ref{fig:grid_srch_cv_Sonora232} show that, for the Sonora232 site, we obtain optimal $tss$ performance when we fix the consecutive number of violations of threshold $C$ at five. 
		
\begin{figure}[H]
		  \centering
		  \includegraphics[width=.5\textwidth]{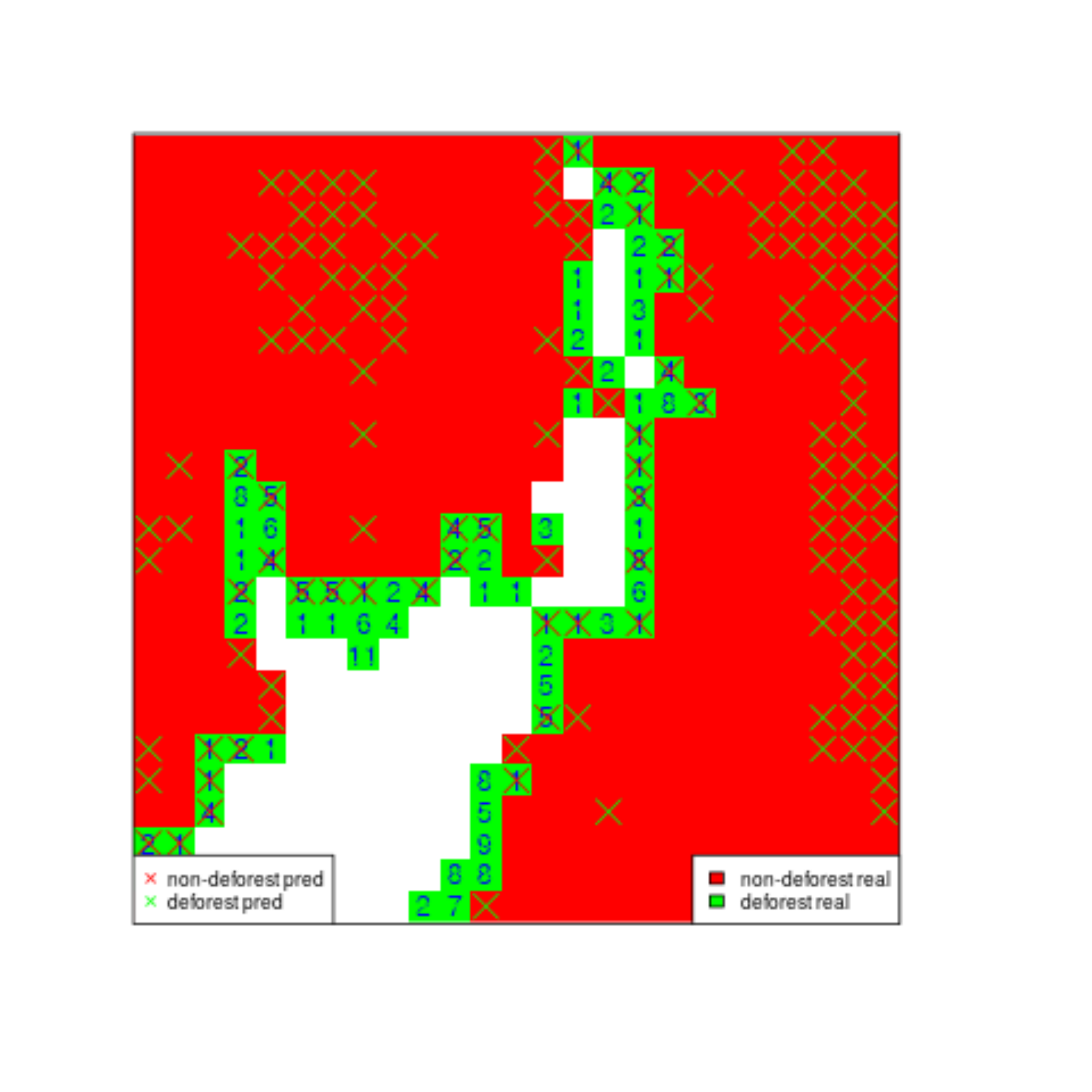} 
		  \caption[Spatial results for Sonora232 site of applying optimal threshold using 3 consecutive violation rule.]
		  {Spatial results for Sonora232 site of applying optimal threshold using 3 consecutive violation rule.}
		  \label{fig:grdSrch_spatial_Sonora232}
		\end{figure}

		Figure \ref{fig:grdSrch_spatial_Sonora232} displays the results of applying the algorithm to the Sonora232 site, for parameter value $C=3$ and the associated optimal NIR reflectance prediction error threshold, spatially. Green and red background color represents deforested and non-deforested pixels while green and red crosses represent the \emph{deforestation} model prediction. The degree of deforestation (the number of 250m resolution deforested pixels, out of a possible 16) is also displayed so that we can explore its impact on deforestation detection. 
		
		Figure \ref{fig:grdSrch_spatial_Sonora232} shows that most of the errors occur in patches. However, there are a few isolated pixels where the algorithm did not classify correctly. In these cases spatial smoothing of classification predictions could help improve the performance. 
	
		\subsection{Forest to urban or cropland sites} \label{ss:grdSrchUrban}
		
		\subsubsection{Yucatan180}
		
\begin{figure}[H]
		  \centering
		  \includegraphics[width=.5\textwidth]{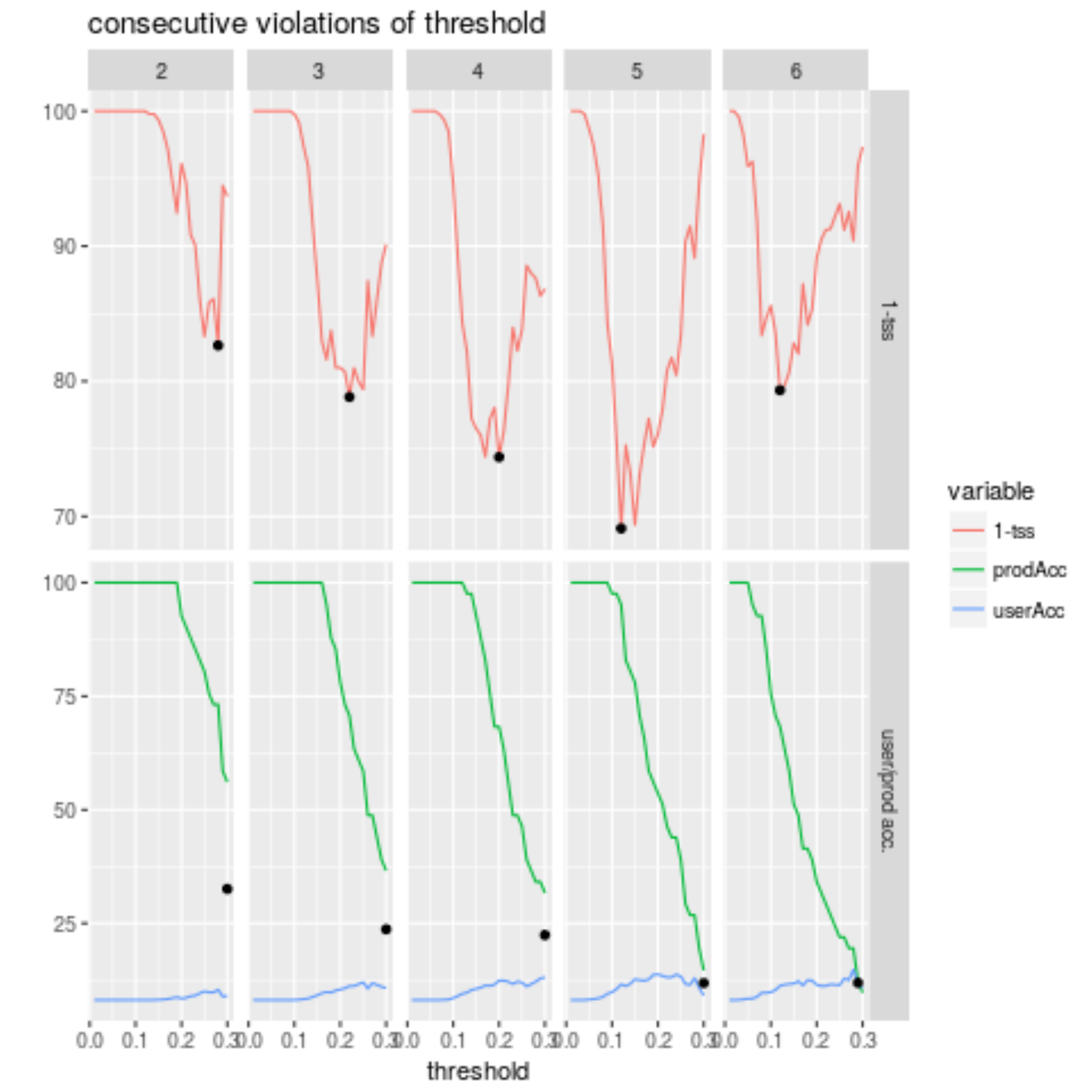} 
		  \caption[Grid search for optimal NDVI threshold for Yucatan180 site (training $1-tss$).]
		  {Grid search for optimal NDVI threshold for Yucatan180 site (training $1-tss$).}
		  \label{fig:grdSrch_Yucatan180}
		\end{figure}

\begin{figure}[H]
	      \begin{subfigure}{.5\textwidth}
	        \centering
	    	\includegraphics[width=1\textwidth]{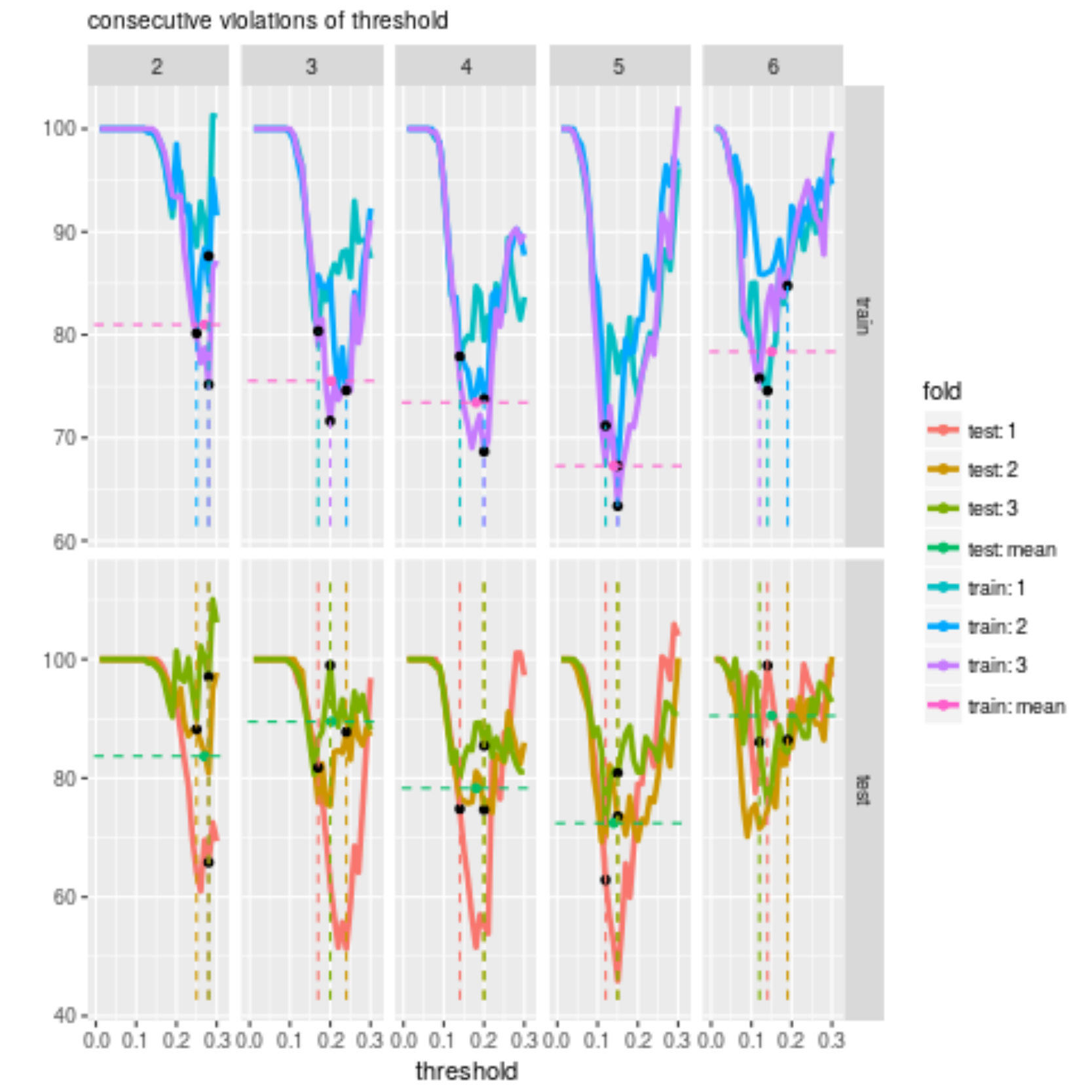} 
	        \caption{$1-tss$}
	        \label{fig:defo_pix}
	      \end{subfigure}%
	      \begin{subfigure}{.5\textwidth}
	        \centering
	        \includegraphics[width=1\textwidth]{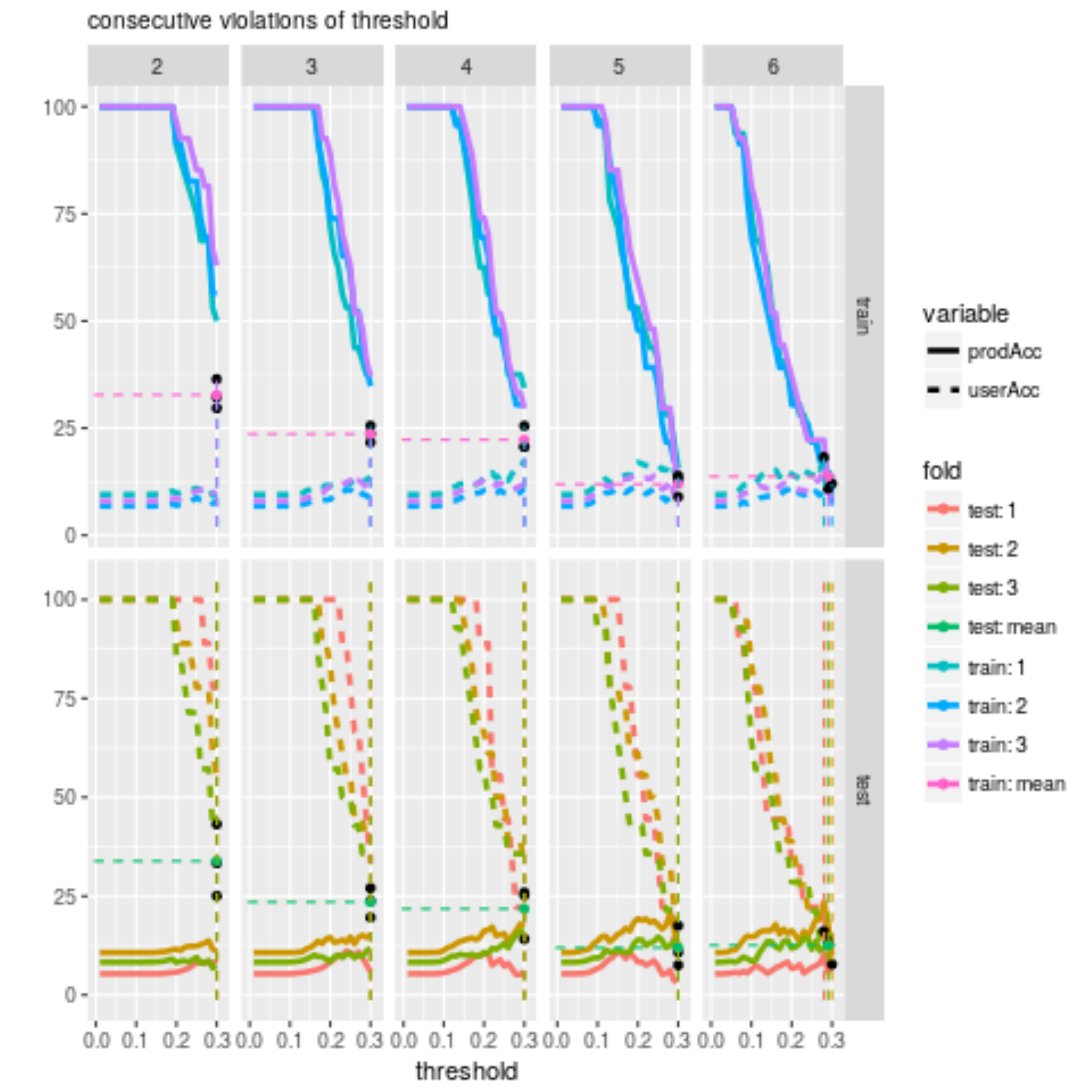}
	        \caption{$a_p$ and $a_u$}
	        \label{fig:non_defo_pix}
	      \end{subfigure}
	      \caption[Grid search for optimal NDVI threshold for Yucatan180 site (cross-validated $1-tss$, $a_p$ and $a_u$).]
	      {Grid search for optimal NDVI threshold for Yucatan180 site (cross-validated $1-tss$, $a_p$ and $a_u$).}
	      \label{fig:grid_srch_cv_Yucatan180}
	    \end{figure}

		Figures \ref{fig:grdSrch_Yucatan180} and \ref{fig:grid_srch_cv_Yucatan180} illustrate the grid-search procedure used to find the optimal NDVI prediction error threshold, for different values of $C$, in the Yucatan180 site. 
		
		Figures \ref{fig:grdSrch_Yucatan180} and \ref{fig:grid_srch_cv_Yucatan180} show that, for the Yucatan180 site, we obtain optimal $tss$ performance when we fix the consecutive number of violations of threshold $C$ at five. 
		
\begin{figure}[H]
		  \centering
		  \includegraphics[width=.5\textwidth]{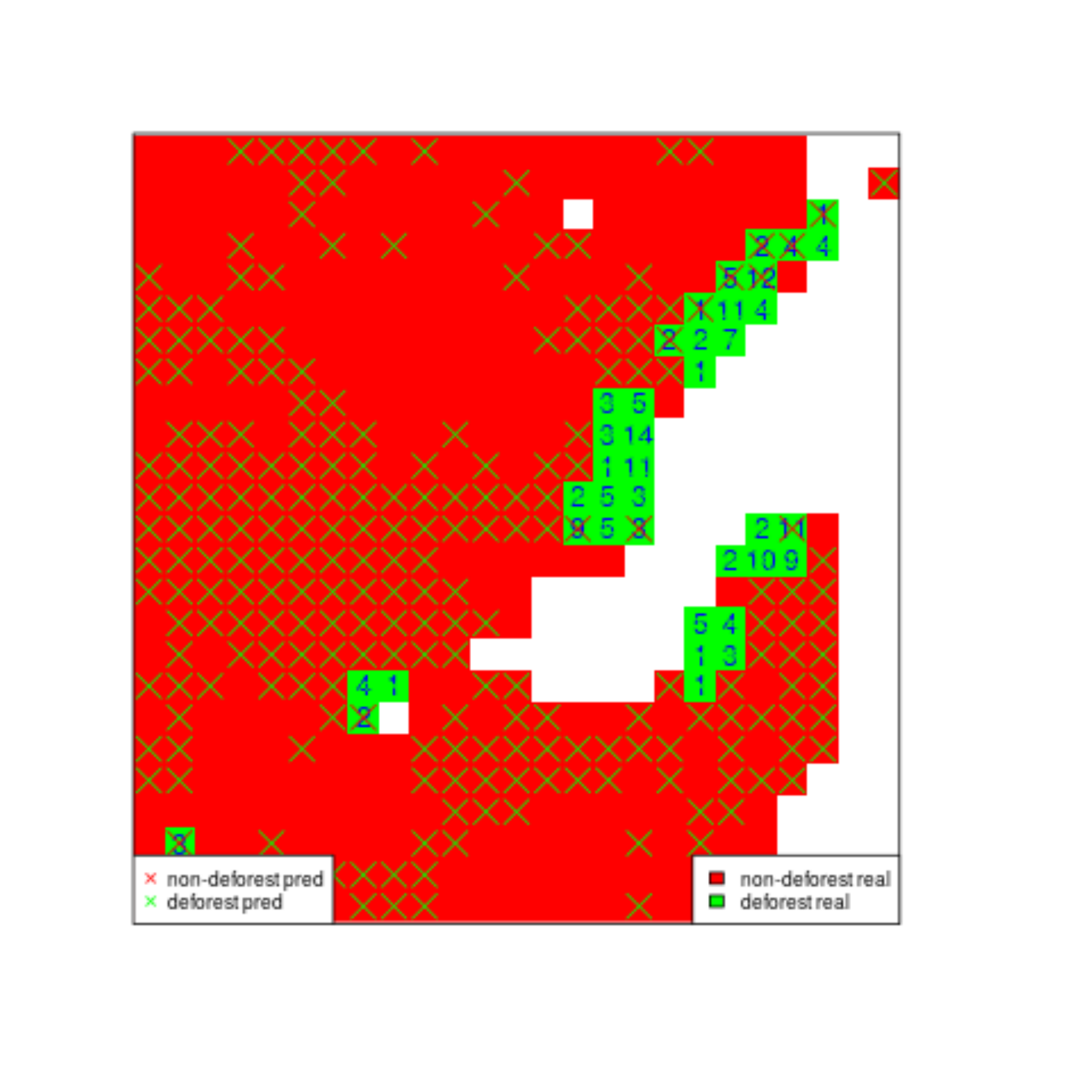} 
		  \caption[Spatial results for Yucatan180 site of applying optimal threshold using 3 consecutive violation rule.]
		  {Spatial results for Yucatan180 site of applying optimal threshold using 3 consecutive violation rule.}
		  \label{fig:grdSrch_spatial_Yucatan180}
		\end{figure}

		Figure \ref{fig:grdSrch_spatial_Yucatan180} displays the results of applying the algorithm to the Yucatan180 site, for parameter value $C=3$ and the associated optimal NDVI prediction error threshold, spatially. Green and red background color represents deforested and non-deforested pixels while green and red crosses represent the \emph{deforestation} model prediction. The degree of deforestation (the number of 250m resolution deforested pixels, out of a possible 16) is also displayed so that we can explore its impact on deforestation detection.
		
		Figure \ref{fig:grdSrch_spatial_Yucatan180} shows that most of the errors occur in patches. However, there are a few isolated pixels where the algorithm did not classify correctly. In these cases spatial smoothing of classification predictions could help improve the performance. The performance of the algorithm seems quite poor especially for detecting non-deforestation events. If we look at table \ref{tab:grdSrch_consecVaries_CV} we see that the  $1-tss$ value for $C=3$ is 87.23\% corresponding to a $tss$ of 12.77\%. Recall, that the $tss$ gives equal weight to detecting deforestation and non-deforestation events. In this case the algorithm is quite good at detecting deforestation events, it detects 69.2\% of the deforested pixels, but quite poor at detecting non-deforestation events, it detects only 13.08\% of non-deforestation events.

	\section{Multivariate thresholds} \label{s:simAn} 
	
	If there was every kind of deforestation in every 25 by 25km site then we could train thresholds on a site by site basis. However most 25 by 25km sites have not had deforestation for periods where reflectance information is available and most of those that have, only experienced one type of deforestation. This is why we chose the nine 25 by 25km sites, in the hope they are a sample of pixels where the amount and type of deforestation that has occurred is representative of the kind that could happen in Mexico on any given forest pixel going forward. 
	
	Although we don't know what type of deforestation happened we could simply apply both thresholds separately on the reflectance data of a given pixel to check if the algorithm flags the pixel for \emph{forest to water} deforestation, \emph{forest to urban or cropland} deforestation, neither, or both. Of course before doing this we would need to aggregate the thresholds of each site from each respective type of deforestation into a single threshold. This could be done by taking a weighted average of the thresholds or by re-training the thresholds on all the pixels that suffered a given type of deforestation together, regardless of which site they came from. In order to avoid having the potentially troublesome possibility of having a pixel flagged for more than one type of deforestation and in order to enable us to pool all the data for the training of the thresholds, we try two approaches which both allow for the training of thresholds using both time series and the pixels from all nine sites:
	
	\begin{enumerate}
	\item \textbf{Multivariate thresholds}: this means training both thresholds simultaneously.  
	\item \textbf{Mahalanobis distance}: constructing and index from both time series which captures most of the information available from each and then thresholding this index. 
	\end{enumerate}
	
	These two approaches are the subject of section \ref{s:simAn} and \ref{s:mahl}.

\begin{figure}[H]
	  \centering
	  \includegraphics[width=1\textwidth]{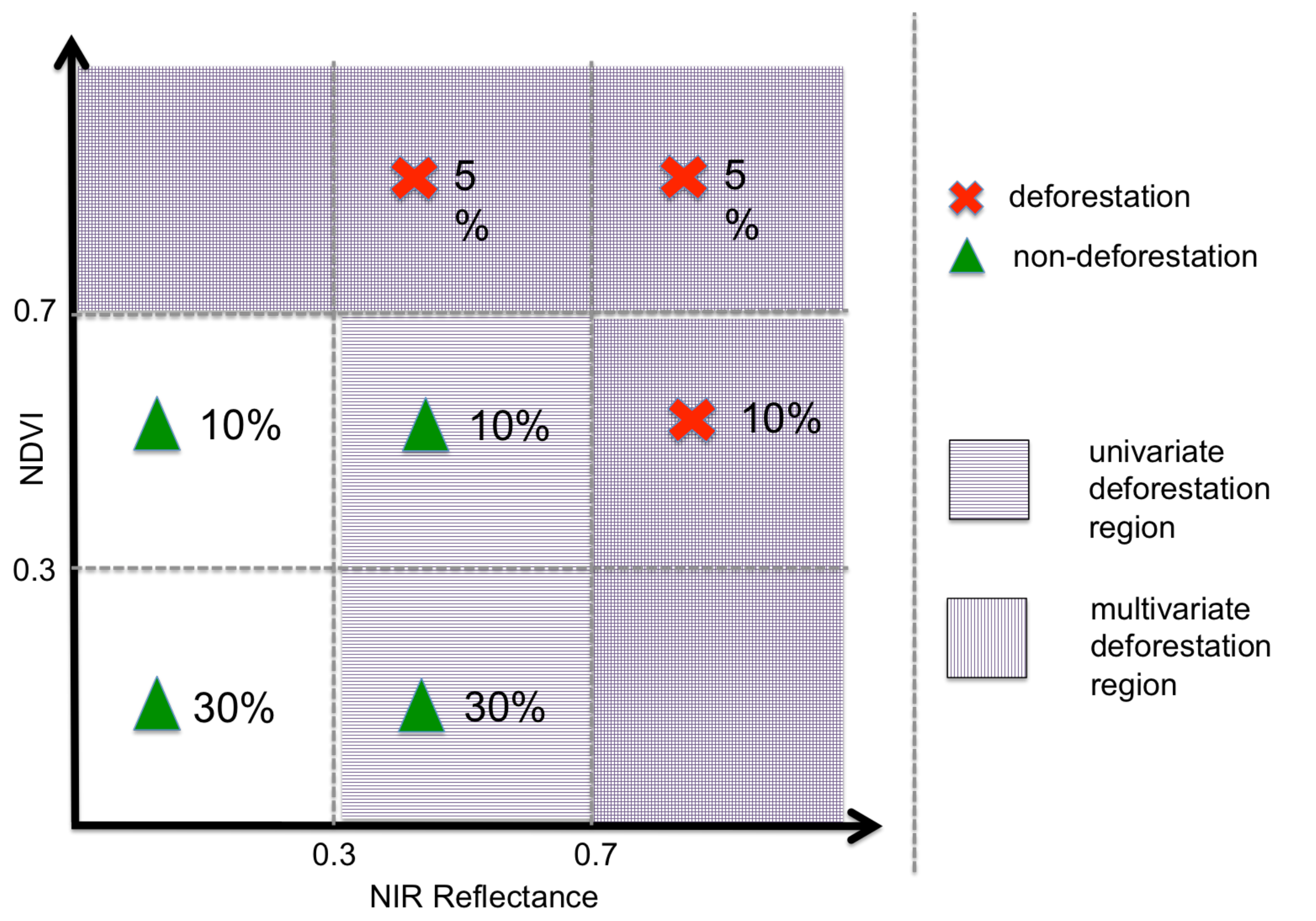} 
	  \caption[Simplified example of multivariate thresholding approach. ]
	  {Simplified example of multivariate thresholding approach. Absolute predicted error of NIR reflectance and NDVI bands is plotted and deforestation and non-deforestation cases distinguished.}
	  \label{fig:multivar_approach}
	\end{figure}

	Although each type of deforestation seems to affect one of the time series more than the other, in reality they will both be affected. Perhaps a difference of 0.3 between observed and predicted values was not enough to trigger a deforestation flag when looked at separately, but the fact that both time series present this change could be sufficient evidence to trigger a flag. Figure \ref{fig:multivar_approach} shows a simplified case where it is desirable to follow a multivariate approach.
	
	In the example of figure \ref{fig:multivar_approach} we only consider two possible thresholds: $C \in \{0.3, 0.7\}$. We consider a simplified version of the algorithm where the number of consecutive violations $C$ is equal to one. If we optimize our $tss$ only looking at each of the two variables separately we would choose a threshold of 0.7 for the NIR reflectance absolute predicted error and 0.3 for the NDVI absolute predicted error. If we apply both of these rules simultaneously, predicting deforestation when any of the two thresholds are breached, we obtain a $tss$ of 75 \%. However, by using the multivariate threshold (0.7, 0.7), i.e. a threshold of 0.7 for both variables, we can get a $tss$ of 100\%. So if we optimize simultaneously on both variables we can obtain a better solution. A limitation of this approach is that the deforestation discrimination region which we generate is quite inflexible: we can only obtain inverted L-shaped regions as in \ref{fig:multivar_approach}.
	
	The utility function we would like to maximize in the multivariate case changes to:
	
	\begin{align} \label{utility_multivariate}
		& tss(L; C, \{z_p\}_{p \in \mathcal{S}_1 \cup \cdot \cdot \cdot \cup \mathcal{S}_9},\{\hat{\epsilon}_{b,p,t}^j\}_{b \in \{2,8\},p \in \mathcal{S}_1 \cup \cdot \cdot \cdot \cup \mathcal{S}_9,j \in \{1,...,5\},t \in \mathcal{P}(j,p)} ) =  \nonumber \\
		& \frac{S(L; C, \{\hat{\epsilon}_{b,p,t}^j\}_{b \in \{2,8\},p \in \mathcal{S}_1 \cup \cdot \cdot \cdot\cup \mathcal{S}_9,j \in \{1,...,5\},t \in \mathcal{P}(j,p)} )}{N_1(\{z_p\}_{p \in \mathcal{S}_1 \cup \cdot \cdot \cdot \cup \mathcal{S}_9})} + \nonumber \\ 
		& \frac{U(L; C, \{\hat{\epsilon}_{b,p,t}^j\}_{b \in \{2,8\}, p \in \mathcal{S}_1 \cup \cdot \cdot \cdot \cup \mathcal{S}_9,j \in \{1,...,5\},t \in \mathcal{P}(j,p)} )}{N_0(\{z_p\}_{p \in \mathcal{S}_1 \cup \cdot \cdot \cdot \cup \mathcal{S}_9})}-1 = \nonumber \\
		& \frac{\sum_{i=1}^9 \sum_{p \in \mathcal{S}_i} z_p \bigg(\prod_{j=1}^5 \prod_{t \in \mathcal{P}_1(j,p)} h_L(\{\hat{\epsilon}_{b,p,s}^j\}_{b \in \{2,8\},s \in \{t, t+1, t+C-1\}}) \bigg)}{\sum_{i=1}^9 \sum_{p \in \mathcal{S}_i} z_p} + \nonumber \\
		& \frac{\sum_{i=1}^9 \sum_{p \in \mathcal{S}_i} (1-z_p) \bigg(1-\prod_{j =1}^5 \prod_{t \in \mathcal{P}_1(j,P)} h_L(\{\hat{\epsilon}_{b,p,s}^j\}_{b \in \{2,8\}s \in \{t, t+1, t+C-1\}}) \bigg)}{\sum_{i=1}^9|\mathcal{S}_i|-\sum_{i=1}^9 \sum_{p \in \mathcal{S}_i} z_p} - 1
	\end{align}	
	
	where
	\begin{itemize}
		\item $\mathcal{S}_i$ is site number $i$. 
		\item $\mathcal{P}(j,p)$ is the set of dates with clear reflectance observations for pixel $p$ and prediction \textbf{window}  $j$ i.e. as described in section \ref{s:adapt} and in figure \ref{fig:trainScheme_adapt}.
		\item $\mathcal{P}_1(j,p)$ is the set of dates with clear reflectance observations for pixel $p$ and  prediction \textbf{year} $j$,
		\item the super index $j$ indicates the training window $\mathcal{T}_{j,p}$ used to train the reflectance model parameters and obtain the errors $\hat{\epsilon}_{b,p,t}^j$, and
		\item $b_2$ and $b_8$ refer to the NIR reflectance and NDVI bands respectively, and,
		\item the thresholding function $h_L(\cdot)$ with \textbf{multivariate} threshold $L$ is defined as follows:
		
		\begin{align} \label{thrs_multivariate}
			h_L(\{\hat{\epsilon}_{b,p,s}^j\}_{b \in \{2,8\},s \in \{t, t+1, t+C-1\}}) &:= \mathbbm{1}_{\{h'_L(\{\hat{\epsilon}_{b,p,s}^j\}_{b \in \{2,8\},s \in \{t, t+1, t+C-1\}}) > 0 \}}  \\ 
			h'_L(\{\hat{\epsilon}_{b,p,s}^j\}_{b \in \{2,8\},s \in \{t, t+1, t+C-1\}}) &:= \sum_{b \in \{2,8\}} \bigg(\prod_{r=0}^{C-1} \mathbbm{1}_{\{\hat{\epsilon}_{b,p,t+r}^j > L\}} + \prod_{r=0}^{C-1} \mathbbm{1}_{\{\hat{\epsilon}_{b,p,t+r}^j < -L\}} \bigg) 
		\end{align}	
	\end{itemize}
	
	Since a grid-search in the multivariate case is very expensive and since the utility function is not even continuous we chose to optimize it using  \emph{simulated annealing} because this technique is useful for exploring a large part of the parameter (threshold) space to find a solution close to the global optimum and doesn't require smooth or even continuous utility surfaces. We restricted the solution space to a grid consisting of the cartesian product of equally spaced sequences of values for the NIR and NDVI prediction error thresholds. The initial solution used for the NIR and NDVI thresholds,  corresponds to the average of the optimal thresholds found in the site-by-site optimization results of section \ref{s:gridSearch}. For the NIR threshold, the results of \emph{forest to water} sites are averaged and for the NDVI threshold those of \emph{forest to urban or cropland} sites are averaged.
		
		\subsection{Results} \label{ss:simAnRes}
		
		Tables \ref{tab:simAnn_consecVaries_thrs} and \ref{tab:simAnn_consecVaries} show the optimal multivariate thresholds and $1-tss$ performance results, for various values of $C$,  of the simulated annealing optimization.
	
	\begin{center}
		\footnotesize\addtolength{\tabcolsep}{-2pt}
\begin{table}[ht]
\centering
\begin{tabular}{l|ccccc}
 threshold & 2 & 3 & 4 & 5 & 6 \\ 
  \hline
NIR\_reflectance & 0.120 & 0.101 & 0.082 & 0.062 & 0.053 \\ 
   \hline
NDVI & 0.235 & 0.193 & 0.182 & 0.134 & 0.129 \\ 
   \hline
\end{tabular}
\caption[optimal multivariate thresholds trained on 9 sites using simulated annealing]{optimal multivariate thresholds trained on 9 sites using simulated annealing} 
\label{tab:simAnn_consecVaries_thrs}
\end{table}

	\end{center}

	\begin{center}
		\footnotesize\addtolength{\tabcolsep}{-2pt}
\begin{table}[ht]
\centering
\begingroup\tiny
\begin{tabular}{l|ccccc}
 site & 2 & 3 & 4 & 5 & 6 \\ 
  \hline
all & 77.466 & 72.125 & 62.989 & 66.058 & 65.678 \\ 
   \hline
Sonora232 & 99.650 & 102.536 & 75.960 & 84.572 & 74.407 \\ 
  Jalisco164 & 71.560 & 52.394 & 40.860 & 39.521 & 28.227 \\ 
  Nayarit151 & 51.028 & 28.436 & 25.244 & 43.723 & 35.958 \\ 
  Nayarit109 & 36.111 & 41.468 & 49.206 & 53.968 & 47.421 \\ 
   \hline
Yucatan180 & 98.628 & 94.454 & 83.962 & 88.057 & 91.053 \\ 
  QuintanaRoo100 & 93.893 & 94.132 & 92.717 & 85.091 & 91.912 \\ 
  Michoacan98 & 86.865 & 72.108 & 79.563 & 68.487 & 64.056 \\ 
  QuintanaRoo77 & 89.716 & 86.418 & 82.658 & 71.016 & 78.511 \\ 
  Sonora74 & 67.119 & 100.339 & 77.627 & 91.356 & 97.797 \\ 
   \hline
\end{tabular}
\endgroup
\caption[optimal 1-tss for each site for general thresholds trained with simulated annealing]{optimal 1-tss for each site for general thresholds trained with simulated annealing} 
\label{tab:simAnn_consecVaries}
\end{table}

	\end{center}
	
	By observing the results of table  \ref{tab:simAnn_consecVaries} we can see that applying a single (multivariate) threshold to all sites causes the performance to deteriorate when compared to the site-by-site trained performance results of section \ref{s:gridSearch}. 
	
\begin{figure}[H]
		  \centering
		  \includegraphics[width=.5\textwidth]{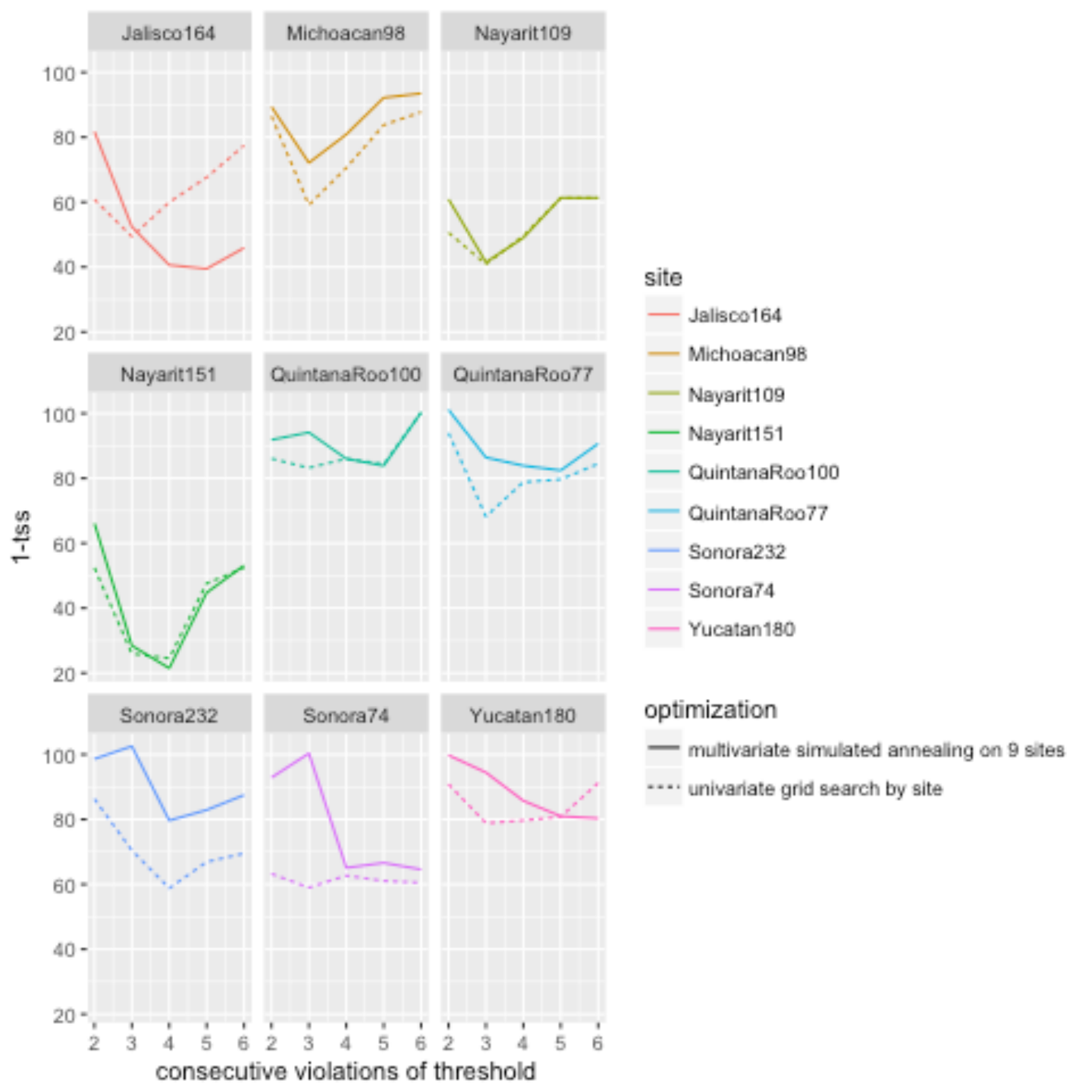} 
		  \caption[$1-tss$ results for multivariate training of thresholds with all 9 sites. Results for optimal threshold for 3 consecutive violations. Comparison with site by site grid-search.]
		  {$1-tss$ results for multivariate training of thresholds with all 9 sites. Results for optimal threshold for 3 consecutive violations. Comparison with site by site grid-search.}
		  \label{fig:simAnn_consecFix}
		\end{figure}

	In order to visualize how the performance changes as we vary the parameter $C$, but while keeping the threshold fixed, we applied the optimal threshold found for $C=3$ accross the board, i.e. to values of $C$ in the 2-6 range. Table \ref{tab:simAnn_consecFix} and figure \ref{fig:simAnn_consecFix} show the results. In figure \ref{fig:simAnn_consecFix} we also compare these performance results to those obtained using site-by-site grid search in section \ref{s:gridSearch}. 
	
		\begin{center}
			\footnotesize\addtolength{\tabcolsep}{-2pt}
\begin{table}[ht]
\centering
\begingroup\tiny
\begin{tabular}{lccl|ccccc}
 site & threshold NIR & threshold NDVI & variable & 2 & 3 & 4 & 5 & 6 \\ 
  \hline
all & 0.101 & 0.193 & both & 85.959 & 72.125 & 63.866 & 70.102 & 75.060 \\ 
   \hline
Sonora232 & 0.101 & 0.193 & both & 98.632 & 102.536 & 79.659 & 82.903 & 87.504 \\ 
  Jalisco164 & 0.101 & 0.193 & both & 81.711 & 52.394 & 40.621 & 39.459 & 45.887 \\ 
  Nayarit151 & 0.101 & 0.193 & both & 66.098 & 28.436 & 21.483 & 44.751 & 53.139 \\ 
  Nayarit109 & 0.101 & 0.193 & both & 60.913 & 41.468 & 49.008 & 61.310 & 61.310 \\ 
   \hline
Yucatan180 & 0.101 & 0.193 & both & 99.784 & 94.454 & 85.734 & 80.971 & 80.314 \\ 
  QuintanaRoo100 & 0.101 & 0.193 & both & 91.853 & 94.132 & 86.014 & 83.870 & 100.268 \\ 
  Michoacan98 & 0.101 & 0.193 & both & 89.356 & 72.108 & 80.872 & 92.160 & 93.489 \\ 
  QuintanaRoo77 & 0.101 & 0.193 & both & 101.116 & 86.418 & 83.895 & 82.440 & 90.662 \\ 
  Sonora74 & 0.101 & 0.193 & both & 92.881 & 100.339 & 65.085 & 66.610 & 64.576 \\ 
   \hline
\end{tabular}
\endgroup
\caption[1-tss for all sites using thresholds trained simultaneously with simulated annealing]{1-tss for all sites using thresholds trained simultaneously with simulated annealing} 
\label{tab:simAnn_consecFix}
\end{table}

		\end{center}
	
	Figure \ref{fig:simAnn_consecFix} shows that especially for some sites such as Sonora232, Sonora74 and Michoacan98 applying a general threshold causes a deterioration of the performance. 
		
	\section{Index approach} \label{s:mahl}

	As we saw in section \ref{s:simAn} when we threshold the prediction errors of the reflectance bands directly we generate deforestation discrimination regions that are quite inflexible: in the two-dimensional, NIR reflectance and NDVI case, they have an inverted L-shape as in figure \ref{fig:multivar_approach}. By constructing an index as a function of the relevant reflectance bands and then thresholding the prediction error of this index we can generate more flexible deforestation discrimination regions. Additionally, if we use knowledge of what distinguishes forests from other landcovers, from a reflectance point of view, we can construct indices which measure these characteristics. This is what is done in \cite{CMFDA} as the \emph{forest} index they use is in turn constructed with brightness, greenness and wetness indices, features which are useful in distinguishing forests. The modis 13A2 data set does not have the necessary reflectance information to construct these indices so we are unable to try this specific approach. Instead we build an index based on the local Mahalanobis distance between the predicted errors of the NIR reflectance and NDVI bands. The idea is to build an index which takes into account the variance of predicted errors of NIR reflectance and NDVI and the correlation between them. This index is \emph{local} in the sense that it will use information on the correlation between the predicted errors of NIR reflectance and NDVI, for pixels that are \emph{close} together, and using observations from a certain part of the year. 
	
		\subsection{Mahalanobis distance} \label{ss:mahlDist}

\begin{figure}[H]
		  \centering
		  \includegraphics[width=.5\textwidth]{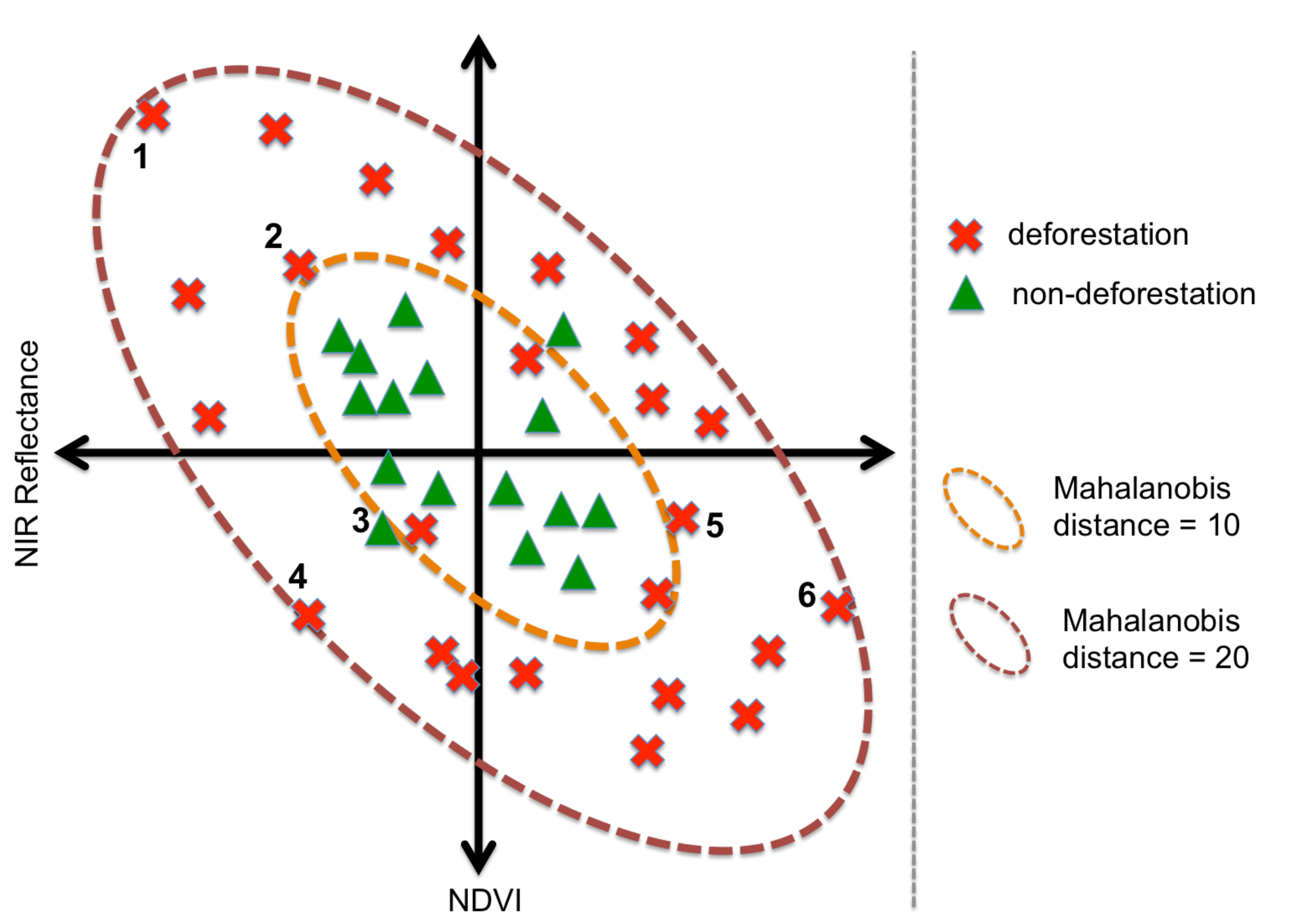} 
		  \caption[Simplified example of Mahalanobis distance thresholding approach. ]
		  {Simplified example of Mahalanobis distance thresholding approach. Predicted error of NIR reflectance and NDVI bands is plotted and deforestation and non-deforestation cases distinguished.}
		  \label{fig:mahl_approach}
		\end{figure}

		The Mahalanobis distance is a multivariate generalization of the idea of standardizing a variable by dividing by its standard deviation such that it is unitless and scale-invariant. Instead of using the original NIR reflectance and NDVI prediction error coordinate system to measure distance between points, we use the principal components as our axes. We relativize the euclidean distance between any two points by the observed variance in a certain direction: the direction defined by the two points. Figure \ref{fig:mahl_approach} illustrates the idea.

		Although the three pairs of points, (1,2), (3,4) and (4,5), illustrated in figure \ref{fig:mahl_approach} are at different euclidean distances, when we take into account the covariance between NIR reflectance and NDVI predicted errors, and relativize the euclidean distances with respect to this covariance we may observe they are the same distance apart. The hope is that the deforested and non-deforested pixels will be distributed somewhat similarly to figure \ref{fig:mahl_approach} such that the Mahalanobis distance can help us discriminate between them.
		
		The utility function we would like to maximize in this case is:
	
		\begin{align} \label{utility_mahalanobis}
			& tss(L; C, \{z_p\}_{p \in \mathcal{S}_1 \cup \cdot \cdot \cdot \cup \mathcal{S}_9},\{\hat{\epsilon}_{b,p,t}^j\}_{b \in \{2,8\},p \in \mathcal{S}_1 \cup \cdot \cdot \cdot \cup \mathcal{S}_9,j \in \{1,...,5\},t \in \mathcal{P}(j,p)} ) =  \nonumber \\
			& \frac{S(L; C, \{\hat{\epsilon}_{b,p,t}^j\}_{b \in \{2,8\},p \in \mathcal{S}_1 \cup \cdot \cdot \cdot \cup \mathcal{S}_9,j \in \{1,...,5\},t \in \mathcal{P}(j,p)} )}{N_1(\{z_p\}_{p \in \mathcal{S}_1 \cup \cdot \cdot \cdot \cup \mathcal{S}_9\}})} + \nonumber \\ 
			& \frac{U(L; C, \{\hat{\epsilon}_{b,p,t}^j\}_{b \in \{2,8\}, p \in \mathcal{S}_1 \cup \cdot \cdot \cdot \cup \mathcal{S}_9,j \in \{1,...,5\},t \in \mathcal{P}(j,p)} )}{N_0(\{z_p\}_{p \in \mathcal{S}_1 \cup \cdot \cdot \cdot \cup \mathcal{S}_9})}-1 = \nonumber \\
			& \frac{\sum_{i=1}^9 \sum_{p \in \mathcal{S}_i} z_p \bigg(\prod_{j=1}^5 \prod_{t \in \mathcal{P}_1(j,p)} h_L(\{I_{\mathcal{M}}(\{\hat{\epsilon}_{b,p,s}^j\}_{b \in \{2,8\}}; \hat{\Sigma}_{p,s})\}_{s \in \{t, t+1, t+C-1\}}) \bigg)}{\sum_{i=1}^9 \sum_{p \in \mathcal{S}_i} z_p} + \nonumber \\
			& \frac{\sum_{i=1}^9 \sum_{p \in \mathcal{S}_i} (1-z_p) \bigg(1-\prod_{j=1}^5 \prod_{t \in \mathcal{P}_1(j,p)} h_L(\{I_{\mathcal{M}}(\{\hat{\epsilon}_{b,p,s}^j\}_{b \in \{2,8\}}; \hat{\Sigma}_{p,s})\}_{s \in \{t, t+1, t+C-1\}}) \bigg)}{\sum_{i=1}^9|\mathcal{S}_i|-\sum_{i=1}^9 \sum_{p \in \mathcal{S}_i} z_p} - 1
		\end{align}	
	
		where
		\begin{itemize}
			\item $\mathcal{S}_i$ is site number $i$. 
			\item $\mathcal{P}(j,p)$ is the set of dates with clear reflectance observations for pixel $p$ and prediction \textbf{window}  $j$ i.e. as described in section \ref{s:adapt} and in figure \ref{fig:trainScheme_adapt}.
			\item $\mathcal{P}_1(j,p)$ is the set of dates with clear reflectance observations for pixel $p$ and  prediction \textbf{year} $j$,
			\item the super index $j$ indicates the training window $\mathcal{T}_{j,p}$ used to train the reflectance model parameters and obtain the errors $\hat{\epsilon}_{b,p,t}^j$, and
			\item $b_2$ and $b_8$ refer to the NIR reflectance and NDVI bands respectively, 
			\item the local Mahalanobis distance index function $I_{\mathcal{M}}(\cdot)$ is defined as follows:
			
			\begin{align}
				I_{\mathcal{M}}(\{\hat{\epsilon}_{b,p,t}^j\}_{b \in \{2,8\}}) &= \left( \left( \begin{array}{cc}
\hat{\epsilon}_{2,p,t} & \hat{\epsilon}_{8,p,t} 
\end{array} \right)
\hat{\Sigma}_{p,t}^{-1}
\left( \begin{array}{c}
\hat{\epsilon}_{2,p,t}  \\
\hat{\epsilon}_{8,p,t}
\end{array} \right) \right)^{\sfrac{1}{2}} \\
			&=\left( \left( \begin{array}{cc}
\hat{\epsilon}_{2,p,t} & \hat{\epsilon}_{8,p,t} 
\end{array} \right)
\left( \begin{array}{cc}
\hat{\sigma}_{2,c(p,t)}^2 & \hat{\sigma}_{28,c(p,t)} \\
\hat{\sigma}_{28,c(p,t)} & \hat{\sigma}_{8,c(p,t)}^2 
\end{array} \right)^{-1}
\left( \begin{array}{c}
\hat{\epsilon}_{2,p,t}  \\
\hat{\epsilon}_{8,p,t}
\end{array} \right) \right)^{\sfrac{1}{2}}, 
			\end{align}	
			where
			\begin{itemize}
				\item $\hat{\sigma}_{2,c(p,t)}^2$ and $\hat{\sigma}_{8,c(p,t)}^2$ are the historical sample variances of NIR reflectance and NDVI prediction errors belonging to cube $c(p,t)$,
				\item  $\hat{\sigma}_{28,c(p,t)}$ is the historical sample covariance of NIR reflectance and NDVI prediction errors belonging to cube $c(p,t)$
				\item cube $c(p,t)$  consists of observations for the 25 pixels within one of 25, 5 by 5km, squares within each site, and, which occurred on dates corresponding to a specific 5 day period of the year. 
			\end{itemize}	
			
			\item the thresholding function $h_L(\cdot)$ with \textbf{univariate} threshold $L$ is defined as follows:
		
			\begin{align} \label{thrs_mahalanobis}
				& h_L(\{I_{\mathcal{M}}(\{\hat{\epsilon}_{b,p,s}^j\}_{b \in \{2,8\}}; \hat{\Sigma}_{p,s})\}_{s \in \{t, t+1, t+C-1\}}) &:= \nonumber \\
				& \prod_{r=0}^{C-1} \mathbbm{1}_{\{I_{\mathcal{M}}(\{\hat{\epsilon}_{b,p,t+r}^j\}_{b \in \{1,2\}}; \hat{\Sigma}_{p,t+r}) > L\}} + \prod_{r=0}^{C-1} \mathbbm{1}_{\{I_{\mathcal{M}}(\{\hat{\epsilon}_{b,p,t+r}^j\}_{b \in \{1,2\}}; \hat{\Sigma}_{p,t+r}) < -L\}}
			\end{align}	
		\end{itemize}

		\subsection{Results} \label{ss:mahlRes}
		
		Tables \ref{tab:Mahldist_consecVaries_thrs_all_mahl} and \ref{tab:Mahldist_consecVaries_all_mahl} show the optimal thresholds and $1-tss$ performance results, for various values of $C$,  of the grid-search optimization.

			\begin{center}
				\footnotesize\addtolength{\tabcolsep}{-2pt}
\begin{table}[ht]
\centering
\begin{tabular}{l|ccccc}
 variable & 2 & 3 & 4 & 5 & 6 \\ 
  \hline
Mahl. dist. & 17.471 & 16.092 & 11.724 & 8.506 & 6.667 \\ 
   \hline
\end{tabular}
\caption[optimal Mahalanobis distance thresholds trained on 9 sites using grid search]{optimal Mahalanobis distance thresholds trained on 9 sites using grid search} 
\label{tab:Mahldist_consecVaries_thrs_all_mahl}
\end{table}

			\end{center}

			\begin{center}
				\footnotesize\addtolength{\tabcolsep}{-2pt}
\begin{table}[ht]
\centering
\begingroup\tiny
\begin{tabular}{l|ccccc}
 site & 2 & 3 & 4 & 5 & 6 \\ 
  \hline
all & 72.428 & 69.240 & 68.840 & 69.135 & 70.604 \\ 
   \hline
Sonora232 & 99.390 & 88.612 & 95.648 & 102.464 & 98.080 \\ 
  Jalisco164 & 38.316 & 38.910 & 38.449 & 32.988 & 38.449 \\ 
  Nayarit151 & 49.892 & 44.372 & 40.936 & 43.966 & 38.068 \\ 
  Nayarit109 & 45.833 & 56.151 & 51.190 & 41.667 & 46.429 \\ 
   \hline
Yucatan180 & 97.892 & 68.934 & 73.812 & 70.227 & 71.883 \\ 
  QuintanaRoo100 & 90.110 & 94.981 & 96.321 & 86.819 & 97.930 \\ 
  Michoacan98 & 99.482 & 98.884 & 91.602 & 94.279 & 87.383 \\ 
  QuintanaRoo77 & 72.908 & 75.333 & 65.074 & 61.654 & 70.143 \\ 
  Sonora74 & 70.169 & 75.085 & 70.000 & 79.831 & 91.864 \\ 
   \hline
\end{tabular}
\endgroup
\caption[optimal 1-tss for each site for general Mahalanobis distance thresholds trained with grid search]{optimal 1-tss for each site for general Mahalanobis distance thresholds trained with grid search} 
\label{tab:Mahldist_consecVaries_all_mahl}
\end{table}

			\end{center}
			
			By observing the results of table  \ref{tab:Mahldist_consecVaries_thrs_all_mahl} we can see that, as was the case in section \ref{s:simAn}, applying a single threshold to all sites causes the performance to deteriorate when compared to the site-by-site trained performance results of section \ref{s:gridSearch}. 
	
\begin{figure}[H]
				  \centering
				  \includegraphics[width=.5\textwidth]{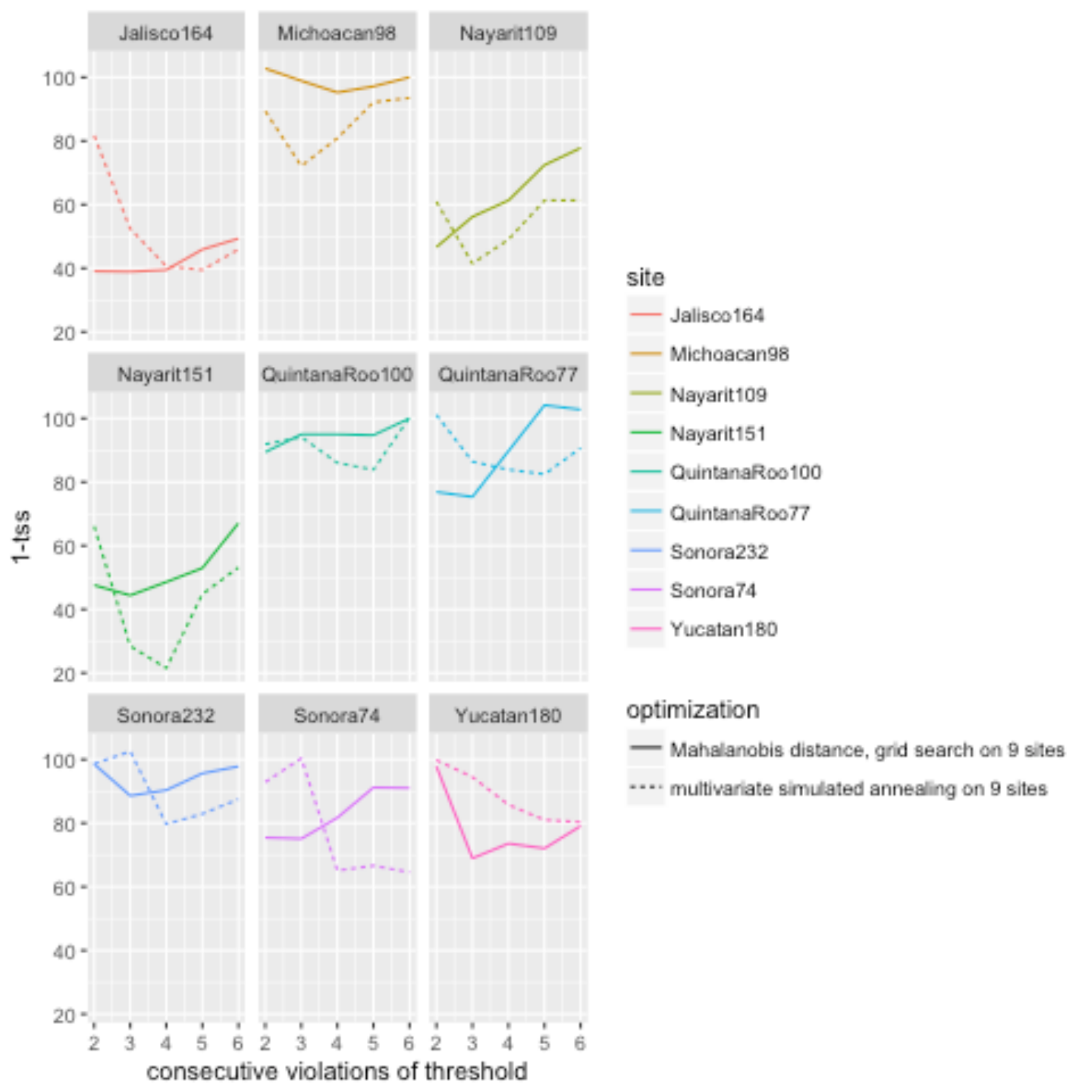} 
				  \caption[$1-tss$ results for grid search training of Mahalanobis distance thresholds with all 9 sites. Results for optimal threshold for 3 consecutive violations. Comparison with equivalent multivariate thresholds.]
				  {$1-tss$ results for grid search training of Mahalanobis distance thresholds with all 9 sites. Results for optimal threshold for 3 consecutive violations. Comparison with equivalent multivariate thresholds}
				  \label{fig:mahl_consecFix}
				\end{figure}

			In order to visualize how the performance changes as we vary the parameter $C$, but while keeping the threshold fixed, we applied the optimal threshold found for $C=3$ accross the board, i.e. to values of $C$ in the 2-6 range. Table \ref{tab:Mahldist_consecFix} and figure \ref{fig:mahl_consecFix} show the results. In figure \ref{fig:mahl_consecFix} we also compare these performance results to those obtained using multivariate thresholds in section \ref{s:simAn}.

				\begin{center}
					\footnotesize\addtolength{\tabcolsep}{-2pt}
\begin{table}[ht]
\centering
\begingroup\tiny
\begin{tabular}{lcl|ccccc}
 site & threshold & variable & 2 & 3 & 4 & 5 & 6 \\ 
  \hline
all & 16.092 & Mahl. dist. & 72.393 & 69.240 & 73.218 & 78.612 & 83.012 \\ 
   \hline
Sonora232 & 16.092 & Mahl. dist. & 98.525 & 88.612 & 90.356 & 95.616 & 97.808 \\ 
  Jalisco164 & 16.092 & Mahl. dist. & 39.060 & 38.910 & 39.459 & 45.887 & 49.309 \\ 
  Nayarit151 & 16.092 & Mahl. dist. & 47.592 & 44.372 & 48.566 & 52.949 & 67.045 \\ 
  Nayarit109 & 16.092 & Mahl. dist. & 46.627 & 56.151 & 61.310 & 72.421 & 77.778 \\ 
   \hline
Yucatan180 & 16.092 & Mahl. dist. & 97.824 & 68.934 & 73.575 & 72.056 & 79.079 \\ 
  QuintanaRoo100 & 16.092 & Mahl. dist. & 89.380 & 94.981 & 94.981 & 94.713 & 100.000 \\ 
  Michoacan98 & 16.092 & Mahl. dist. & 102.757 & 98.884 & 95.223 & 97.130 & 100.000 \\ 
  QuintanaRoo77 & 16.092 & Mahl. dist. & 76.862 & 75.333 & 89.740 & 104.147 & 102.765 \\ 
  Sonora74 & 16.092 & Mahl. dist. & 75.424 & 75.085 & 81.695 & 91.186 & 91.017 \\ 
   \hline
\end{tabular}
\endgroup
\caption[1-tss for all sites using Mahalanobis distance thresholds trained with grid search]{1-tss for all sites using Mahalanobis distance thresholds trained with grid search} 
\label{tab:Mahldist_consecFix}
\end{table}

				\end{center}
	
	Figure \ref{fig:mahl_consecFix} shows that when comparing the overall grid-search optimized performance when using the Mahalanobis distance with the performance when using multivariate thresholds, the Mahalanobis distance version of the algorithm performs better for 2-3 consecutive violations of the threshold while the multivariate version of the algoritm performs better for 4-6 consecutive violations of the threshold. This means if we want to detect deforestation faster it is better to use the Mahalanobis distance of the prediction errors. 		
		
\chapter{Model heterogeneity} \label{ch:modhet}

As we saw in chapter \ref{ch:meth} the underlying model used for the reflectance (NIR reflectance and NDVI) time series is:

\begin{align}
	y_{b,p,t} = g_{b,p}(d) + \epsilon_{b,p,t} = a_{b,p}^T X + \epsilon_{b,p,t}
\end{align}	

where,

\begin{itemize}
	\item $y_{b,p,t}$ is the surface reflectance for spectral band $b$, pixel $p$ and date $t$,
	\item $d$ is the composite day of the year,
	\item $a_{b,p}=(\alpha_{0,b,p},\alpha_{1,b,p},\beta_{1,b,p},\alpha_{2,b,p},\beta_{2,b,p})^T$, and,
	\item $X = (1, \cos(\frac{2\pi d}{T}),\sin(\frac{2\pi d}{T}),\cos(\frac{2\pi d}{0.5T}),\sin(\frac{2\pi d}{0.5T}))^T$
\end{itemize}	

If model assumptions (zero expected value, constant variance and uncorrelatedness for the errors $\epsilon_{b,p,t}$ for $t \in \mathcal{T}$) are not met then OLS may not be the best method for esimating $a_{b,p}$. However since we wish to fit this model for every forest pixel in Mexico, we do not carry out diagnostic analysis to see if the model assumptions are satisfied, because:

\begin{enumerate}
	\item It is unfeasible to find a custom model for each pixel which satisfies assumptions, and,
	\item Fitting methods other than ordinary least square (OLS) may not be as computationally feasible given the amount of forest pixels to be fitted.
\end{enumerate}  

However, it is clear that since surface reflectance time series depend on seasonal factors such as the amount of sun radiance which reaches the surface and the phenological cycle of vegetation, it is likely the errors $\epsilon_{b,p,t}$ will be time correlated i.e. $Corr(\epsilon_{b,p,t},\epsilon_{b,p,s}) \neq 0$ for $t \neq s$. 

Additionally, although we followed a pixel-by-pixel modeling approach, and given that surface reflectance depends on factors which are spatially correlated such as the type of land cover at the surface, it is likely that both the model parameters $a_{b,p}$ and the error distribution $\epsilon_{b,p,t}$ will be spatially correlated i.e. $Corr(a_{b,p}, a_{b,q}) \neq 0$ and $Corr(\epsilon_{b,p,t},\epsilon_{b,q,t}) \neq 0$ for $p \neq q$. 

In section \ref{s:coef} we explore the spatial correlation of $a_{b,p}$ for different pixels $p$ by plotting the fitted $a_{b,p}$ for different pixels. 

In section \ref{s:var} we explore the spatial and time correlation of the errors, $\epsilon_{b,p,t}$, by comparing the distribution of prediction errors for different pixels and accross different periods. Since the adapted CMFDA algorithm is built on the prediction error, the difference between predicted and observed reflectance, $\hat{\epsilon}_{b,p,t}= \hat{g}_{b,p}(d) - y_{b,p,t}$, it is important to understand their spatial and time correlation. The heterogeneity in the distribution of errors $\epsilon_{b,p,t}$ accross pixels is likely the cause of variance in the optimal thresholds which we observed in chapter \ref{ch:meth}, and which complicates the algorithm by not allowing us to to find a single threshold (univariate or multivariate) which works reasonably well for different sites. The heterogeneity in the distribution of errors $\epsilon_{b,p,t}$ accross periods of the year could be leading to poor performance of the algorithm for periods of the year which have higher error variance.  In the rest of the chapter, sections \ref{s:homStdr}-\ref{s:genRes}, we explore transformations that can be applied to the prediction error so that its distribution accross pixels and periods is homogenized. The ultimate goal is to obtain transformed prediction errors which behave similarly accross sites so that applying one threshold to all yields reasonably good performance. 

	\section{Model coefficients} \label{s:coef}
	
\begin{figure}[H]
	  \centering
	  \includegraphics[width=.5\textwidth]{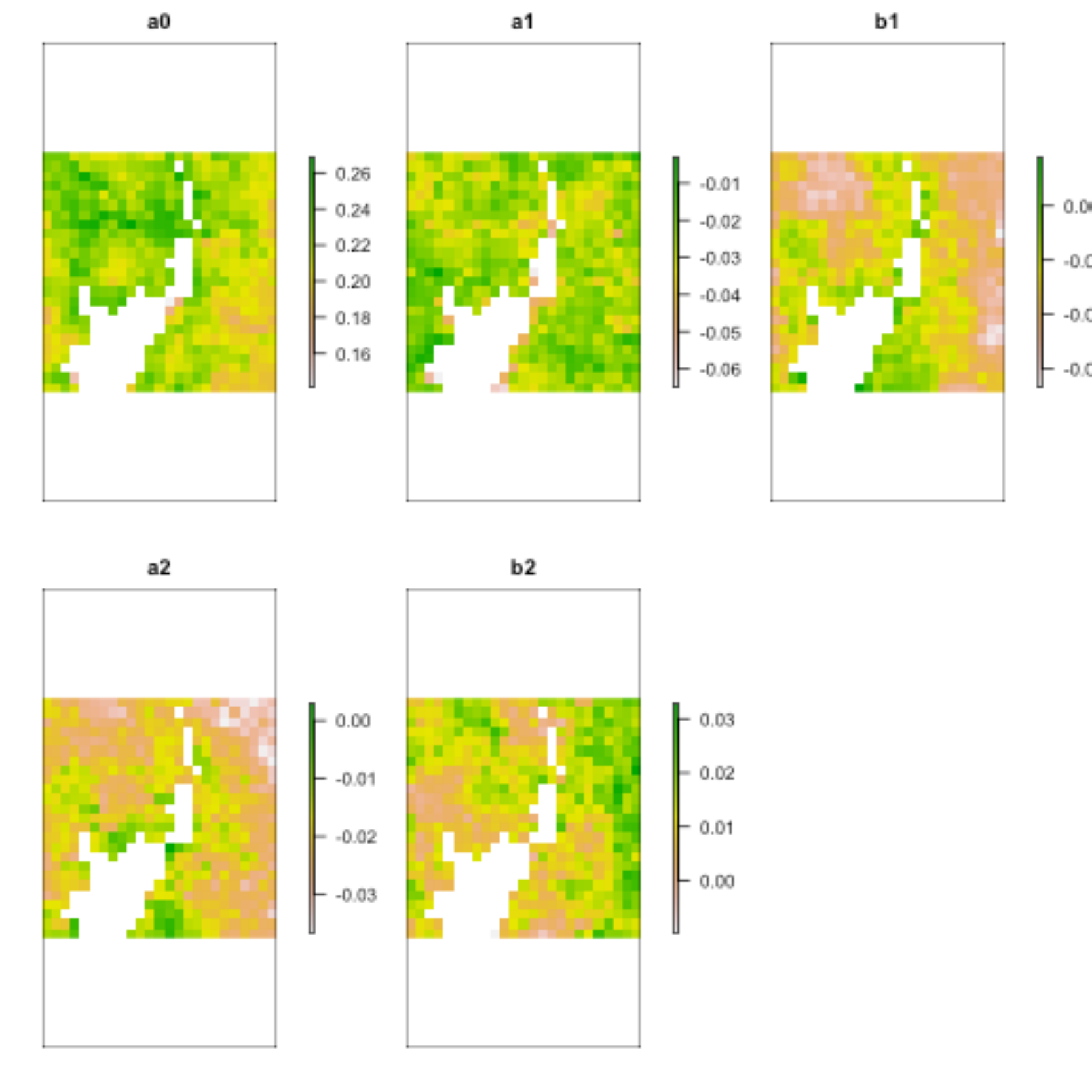} 
	  \caption[Model coefficients for NIR reflectance model for Sonora232 site and 2003-2004 training period.]
	  {Model coefficients for NIR reflectance model for Sonora232 site and 2003-2004 training period.}
	  \label{fig:modelCoef_Sonora232}
	\end{figure}

\begin{figure}[H]
	  \centering
	  \includegraphics[width=.5\textwidth]{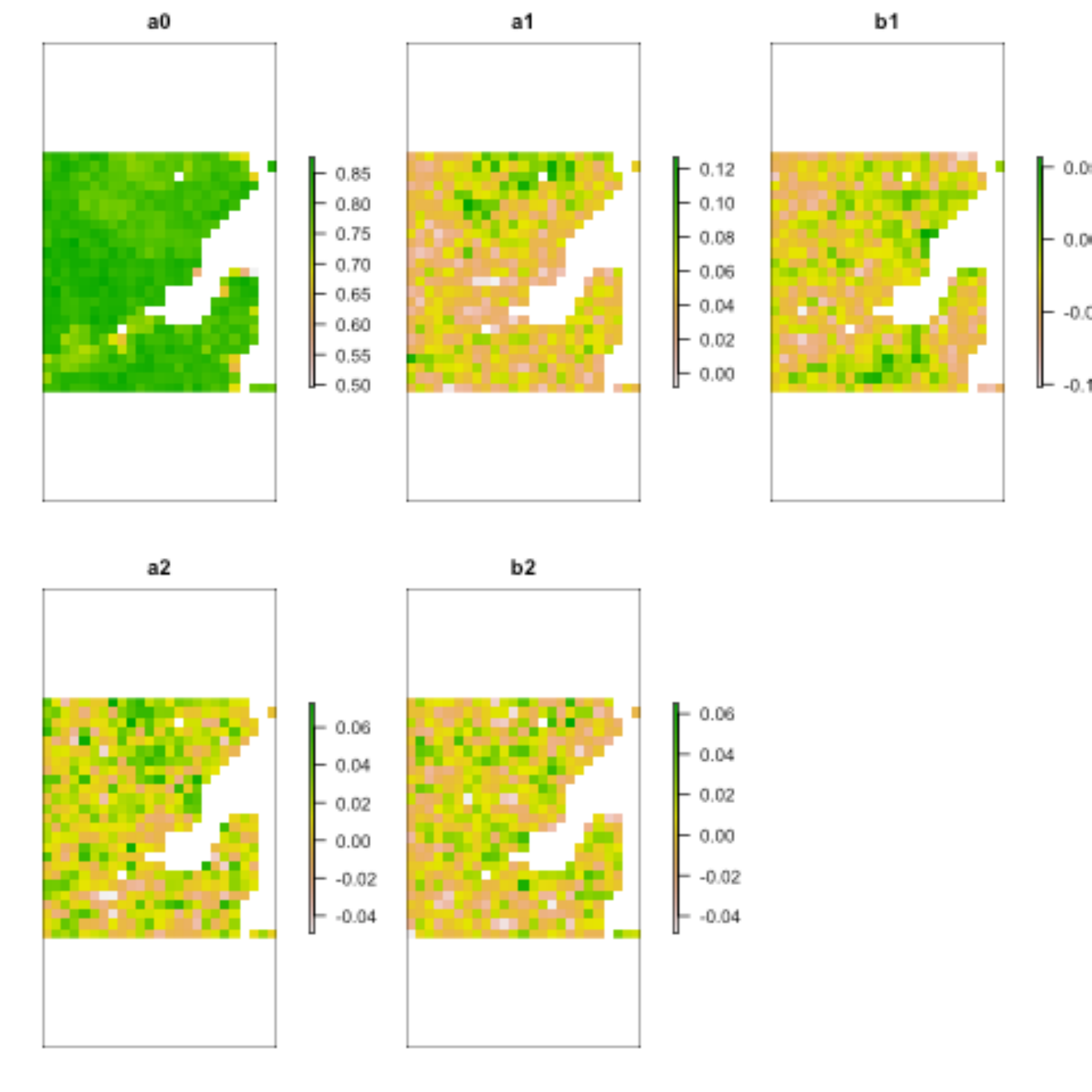} 
	  \caption[Model coefficients for NDVI model for Yucatan180 site and 2003-2004 training period.]
	  {Model coefficients for NDVI model for Yucatan180 site and 2003-2004 training period.}
	  \label{fig:modelCoef_Yucatan180}
	\end{figure}

	Figures \ref{fig:modelCoef_Sonora232} and \ref{fig:modelCoef_Yucatan180}  show the fitted model coefficients for the Sonora232 and Yucatan180 sites, for the 2003-2004 training period. For Sonora232 NIR reflectance was modeled and for Yucatan180 the NDVI. 
	
	As we can see there is a clear spatial correlation between the fitted model coefficients. If we use a spatial model, one where the spatial dependence of the coefficients $a(x,y)$ is taken into account, we can pool the data locally so that the fitted coefficients for a given pixel also use information from nearby pixels. This is not explored further in this work. 

	\section{Variance of errors} \label{s:var}
\begin{figure}[H]
	      \begin{subfigure}{.5\textwidth}
	        \centering
	    	\includegraphics[width=0.9\textwidth]{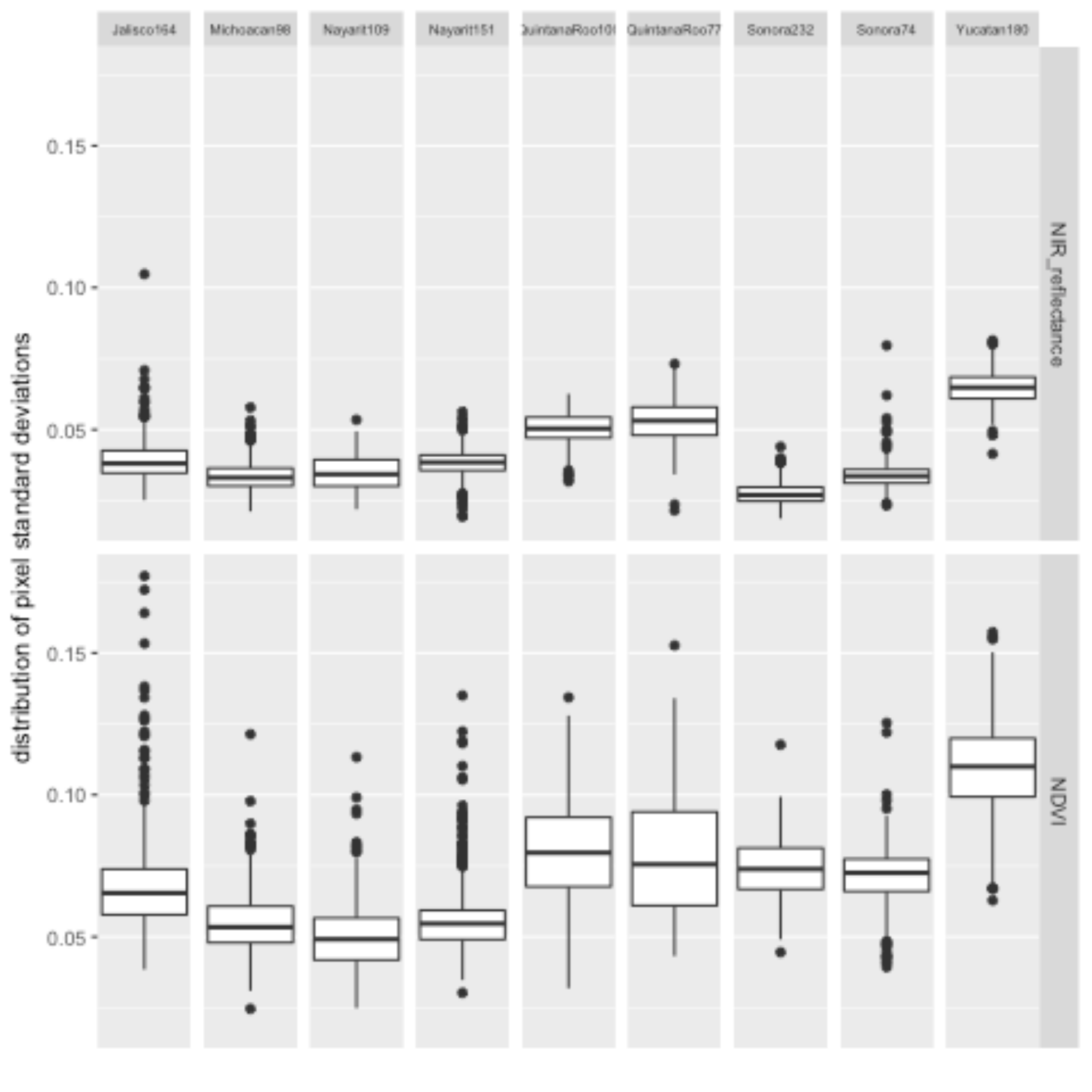} 
	        \caption{Boxplot of pixel standard deviations by site}
	        \label{fig:varDep_site}
	      \end{subfigure}%
	      \begin{subfigure}{.5\textwidth}
	        \centering
	        \includegraphics[width=1\textwidth]{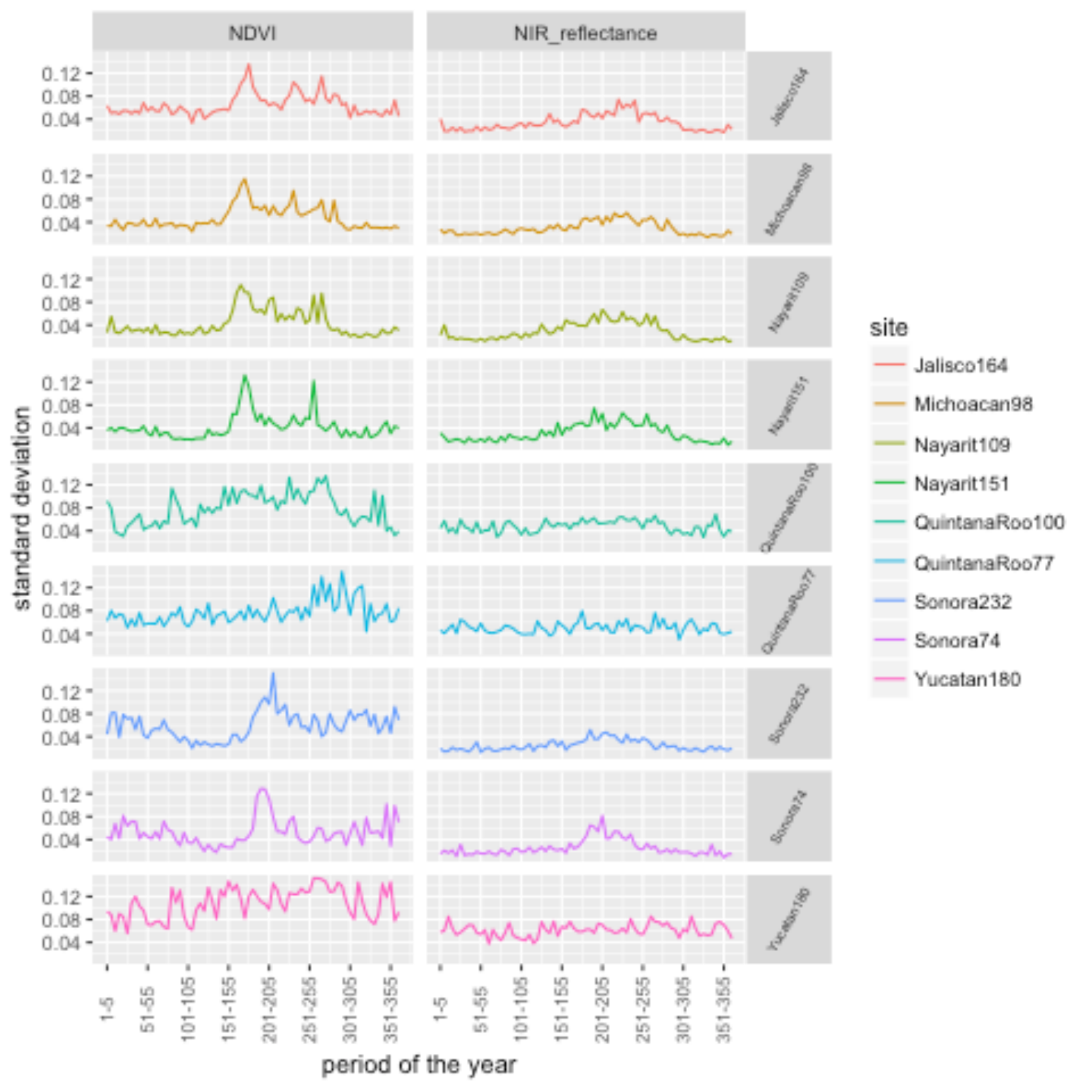}
	        \caption{Line graph of period standard deviations by site}
	        \label{fig:varDep_period}
	      \end{subfigure}
	      \caption[Standard deviation of prediction error by site and time of the year]
	      {Standard deviation of prediction error by site and time of the year}
	      \label{fig:varDep}
	    \end{figure}

		As figure \ref{fig:varDep} shows the variance of the prediction errors depends both on the location of the pixel and on the period of the year.   Figure \ref{fig:varDep_site} shows the distribution of the pixel-wise prediction error standard deviations for each site. Figure \ref{fig:varDep_period} shows the sample standard deviation for each site and each one of the 73, five day periods of the year.
		
	\section{Homogenizing error variance: standardization} \label{s:homStdr}	
		
		In order to homogenize the distribution of prediction errors accross pixels and period of the year we tried dividing it by an estimate of its standard deviation $\sigma$. We tried doing this in several ways following different assumptions about the way the variance depends on space and/or time:
		
		 \begin{enumerate}[1.]
			 \item \textbf{Depends only on space}: divide prediction error of a pixel by the sample standard deviation of prediction errors observed for that pixel,
			 \item \textbf{Depends only on time}: divide prediction error observed at a given date by the sample standard deviation of prediction errors observed on all 9 sites during the 5 day period of the year corresponding to the date of the observation. The year is divided into following 73 periods: $\{1-5, 6-10,...,356-360, 361-366 \}$,
			 \item \textbf{Depends on time and on space globally}: divide prediction error observed at a given site and date by the sample standard deviation of prediction errors observed on that site during the 5 day period of the year corresponding to the date of the observation, 
			 \item \textbf{Depends on space and time}. For this assumption we tried several variants,
			 	\begin{enumerate}[a.]
					\item Apply 2 then 1 to prediction errors,
					\item Apply 3 then 1 to prediction errors, 
					\item \textbf{Space-time cubes}: divide prediction error observed at a given pixel and date by the sample standard deviation of prediction errors observed on the corresponding space-time \emph{cube}. A cube of observations consists of the observations for the 25 pixels within one of 25, 5 by 5km, squares of which a site consists, and, which occurred on dates corresponding to a specific 5 day period of the year. 
				\end{enumerate}	
		 \end{enumerate}

	\section{Homogenizing error variance: empirical cdf} \label{s:homECDF}	
		
		If the prediction error distributions for different sites had similar shape distributions and only differed in variance then the standardization procedure described in sections \ref{s:homStdr} would, under certain other assumptions, lead to similar thresholds for the standardized differences, accross sites. The following graphs show the distribution of standardized prediction errors by site.
		
		 \begin{figure}[H]
	      \begin{subfigure}{.5\textwidth}
	        \centering
	    	\includegraphics[width=0.9\textwidth]{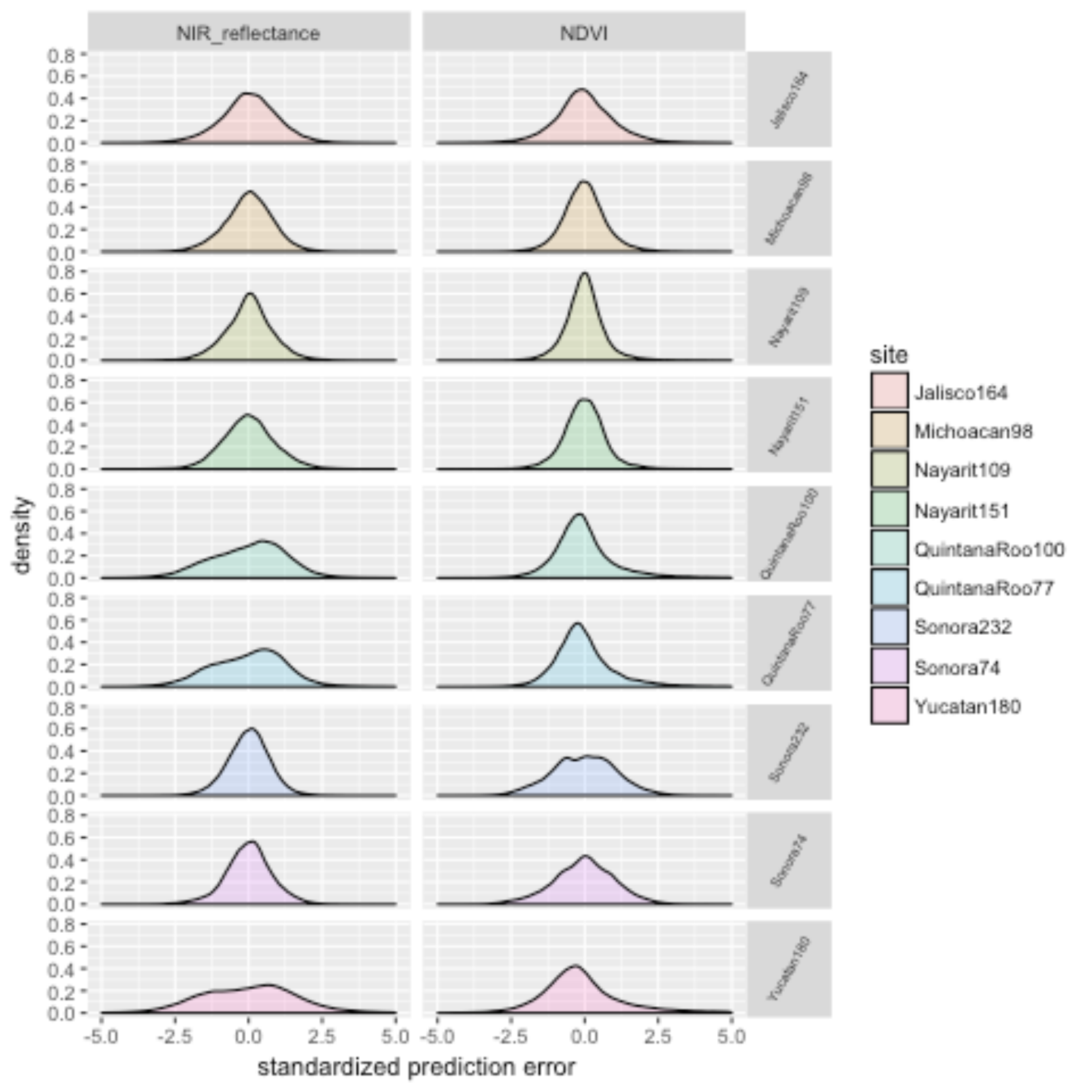} 
	        \caption{Density of standardized prediction errors by site}
	        \label{fig:distErrorStdr_dens}
	      \end{subfigure}%
	      \begin{subfigure}{.5\textwidth}
	        \centering
	        \includegraphics[width=1\textwidth]{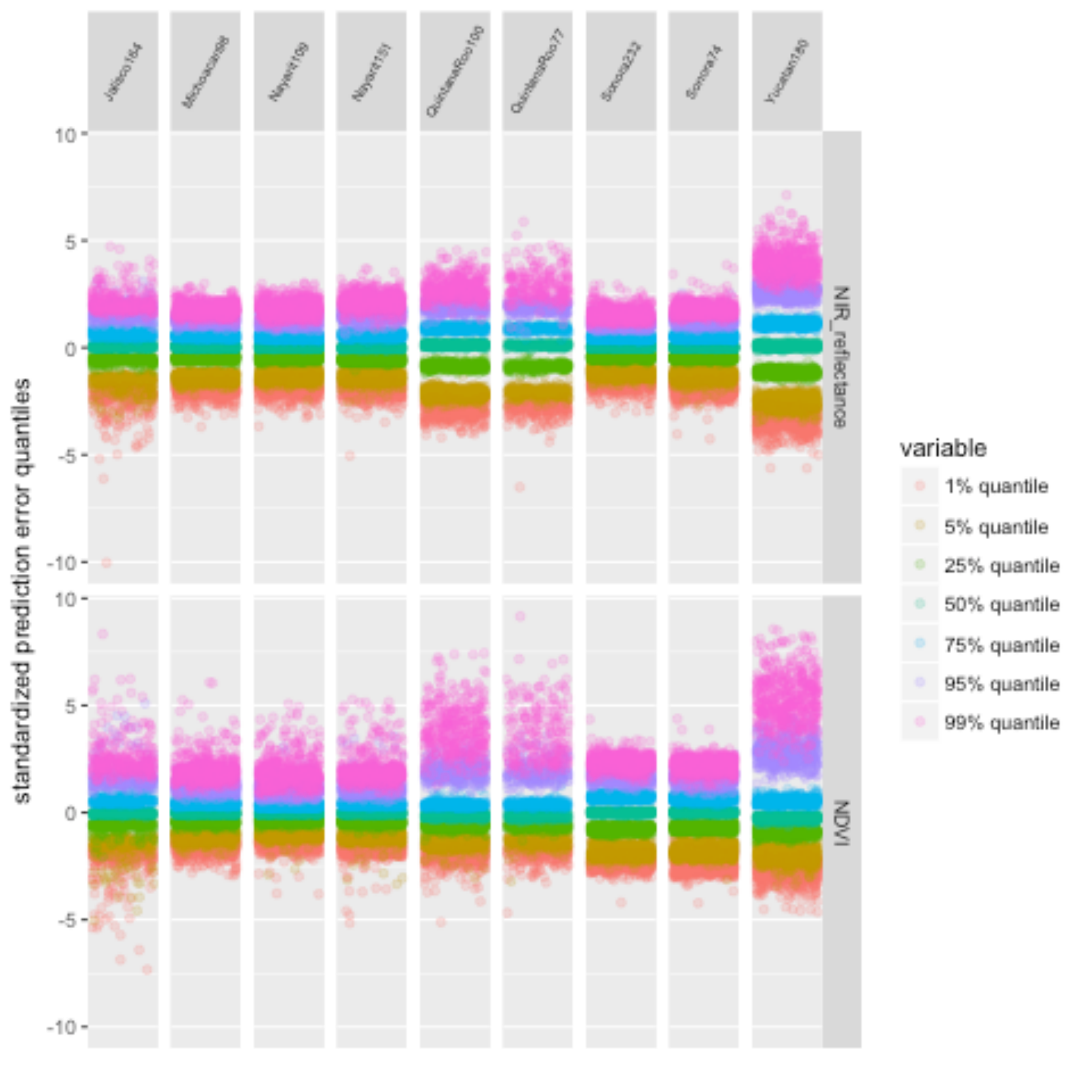}
	        \caption{Quantiles of standardized prediction errors by pixel organized by site}
	        \label{fig:distErrorStdr_quants}
	      \end{subfigure}
	      \caption[Distribution of standardized prediction error by site]
	      {Distribution of standardized prediction error by site}
	      \label{fig:distErrorStdr}
	    \end{figure}
		
		We can see that the high quantiles of the prediction error distribution can vary a lot accross pixels, and, especially for NDVI, even more accross sites. This may explain why the optimal thresholds for the NDVI sites were so different (see figure \ref{fig:opt_thrs}). 
				
		The adapted CMFDA algorithm relies on finding a \emph{high} absolute predicted error to determine deforestation events. If we knew the exact distribution of predicted error we could simply define a deforestation event in terms of a certain quantile of that distribution and even calculate the probability of the deforestation flag actually corresponding to a deforestation event.  Figure \ref{fig:distErrorStdr_quants} shows that the high quantiles of the empirical standardized predicted error distributions by pixel vary a lot. In order to address this issue we tried \emph{standardizing} the predicted error distributions using the empirical cumulutive distribution function (ecdf). We did this in several ways as before: 
		
	 \begin{enumerate}[I.]
		 \item \textbf{Depends only on space}: calculate ecdf function for each pixel using past prediction errors. Apply corresponding ecdf to new predicted errors according to pixel. Subtract 0.5 so that \emph{standardized} errors by pixel now have a uniform distribution from -0.5 to 0.5 and we can threshold using the absolute error as before. 
		 \item \textbf{Depends only on time}: calculate ecdf function for prediction errors observed on \textbf{all} 9 sites during a given 5 day period of the year. The year is divided into following 73 periods: $\{1-5, 6-10,...,356-360, 361-366 \}$. Apply corresponding ecdf to new predicted errors according to the observation date. Subtract 0.5 so that \emph{standardized} errors by period now have a uniform distribution from -0.5 to 0.5 and we can threshold using the absolute error as before,
		 \item \textbf{Depends on time and on space globally}: calculate ecdf function for prediction errors observed on \textbf{each} of the 9 sites during a given 5 day period of the year. The year is divided into following 73 periods: $\{1-5, 6-10,...,356-360, 361-366 \}$. Apply corresponding ecdf to new predicted errors according to observation date and site. Subtract 0.5 so that \emph{standardized} errors by period and site now have a uniform distribution from -0.5 to 0.5 and we can threshold using the absolute error as before. 
		 \item \textbf{Depends on space and time}. For this assumption we tried several variants,
		 	\begin{enumerate}[i.]
				\item Apply II then I to prediction errors (with variant): calculate ecdf function of prediction errors observed on \textbf{all} 9 sites during a given 5 day period of the year. Apply corresponding ecdf to new predicted errors according to the observation date. Apply inverse standard normal cdf so that prediction errors by period are distributed normally. Now calculate ecdf function for each pixel using  prediction errors standardized with ecdf by date. Apply corresponding ecdf to new prediction errors according to pixel. Subtract 0.5 so that \emph{standardized} errors by pixel now have a uniform distribution from -0.5 to 0.5 and we can threshold using the absolute error as before. 
				\item Apply III then I to prediction errors: calculate ecdf function for prediction errors observed on \textbf{each} of the 9 sites during a given 5 day period of the year. Apply corresponding ecdf to new prediction errors according to observation date and site. Apply inverse standard normal cdf so that prediction errors by period and site are distributed normally. Now calculate ecdf function for each pixel using  prediction errors standardized with ecdf by date and site. Apply corresponding ecdf to new predicted errors according to pixel. Subtract 0.5 so that \emph{standardized} errors by pixel now have a uniform distribution from -0.5 to 0.5 and we can threshold using the absolute error as before.
				\item \textbf{Space-time cubes}: calculate ecdf function of prediction errors for each cube. Apply corresponding ecdf to new predicted errors according to cube. Subtract 0.5 so that \emph{standardized} errors by cube have a uniform distribution from -0.5 to 0.5 and we can threshold with absolute error as before.  
			\end{enumerate}	
	 \end{enumerate}

	\section{Results}	\label{s:genRes}
	
	Since we automated the implementation of the different \emph{standardization} schemes, we also tried combinations where, for example,
	we standardized predicted errors by date by dividing by the corresponding standard deviation, and then standardized the resulting errors by pixel by applying the prediction error ecdf by pixel. In total we tried 17 different standardization schemes summarized in the following table:
	
	 \begin{table}[H] 
	\begin{center}
	 \begin{tabular}{|l|l|c|p{2cm}|p{2cm}|}
		 \hline
	 \textbf{Name} & \textbf{Code}  & by \textbf{date (time)} &  by \textbf{pixel (space)} & by \textbf{cube (space-time)} \\ 
	 \hline                                                                                                                                    
	 ---                       &        &                &       &         \\
	 \hline
	  day-sd.overall--         &  2     &  sd all sites   &      &         \\
	  day-sd.site--            &  3     &  sd by site     &      &         \\
	  day-qs.overall--         &  II    &  ecdf all sites &      &         \\
	  day-qs.site--            &  III   &  ecdf by site   &      &         \\
	  \hline
	  pixel--sd-               &  1     &                 & sd   &         \\
	  pixel--qs-               &  I     &                 & ecdf &         \\
	  \hline
	  day\_pixel-sd.overall-sd- &  4a    &  sd all sites   & sd   &        \\
	  day\_pixel-sd.site-sd-    &  4b    &  sd by site     & sd   &        \\
	  day\_pixel-qs.overall-sd- &  II.1  &  ecdf all sites & sd   &        \\
	  day\_pixel-qs.site-sd-    &  III.1 &  ecdf by site   & sd   &        \\
	  day\_pixel-sd.overall-qs- &  2.I   &  sd all sites   & ecdf &        \\
	  day\_pixel-sd.site-qs-    &  3.I   &  sd by site     & ecdf &        \\
	  day\_pixel-qs.overall-qs- &  IVi   &  ecdf all sites & ecdf &        \\
	  day\_pixel-qs.site-qs-    &  IVii  &  ecdf by site   & ecdf &        \\
	  \hline
	  cube---sd                &  4c    &                  &       & sd      \\
	  cube---qs                &  IViii &                  &       & ecdf    \\
	\hline                                                                                       
	 \end{tabular}\\
	 \caption{Summary of different prediction error standardization schemes used}
	 \label{stdrSchemes}
	\end{center}
	 \end{table}
	 
\begin{figure}[H]
      \begin{subfigure}{.5\textwidth}
        \centering
    	\includegraphics[width=0.9\textwidth]{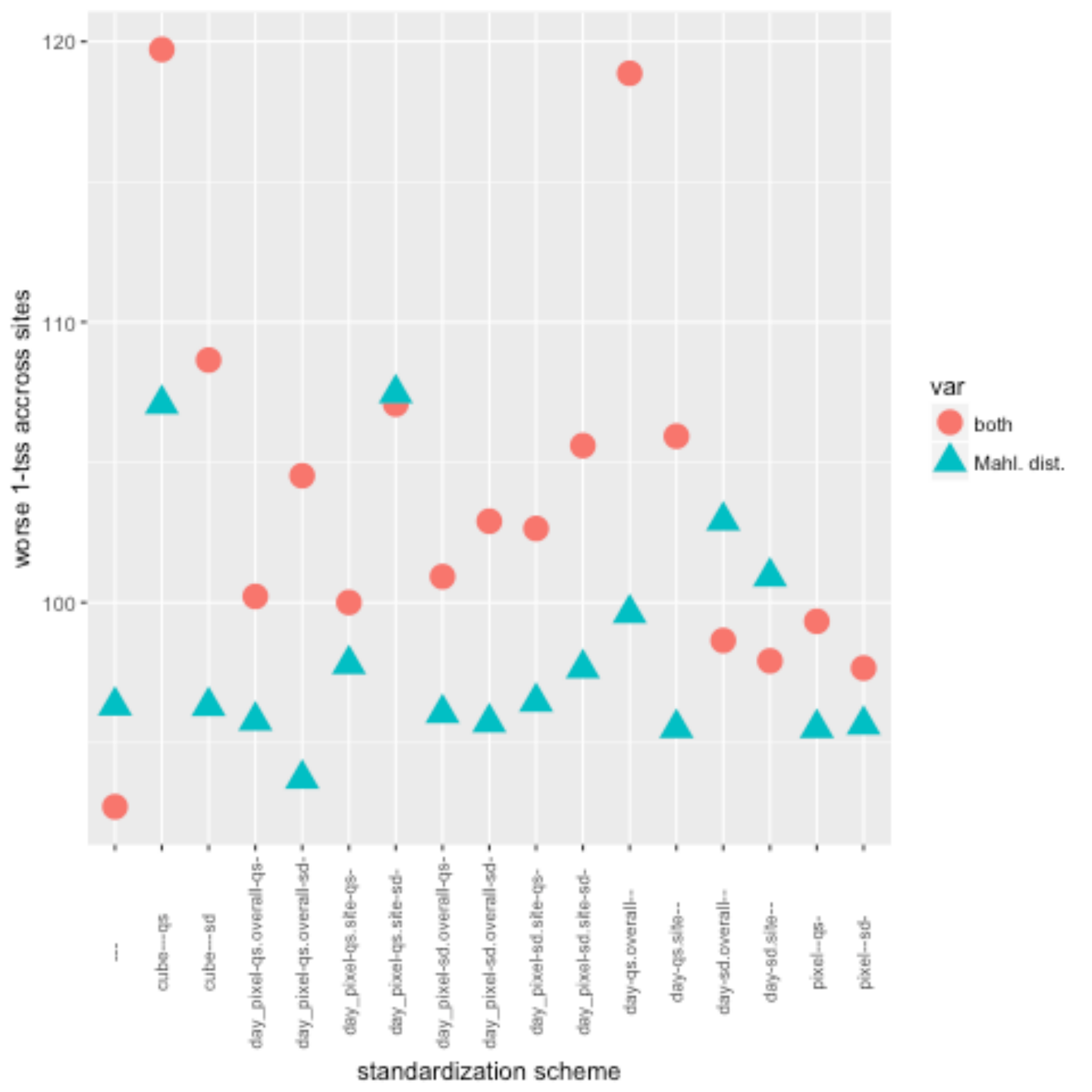} 
        \caption{$1-tss$ of worse performing site for each standardization scheme}
        \label{fig:stdrScheme_max}
      \end{subfigure}%
      \begin{subfigure}{.5\textwidth}
        \centering
        \includegraphics[width=1\textwidth]{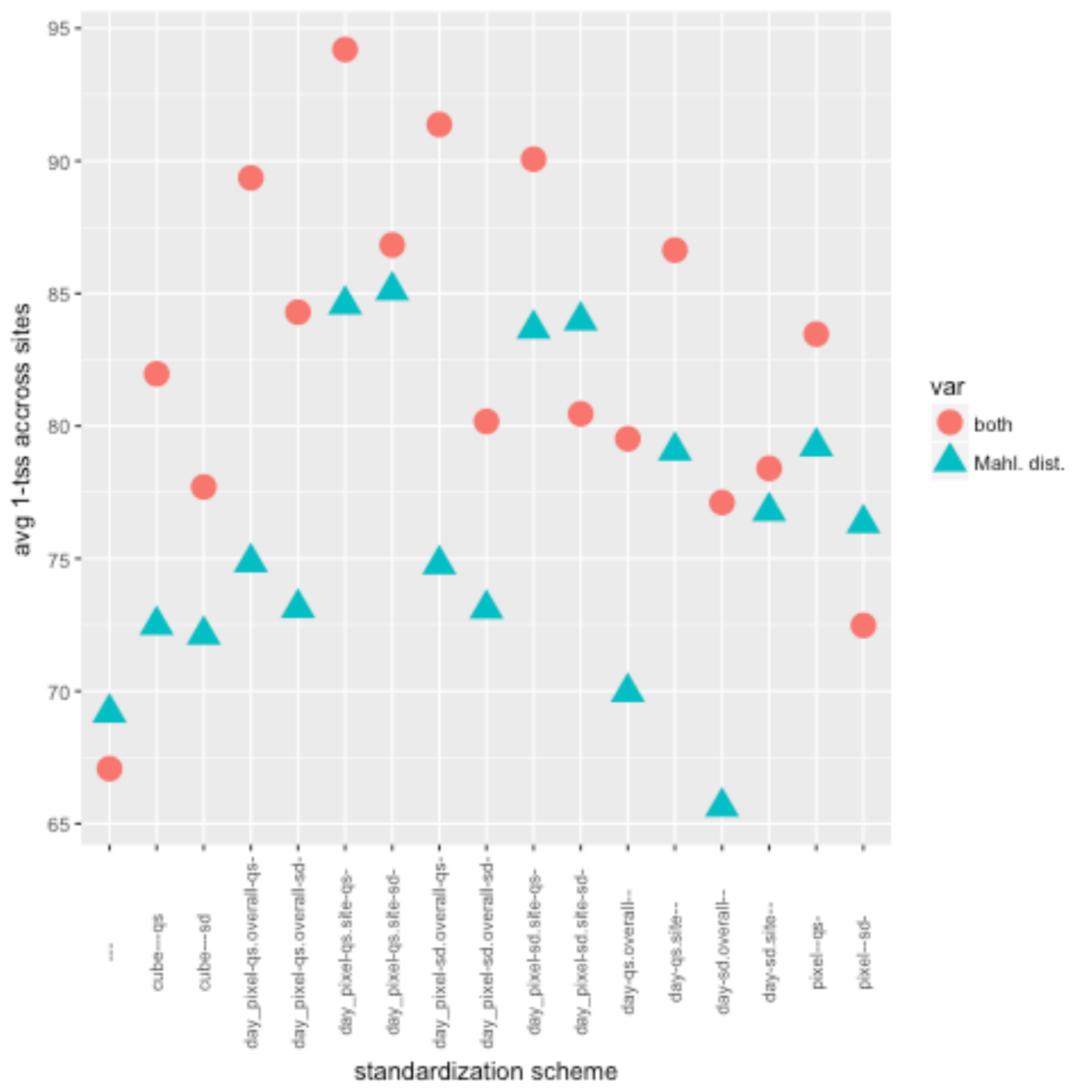}
        \caption{average $1-tss$ accross sites for each standardization scheme}
        \label{fig:stdrScheme_avg}
      \end{subfigure}
      \caption[Summary of performance of difference standardization schemes]
      {Summary of performance of difference standardization schemes}
      \label{fig:stdrScheme}
    \end{figure}

	 Figures \ref{fig:stdrScheme_max} and \ref{fig:stdrScheme_avg} show the worse and average  performing $1-tss$ accross sites for each of the standardization schemes. To simplify the analysis we only took into account the optimal $1-tss$ resulting from applying the adapted CMFDA algorithm with 4 consecutive violations of threshold. We took into account the multivariate approach, optimizing the multivariate thresholds using simulated annealing, and the index approach, using grid-search to optimize the Mahalanobis threshold. In both cases threshold optimization was done using all 9 sites. 
	 
	Although on average using the Mahalanobis distance of prediction errors and standardizing by dividing them by the overall period-wise standard deviation (standardization scheme \emph{day-sd-.overall--}) performs better than the non-standardized multi-variate thresholding version, in terms of the worse performing site, the latter performs better. Since the standardization schemes were implemented so as to improve the generalization of thresholds accross different locations we conclude that we have not found an appropriate standardizations scheme. More study needs to be made regarding the prediction error distributions and how they vary accross time and space. 
	
\chapter{Implementation} \label{ch:imp}

Although we were not able to find a satisfactory solution to the problem of heterogeneity in the prediction errors accross pixels and period of the year the results of chapter \ref{ch:meth} indicate we already have an algorithm that can achieve an overall $1-tss$ of 62\% (see table \ref{tab:simAnn_consecVaries}). Such a score could, for example, be obtained if the algorithm detects 68\% of deforestation events and 68\% of non-deforestation events to put it into context. The algorithm in question is the adaptation of the CMFDA algorithm described in sections \ref{s:CMFDA} and \ref{s:adapt} using the multivariate thresholding function described in section \ref{s:simAn}. We recommend this thresholding function be used with four consecutive violations of the threshold i.e. $C=4$. The algorithm can be implemented online with the following steps:

If a new batch of 13A2 modis, 1km resolution surface reflectance data observations which have a nominal date $t$ (recall this means with composite dates 0-15 days after $t$) carry out the following steps:
\begin{enumerate}[1.]
	\item \textbf{Forest mask:} Identify the set of forest pixels $\mathcal{F}_t$ by using latest available NALCMS landcover classification excluding any pixels already flagged as deforested in the time between the latest available classification and the date of the previous \emph{clear} observations.
	\item \textbf{Filter clouds:} Identify the set of \emph{clear} pixels $\mathcal{C}_t$ for nominal date $t$. They consist of all the pixels that have a \emph{pixel reliability} flag taking values \emph{good data} or \emph{marginal data}.
	\item For each pixel $p$ such that $p \in \mathcal{A}_t = \mathcal{F}_t \cap \mathcal{C}_t$ carry out following steps:
	\begin{enumerate}[a.]
		\item Obtain \textbf{three} \emph{clear} NIR reflectance and NDVI observations prior to new \emph{clear} observation, to conform prediction set $Y_p^\mathcal{P} = \{y_{b,p,s}\}_{b \in \{2,8\}, s \in \{t, t-1, t-2, t-3\}}$ 
		\item Obtain all  \emph{clear} observations from two year period prior to $t-3$ to conform training set $Y_p^\mathcal{T} = \{y_{b,p,s}\}_{b \in \{2,8\}, s \in \{t-4, t-5,..., T(t-3)\}}$ where  $T(t)$  is the first available clear observation in the two year period starting two years before $t$.
		\item Fit Fourier model \ref{fourierModel} to training set $Y_p^\mathcal{T}$ using ordinary least squares, i.e. obtain $\hat{g}_{b,p}$ for $b \in \{2,8\}$.
		\item Predict reflectance values for prediction set dates using fitted Fourier model, i.e. obtain $\hat{Y}_p^\mathcal{P} = \{\hat{y}_{b,p,s}\}_{b \in \{2,8\}, s \in \{t, t-1, t-2, t-3\}}$ where $\hat{y}_{b,p,t}=\hat{g}_{b,p}(d(t))$ and $d(t)$  maps the date $t$ to the day of the year.
		\item Obtain predicted error set $\hat{E}_p = \{\hat{\epsilon}_{b,p,s}\}_{b \in \{2,8\}, s \in =\{t, t-1, t-2, t-3\}}$ where $\hat{\epsilon}_{b,p,t}=y_{b,p,t}-\hat{y}_{b,p,t}$.
		\item Apply threshold function \ref{thrs_multivariate}, to error set with $L=(0.082, 0.182)$ (see table \ref{tab:simAnn_consecVaries_thrs}) to obtain deforestation prediction. 
	\end{enumerate}	
\end{enumerate}

\chapter{Summary} \label{ch:Summary}

In chapter \ref{ch:intro} we introduced the problem of deforestation and mentioned two approaches to deforestation detection using satellite images: comparison of two images and use of multiple images during the growing season to detect deforestation. We gave a brief conceptual overview of the CMFDA algorithm, a method which builds on the approach of the latter category by implementing continuous monitoring of images throughout the year. This algorithm works on 30m resolution Landsat images. We also describe the adaptations we make to this algorithm for it to work on 1km resolution Terra images. We saw that both the original and adapted CMFDA algorithms are composed of two submodels: a \emph{reflectance model} and a \emph{deforestation model}. The reflectance model estimates the amount of surface reflectance for different spectral bands for a given forest pixel, as a function of the day of the year. The deforestation model estimates the ocurrence of a deforestation event by comparing the last few (2-6) predicted and real reflectance values. 

In chapter \ref{ch:data} we explored the data we used to train and test the algorithm. In section \ref{s:landuse} we took a look at the data used to construct the deforestation flag, which is what we ultimateley want to model, and which is necessary to fit the deforestation model. This data is not \emph{ground truth} deforestation data, rather its the output of a land-use classification model which works with 250m resolution reflectance data. We compared the output of the classification model for 2005 and 2010 to build a \emph{deforestation} 2005-2010 flag. We explain the input data, statistical techniques and ground truth data that were used to develop this model. We learned that according to this model the net loss of forest rate from 2005 to 2010 is only 0.032\% compared to \cite{InformeMexico} which estimates it at 0.87\%. By choosing study areas with high deforestation, such that the new land-use type is widespread, we hope the classification model has reasonable accuracy: the idea is that this new type of land-use will be easily detected by the classification model since it is not spatially isolated. In section \ref{s:refl} we study the modis 13A2 dataset. This data set includes \emph{quality}, \emph{sun-sensor geometry}, \emph{surface reflectance} and \emph{vegetation index} variables. We first give a detailed account of the measurement and data processing that goes into transforming the raw radiance data obtained by the modis sensor aboard the Terra into gridded, composited surface reflectance data. The particular circumstances in which a reflectance value for a given pixel was recorded (angle between sun, surface and sensor, presence of clouds, shadows or snow, etc) and the processing involved (whether the reflectance value was atmosphere corrected or not) are described by the quality and sun-sensor geometry variables. We explore their distribution and behavior accross pixels and time. Although a lot of this analysis is not used further in this work we include it as we think it helps the understanding of the underlying physical causal processes involved. This in turn can inform future improvements to the methodology described in this work. The \emph{pixel reliability} flag turns out to be a good summary of the rest of the quality variables. Measurements flagged as \emph{good data} have no undesirable features and make up about 65\% of the data. Measurements flagged as \emph{marginal data} make up around 31\% of the data and, in general, are flagged as such because it is possible that a cloud or shadow influenced the measurement. The remaining 4\% of the data has bad quality. We use this flag in order to filter poor measurements as is described later in section \ref{s:adapt}. We also learn that the quality of the data goes down during the summer months due to the rainy season that occurs throughout most of Mexico. The sun-sensor geometry of recordings is described by three variables, \emph{sun zenith angle}, \emph{view zenith angle} and \emph{relative azimuth angle}. We see that \emph{sun zenith angle} is mostly a function of the day of the year, \emph{view zenith angle} has no seasonal component and mostly occurs at discrete intervals and \emph{relative azimuth angle} is mostly a function of the \emph{sun zenith angle}. The 13A2 modis dataset includes surface reflectance data for four spectral bands: red, near infra-red (NIR), blue and mid infra-red (MIR) reflectance. All four bands have a seasonal pattern peaking in the summer months although NIR reflectance peaks a little later, in the months of July and August, as opposed to May and June. All four bands have positively-skewed distributions indicating that it may be necessary to use a transformation of the surface reflectance instead of the surface reflectance itself, with regard to the \emph{reflectance} model (we didn't explore this, but it may provide better results). We explored why using the red, NIR and blue spectral bands may be useful in detecting deforestation given that the red and blue bands are mostly absorbed and the NIR band mostly reflected or transmited by vegetation cover. However, we also observed that it is difficult to distinguish landcovers by their seasonal median surface reflectance pattern, at least visually. This is probably due to local factors, such as the local solar time at which observations are made for different pixels. The 13A2 modis dataset includes two vegetation indices, the normalized difference vegetation index (NDVI) and the enhanced vegetation index (EVI) which are constructed as ratios of the four spectral bands. In essence, both seek to measure the amount of vegetation by contrasting, through a ratio, the amount of red and NIR reflectance. The EVI seeks to improve the sensitivity of the NDVI by adjusting for atmospheric and soil factors. Both indices decrease slowly from January to around May and June when they reach their minimum. They then increase drastically peaking around July and August, and decreasing steadily for the rest of the year. It is also difficult to distinguish vegetation land cover types using their  seasonal median vegetation index pattern. This hints at the possibility that we can only use surface reflectance models to detect deforestation locally. Indeed, as we see in chapter \ref{ch:meth} the methodology we implement works locally on a pixel-by-pixel basis.

In chapter \ref{ch:meth} we give a detailed description of the CMFDA methodology for 30m resolution Landsat images, 16-day frequency, (top-of-atmosphere reflectance) and its adaptation to 1km resolution, 16-day frequency, Terra images (surface reflectance). We select the study area for training and testing the algorithm and implement two variants of the algorithm, giving performance results. In section \ref{s:CMFDA} we describe the CMFDA algorithm described in \cite{CMFDA}. It consists of seven steps geared toward estimating two models: a \emph{reflectance} model in which surface reflectance for each band is modeled as a function of the day of the year; and a \emph{deforestation} model which models the occurrence of a deforestation event as a function of the size of recent prediction errors. The idea is that if observed reflectance deviates a lot from model predicted reflectance it is likely due to a landcover change. Steps 1 and 2 are filtering steps designed to detect and eliminate reflectance observations taken under cloud, shadow or snow conditions. Step 3 is an atmosphere correction step to transform top-of-atmosphere reflectance to an estimate of the surface reflectance. In step 4 surface reflectance is modeled in terms of harmonic functions of the day of the year. A two year training window is used to estimate this model.  This model is fitted to any pixel that could potentially be a forest pixel because in step 5 forest pixels are identified by using model coeficients and knowledge about the range of values these may take when the surface is forest. In step 6 the fitted models (one for each pixel and spectral band) are used to obtain predictions for the surface reflectance, at each pixel and band, during a prediction window spanning three \emph{clear} observations. In step 7 the real and predicted surface reflectance values for different bands are combined into a \emph{forest} index. Real and predicted values are compared for the three dates  and if the difference exceeds a threshold of 0.12 for each then the deforestation flag is triggered. Training of the threshold was carried out using information from a single site along the Savannah river. The harmonic models were fitted using reflectance data from 2001-2002 and were then used to predict the data for 2003. Ground truth data was available for 2003 so an optimal threshold was determined using this data. In order to evaluate the performance of the algorithm and train the thresholds the CMFDA algorithm uses a combination of the producer's and user's accuracy. 

There are three main differences between the Landsat images used in \cite{CMFDA} and the Terra 13A2 images we use in this work: the resolution is 30m for the former and 1km for the latter; the Landsat dataset used in \cite{CMFDA} consists of top-of-atmosphere reflectance whereas the 13A2 Terra dataset consists of surface reflectance; and, the spectral bands available are different in each dataset. In section \ref{s:adapt} we describe the changes we made to the algorithm to adapt it to this new dataset. In essence these changes consist of a simplification of the algorithm. The cloud, shadow and snow observation filtering steps are replaced with the use of data quality flag since the original top-of-atmosphere reflectance data has already been labeled for cloud, shadow and snow conditions and also atmosphere corrected. To define the forest mask the CMFDA uses the model coefficients for the surface reflectance at certain spectral bands, such as short wave infra-red (SWIR), which are not available in the modis 13A2 dataset. To obtain a forest mask for which the deforestation detection algorithm should be applied we simply use the NALCMS land-use classification model described in section \ref{s:landuse}. Since the 13A2 data set does not include the necessary bands in order to construct  the \emph{forest} surface reflectance index we were forced to adapt the way we compare model predicted and real surface reflectance values. In \ref{s:adapt} we merely identified two different approaches, a \emph{multivariate} and an \emph{index} approach, and left it for sections \ref{s:simAn} and \ref{s:mahl} to implement each in order to determine which has better performance. In the multivariate approach a subset of the reflectance bands are first predicted and if the prediction error of each exceeds a certain threshold, particular to each band, then a band flag is triggered. The flags for the subset of bands are then combined using an \emph{or} rule to obtain the general deforestation flag. In the index approach, as is the case in CMFDA, the subset of reflectance bands are first combined to form a reflectance index, and the prediction error for the index then checked against a threshold. Since we did not have the necessary bands to construct the \emph{forest} index we constructed a local Mahalanobis distance index which helps to detect truly unusual prediction errors by taking into account the covariance between the errors for the different bands.  The subset of bands was chosen in section \ref{s:studyArea} by visually inspecting the time-series of the different bands for pixels where deforestation took place. For training the different thresholds we used a more extensive study period and study area than that used in \cite{CMFDA}, so that we could verify that the methodology generalizes to sites with different characteristics and accross time. This was possible, in part, because the data we used has lower resolution, meaning we could use a larger overall area. The fact that we used a classification model to obtiain our \emph{ground truth} deforestation flag also meant we were not restricted to sites where data has been \emph{manually} collected. The harmonic models were fitted using reflectance data from two year windows in the 2003-2009 period (i.e. 2003-2004, 2004-2005,...,2008-2009). Models associated to each respective training window were then used to predict reflectance for the following year (i.e. 2005, 2006,...,2010 respectively). The data came from nine sites selected in section \ref{s:studyArea}. Thresholds were trained using the 2005-2010 deforestation flag described in section \ref{s:landuse}. In order to evaluate the performance of the algorithm and train the thresholds we chose to use the true skill statistic ($tss$), which gives equal weight to detecting deforestation and non-deforestation events.  

In section \ref{s:studyArea} we selected the nine sites we would use for threshold training.  We looked for 25 by 25km sites with a high amount of deforestation and such that the type of deforestation (i.e. the type of new landcover) was representative of the types that can occur going forward. We identified two types of deforestation sites, four \emph{forest to water} sites  and five \emph{forest to urban or cropland} sites. The deforestation at \emph{forest to water} sites is more easily detected by looking at the time series of the NIR reflectance band, while the deforestation at \emph{forest to urban or cropland} sites is more easily detected by looking at the time series of NDVI. Based on this exploration we chose the subset of bands, to be used in the multivariate and index thresholding approaches, to be NIR and NDVI. Although the NDVI is not strictly speaking a band, but a function of the red and NIR bands, we treated it as such. 

In section \ref{s:gridSearch} we trained univariate (the prediction errors of only one reflectance band used) thresholds on a site-by-site basis. The spectral band for each site was chosen depending on the deforestation type that predominantly occurred there: NIR for \emph{forest to water} site and NDVI for \emph{forest to urban or cropland} sites. There are two reasons why this exercise is not implementable: we don't know beforehand which type of deforestation will occur and so have to combine the prediction errors of both bands; and, our goal is to apply one threshold (univariate or multivariate) to all of Mexico. Although in theory we could localize thresholds so that they would vary according to local conditions (type of forest, reflectance ranges, etc) this is problematic since we do not have deforestation information available, for all sites, with which to train these local thresholds, even when using \emph{ground truth} from a classification model as we do in this work. Despite these shortcomings, this site-by-site training provides a performance benchmark with which to compare the results for the implementable methods of sections \ref{s:simAn} and \ref{s:mahl}. We introduce the univariate form of the utility function, $tss$, for the site-by-site univariate thresholding case. The threshold values were trained using grid-search on an equally spaced sequence of values. In general, we found that we obtained better $tss$ performance for the \emph{forest to water} sites. We learn that there is a lot of variance in the optimal threshold accross sites especially for \emph{forest to urban or cropland} sites where the NDVI is used. This gave us an early indication that when, in sections \ref{s:simAn} and \ref{s:mahl}, we attempt to find a single optimal threshold for all 9 sites, the performance results will deteriorate, given that for most sites the overall threshold will to be far from the site-specific optimal value. The performance accross sites also showed a lot of variance. For the \emph{forest to water} sites the best and worse performing site, according to  $tss$,  was Nayarit151 and Sonora232, with a cross-validated $tss$ of 73.65 \% and 45.62 \% respectively ($1-tss$ of 26.35 \% and 54.38 \%), achieved with a threshold of 0.08 and 0.06 respectively, for the NIR predicted error, and, applied 4 and 5 consecutive times respectively ($C=4$ and $C=5$).  For the \emph{forest to urban or cropland} sites the best and worse performing site, according to $tss$, was QuintanaRoo77 and QuintanaRoo100, with a cross-validated $tss$ of  42.24 \% and 19.23 \% respectively ($1-tss$ of 57.76 \% and 80.77 \%), achieved with a threshold of 0.13 and 0.24 respectively, for the NDVI predicted error, and, applied 5 and 2 consecutive times respectively ($C=5$ and $C=2$). These results refer to table \ref{tab:grdSrch_consecVaries_CV}.

In section \ref{s:simAn} we implemented the multivariate approach to thresholding predicted errors which consists of thresholding the predicted errors of each band separatley, with independent threshold values, and then combining the results with an \emph{or} rule. The resulting flag indicates whether the predicted errors, of either NIR or NDVI, violates the respective threshold a number of consecutive times $C$ or not. We introduce the multivariate form of the utility function, $tss$, for optimizing the multivariate threshold of all nine sites. We trained the thresholds using simulated annealing on a grid of NIR and NDVI threshold values. In this case grid-search was not used because the computing cost of training a multivariate threshold with this exhaustive approach was considered too high. The starting threshold used for NIR and NDVI,  in the simulated annealing algorithm, corresponds to the average of the optimal thresholds found in the site-by-site optimization results of section \ref{s:gridSearch}. For the NIR threshold, the results of \emph{forest to water} sites are averaged and for the NDVI threshold those of \emph{forest to urban or cropland} sites are averaged. The best overall $tss$ performance of 37.01 \% ($1-tss$ of 62.99 \%) was achieved using a NIR and NDVI threshold of 0.082 and 0.182 respectively, applied 4 consecutive times ($C=4$). As the variance of the optimal site-by-site thresholds indicated, when we break these overall, multivariate threshold, $tss$ results down by site we find that the performance, in general, deteriorates when we compare it to the performance of the site-specific univariate thresholds. This is especially the case for some sites such as Sonora232 and Michoacan98 (see figure \ref{fig:simAnn_consecFix}). 

In section \ref{s:mahl} we implemented the index approach to thresholding predicted errors which consists of combining the predicted errors of the NIR ($\hat{\epsilon}_{2,p,t}$) and NDVI ($\hat{\epsilon}_{8,p,t}$) bands, and then thresholding the resulting predicted error index. We chose the local Mahalanobis distance of the predicted errors as our index since it can be thought of as a measure of how atypical a given bivariate, NIR and NDVI, predicted error observation is with respect to the historical pattern of observations. At least, this is the case, if we can assume NIR and NDVI predicted errors to have a more or less linear relationship. The Mahalanobis distance between two variables is calculated using the covariance matrix between them. We assumed the covariance matrix between the predicted NIR and NDVI errors is a function of the period of the year and the location. With this assumption in mind we estimated a different covariance matrix for each space-time \emph{cube}. We then calculated the Mahalanobis distance of the predicted errors $(\hat{\epsilon}_{2,p,t},\hat{\epsilon}_{8,p,t})$,  using the covariance matrix of the cube corresponding to the pixel $p$ and day of the year $d(t)$. We defined the set of space-time cubes by dividing each 25 by 25km site into five 5 by 5km squares and by dividing the days of the year into 73 five day-intervals. The set of cubes is then defined by the cartesian product of the set of squares and the set of day-intervals. We introduce the univariate, index form of the utility function, $tss$, for optimizing the univariate, index threshold of all nine sites, including the formula for calculating the Mahalanobis distance and the definition of a space-time \emph{cube}. We trained the thresholds using grid-search on an equally spaced sequence of values, first on a site-by-site basis and then for all nine sites together. The site-by-site optimized performance  when using the Mahalanobis distance of the prediction errors was worse than the performance when using one of NIR prediction errors or NDVI prediction errors (depending on the type of deforestation at the site). However for some sites, such as Jalisco164, Nayarit109 and Yucatan180 using the Mahalanobis distance provided better performance. This despite the fact that for the optimization of NIR or NDVI prediction error thresholds by site, used information about the type of deforestation that would occur, making it an unfair comparison. In any case, even though we do not use this information in the case of the Mahalanobis distance algorithm, it is not feasible to train thresholds locally since deforestation information is not available for many regions. The optimal Mahalanobis distance thresholds accross sites also have lot of variance indicating that applying a single threshold accross sites will deteriorate the performance. In comparing the overall grid-search optimized performance when using the Mahalanobis distance of the prediction errors with the performance when using the NIR and NDVI prediction errors together with a multivariate threshold we noticed something interesting: the Mahalanobis distance version of the algorithm performs better for 2-3 consecutive violations of the threshold while the multivariate version of the algoritm performs better for 4-6 consecutive violations of the threshold. This means if we want to detect deforestation faster it is better to use the Mahalanobis distance of the prediction errors. The best overall $tss$ performance of 31.16 \% ($1-tss$ of 68.84 \%) was achieved using a Mahalanobis distance threshold of 11.72, applied 4 consecutive times ($C=4$). As the variance of the optimal site-by-site thresholds indicated, when we break these overall, index threshold, $tss$ results down by site we find that the performance, in general, deteriorates when we compare it to the performance of the site-specific  thresholds. This is especially the case for some sites such as Sonora232 and Michoacan98 (see figure \ref{fig:mahl_consecFix}).

In chapter \ref{ch:modhet} we explore how the components of the reflectance model $g_{b,p}(d)$ and $\epsilon_{b,p,t}$ vary accross space (accross pixels $p$) and accross time (accross the day of the year). In section \ref{s:coef} we verify visually that the expected surface reflectance, $g_{b,p}(d)$ is spatially correlated. In section \ref{s:var} we verify visually that the variance $\sigma_{p,t}^2$ of the errors $\epsilon_{b,p,t}$ varies accross pixels $p$ and day of the year $d$.  Since the algorithm described in chapter \ref{ch:meth} is based on applying a single threshold (multivariate or index based) to the prediction errors, regardless of the location of the pixel $p$ or the time of the year $d$, the tacit assumption is that the errors $\epsilon_{b,p,t}$ have the same distribution accross pixels $p$ and day of the year $d(t)$. However, despite the fact that the algorithm obtains reasonable results, we have seen that this is not the case. In sections \ref{s:homStdr}-\ref{s:homECDF} we detail two approaches for transforming the prediction errors $\hat{\epsilon}_{b,p,t}$ such that they no longer depend on space and time. In section \ref{s:homStdr} the approach detailed involves dividing the errors by an estimate of the standard deviation. We estimate the standard deviation under various assumptions about how it depends on either space (pixel), time (day of the year) or both. In section \ref{s:homECDF} we explore the distribution of standardized predicted errors, where the standard deviation of a predicted error is estimated under the assumption that it depends on the space-time \emph{cube} to which it belongs (i.e. standard deviation depends on pixel and day of the year). We learn that the high quantiles of the standardized prediction error distribution vary a lot accross pixels, and, especially for NDVI, even more accross sites. Although dividing by the standard deviation normalizes the variance of the predicted error distribution of a given cube to one, it does not necesaily normalize the high quantiles. The adapted CMFDA algorithm relies on finding a high absolute predicted error to determine deforestation events. Conceptually, it finds a high enough quantile of the predicted error distribution, such that the probability that errors of that magnitude or higher occurred $C$ consecutive times is very low. The fact that the high quantiles vary a lot accross pixels and sites means that we have not achieved our goal of normalizing the optimal thresholds. To address this problem, in section \ref{s:homECDF} we propose using the empirical cumulative distribution function (ecdf) to transform and homogenize the predicted error distributions. Again, we estimate the ecdf under various assumptions about how it depends on either space (pixel), time (day of the year) or both.

In section \ref{s:genRes} we compare all the different variance/distribution homogenization schemes, including when we leave the prediction errors as they are, for both the multivariate and the Mahalanobis distance index approach. In total there are 17 different variance/distribution homeginization schemes applied to both thresholding approaches. To simplify the analysis we focus on the results when the number of consecutive violations of the threshold is fixed at 4 ($C=4$). We hope that transforming the prediction errors will lead to obtaining similar thresholds accross sites such that the overall optimal threshold for the nine sites is close to the site-specific optimal thresholds. This in turn should lead to improved $tss$ performance for each site since we would no longer be applying a general threshold  to all sites that is significantly sub-optimal when compared to the optimal site-by-site threshold. To compare the homogenization schemes, then, we looked at the $tss$ performance of the worse performing site. It turned out that with this measure the best performing scheme was the \emph{vanilla} scheme,  where the errors were left untransformed. When looking at the average $tss$ performance accross sites the variance homogenization scheme consisting of dividing the Mahalanobis distance prediction errors by their standard deviation, under the assumption that it depends on the day of the year only, obtained slightly better $tss$ performance than the vanilla method, however we think the first \emph{minimax} type measure is best in this case. We conclude that the assumptions made do not reflect the way the prediction error variance really behaves. 

Finally in chapter \ref{ch:imp} we give a detailed description of how to implement the adapted CMFDA algorithm described in section \ref{s:CMFDA} and \ref{s:adapt} with the multivariate approach described in section \ref{s:simAn}, since it obtained slightly better performance than the index approach. We do not transform the prediction errors since in chapter \ref{ch:modhet} we did not find evidence that this improves performance.

\section{Future Work}\label{s:FutureWork}

The following is a list of possible ways to extend or improve the work presented here:

\begin{enumerate}[1.]
	
	\item \textbf{Reflectance model}
		\begin{itemize}

			\item \textbf{Local factors and sun-sensor geometry variables}. In some of the exploratory analysis not included in this work we tried to model surface reflectance globally, as opposed to on a pixel-by-pixel basis. The predictive power was much lower than that achieved with the pixel-by-pixel models however we noticed two interesting behaviors: sun sensor geometry variables such as sun zenith angle were statistically significant, even while also including other seasonal factors. In addition, the land-use type was only significant when we tried modeling reflectance locally, i.e. for a cluster of pixels that are close together. This suggests the following direction for future research: why can't surface reflectance be modeled globally? What factors which capture local conditions would we need to include? Are sun-sensor geometry variables one of these local factors? How can we incorporate sun-sensor geometry variables into the reflectance model? Can it be done at pixel-by-pixel modeling level for example?
						
			\item \textbf{Reflectance vs. landuse type}. In section \ref{s:refl} we saw it is not easy to distinguish between landuse covers by simply looking at the median reflectance or vegetation index for the the different months of the year. This is possibly due to local factors influencing reflectance. We think it is worth exploring in this direction further to get an understanding of how we can learn about the landuse type by looking at reflectance patterns. One approach would be to fit Fourier models such as \ref{fourierModel} to pixels that don't change landcover in 2005-2010 period, for example those of the sample that was described in section \ref{s:refl}. We could then try and relate landcover type to the fitted coefficients at a global and local level. 
			
			\item \textbf{Error analysis}. We did not carry out formal error analysis on the pixel-by-pixel Fourier models because of the amount of models and impossibility of fitting customized models for each pixel. However, more analysis needs to be made in understanding the distribution of the errors and how they depend on space and time. The improvement of the algorithm depends on being able to homogenize the errors accross time and space so that it is reasonable to apply a single threshold to the prediction errors, regardless of the location and time of the year. In chapter \ref{ch:modhet} we tried several approaches for homogenizing the prediction error distribution but found none that worked. It may be necessary to apply transformations to the surface reflectance variables in order to obtain better behaved errors. Or it may be advisable to work with the NIR and red reflectance variables instead of NIR and NDVI. The NDVI is a function of the NIR and red bands but the fact that NDVI is a ratio suggests a different treatment may be necessary. 
			
			\item \textbf{Model space and time dependance}. Our current approach deals with dependence of reflectance on local factors by simply fitting one model per pixel. It would be beneficial to estimate a model that takes into account spatial and time dependence, so as to better understand surface reflectance and how it is affected by landcover type, aswell as to allow for the pooling of data accross pixels. The current model takes time dependence into account, for the mean component of the model, but not for the variance component. Such a model could be generally described in the following way:

			\begin{align}
				y_{b,p,t} = g_{b}(d,x(p),y(p)) + \epsilon_{b}(d, x(p), y(p)),
			\end{align}	
			
			where $b \in \{1,2,..,B\}$, $p \in \mathcal{P}$,$t \in \mathcal{T}$ and $d \in \{1,2,...,366\}$ refer to the band, pixel, date and day of the year respectively, and $(x(p), y(p))$ refer to the x-y coordinates of the pixel $p$. The assumption is that $\epsilon$ has a distribution $D$ such that $\mathbb{E}[D]=0$ and $\mathbb{V}[D]=\sigma(d, x(p), y(p))^2$.

		\end{itemize}
	
	\item \textbf{Deforestation model}
	
		\begin{itemize}
			
			\item \textbf{Ground truth data}. The deforestation data we used to relate reflectance prediction errors to deforestation events was not actual ground truth data, rather the output of a classification model. This was somewhat justified in that the classification model worked with higher resolution reflectance data and that it allowed us to use a large and diverse study area. Although we probably lost accuracy by training on noisy data, our conclusions on how well the methodology generalizes accross regions are richer. In any case, it is necessary to test the methodology on real ground truth data to better assess its accuracy. Bearing in mind we know that, when comparing to \cite{InformeMexico} the classification model underestimates the amount of net forest loss that occurred between 2005-2010, there is even the possibility that the CMFDA methodology works better than our performance measures suggest. Figure \ref{fig:grdSrch_spatial_Yucatan180} , for example, which shows the spatial results of implementing a certain version of the CMFDA methodology for the Yucatan180 site, suggests that perhaps deforestation in that site is more extended than the classification model implies. This is merely a possibility, but one that is worth exploring.  
	
			\item \textbf{Mahalanobis distance}. We chose to use the Mahalanobis distance of the NIR and NDVI predicted errors to build an index that takes into account the predicted errors of the different bands. Although this approach did not obtain as good results as the multivariate approach, it showed promise since its performance was reasonably close in general, and in particular proved better at detecting deforestation events quickly (with only 2-3 observations). However, we did not sufficiently explore the relationship between NIR and NDVI accross pixels and time of the year. If the dependence between the two is not linear then perhaps the variables need to be transformed first, or a different set of bands (again perhaps NIR and red reflectance) used, in order to obtain better performance with this approach. Actually, we know the relationship between NIR and NDVI is not likely to be well approximated with a linear model for the dependence since NDVI is a non-linear function of NIR and red reflectance. 
	
			\item \textbf{Monitor ratio instead of difference.} As we saw in figure \ref{fig:varDep_period} the predicted errors have a seasonal component. If instead of monitoring the difference between predicted and real observations we monitor the ratio it is possible the seasonal component could dissappear. 
			
			\item \textbf{Complex discrimination regions.} Observing figures \ref{fig:multivar_approach} and \ref{fig:mahl_approach} which give a simplified, schematic representation of the deforestation discrimination regions on the NIR and NDVI predicted error space, we see that these regions are quite simple. This is in part due to the thresholding approach followed by the CMFDA methodology. We could instead seek to find more complex discrimination regions which better exploit the multivariate nature of predicted reflectance error observations.  To do this we would simply need to identify, for each pixel, during the prediction window, the 2-6 consecutive observations which are most likely to correspond to a deforestation event. We could perhaps do this using a moving average of the Mahalanobis distance of predicted errors (for all bands this time). We could then relate the reflectance observations for the different bands, and for the 2-6 observation dates, to the occurrence of deforestation within the prediction window. In turn, this would allow the use of one of many available regression techniques to estimate a more complex discrimination region (i.e. a more complex \emph{deforestation} model): logistic regression, support vector machines, random forest or neural networks to name a few. We could still implement the deforestation detection online since every time a new clear observation became available we can calculate prediction errors for the different bands and use the last 2-6 consecutive prediction errors to build our features and predict whether deforestation took place or not.

		\end{itemize}

	\item \textbf{Scaling algorithm up}
	
		\begin{itemize}
			\item \textbf{Sharper resolution}. The surface reflectance information used from the Terra satellite is also available at 250m resolution. Additionally, 30m resolution reflectance data is available from the Landsat satellite. Higher resolution data allows more accurate and reliable monitoring of deforestation but represents computational challenges as the amount of data increase quadratically with the increase in resolution.  
 		\end{itemize}

\end{enumerate}	



\addtocontents{toc}{\vspace{.5\baselineskip}}
\cleardoublepage
\phantomsection
\addcontentsline{toc}{chapter}{\protect\numberline{}{Bibliography}}
\bibliography{8_myReferences}
\end{document}